\documentclass[11pt,a4paper]{article} 
\usepackage{jheppub}
\pdfoutput=1
\usepackage{epsfig}
\usepackage{graphicx}
\usepackage{amsmath}
\usepackage{amsfonts}
\usepackage{amssymb}
\usepackage{slashed}
\usepackage{natbib}
\usepackage{xcolor}
\usepackage{subfigure}
\usepackage{placeins} 

\def\beq{\begin{equation}}
\def\eeq{\end{equation}}
\def\beqa{\begin{eqnarray}}
\def\eeqa{\end{eqnarray}}
\newcommand{\nn}{\nonumber}

\def\eqn#1{eq.~(\ref{#1})}
\def\eqns#1#2{eqs.~(\ref{#1}) and~(\ref{#2})}

\def\sec#1{section~{\ref{#1}}}
\def\app#1{appendix~\ref{#1}}

\def\bsp#1\esp{\begin{split}#1\end{split}}

\newcommand{\ord}{{\cal O}}
\def\cM{{\cal M}}

\def\cN{{\cal N}}

\newcommand{\IM}{{\rm Im}}

\def\ifm{\ifmmode}

\def \tr {\mbox{tr\,}}

\newcommand{\Li}{{\rm Li_2}}
\newcommand\real[1]{{\mathrm{Re}\left[{#1}\right]}}
\newcommand\imag[1]{{\mathrm{Im}\left[{#1}\right]}}
\newcommand\disp[1]{{\mathrm{Disp}\left[{#1}\right]}}
\newcommand\absorp[1]{{\mathrm{Absorp}\left[{#1}\right]}}
\newcommand{\NMRK}{{ \mathrm{NMRK}}}

\newcommand{\MRKx}{{ \mathrm{MRK}^\prime}}
\newcommand{\sigmaAt}{\overset{\leftrightarrow}{\sigma_A}}
\newcommand{\sigmaAb}{\underset{\leftrightarrow}{\sigma_A}}
\newcommand{\sigmaAtb}{\overset{\leftrightarrow}{\underset{\leftrightarrow}{\sigma_A}}}
\newcommand{\sigmaBt}{\overset{\leftrightarrow}{\sigma_B}}
\newcommand{\sigmaBb}{\underset{\leftrightarrow}{\sigma_B}}
\newcommand{\sigmaBtb}{\overset{\leftrightarrow}{\underset{\leftrightarrow}{\sigma_B}}}

\newcommand{\tocol}{\xrightarrow[p_4 || p_5]{}}
\newcommand{\tosoft}{\xrightarrow[p_4 \to 0]{}}
\newcommand{\toMRK}{\xrightarrow[\mathrm{MRK}]{}}
\newcommand{\toNMRK}{\xrightarrow[\mathrm{NMRK}]{}}
\newcommand{\toMRKx}{\xrightarrow[\mathrm{MRK}^\prime]{}}

\def\cg{\kappa_\Gamma}
\newcommand\sss{\scriptscriptstyle}
\newcommand\as{\alpha_{\sss S}} 
\newcommand\gs{g_{\sss S}}

\newcommand\Soft{\mathrm{Soft}}

\def \al #1 {\frac {\as({#1})}{\pi} }
\def \ds #1 {\ooalign{$\hfil/\hfil$\crcr$#1$}}

\def \QCD {\mbox{{\tiny QCD}}}

\def\slash#1{\rlap{\hbox{$\mskip 1 mu /$}}#1}      

\allowdisplaybreaks[1]

\preprint{SLAC--PUB--17654}

\title{One-loop central-emission vertex for two gluons in $\mathcal{N}=4$ super Yang-Mills theory}

\author[a]{Emmet P. Byrne,}
\author[b,c,1]{Vittorio Del Duca,}
\author[d]{Lance J. Dixon,}
\author[a]{Einan Gardi} 
\author[a]{and~Jennifer~M.~Smillie}
\note{On leave from INFN, Laboratori Nazionali di Frascati, Italy.}
\affiliation[a]{Higgs Centre for Theoretical Physics, School of Physics and Astronomy, The University of Edinburgh, Edinburgh EH9 3FD, Scotland, UK}
\affiliation[b]{Institute for Theoretical Physics, ETH Z\"urich, 8093 Z\"urich, Switzerland}
\affiliation[c]{Physik-Institut, Universit\"at Z\"urich, 8057 Z\"{u}rich, Switzerland}
\affiliation[d]{SLAC National Accelerator Laboratory, Stanford University, Stanford, CA 94309, USA}

\emailAdd{Emmet.Byrne@ed.ac.uk}
\emailAdd{delducav@itp.phys.ethz.ch}
\emailAdd{lance@slac.stanford.edu}
\emailAdd{Einan.Gardi@ed.ac.uk}
\emailAdd{J.M.Smillie@ed.ac.uk}

\abstract{A necessary ingredient for extending the BFKL equation to next-to-next-to-leading logarithmic (NNLL) accuracy is the one-loop central emission vertex (CEV) for two gluons which are not strongly ordered in rapidity. Here we consider the one-loop six-gluon amplitude in $\mathcal{N}=4$ super Yang-Mills (SYM) theory in a central next-to-multi-Regge kinematic (NMRK) limit, we show that its dispersive part factorises in terms of
the two-gluon CEV, and we use it to extract the 
one-loop two-gluon CEV for any helicity configuration within this theory. This is a component of the two-gluon CEV in QCD.
Although computed in the NMRK limit, both the colour structure and the kinematic dependence of the two-gluon CEV capture much of the complexity of the six-gluon amplitudes in general kinematics. In fact, the transcendental functions of the latter can be conveniently written in terms of impact factors, trajectories, single-emission CEVs and a remainder, which is a function of the conformally invariant cross ratios which characterise the six-gluon amplitudes in planar $\mathcal{N}=4$ SYM.
Finally, as expected, in the MRK limit the two-gluon CEV neatly factorises in terms of two single-emission CEVs.
}


\begin{document}

\maketitle

\section{Introduction}
\label{sec:intro}

In the Regge limit, in which the squared centre-of-mass energy $s$ is much larger than the momentum transfer $|t|$, $s\gg |t|$,  any $2\to 2$ scattering process is dominated by the exchange in the $t$ channel of the highest-spin particle. In the case of QCD or ${\cal N}=4$ Super Yang-Mills (SYM) theory, that implies the exchange of a gluon in the $t$ channel. Contributions that do not feature gluon exchange in the $t$ channel are power suppressed in $t/s$.
Building upon this feature, the Balitsky-Fadin-Kuraev-Lipatov (BFKL) equation describes
strong-interaction processes with two large and disparate scales, $s\gg |t|$, by resumming the radiative corrections to parton-parton scattering to all orders at leading logarithmic (LL)~\cite{Lipatov:1976zz,Kuraev:1976ge,Kuraev:1977fs,Balitsky:1978ic} and next-to-leading logarithmic (NLL) accuracy~\cite{Fadin:1998py,Ciafaloni:1998gs,Kotikov:2000pm,Kotikov:2002ab} in $\log(s/|t|)$. The resummation of large energy logarithms extends to higher-multiplicity final states in special kinematic limits and improves the description of scattering events where large rapidity intervals are spanned~\cite{Mueller:1986ey,DelDuca:1993mn,Stirling:1994he,Andersen:2001kta,Colferai:2010wu,Andersen:2011hs,Ducloue:2013hia}. 
In the last decade or so, the Regge limit has been explored extensively in both ${\cal N}=4$ SYM~\cite{Bartels:2008ce,Dixon:2012yy,Basso:2014pla,DelDuca:2016lad,DelDuca:2019tur,Caron-Huot:2020vlo} and QCD~\cite{Caron-Huot:2013fea,Caron-Huot:2017fxr,Caron-Huot:2017zfo,Caron-Huot:2020grv,Falcioni:2021buo} amplitudes and cross sections~\cite{DelDuca:2013lma,DelDuca:2017peo}. Furthermore, owing to its relative simplicity, it has been used to constrain, compute or validate amplitudes in general kinematics~\cite{DelDuca:2009au,DelDuca:2010zg,Dixon:2014iba,Henn:2016jdu,Caron-Huot:2016owq,Almelid:2017qju,Caron-Huot:2019vjl,Falcioni:2021buo,Caola:2021izf}.
\par
\begin{figure}[htb]
\centering
     \subfigure[]{\label{fig:alpha1_alpha0}\includegraphics[scale=0.5]{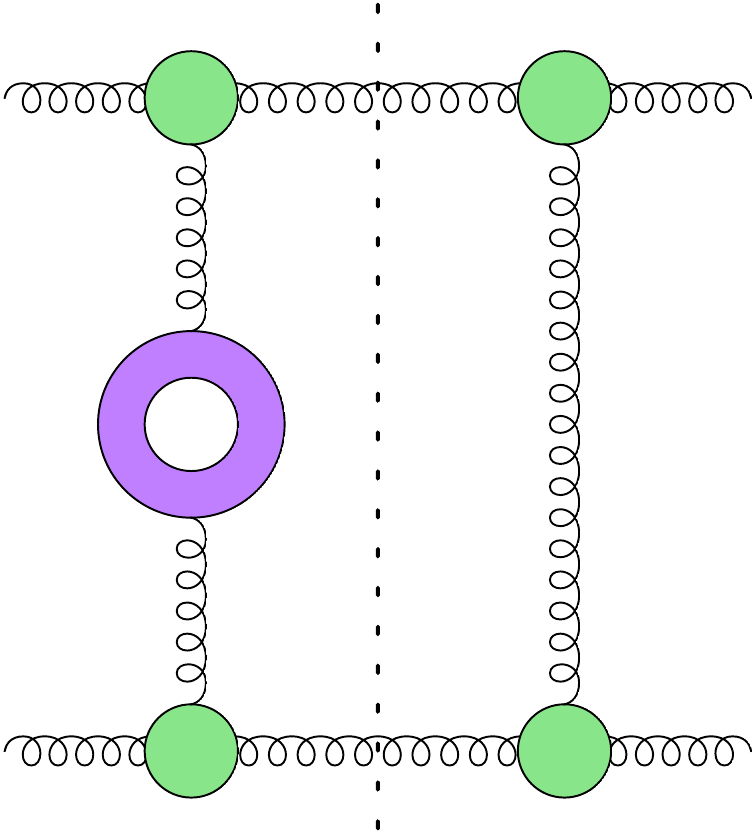}}
    \hspace*{20pt}
    \subfigure[]{\label{fig:V0g_V0g}\includegraphics[scale=0.5]{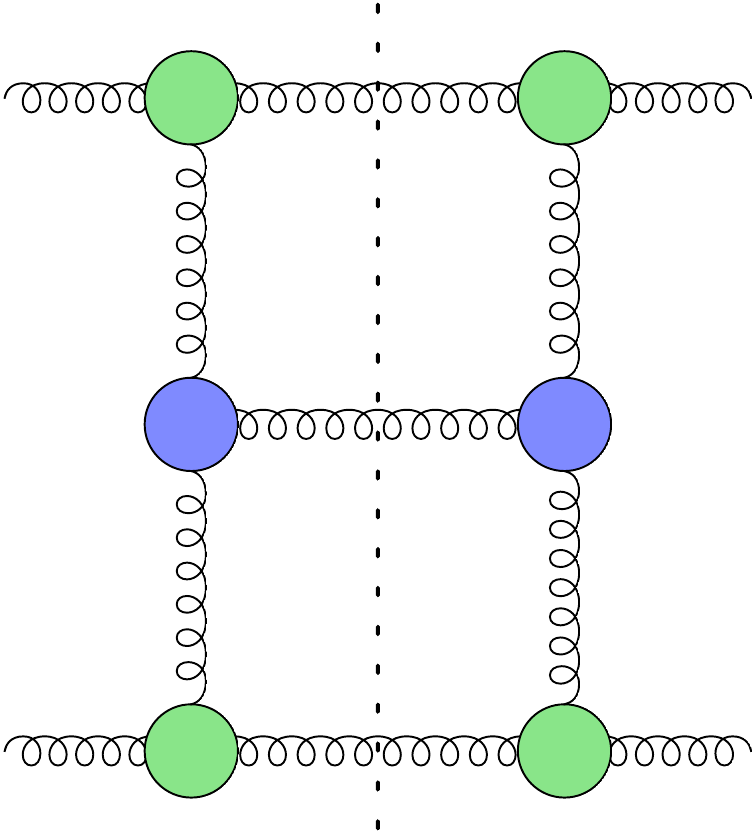}}
    \caption{Interference diagrams which contain the building blocks required for constructing the BFKL equation at LL accuracy: (a) gluon Regge trajectory at one loop;  (b) one-gluon central emission vertex at tree level. Tree-level impact factors (shown in green) are required for the computation of cross sections, but are not required to determine the evolution of the $t$-channel Reggeon.}
    \label{fig:BFKL_LL_Building_blocks}
\end{figure}

The BFKL equation~\cite{Lipatov:1976zz,Kuraev:1976ge,Kuraev:1977fs,Balitsky:1978ic} 
describes rapidity evolution in terms of an integral of the gluon propagator exchanged in the $t$ channel over all transverse momenta. The kernel of this integral equation consists of two elements, which are described schematically in fig.~\ref{fig:BFKL_LL_Building_blocks} and
which we review in section~\ref{sec:bfklblocksll}: 
a virtual component corresponding to the Regge trajectory, which first occurs in the one-loop $2\to 2$ scattering amplitude at LL accuracy in $\log(s/|t|)$ (fig.~\ref{fig:BFKL_LL_Building_blocks}(a)),
and a real component involving the emission of a gluon into the final state. The latter involves a vertex termed 
the Lipatov or \emph{central-emission vertex} (CEV), which first occurs in the tree-level $2\to 3$ amplitude in multi-Regge kinematics (MRK) (fig.~\ref{fig:BFKL_LL_Building_blocks}(b)), where the outgoing partons are strongly ordered in light-cone momentum, or equivalently in rapidity.
Upon integrating the BFKL evolution equation one thus resums corrections in which all emissions are ordered in rapidity, forming the famous ladder graphs. At LL accuracy, each gluon emission along the ladder introduces a factor of ${\cal O}(\as \log(s/|t|))$ after rapidity integration.
The impact factors, \eqn{centrc}, are not part of the gluon ladder, but sit at its ends. In fig.~\ref{fig:BFKL_LL_Building_blocks}, they are represented by the green blobs. When they are squared, they constitute the jet impact factors and contribute to jet cross sections at LL accuracy~\cite{Mueller:1986ey,DelDuca:1993mn,Stirling:1994he,Andersen:2001kta,Andersen:2011hs}. 
\par
The $2\to 2$ scattering amplitudes at LL accuracy are real-valued, and the terms of ${\cal O}(\as^n \log^n(s/|t|))$ exponentiate, manifesting the phenomenon of \emph{gluon Reggeization}~\cite{Fadin:2020lam}. The hard scattering is effectively mediated by the $t$-channel exchange of a \emph{single} Reggeized gluon, or Reggeon, carrying an octet colour charge, just as at tree level, and it gives rise to a Regge pole in the complex angular momentum plane.

\begin{figure}[bht]
\centering
    \subfigure[]{\label{fig:alpha2_alpha0}\includegraphics[scale=0.5]{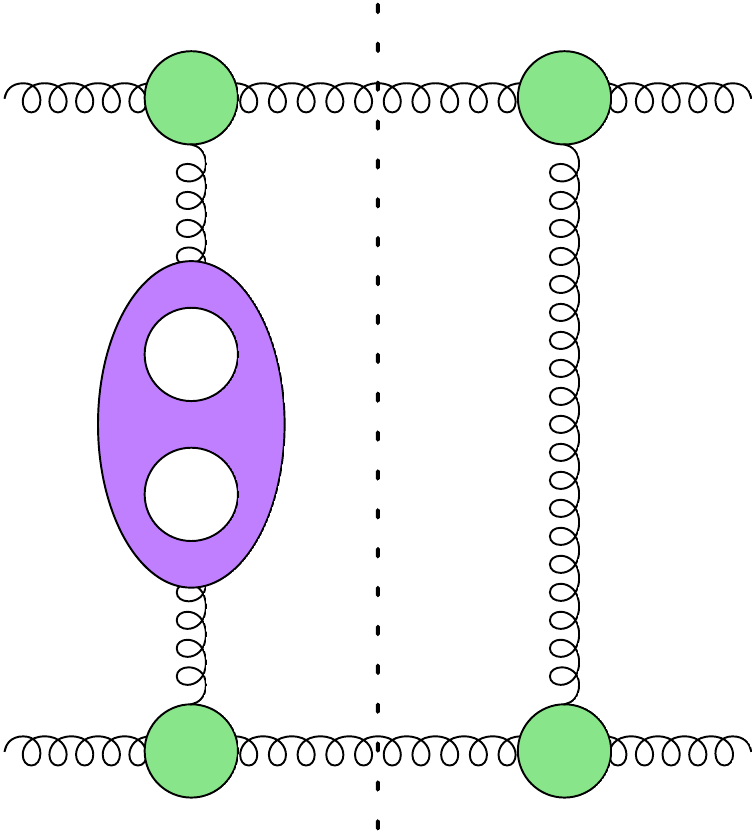}}
    \hspace*{20pt}
    \subfigure[]{\label{fig:V1g_V0g}\includegraphics[scale=0.5]{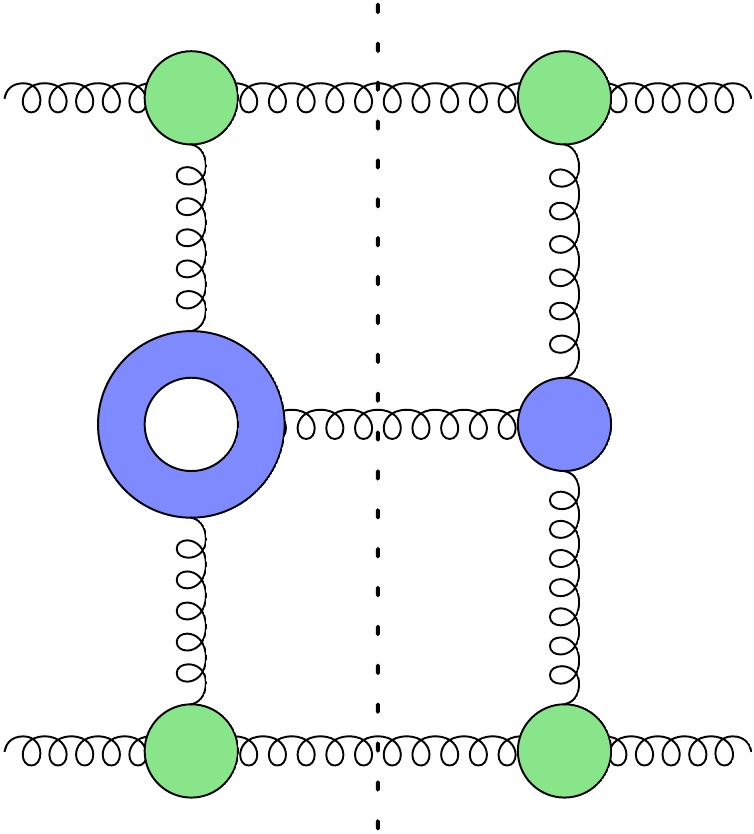}}
    \hspace*{20pt}
    \subfigure[]{\label{fig:V0gg_V0gg}\includegraphics[scale=0.5]{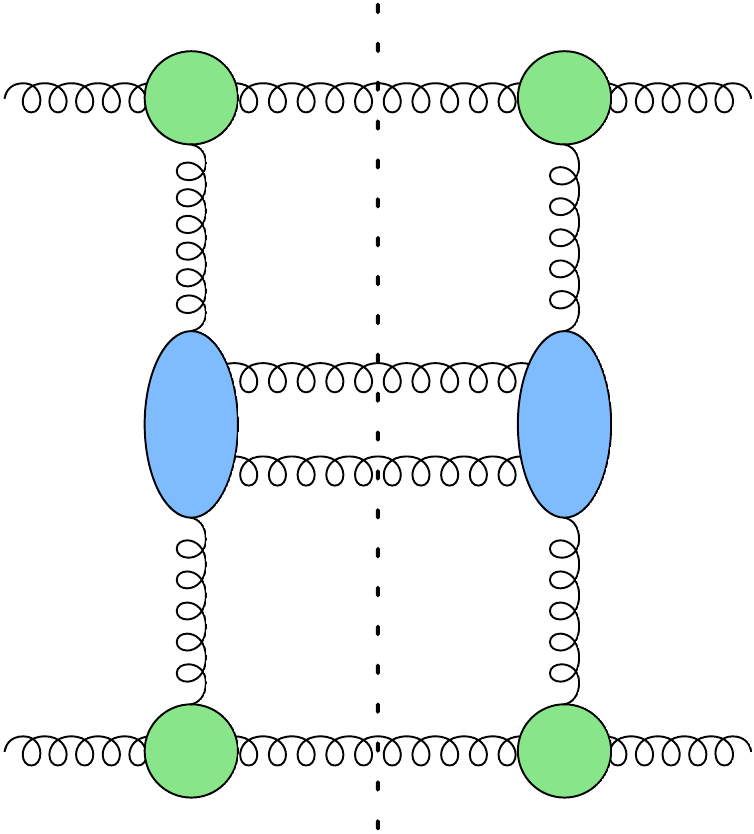}}
    \caption{Interference diagrams which contain the building blocks required for constructing the BFKL equation at NLL accuracy: (a) gluon Regge trajectory at two loops;  (b) one-gluon central-emission vertex at one loop; (c) two-gluon central emission vertex at tree level.}
    \label{fig:BFKL_NLL_Building_blocks}
\end{figure}

The BFKL equation has been extended to next-to-leading logarithmic (NLL) accuracy~\cite{Fadin:1998py,Ciafaloni:1998gs,Kotikov:2000pm,Kotikov:2002ab}, allowing one to resum the terms of ${\cal O}(\as^n \log^{n-1}(s/|t|))$~\cite{Fadin:2006bj,Fadin:2015zea}. 
This generalization involves next-to-leading-order corrections to the BFKL kernel, consisting of three elements, which are depicted schematically in fig.~\ref{fig:BFKL_NLL_Building_blocks}: 
two-loop corrections to the Regge trajectory~\cite{Fadin:1995xg,Fadin:1996tb,Fadin:1995km,Blumlein:1998ib,DelDuca:2001gu} (fig.~\ref{fig:BFKL_NLL_Building_blocks}(a)), one-loop corrections to the central emission vertex ~\cite{Fadin:1993wh,Fadin:1994fj,Fadin:1996yv,DelDuca:1998cx,Bern:1998sc} (fig.~\ref{fig:BFKL_NLL_Building_blocks}(b)), which we review in section~\ref{sec:bfklblocksnll},
and a new tree-level vertex function for the emission of two gluons (fig.~\ref{fig:BFKL_NLL_Building_blocks}(c)) (or of a quark-antiquark pair) that are \emph{not} strongly ordered in rapidity~\cite{Fadin:1989kf,DelDuca:1995ki,Fadin:1996nw,DelDuca:1996nom,DelDuca:1996km}, which we review in section~\ref{sec:tcev}.
The latter requires the evaluation of a six-parton tree-level amplitude in the so-called next-to-multi-Regge kinematics (NMRK), in which the outgoing partons are strongly ordered in rapidity except for two gluons (or a quark-antiquark pair) emitted centrally off the gluon ladder. When these are integrated in rapidity, they yield a factor of ${\cal O}( \as^2 \log(s/|t|) )$, thus contributing at NLL accuracy.
Crucial to carrying out the BFKL program at NLL accuracy is the fact that the real part -- or more precisely, the $s\leftrightarrow u$ signature-odd part -- of the $2\to 2$ amplitude is still described by the exchange of a single Reggeized gluon~\cite{Fadin:2006bj,Fadin:2015zea}. Note that the same is not true for
the imaginary (signature even) part of the amplitude, which is governed by the exchange of a pair of Reggeized gluons~\cite{Caron-Huot:2013fea,Caron-Huot:2017zfo,Caron-Huot:2020grv}.

In order to compute jet cross sections at NLL accuracy~\cite{Colferai:2010wu,Ducloue:2013hia} through the BFKL equation, jet impact factors at next-to-leading order (NLO) in $\as$~\cite{Bartels:2001ge,Bartels:2002yj} are required. They are based on the one-loop impact factor~\cite{Fadin:1993wh,Fadin:1992zt,Fadin:1993qb,DelDuca:1998kx,Bern:1998sc}, \eqn{eq:if1}, and the impact factor for the emission of two gluons or of a quark-antiquark pair~\cite{Fadin:1989kf,DelDuca:1995ki,Fadin:1996nw,DelDuca:1996nom,DelDuca:1996km,Duhr:2009uxa}, evaluated in the NMRK in which the gluons are strongly ordered in rapidity except for two partons emitted at one end of the ladder.

Extending the BFKL equation to next-to-next-to-leading logarithmic (NNLL) accuracy faces additional challenges\footnote{The BFKL equation has been extended to NNLL~\cite{Gromov:2015vua,Velizhanin:2015xsa} and to ${\rm N}^3$LL accuracy~\cite{Velizhanin:2021bdh} in planar $\mathcal{N}=4$ SYM theory.}. Chief amongst these is the fact that starting at this logarithmic order, the real part of $2\to 2$ amplitudes is not anymore governed by a single Reggeized gluon. Instead, it also involves contributions from three Reggeon exchange~\cite{DelDuca:2001gu,DelDuca:2014cya,Fadin:2017nka,Caron-Huot:2017zfo,Falcioni:2020lvv,Falcioni:2021buo} and the mixing of the latter with a single Reggeon. These give rise to a Regge cut as well as a Regge pole in the complex angular momentum plane. This makes the determination of the three-loop Regge trajectory, characterizing the Regge-pole term, a subtle problem, as it requires to disentangle Regge pole and cut contributions, see e.g.~\cite{DelDuca:2014cya,Caron-Huot:2017zfo,Fadin:2021csi}. Recently this problem has been resolved~\cite{Falcioni:2021buo,Falcioni:2021dgr} 
and the three-loop Regge trajectory in QCD has been determined~\cite{DelDuca:2021vjq,Falcioni:2021dgr,Caola:2021izf}. This gives some hope that the analytic structure of amplitudes in the Regge limit can be understood beyond NLL accuracy also in the multi-leg case, and indeed that the BFKL programme could eventually be extended to NNLL and beyond.

\begin{figure}[htb]
\centering
    \subfigure[]{\label{fig:alpha3_alpha0}\includegraphics[scale=0.5]{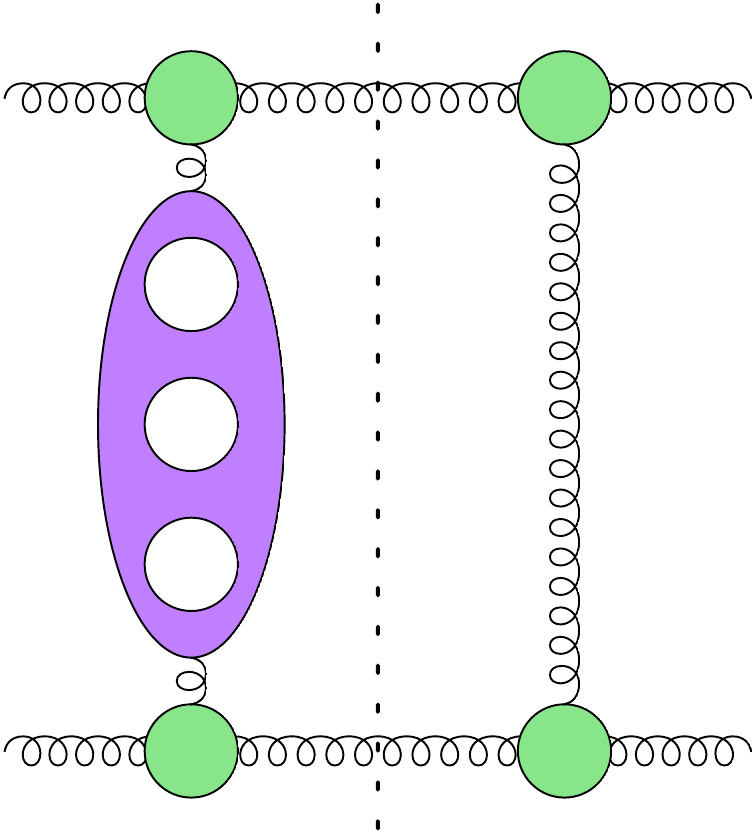}}
    \hspace*{20pt}
    \subfigure[]{\label{fig:V2g_V0g}\includegraphics[scale=0.5]{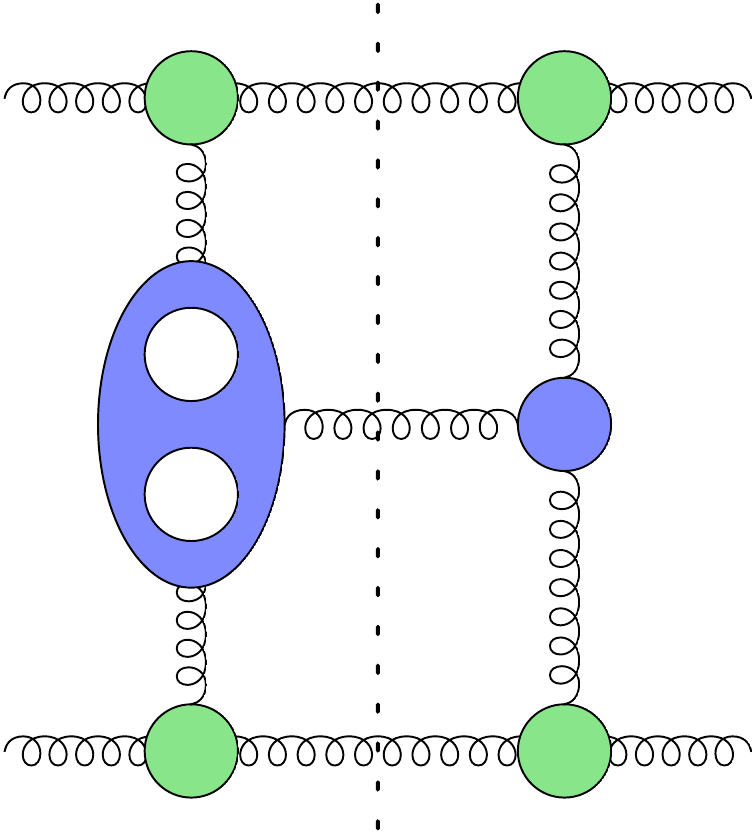}}
    \hspace*{20pt}
    \subfigure[]{\label{fig:V1g_V1g}\includegraphics[scale=0.5]{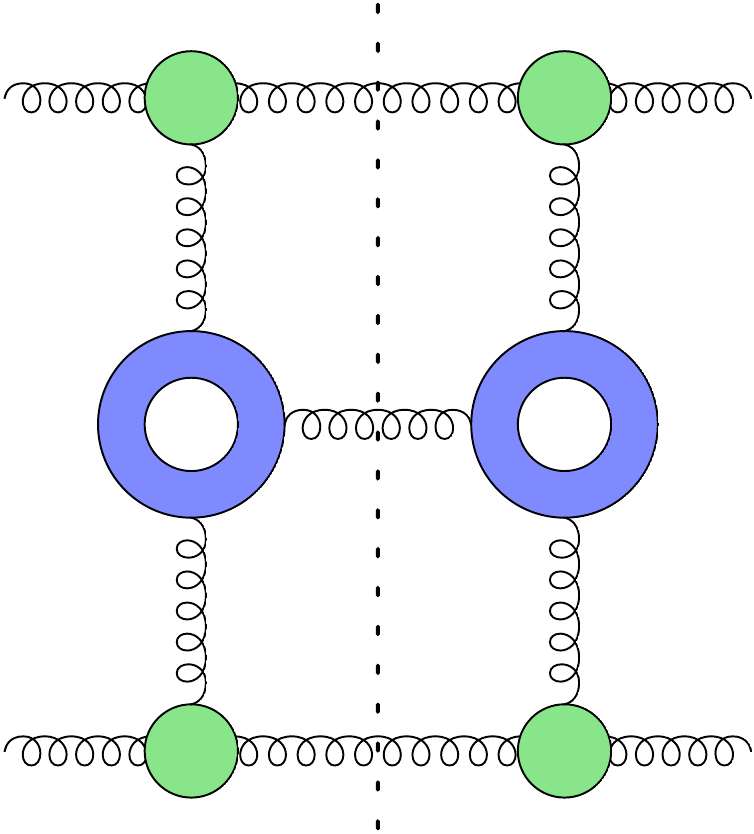}}
    \hspace*{20pt}
    \subfigure[]{\label{fig:V1gg_V0gg}\includegraphics[scale=0.5]{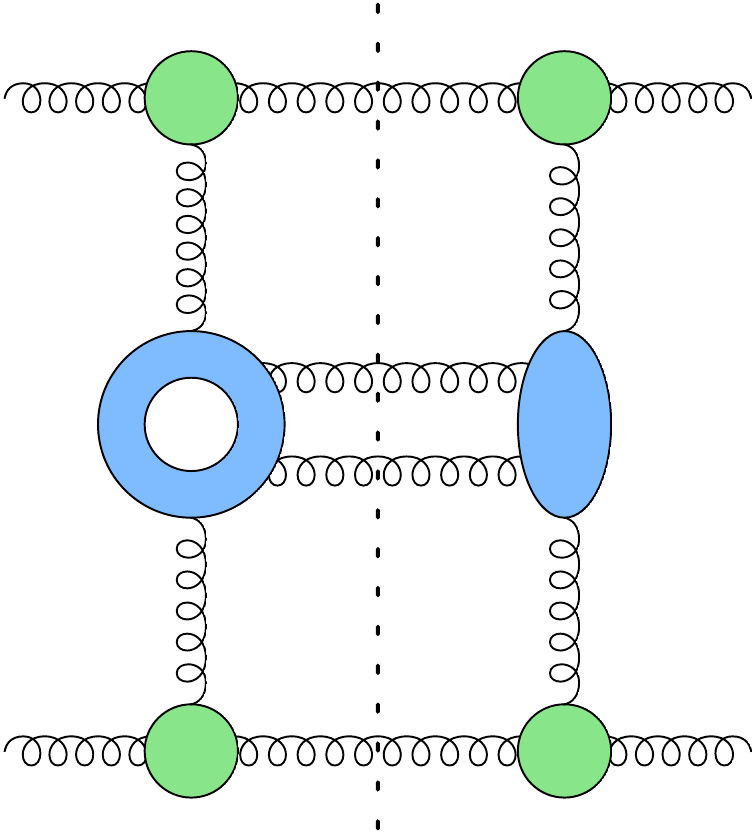}}
    \hspace*{20pt}
    \subfigure[]{\label{fig:V0ggg_V0ggg}\includegraphics[scale=0.5]{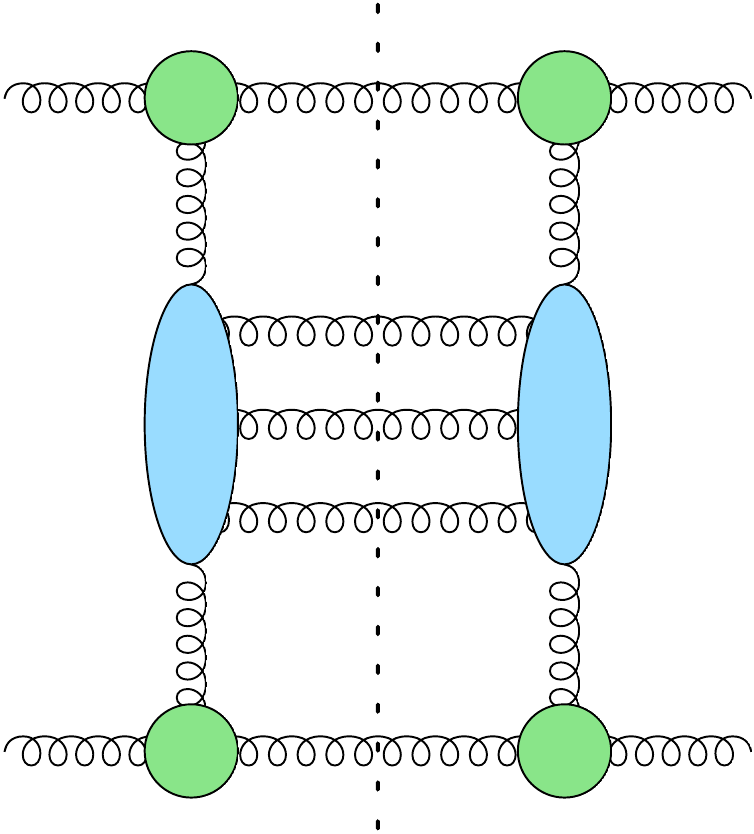}}
    \caption{Interference diagrams which contain the building blocks required for constructing the BFKL equation at NNLL accuracy: (a) gluon Regge trajectory at three loops;  (b) one-gluon CEV at two loops; (c) self-interference of one-gluon CEV at one loop; (d) two-gluon CEV at one loop; (e) three-gluon CEV at tree level.}
    \label{fig:BFKL_NNLL_Building_blocks}
\end{figure}

Carrying out the BFKL program at NNLL accuracy one needs,
in addition to the three-loop Regge trajectory (fig.~\ref{fig:BFKL_NNLL_Building_blocks}~(a)), which has recently been determined~\cite{DelDuca:2021vjq,Falcioni:2021dgr,Caola:2021izf}, 
various corrections to the central emission vertex. These are described schematically in fig.~\ref{fig:BFKL_NNLL_Building_blocks} and include the following elements: the emission vertex of three partons along the gluon ladder~\cite{DelDuca:1999iql,Antonov:2004hh,Duhr:2009uxa} (fig.~\ref{fig:BFKL_NNLL_Building_blocks}~(e)), 
which have been obtained by evaluating seven-parton tree-level amplitudes in next-to-next-to-multi-Regge kinematics (NNMRK);
the square of the one-loop correction to the emission vertex of one gluon (fig.~\ref{fig:BFKL_NNLL_Building_blocks}~(c));
the one-loop correction to the emission vertex of two gluons~(fig.~\ref{fig:BFKL_NNLL_Building_blocks}~(d)) or of a quark-antiquark pair along the gluon ladder;
and the two-loop correction to the central-emission vertex of one gluon (fig.~\ref{fig:BFKL_NNLL_Building_blocks}~(b)). The latter two contributions are yet to be determined. In this paper, we compute the one-loop corrections to the central-emission vertex of two gluons in $\mathcal{N}=4$ SYM. Utilizing the supersymmetric decomposition of QCD loop amplitudes~\cite{Bern:1993mq,Bern:1994zx,Bern:1994cg}, the $\mathcal{N}=4$ result is a first step toward the determination of the same quantity in QCD. To this end we use the one-loop six-gluon amplitudes in $\mathcal{N}=4$ SYM based on refs.~\cite{Bern:1994zx,Bern:1994cg}. By considering the appropriate NMRK kinematic limit of these amplitudes and comparing the result with a Regge-factorized form, we extract the two-gluon central-emission vertex in both the same helicity and opposite helicity configurations.

An important motivation for our investigation, besides the extension of the BFKL program, stems from the potential of the NMRK limit to unravel the analytic structure (as well as the colour structure) of multi-leg amplitudes. Indeed, direct analytical evaluation of multi-leg amplitudes in general kinematics at one-loop level is highly demanding, and the extension to two loops and beyond is ever more so. Even when such amplitudes are known, bringing them to a compact form and evaluating them efficiently are in general unsolved problems, which are of high interest both from a theoretical perspective and because of their immediate relevance to precision collider physics. Understanding the behaviour of amplitudes in various limits, where drastic simplifications occur, is a key strategy towards understanding them in general kinematics. Specifically, in the MRK limit one-loop amplitudes are very simple indeed: they are proportional to the tree amplitude, and involve simple kinematic dependence on the transverse momenta of the emitted gluons.  We shall see that the NMRK limit of six gluon amplitudes (already in ${\cal N}=4$ SYM) features several new aspects compared to MRK, including new colour structures as well as rational functions that depart from the tree-level amplitude, making it closer in complexity to the amplitude in general kinematics, while still more tractable.

We find that a useful way to organise the one-loop amplitudes in ${\cal N}=4$ SYM is by separately considering the dispersive and absorptive parts of the amplitude. We find that in the NMRK limit, the dispersive part of the one-loop amplitude factorises in terms of a one-loop two-gluon CEV, while the absorptive part admits no such simple factorisation. However, upon considering the helicity-summed interference of the one-loop amplitude with the tree-level amplitude, we observe that only the dispersive part of the amplitude contributes at NNLL accuracy, which is a promising result for the prospect of extending the BFKL kernel to this logarithmic accuracy. We also find that simplifications occur when considering the interference of the dispersive part of the one-loop amplitude with the tree-level amplitude: While, as mentioned above, the one-loop two-gluon CEV has an additional, fully symmetric colour structure which does not feature at tree level, this new colour structure does not contribute at NNLL accuracy. It does however, contribute to cross sections at N$^3$LL accuracy, where one must consider the square of the one-loop two-gluon CEV, giving rise to a quartic Casimir invariant.

The structure of this paper is as follows. 
In section \ref{sec:bfklblocks} we review the factorised structure of amplitudes at LL and NLL accuracy. In section \ref{sec:tcev} we review the properties of the tree-level two-gluon central-emission vertex and we obtain new representations of this vertex.
In section \ref{sec:6g1lamp} we take the central NMRK limit of the one-loop six-gluon amplitude in $\cN=4$ SYM. We find that the dispersive part of the amplitude factorises in terms of a one-loop two-gluon CEV. We obtain this CEV for the case of same-helicity gluons in section \ref{sec:N4same-hel} and for the case of opposite-helicity gluons in section \ref{sec:N4opp-hel}. In section~\ref{sec:NLO_xs} we consider the NLO $gg \to gggg$ squared matrix element and show that several simplifications occur in this context. We discuss the implications of these results for the NNLL BFKL kernel. In section \ref{sec:all-orders} we speculate how the behaviour we have observed for six-gluon one-loop amplitudes may be extended to higher multiplicity and loop order. Finally, we conclude in section \ref{sec:conclusions}. Several appendices are included:
in appendices~\ref{sec:appkin} and \ref{sec:min-invts} we display the kinematic regions and invariants of interest; in appendix~\ref{sec:power-suppressed} we list the power-suppressed helicity configurations; in appendix~\ref{sec:nmhv6gamp} we list the tree-level and one-loop six-gluon NMHV amplitudes in $\cN=4$ SYM in general kinematics; in appendix~\ref{sec:altern} we include
alternative forms of the tree-level opposite-helicity CEV; in appendix~\ref{sec:absorp} we display the absorptive part of the one-loop six-gluon amplitude in \texorpdfstring{$\cN=4$}{N=4} SYM in
the NMRK limit; in appendix~\ref{sec:lim} we compute the MRK, soft and collinear limits of the one-loop two-gluon CEV.

\section{The building blocks of the BFKL equation}
\label{sec:bfklblocks}

In this section we review the factorized structure of four and five gluon amplitudes in the Regge limit.
We define the elements which appear in this factorization, which are also the building blocks of the BFKL equation, as shown schematically in figs. \ref{fig:BFKL_LL_Building_blocks}, \ref{fig:BFKL_NLL_Building_blocks}
and \ref{fig:BFKL_NNLL_Building_blocks}. We will need these when discussing the more involved two-gluon central emission vertex in ${\cal N}=4$ SYM, which we determine from the six-gluon amplitude in the following sections. As high-energy factorization properties are universal, that is independent of the particular gauge theory being considered, our discussion in the present section will be entirely general. Note however that the functions appearing in the factorized amplitudes, such as impact factors, Regge trajectory and central emission vertices, do depend on the theory under consideration starting from one loop.

\subsection{Building blocks at LL accuracy}
\label{sec:bfklblocksll}

In the Regge limit, one can write the tree-level $2\to 2$ amplitudes in a factorised way. For example, the amplitude for gluon-gluon scattering $g_1\, g_2\to g_3\,g_4$ at tree level and at fixed helicities may be written as \cite{Lipatov:1976zz,Kuraev:1976ge}, 
\begin{equation}
\cM_{4g}^{(0)} = 
\left[\gs (F^{a_3})_{a_2c}\, C^{g(0)}(p_2^{\nu_2}, p_3^{\nu_3}) \right]
\frac{s}{t} \left[\gs (F^{a_4})_{a_1c}\, C^{g(0)}(p_1^{\nu_1}, p_4^{\nu_4}) \right] \, ,
\label{elas}
\end{equation}
with the incoming momenta $p_1$ and $p_2$ parametrised as in \eqn{in}, and with $s=(p_1+p_2)^2$, $q = -(p_2 + p_3)$, $t=q^2  \simeq -|q_\perp|^2$ 
and $(F^c)_{ab} = i\sqrt{2} f^{acb}$, 
and where the superscript $\nu_i$ labels the helicity of gluon~$g_i$. We consider all the momenta as outgoing, so the helicities for incoming partons are reversed, see \app{sec:appa}. 
As is apparent from the colour coefficient  $(F^{a_3})_{a_2c} (F^{a_4})_{a_1c}$, in \eqn{elas} only the antisymmetric octet ${\bf 8}_a$ is exchanged in the $t$ channel.

Since the four-gluon amplitude is a maximally helicity-violating (MHV) amplitude, \eqn{elas} describes $\binom{4}{2} = 6$ helicity configurations. However, at leading power in $t/s$, helicity is conserved along the $s$-channel direction, so in \eqn{elas} four helicity configurations are leading, two for each tree impact factor, $g^*\, g \rightarrow g$, with $g^*$ an off-shell gluon~\cite{DelDuca:1995zy},
\begin{equation}
C^{g(0)}(p_2^\ominus, p_3^\oplus) = 1\,, \qquad C^{g(0)}(p_1^\ominus, p_4^\oplus) = \frac{p_{4\perp}^*}{p_{4\perp}}\, ,\label{centrc}
\end{equation} 
with complex transverse coordinates $p_{\perp}$ as in \eqn{eq:scalprod}. At this order, the impact factors are just overall phases, and they transform under parity into their complex conjugates,
\begin{equation}
[C^{g}(p^\nu,p'^{\nu'})]^* = C^{g}(p^{-\nu},p'^{-\nu'})\, . 
\label{eq:ifconj}
\end{equation} 
The helicity-flip impact factor $C^{g(0)}(p^\oplus,p'^\oplus)$ and its parity conjugate $C^{g(0)}(p^\ominus,p'^\ominus)$ are power suppressed in $t/s$.
In~\eqn{centrc}, and in what follows, we encircle the helicity index to avoid any confusion with the light-cone direction labelling.

The tree amplitudes for quark-gluon or quark-quark scattering have the same form as \eqn{elas}, up to replacing one or both gluon impact factors $C^{g(0)}$ (\ref{centrc}) with quark impact factors $C^{q(0)}$, and the colour factors $(F^c)_{ab}$ in the adjoint representation with the colour factors $T^c_{ij}$ in the fundamental representation of SU($N_c$), which we normalise as $\tr(T^aT^b)= T_F \delta^{ab}$, with $T_F=1$. So in the Regge limit, $2\to 2$ scattering amplitudes factorise into gluon or quark impact factors and a gluon propagator in the $t$ channel, and are uniquely determined by them.

At LL accuracy in $\log(s/|t|)$, the four-gluon amplitude is given to all orders in $\as$ by Reggeizing the gluon, i.e. by dressing the gluon propagator in \eqn{elas} as~\cite{Lipatov:1976zz,Kuraev:1976ge}
\begin{equation}
\frac1{t} \to \frac1{t} \left(\frac{s}{\tau}\right)^{\alpha(t)}\,,
\label{eq:Reggeize}
\end{equation}
where $\tau > 0$ is a Regge factorisation scale, which is of order of~$|t|$, and much smaller than~$s$. 
The function $\alpha(t)$ is the gluon Regge trajectory which is related to the loop transverse-momentum integration. The latter is infrared divergent, and in dimensional regularization with $d=4-2\epsilon$, it can be written as
\begin{equation}
\alpha(t) =\frac{\as}{4\pi}\alpha^{(1)}(t)+\mathcal{O}(\as^2) \,, \qquad {\rm with} \qquad 
\alpha^{(1)}(t) =  N_c\, \frac{2}{\epsilon} 
\left(\frac{\mu^2}{-t}\right)^{\epsilon} \kappa_{\Gamma}\, ,\label{alph1}
\end{equation}
where for brevity we omit the dependence of $\alpha(t)$ on $\mu^2$. In \eqn{alph1}, $\as \equiv \gs^2/(4\pi)$ is the bare coupling, and the $\mathcal{O}(\as^2)$ term indicates the presence of higher corrections which become relevant beyond LL accuracy, $N_c$ is the number of colours, and 
\begin{equation}
\kappa_\Gamma \equiv (4\pi)^\epsilon\, \frac{\Gamma(1+\epsilon)\,
\Gamma^2(1-\epsilon)}{\Gamma(1-2\epsilon)}\, .\label{cgam}
\end{equation}
It is said that in the Regge limit, the four-gluon amplitude of \eqn{elas} with the dressing in \eqn{eq:Reggeize}, is determined by the $t$-channel exchange of a single Reggeized gluon, or Reggeon. 

In five-gluon scattering $g_1\, g_2\to g_3\,g_4\, g_5$, the multi-Regge kinematics (MRK) are defined as a strong ordering in the light-cone momenta,
\beq
p_3^+\gg p_4^+\gg p_5^+\,, \qquad |p_{3\perp}| \simeq |p_{4\perp}| \simeq |p_{5\perp}| \,.
\label{eq:mrk}
\eeq
or equivalently in the rapidities, of the produced gluons, \eqn{eq:mrkapp}.
In MRK, appendix~\ref{sec:appb}, the tree five-gluon amplitude takes the factorised form (fig.~\ref{fig:BFKL_LL_Building_blocks} (b)),
\begin{eqnarray}
\cM_{5g}^{(0)}  &=& 
s \left[\gs (F^{a_3})_{a_2c_1}\, C^{g(0)}(p_2^{\nu_2}, p_3^{\nu_3})) \right]\, 
\frac{1}{t_1} \label{three}\\ &&\times \left[\gs (F^{a_4})_{c_1c_2}\, 
V^{g(0)}(q_1,p^{\nu_4}_4,q_2) \right]\, \frac{1}{t_2}\, 
\left[\gs (F^{a_5})_{a_1c_2}\, C^{g(0)}(p_1^{\nu_1}, p_5^{\nu_5}) \right] \,, \nonumber
\end{eqnarray}
with $q_1 = -(p_2+p_3)$, $q_2= q_1-p_4$, and $t_i= q_i^2\simeq - q_{i\perp} q_{i\perp}^\ast$ ($i = 1, 2$). The impact factors are given in \eqn{centrc}, while the emission vertex of a real gluon from the $t$-channel gluon, which is termed the CEV, is~\cite{Lipatov:1976zz,Lipatov:1991nf}
\beq
V^{g(0)}(q_1,p^{\oplus}_4,q_2) = \frac{q_{1\perp}^\ast q_{2\perp}}{p_{4\perp}}\,,
\label{eq:centrv}
\eeq
which transforms under parity into its complex conjugate,
\beq
\left[ V^{g(0)}(q_1,p^{\nu},q_2) \right]^\ast = V^{g(0)}(q_1,p^{-\nu},q_2)\,.
\eeq

\begin{figure}[hbt]
\centering
     \includegraphics[scale=0.35]{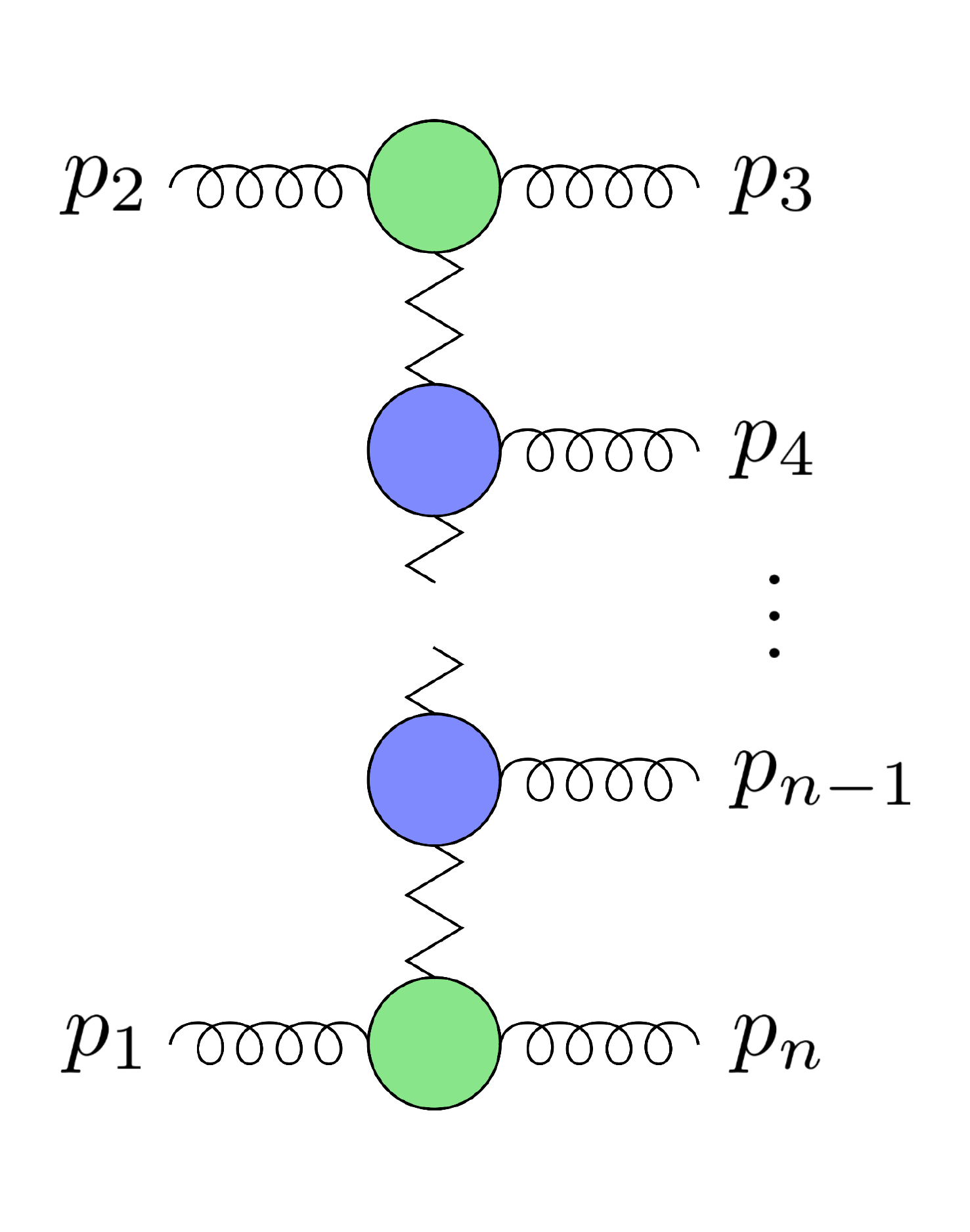}
    \caption{The factorised structure of gluon amplitudes in MRK. The zigzag line represents the Reggeized gluon propagator given in \eqn{eq:Reggeize}.
    The coloured blobs represent impact factors, and one-gluon CEVs. At LL accuracy these elements are only required at tree level, but much like the gluon Regge trajectory $\alpha(t)$, they can be defined to all orders in perturbation theory (note that this interpretation departs from the previous figures where the blobs are specific to a given loop order).
    }
    \label{fig:ng_MRK}
\end{figure}

The one-loop gluon Regge trajectory,~\eqn{alph1}, and the tree-level central-emission vertex (CEV),~\eqn{eq:centrv}, are the building blocks of an iterative structure, in the sense that a tree six-gluon amplitude in MRK, $g_1g_2\to g_3 g_4 g_5 g_6$, will display two 
CEVs along the gluon ladder, with the four outgoing gluons separated by three large rapidity gaps. At LL accuracy, each of the corresponding $t$-channel gluons Reggeizes according to eq.~(\ref{eq:Reggeize}) with the one-loop trajectory of eq.~(\ref{alph1}). Similarly, a seven-gluon amplitude in MRK will display three CEVs, and at LL accuracy it will feature four Reggeized $t$-channel gluons across each of the rapidity gaps, and so on for higher-multiplicity amplitudes. This iterative structure is captured by the leading-order BFKL equation. It is depicted in figure~\ref{fig:ng_MRK}, which is suggestive of an all-multiplicity, all-order generalization. 

\subsection{Building blocks at NLL accuracy}
\label{sec:bfklblocksnll}

As mentioned in the introduction, in order to extend the BFKL equation to NLL accuracy, as done in refs.~\cite{Fadin:1998py,Ciafaloni:1998gs,Kotikov:2000pm,Kotikov:2002ab}, one needs three ingredients, which are described schematically by fig.~\ref{fig:BFKL_NLL_Building_blocks}. First, one needs one-loop corrections to the single-gluon CEV of eq.~(\ref{eq:centrv}), which can be extracted from a five-gluon one-loop amplitude~\cite{Fadin:1993wh,Fadin:1994fj,Fadin:1996yv,DelDuca:1998cx,Bern:1998sc} (fig.~\ref{fig:BFKL_NLL_Building_blocks} (b)), as we show later in this section.
Second, one requires the tree-level CEV for two gluons, which are not ordered in rapidity. This object can be extracted from a tree-level six-gluon amplitude at NMRK~\cite{Fadin:1989kf,DelDuca:1995ki,Fadin:1996nw,DelDuca:1996km,Duhr:2009uxa} (fig.~\ref{fig:BFKL_NLL_Building_blocks} (c)) and it will be again extracted and analysed in some detail in section~\ref{sec:tcev} below.
Finally, one needs two-loop corrections to the Regge trajectory~$\alpha(t)$~\cite{Fadin:1995xg,Fadin:1996tb,Fadin:1995km,Blumlein:1998ib,DelDuca:2001gu}.
The latter may be extracted from $2\to 2$ amplitudes at two loops (fig.~\ref{fig:BFKL_NLL_Building_blocks} (a)). This requires an operative prescription to disentangle the Regge trajectory from the impact factor. In what follows, we briefly present such a prescription following ref.~\cite{Fadin:1993wh}.

We recall that in the Regge limit, $2\to 2$ scattering amplitudes $g_1\, g_2\to g_3\,g_4$, can be conveniently split into odd ${\cal M}_{4g}^{(-)}$ and even ${\cal M}_{4g}^{(+)}$ components under so-called signature symmetry, corresponding to a kinematic $s\leftrightarrow u$ interchange,
\begin{equation}
{\cal M}_{4g}(s,t)= {\cal M}_{4g}^{(-)}(s,t)+{\cal M}_{4g}^{(+)}(s,t)\,,\quad
\qquad\cM_{4g}^{(\pm)}(s,t) = \frac{ \cM_{4g}(s,t) \pm \cM_{4g}(u,t) }{2}\,,
\label{eq:sucross}
\end{equation}
with $u=-s-t\simeq -s$. 
In order to resum energy logarithms in ${\cal M}_{4g}(s,t)$ beyond LL accuracy it is useful to define a signature-symmetric logarithm,
\beq
\label{eq:lcross}
L\equiv \log\left(\frac{s}{-t}\right)-\frac{i\pi}{2}=\frac12\left[\log\left(\frac{-s-i0}{-t}\right)+\log\left(\frac{-u}{-t}\right)\right],
\eeq
and expand the amplitude at leading power in the Regge limit as 
\begin{equation}\label{eq:expansionDef}
	{\cal M}_{4g}^{(\pm)} = \gs^2 \sum_{n=0}^\infty 
	\left(\frac{\as}{4\pi}\right)^n \sum_{m=0}^nL^m{\cal M}_{4g}^{(\pm,n,m)}\,,
\end{equation}
with $m=n$ yielding the coefficients at LL accuracy, $m=n-1$ the coefficients at NLL accuracy, and so on.
With \eqns{eq:lcross}{eq:expansionDef} at work, one can show~\cite{Caron-Huot:2017fxr} that the coefficients of the odd (even) amplitude, ${\cal M}_{4g}^{(\mp,n,m)}$, are purely real (imaginary). Moreover, ${\cal M}_{4g}^{(-)}$ only receives contributions from $t$-channel exchange of an odd number of Reggeons, while ${\cal M}_{4g}^{(+)}$, from an even number.

Owing to Bose symmetry, also the colour factors of ${\cal M}_{4g}^{(-)}$ and ${\cal M}_{4g}^{(+)}$ are respectively odd and even under the interchange of the colour indices of the corresponding gluons $g_2$ and $g_3$ (or, instead, those of $g_1$ and $g_4$). The tree amplitude in eq.~(\ref{elas}) is entirely odd under signature, and indeed, its colour structure is antisymmetric under interchanging~$a_2$ and $a_3$, corresponding to a pure $t$-channel octet exchange, ${\bf 8}_a$.
Higher-order contributions to this process
involve in general additional colour structures, as dictated by the decomposition of the product ${\bf 8}_a \otimes {\bf 8}_a$ into irreducible representations,
\beq
  {\bf 8}_a \otimes {\bf 8}_a \, = \, \{ {\bf 1} \oplus {\bf 8}_s \oplus {\bf 27} \} 
  \oplus [ {\bf 8}_a \oplus {\bf 10} \oplus \overline{{\bf 10}} ]\ \, ,
\label{8*8}
\eeq
where the curly (square) brackets contain the representations which are even (odd) under signature.  However, at NLL accuracy, the signature-odd component of the amplitude, i.e. $\cM_{4g}^{(-,n,n-1)}$ for all $n$ in eq.~(\ref{eq:expansionDef}), is still governed by the exchange of a single Reggeon in the~$t$ channel, and~\eqn{eq:Reggeize} generalises in a straightforward way to~\cite{Fadin:1993wh},
\begin{align}
{ \cM^{(-)[8_a]}_{4g} } \label{sudall}
&= \frac1{2}  
\left[\gs (F^{a_3})_{a_2c}\, C^{g}(p_2^{\nu_2}, p_3^{\nu_3})) \right] \frac{s}{t} 
\left[ \left(\frac{s}{\tau}\right)^{\alpha(t)} + \left(\frac{-s}{\tau}\right)^{\alpha(t)} \right]
\, \left[\gs (F^{a_4})_{a_1c}\, C^{g}(p_1^{\nu_1}, p_4^{\nu_4}) \right]\,, 
\end{align}
where the colour and kinematic parts of the amplitude are each odd under $g_2\leftrightarrow g_3$ interchange. 

In \eqn{sudall}, we expand the gluon Regge trajectory in $\as$ as
\begin{equation}
\alpha(t) = \frac{\as}{4\pi} \alpha^{(1)}(t) + 
\left(\frac{\as}{4\pi}\right)^2 \alpha^{(2)}(t) + \ord(\as^3)\,
,\label{alphb}
\end{equation}
with $\alpha^{(1)}$ in \eqn{alph1}, and $\alpha^{(2)}$ is the two-loop Regge trajectory which enters the BFKL equation at NLL accuracy. Likewise, we expand the helicity-conserving impact factor, 
\begin{equation}
C^g(p^\oplus,p'^\ominus)=
C^{g(0)}(p^\oplus,p'^\ominus)+\frac{\as}{4\pi}\,C^{g(1)}(p^\oplus,p'^\ominus)+\mathcal{O}\left(\as^2\right)
\, ,\label{fullv}
\end{equation}
with one-loop corrections,
\begin{equation}
C^{g(1)}(p^\oplus,p'^\ominus) = C^{g(0)}(p^\oplus,p'^\ominus)\,
c^{g(1)}(t)\,.
\label{eq:if1}
\end{equation}
The same applies to the opposite helicity-conserving impact factor $C^g(p^\ominus,p'^\oplus)$. Note that the one-loop coefficients $c^{g(1)}$ are real and independent of the helicity configuration. The coefficient $c^{g(1)}$, and therefore also $C^{g}$, depend on the renormalisation and Regge-factorisation scales $\mu^2$ and $\tau$, but we omit this dependence for brevity.

The $2\to 3$ scattering amplitudes, $g_1\, g_2\to g_3\,g_4\,g_5$, can be likewise split into odd and even components under signature symmetry,
\begin{equation}
    {\cal M}_{5g}={\cal M}_{5g}^{(+,+)}+
    {\cal M}_{5g}^{(+,-)}+{\cal M}_{5g}^{(-,+)}+{\cal M}_{5g}^{(-,-)}\,,
\label{eq:signsymm}
\end{equation}
with
\beqa
    \lefteqn{ {\cal M}_{5g}^{(\sigma_a,\sigma_b)}
    (p_1^{\nu_1},p_2^{\nu_2},p_3^{\nu_3},p_4^{\nu_4},p_5^{\nu_5}) } \nn\\
    &=&
   \frac14 {\cal M}_{5g}(p_1^{\nu_1},p_2^{\nu_2},p_3^{\nu_3},p_4^{\nu_4},p_5^{\nu_5})
    +  \frac{\sigma_a}{4} {\cal M}_{5g}(p_1^{\nu_1},p_3^{\nu_3},p_2^{\nu_2},p_4^{\nu_4},p_5^{\nu_5})
    \label{eq:s34s45cross}\\
   &+& \frac{\sigma_b}{4} {\cal M}_{5g}(p_5^{\nu_5},p_2^{\nu_2},p_3^{\nu_3},p_4^{\nu_4},p_1^{\nu_1})
    +\frac{\sigma_a\sigma_b}{4} {\cal M}_{5g}(p_5^{\nu_5},p_3^{\nu_3},p_2^{\nu_2},p_4^{\nu_4},p_1^{\nu_1}) \,, \nn
\eeqa
where the two signature indices $\sigma_a$ and $\sigma_b$ can separately be $+1$ or $-1$.
By Bose symmetry, the upper and lower vertices (see fig.~(\ref{fig:momenta})) then acquire a corresponding colour symmetry: the upper vertex is symmetric/antisymmetric upon interchanging the colour indices $a_2$ and $a_3$ for $\sigma_a=+$ and $\sigma_a=-$, respectively. Similarly, the lower one is symmetric/antisymmetric upon interchanging  $a_1$ and $a_5$ for $\sigma_b=+$ and $\sigma_b=-$, respectively.
Specifically, the $\cM^{(-,-)}$ amplitude, which we discuss below, is 
antisymmetric in colour in both the upper and lower vertices.

In MRK, \eqn{eq:s34s45cross} is understood to involve the exchanges $s_{34}\leftrightarrow s_{24}=-s_{34}$ and $s_{45}\leftrightarrow s_{41}=-s_{45}$ along with the associated change in sign of $s=s_{12}=-s_{13}=-s_{52}=s_{35}$.
At NLL accuracy, the signature odd-odd component of the amplitude, $\cM^{(-,-)}$, is governed
by an antisymmetric octet exchange, ${\bf 8}_a$, in the $t_1$ and $t_2$ channels,
corresponding to a single Reggeon. One can then generalise \eqn{three},
\begin{eqnarray}
\cM^{(-,-)[8_a]}_{5g}  &=& \frac1{4}
s \left[\gs (F^{a_3})_{a_2c_1}\, C^{g}(p_2^{\nu_2}, p_3^{\nu_3})) \right]\, 
\frac{1}{t_1} \left[ \left(\frac{s_{34}}{\tau}\right)^{\alpha(t_1)} + \left(\frac{-s_{34}}{\tau}\right)^{\alpha(t_1)} \right]
\label{threeall}\\ &&\times \left[\gs (F^{a_4})_{c_1c_2}\, 
V^{g}(q_1,p^{\nu_4}_4,q_2) \right]\, \frac{1}{t_2}
\left[ \left(\frac{s_{45}}{\tau}\right)^{\alpha(t_2)} + \left(\frac{-s_{45}}{\tau}\right)^{\alpha(t_2)} \right] \nn\\
&&\times
\left[\gs (F^{a_5})_{a_1c_2}\, C^{g}(p_1^{\nu_1}, p_5^{\nu_5}) \right]\,, \nonumber
\end{eqnarray}
where the CEV now includes one-loop corrections
\begin{align}
	V^g(q_1,p_4^{\nu_4},q_2)=V^{g(0)}(q_1,p_4^{\nu_4},q_2)+\frac{\as}{4\pi}V^{g(1)}(q_1,p_4^{\nu_4},q_2)+\mathcal{O}\left(\as^2\right)\,, \label{eq:allcev}
\end{align}
which we express as 
\begin{equation}
V^{g(1)}(q_1,p_4^{\nu_4},q_2)
= V^{g(0)}(q_1,p_4^{\nu_4},q_2)\, 
v^{g(1)}(t_1,|p_{4\perp}|^2,t_2) \,.
\label{eq:oneloopcev}
\end{equation}
where $v^{g(1)}$ is real and helicity-independent. Similarly to $c^{g(1)}$, we omit the dependence on the scales $\mu^2$ and $\tau$ in $v^{g(1)}$ and $V^{g}$.

Finally, we note that
while the signature-odd NLL $gg\to gg$ amplitude is governed by a pure $t$-channel ${\bf 8}_a$ exchange, leading to the simple resummation formula in eq.~(\ref{sudall}), the signature-even  NLL amplitude ${\cal M}_{4g}^{(+)}$, which is purely imaginary, is governed by two-Reggeon exchange, leading to a more complex structure involving all $t$-channel representations that are compatible with the even signature in (\ref{8*8}), namely the singlet ${\bf 1}$, the symmetric octet ${\bf 8}_s$ and the ${\bf 27}$. Such a contribution occurs already in the non-logarithmic signature-even one-loop term~${\cal M}_{4g}^{(+,1,0)}$, which is proportional to $i\pi$~\cite{DelDuca:1998kx}. More generally, the signature-even amplitude obeys the BFKL equation~\cite{Kuraev:1976ge,Kuraev:1977fs,Balitsky:1978ic}: it is described by ladder graphs formed between two Reggeized gluons, which admits an all-order iterative solution~\cite{Caron-Huot:2013fea,Caron-Huot:2017zfo,Caron-Huot:2020grv}, rendering both infrared-singular and finite corrections
${\cal M}_{4g}^{(+,n,n-1)}$  computable for all $n$ for general colour. 
The signature-even gluon-gluon amplitude at NLL accuracy can only contribute to a squared amplitude, and thus to a cross section, at NNLL accuracy.

\subsection{Building blocks at NNLL accuracy}
\label{sec:bfklblocksnnll}

As outlined in the introduction, computing the next-to-next-to-leading-order corrections to the BFKL kernel requires to determine five elements:
the three-loop correction to the Regge trajectory~\cite{DelDuca:2021vjq,Falcioni:2021dgr,Caola:2021izf} (fig.~\ref{fig:BFKL_NNLL_Building_blocks}~(a));
the two-loop correction to the single-gluon CEV\footnote{The two-loop correction to the single-gluon CEV is known in planar $\mathcal{N}=4$ SYM theory~\cite{DelDuca:2009ae}.} (fig.~\ref{fig:BFKL_NNLL_Building_blocks}~(b));
the square of the one-loop correction to the single-gluon CEV\footnote{The one-loop correction to the single-gluon CEV is known in planar $\mathcal{N}=4$ SYM theory to all orders 
of~$\epsilon$~\cite{DelDuca:2009ae}. Beyond $\ord(\epsilon^0)$, the 
one-loop single-gluon CEV requires the one-loop pentagon in $6-2\epsilon$ dimensions~\cite{DelDuca:2009ac,Kniehl:2010aj,Kozlov:2015kol,Syrrakos:2020kba}.}  (fig.~\ref{fig:BFKL_NNLL_Building_blocks}~(c));
the one-loop correction to the two-parton (two gluons or a
quark-antiquark pair) CEV (fig.~\ref{fig:BFKL_NNLL_Building_blocks}~(d));
the tree-level three-parton CEV (fig.~\ref{fig:BFKL_NNLL_Building_blocks}~(e)), which is known for the case of three gluons~\cite{DelDuca:1999iql,Antonov:2004hh,Duhr:2009uxa}. We note in passing that the tree-level four-gluon CEV, which would occur at ${\rm N^3}$LL accuracy, is also known~\cite{Duhr:2009uxa}.

Computing the three-loop gluon Regge trajectory has required to tackle a new hurdle: starting at NNLL accuracy the signature-odd part of the $gg\to gg$ amplitude (namely its real part) cannot be described purely in terms of a single-Reggeon exchange~\cite{DelDuca:2001gu,DelDuca:2011wkl,DelDuca:2011ae,DelDuca:2013ara,DelDuca:2014cya,Fadin:2016wso,Caron-Huot:2017fxr,Fadin:2017nka,Falcioni:2020lvv,Falcioni:2021buo,Falcioni:2021dgr}. Instead, contributions from three Reggeon exchange occur, starting from two loops~\cite{Fadin:2016wso,Caron-Huot:2017fxr,Fadin:2017nka}, and furthermore, three-Reggeon mixing with a single Reggeon kicks in starting at three loops~\cite{Caron-Huot:2017fxr}. As a consequence, a resummation formula of the form~\eqn{sudall}, which corresponds to a pure Regge pole in the complex angular momentum plane, is insufficient. It must be supplemented by additional contributions which do not admit such simple factorization and exponentiation properties, and build up a Regge-cut in the complex angular momentum plane.
Since $t$-channel ${\bf 8}_a$ contributions, which are proportional to the tree amplitude, arise from both single and multiple Reggeon exchange, it had not been obvious how to disentangle Regge pole and cut contributions when working at fixed order. 

Infrared factorisation shows that~\cite{Bret:2011xm,DelDuca:2011ae}, in addition to the usual diagonal terms of the colour octet exchange, also non-diagonal terms in the $t$-channel colour basis contribute to the real part of $M^{(-,2,0)}_{4g}$~\cite{DelDuca:2013ara,DelDuca:2014cya}.
More recently, following the computation of NNLL corrections ${\cal M}_{4g}^{(-,n,n-2)}$ to the odd amplitude through four loops~\cite{Falcioni:2020lvv,Falcioni:2021buo}, 
the structure of the NNLL tower has been understood~\cite{Falcioni:2021dgr}, uncovering a deep connection to the non-planar origin of Regge cuts~\cite{Mandelstam:1965zz,Collins:1977jy}.
Specifically, it has become clear that \emph{planar} multi-Reggeon contributions at NNLL only occur at two and three loops. These contribute to the Regge pole, and factorise along with the single-Reggeon contribution as in~\eqn{sudall}, while the \emph{non-planar} multi-Reggeon contributions are precisely those that build up the Regge cut. Accordingly, the three-loop gluon Regge trajectory has been determined~\cite{DelDuca:2021vjq,Falcioni:2021dgr,Caola:2021izf}, providing the fully-virtual input, shown in fig.~\ref{fig:BFKL_NNLL_Building_blocks}~(a), that is necessary for extending the BFKL equation to NNLL accuracy, 

While the two-loop corrections to the single-gluon CEV (fig.~\ref{fig:BFKL_NNLL_Building_blocks}~(b)) are yet to be determined, in this paper, we evaluate the one-loop corrections to the two-gluon CEV  (fig.~\ref{fig:BFKL_NNLL_Building_blocks}~(d)) in $\mathcal{N}=4$ SYM, as a first step toward computing the same quantity in QCD. 

Let us also note that in order to compute jet cross sections at NNLL accuracy through the BFKL 
equation, jet impact factors at next-to-next-to-leading order (NNLO) in $\as$ will be needed. Building blocks to evaluate the latter are the two-loop impact factors~\cite{DelDuca:2014cya};
the one-loop impact factors for the emission of two gluons~\cite{Canay:2021yvm} or of a quark-antiquark pair, evaluated in NMRK; and the tree impact factor for the emission of three partons at one end of the ladder~\cite{DelDuca:1999iql} evaluated in NNMRK. Further, one must include the square of the one-loop helicity-violating impact factor~\cite{DelDuca:1998kx}.
\section{The two-gluon central-emission vertex at tree level}
\label{sec:tcev}
In this section we investigate several equivalent representations of the two-gluon CEV at tree level. In section \ref{sec:minset} we introduce a new set of kinematic variables which, in particular, greatly simplifies the task of showing equivalence between the different representations of the CEV. In section \ref{sec:bcfw} we obtain new representations for the opposite-helicity CEV from amplitudes that were derived via Britto-Cachazo-Feng-Witten (BCFW) recursion relations~\cite{Britto:2004ap,Britto:2005fq}. Not only do these representations make clear the collinear structure of the tree-level CEV, the individual rational terms that comprise the BCFW amplitudes appear also in the one-loop six-gluon NMHV amplitudes of $\cN=4$ SYM. The study of these rational terms therefore also lays the groundwork for section~\ref{sec:6g1lamp}, where we obtain the one-loop two-gluon CEV in $\cN=4$ SYM.

We begin by reviewing the factorisation properties of six-gluon amplitudes in the central NMRK. We consider the production of four gluons of momenta $p_i$, with $i=3, 4, 5, 6$ in the scattering 
of two gluons of momenta $p_1$ and $p_2$, with the momenta parametrised in terms of lightcone coordinates as in \eqn{in}. We suppose that the outgoing gluons are in a central NMRK configuration, as described in appendix~\ref{sec:appd},
where the rapidities and transverse momenta of the outgoing gluons satisfy
\begin{equation}
y_3\gg y_4 \simeq y_5 \gg y_6\,;\qquad |p_{3\perp}|
\simeq |p_{4\perp}| \simeq |p_{5\perp}|  \simeq |p_{6\perp}|\, .
\label{qmrapp1}
\end{equation}
In the NMRK region of \eqn{qmrapp1}, the tree-level amplitude for the $g_1\, g_2\to g_3\, g_4\,g_5\, g_6$ scattering process is~\cite{Fadin:1989kf,DelDuca:1995ki,Fadin:1996nw,DelDuca:1996km}:
\begin{align}
 \left. \cM^{(0) }_{6g} \right|_{\rm NMRK}  
&= \sum_{\sigma \in S_2}\gs^4\, (F^{a_3})_{a_2c_1}
 (F^{a_{\sigma_4}}F^{a_{\sigma_5}})_{c_1c_3} (F^{a_6})_{a_1c_3}\, M^{(0)}_{6g}( p_1^{\nu_1},p_2^{\nu_2}, p_3^{\nu_3}, p^{\nu_{\sigma_4}}_{\sigma_4}, p^{\nu_{\sigma_5}}_{\sigma_5}, p_6^{\nu_6}) \,,
 \label{NLOfactorization}
\end{align}
where the sum is over the two permutations of the gluon labels 4 and 5, and on the right-hand side an NMRK subscript on the colour-ordered amplitudes $M^{(0)}_{6g}$ is understood. The latter factorise in this limit as
\begin{align}
    \label{NLOfactorization2}
 M^{(0)}_{6g}(p_1^{\nu_1},p_2^{\nu_2}, p_3^{\nu_3}, p^{\nu_4}_4, p^{\nu_5}_5, p_6^{\nu_6})
&= s\, C^{g(0)}(p_2^{\nu_2}, p_3^{\nu_3})\, \frac1{t_1}\, A^{gg(0)}(q_1, p^{\nu_4}_4, p^{\nu_5}_5, q_3)\, \frac1{t_3}\, C^{g(0)}(p_1^{\nu_1}, p_6^{\nu_6}) \,, 
\end{align}
with $s=s_{12}=(p_1+p_2)^2$, $q_1 = - ( p_2 + p_3)$, $q_3 = p_1 + p_6$, $s_{23} = t_1 = q_1^2\simeq -|q_{1\perp}|^2$, 
$s_{61} = t_3 = q_3^2\simeq -|q_{3\perp}|^2$.
Here $A^{gg(0)}(q_1, p^{\nu_4}_4, p^{\nu_5}_5, q_3)$ is the tree-level two-gluon central-emission vertex, which is the subject of this section. This quantity has been studied previously~\cite{Fadin:1989kf,DelDuca:1995ki,Fadin:1996nw,DelDuca:1996km} and our analysis here is primarily intended to prepare the grounds for the determination of this quantity at one loop in the next section.

Since the six-gluon amplitude can be in either a maximally helicity-violating (MHV) or next-to-maximally helicity-violating (NMHV)
configuration, \eqn{NLOfactorization} describes a total of $2\binom{6}{2} + \binom{6}{3} = 50$ helicity configurations. 
Out of these, 34 configurations are associated to amplitudes which have at least one helicity-flip impact factor, appendix~\ref{sec:power-suppressed}, and thus are power suppressed in $t_i/s$, with $i=1,3$. Conversely, at leading power in $t_i/s$, helicity is conserved along the $s$-channel direction in Minkowski space, 
so in \eqn{NLOfactorization} 16 helicity configurations are leading. Two are associated to each tree impact factor, $g^*\, g \rightarrow g$, with $g^*$ an off-shell gluon, \eqn{centrc}.

Multiplying the four helicity configurations associated to the impact factors, there are four helicity configurations associated to the
central-emission vertex for the emission of two gluons, $g^*\, g^* \rightarrow g\, g$, which read\footnote{Note that having defined $(F^c)_{ab} = i\sqrt{2} f^{acb}$, the central-emission vertices, eqs.~(\ref{eq:centrv}), (\ref{eq:kosc_same-hel}) and (\ref{kosc}), do not display the usual overall powers of $\sqrt{2}$.}~\cite{Fadin:1989kf,DelDuca:1995ki,Fadin:1996nw,DelDuca:1996km}
\begin{eqnarray}
A^{gg(0)}(q_1, p^\oplus_4, p^\oplus_5, q_3) &=& \frac{q_{1\perp}^\ast q_{3\perp}}{p_{4\perp}} \sqrt{\frac{x_4}{x_5}}\, \frac{1}{\langle 4\, 5\rangle} \,, \label{eq:kosc_same-hel}\\
\left. A^{gg(0)}(q_1, p^\oplus_4, p^\ominus_5, q_3)\right|_1 &=& - \frac{p_{4\perp}^\ast}{p_{4\perp}} 
\left\{ - \frac{1}{s_{45}} \left[ \frac{y_5\, p_{5\perp} |q_{1\perp}|^2 }{ p_{5\perp}^\ast} 
+ \frac{ x_4\,p_{4\perp} |q_{3\perp}|^2 }{p_{4\perp}^\ast} 
+ \frac{s_{234}\, p_{4\perp}p_{5\perp}}{p_4^-p_5^+} \right]\right. \nn\\
&& \qquad\quad + \left. \frac{(q_{3\perp}+p_{5\perp})^2}{s_{234}} -
\frac{q_{3\perp}+p_{5\perp}}{s_{45}}
\left[\frac{p_{4\perp}}{y_4} - \frac{p_{5\perp}}{x_5} \right]\right\} \,, \label{kosc}
\end{eqnarray}
where the kinematic invariants are given in eqs.~(\ref{nmrkinv}), (\ref{eq:s3nmrk}), and (\ref{cnrpro}), and where we use the momentum fractions,
\beq
x_i = \frac{p_i^+}{p_4^+ + p_5^+}\,, \qquad y_i = \frac{p_i^-}{p_4^- + p_5^-}\,, \qquad i=4, 5\,.
\label{sec:x45}
\eeq
In eqs.~(\ref{eq:kosc_same-hel})--(\ref{kosc}), similarly to~\eqn{centrc}, we have encircled the helicity index
and we have appended the index $\left. \right\vert_1$ to the opposite-helicity vertex (\eqn{kosc}) in order to stress that this is 
the first amongst several  equivalent representations we are going to consider for this quantity. 

Note that the impact factors, \eqn{eq:ifconj}, and the central-emission vertex, eqs.~(\ref{eq:kosc_same-hel})--(\ref{kosc}), transform under parity into their complex conjugates,
\beq
\left[ A^{gg(0)}(q_1, p^{\nu_4}_4, p^{\nu_5}_5, q_3)\right]^\ast = A^{gg(0)}(q_1, p^{-\nu_4}_4, p^{-\nu_5}_5, q_3)\,,
\label{eq:compl}
\eeq
allowing one to obtain $A^{gg(0)}(q_1, p^\ominus_4, p^\ominus_5, q_3)$ and 
$A^{gg(0)}(q_1, p^\ominus_4, p^\oplus_5, q_3)$ from those in eq.~(\ref{kosc}). If the further MRK limit is taken, 
i.e.~further requiring $p_4^+\gg p_5^+$, the two-gluon CEV factorises into a product of two Lipatov vertices times an intermediate $t$-channel pole. The MRK limit is discussed in appendix~\ref{sec:mrklim}.

In ref.~\cite{Duhr:2009uxa} an alternative form of the opposite-helicity two-gluon CEV was found by using the CSW rules~\cite{Cachazo:2004kj}, \eqn{eq:AggpmCSW}.
This is a compact form of the vertex, consisting of only four rational terms, but unlike \eqn{kosc}, this representation appears to have unphysical poles e.g. at $p_{4\perp}+p_{5\perp}=0$. This is a feature which is shared with the new representations we will obtain from the BCFW recursion relations in section~\ref{sec:bcfw}. Before obtaining these new representations, we will first introduce a minimal set of independent variables which we will use throughout this paper. In particular, these variables will make it simple to analytically verify the equivalence of all the representations of the opposite-helicity vertex we will study, thereby proving that the apparent unphysical singularites in the CSW and BCFW representations are spurious.

\subsection{A minimal set of variables}
\label{sec:minset}

In MRK the central-emission vertex for the emission of one gluon can be expressed in terms of one independent
transverse momentum, i.e. in terms of two real parameters. We expect that the emission of one more gluon out of the central-emission vertex will require the introduction of the variables of one more on-shell massless particle, i.e. three more real parameters, such that the central-emission vertex for the emission of $n$ gluons will be expressed in terms of a set of $(3n-1)$ independent real parameters. In particular, the central-emission vertex for the emission of two gluons will be given in terms of five real parameters.

We write $p_{4\perp}$ and $p_{5\perp}$ in terms of MRK $t$-channel transverse momenta,
\beq
p_{4\perp}=q_{1\perp}-q_{2\perp}\,, \qquad p_{5\perp}=q_{2\perp}-q_{3\perp}\,.
\label{eq:mrkparam}
\eeq
In the sum of the equations above, $q_{2\perp}$ cancels such that transverse momentum conservation in NMRK is fulfilled. We parametrise the ratios of transverse momenta in~\eqn{eq:mrkparam} in terms of two complex variables, $w$ and $z$,
\beqa
\frac{p_{4\perp}}{q_{1\perp}}&=& \frac{1}{1-z}\,, \qquad \frac{q_{2\perp}}{q_{1\perp}}=\frac{-z}{1-z}\,, \nn\\
\frac{p_{5\perp}}{q_{3\perp}}&=& \frac{-1}{1-w}\,, \qquad \frac{q_{2\perp}}{q_{3\perp}}=\frac{-w}{1-w}\,.
\label{eq:wzparam}
\eeqa
Inverting these relations gives us
\begin{equation}
	z=-\frac{q_{2\perp}}{p_{4\perp}}\,,\qquad w=\frac{q_{2\perp}}{p_{5\perp}}.
	\label{eq:wzparam2}
\end{equation}
Using complex parametrization of the transverse momenta (see appendix \ref{sec:appa}), $z$ and $w$ are also complex. We will use $\bar{z}$ and $\bar{w}$ to denote their complex conjugates.

In summary, we may parametrise the central-emission vertex for the emission of two gluons, involving five real degrees of freedom, in terms of the set $\{w, z, X\}$, where $X$
is a real positive variable which depends on the rapidity difference between $p_4$ and $p_5$,
\beq
X\equiv \frac{p_{4}^+}{p_{5}^+} = \frac{x_4}{x_5} \,.
\label{eq:xparam}
\eeq
The MRK limit is reached by taking $X\to \infty$. When we use the acronym MRK, we will always mean this canonical MRK limit with the strict rapidity ordering $y_3 \gg y_4 \gg y_5 \gg y_6$. The multi-Regge regime with strict rapidity ordering $y_3 \gg y_5 \gg y_4 \gg y_6$ is also contained within the central NMRK limit of appendix \ref{sec:appd}, and we will refer to this kinematic condition as $\MRKx$. The $\MRKx$ limit is reached by taking $X \to 0$.

The leading NMRK behaviour of several Lorentz invariant quantities are listed in terms of the minimal set of variables in appendix \ref{sec:min-invts}.  In terms of this minimal set, eqs.~(\ref{eq:kosc_same-hel})--(\ref{kosc}) become
\begin{align}
		A^{gg(0)}(q_1, p^\oplus_4, p^\oplus_5, q_3) =&\frac{q_{1\perp}^*}{q_{1\perp}}\frac{z(w-1)(z-1)X}{(w+Xz)}\,,\label{eq:koscmin_same-hel}\\
		\begin{split}
		\left. A^{gg(0)}(q_1, p^\oplus_4, p^\ominus_5, q_3)\right|_1=&\frac{\bar{w}^2z^2|z-1|^2X^2}{|w+Xz|^2(|w|^2+X|z|^2)}
		+\frac{z^2X}{(1+X|z|^2)}
		\\
		+& \frac{Xz}{|w+Xz|^2}\bigg(\frac{|w-1|^2X \bar{z}}{(1+X)}+\bar{w}(1+X|z|^2) \\ & \hspace{3cm} -(|w|^2+X|z|^2)-\bar{w}z(1+X)\bigg)\,.	
		\label{eq:koscmin}
		\end{split}
\end{align}

We note that the same-helicity vertex, \eqn{eq:koscmin_same-hel}, contains a remaining phase $q_{1\perp}^*/q_{1\perp}$, which cannot be expressed in terms of the minimal variables. From \eqn{eq:Spa45min} we see that the physical singularity of this vertex corresponds to
\beqa
\langle 4 \, 5\rangle = 0 \quad \leftrightarrow \quad w+X z = 0\,.
\label{eq:physsing}
\eeqa
The opposite-helicity vertex, \eqn{eq:koscmin}, has a more complicated singularity structure. From \eqn{eq:s3nmrkmin} we see that the further physical singularities of this vertex correspond to the vanishing of three-particle Mandelstam invariants,
\begin{equation}
	\begin{array}{rrrrr}
		s_{123} = 0 \quad & \leftrightarrow & \quad 1+X  = 0 \,, \\
		s_{234} = 0 \quad & \leftrightarrow & \quad 1+X |z|^2 = 0\,,\\
		s_{345} = 0 \quad & \leftrightarrow & \quad |w|^2+X|z|^2 = 0  \,. \\ 
	\end{array}
\label{eq:physsing2}
\end{equation}
The right-hand side of \eqn{eq:physsing} and \eqn{eq:physsing2} define the physical-singularity surfaces of the minimal set $\{w, z, X\}$.
However, in NMRK, if we require all transverse momenta to be non-vanishing, the only singularity surface which can be reached is the collinear limit $w= -zX$. This collinear limit is studied in appendix \ref{sec:collim}. This agrees with the intuitive picture of NMRK where, for example, the large rapidity separation $y_3\gg y_4$ means we cannot take the further collinear limit $p_3 || p_4$. Relaxing the requirement of non-vanishing transverse momenta, \eqns{eq:koscmin_same-hel}{eq:koscmin} are singular in the soft limits $p_4\to 0$ and $p_5\to 0$, which are explored in appendix \ref{sec:softlim}.

\subsection{BCFW representations}
\label{sec:bcfw}
In ref.~\cite{DelDuca:1996km}, the opposite-helicity vertex of \eqn{kosc} was obtained from the tree-level six-gluon amplitude given in ref.~\cite{Mangano:1987xk}. Compact expressions for NMHV six-gluon amplitudes were later derived using the BCFW on-shell recursion relations~\cite{Britto:2004ap,Britto:2005fq}. We present one of the amplitudes obtained in ref.~\cite{Britto:2004ap} here, which we will use as an example in the following\footnote{This amplitude is equal to the complex conjugate of \eqn{eq:nmhvmmpmpp} and we have expanded the $D_i$ functions to display the physical and spurious poles of the amplitude.}:
\begin{align}
\begin{split}
		M^{(0)}_{(\ref{eq:BCFWexample})}\left(p_1^\oplus, p_2^\oplus, p_3^\ominus, p_4^\oplus, p_5^\ominus, p_6^\ominus\right) =&\ -\frac{\langle 3|1+2| 4]^{4}}{\langle 1\, 2\rangle\langle 2\, 3\rangle[4\, 5][5\, 6] s_{123}\langle 1|2+3| 4]\langle 3|4+5| 6]}\\
		&-\frac{\langle5\,  6\rangle^{4}[2\, 4]^{4}}{\langle 5\, 6\rangle\langle 6\, 1\rangle[2\, 3][3\, 4] s_{234}\langle 5|6+1| 2]\langle 1|2+3| 4]} \\
		&-\frac{\langle 3\, 5\rangle^{4}[1\, 2]^{4}}{[6\, 1][1\, 2]\langle 3\, 4\rangle \langle 4\, 5\rangle s_{345}\langle 3|4+5| 6]\langle 5|6+1| 2]}.
		\label{eq:BCFWexample}
\end{split}
\end{align}
A complete set of six-gluon NMHV tree amplitudes are listed in eqs.~(\ref{eq:nmhvmmmppp})--(\ref{eq:nmhvmpmpmp}). A second set of representations of the tree-level amplitudes are given by eqs.~(\ref{eq:NMHVbmmmppp})--(\ref{eq:NMHVbmpmpmp}); these are discussed further in appendix~\ref{sec:1loopNMHV}. As mentioned earlier, the individual terms of these BCFW-based amplitudes appear also in the six-gluon one-loop amplitudes in $\cN=4$ SYM. As an example, note that the rational coefficients that appear in the one-loop 
amplitude~(\ref{eq:N4mmpmpp}) consist of the individual rational terms that make up the tree-level amplitudes~(\ref{eq:nmhvmmpmpp}) and (\ref{eq:NMHVbmmpmpp}). In this section we are therefore interested not only in the NMRK limit of the tree-level amplitudes as a whole, but also in the NMRK behaviour of the individual rational terms which will appear outside the context of tree-level amplitudes in section \ref{sec:N4opp-hel}.

BCFW-based representations have the additional advantage of making the collinear singularity structure of the tree amplitudes manifest. 
For example, \eqn{eq:BCFWexample} has separate poles at $\langle4 \ 5 \rangle =0$ and $[45]=0$, while \eqn{kosc} only displays a pole at $s_{45}=0$. The form in \eqn{kosc} entangles the two types of contributions, from different helicity intermediate states; also, individual terms are more singular than the actual amplitude. The BCFW-based representations are therefore more amenable to the study of collinear limits, a property which is utilised in appendix \ref{sec:collim}.
In addition to physical singularities (spinor products of two cyclically adjacent momenta, and Mandelstam invariants of three cyclically adjacent momenta), the individual terms in these BCFW-based representations also contain unphysical singularities (which cancel in the sum of terms appearing in the amplitude). A simple example is provided by the spinor string
\begin{align}
\begin{split}
	\langle  3 | 4+5 |6]\,&\toNMRK\, -q_{3\perp}^*\sqrt{\frac{p_3^+}{p_6^+}}\,\left(p_{4\perp}+p_{5\perp}\right)
	\,=\, |q_{1\perp}|^2\sqrt{\frac{p_3^+}{p_6^+}}\,\frac{\bar{z}(\bar{w}-1)(w-z)}{|w|^2|z-1|^2}
\end{split}
\end{align}
which appears in the denominator of terms in \eqn{eq:BCFWexample}.

For six-particle kinematics there is a simple relationship between the square of such spinor strings and two- and three-particle Mandelstam invariants,
\begin{equation}
\langle 3 |4+5|6]\langle 6| 4+5|3]= s_{123}s_{345}\left(1-\frac{s_{12}s_{45}}{s_{123}s_{345}}\right).
\end{equation}
The quantity appearing on the right-hand side, $s_{12}s_{45}/(s_{123}s_{345})$, is a conformally-invariant cross ratio~\cite{Drummond:2007au}. In this way we see that each of the unphysical singularity surfaces of the rational terms in eqs.~(\ref{eq:nmhvmmmppp})--(\ref{eq:nmhvmpmpmp}) corresponds to one of the following cross ratios equalling unity,
{
\everymath={\displaystyle}
\begin{equation}
	\begin{array}{ccccccc}
		u_A & = & \frac{s_{12}s_{45}}{s_{123}s_{345}} & \toNMRK &
		\frac{s_{45}}{(p_4^++p_5^+)(p_4^-+p_5^-)}
		&=& \frac{|w+Xz|^2}{(1+X)(|w|^2+X|z|^2)} \,,\\
		v_A & = & \frac{s_{23}s_{56}}{s_{234}s_{123}} & \toNMRK & 
		\frac{(-t_1)p_5^+}{(|q_{2\perp}|^2+p_4^-p_5^+)(p_4^++p_5^+)}
		&=& \frac{X|1-z|^2}{(1+X|z|^2)(1+X)} \,,\\
		w_A & = & \frac{s_{34}s_{61}}{s_{345}s_{234}} & \toNMRK &
		\frac{(-t_3)p_4^-}{(p_4^-+p_5^-)(|q_{2\perp}|^2+p_4^-p_5^+)}
		&=& \frac{X|1-w|^2|z|^2}{(|w|^2+X|z|^2)(1+X|z|^2)}
		\,.
	\end{array}
	\label{eq:uvwA}
\end{equation}
}

We use the label $A$ to indicate that these are the cross ratios of the colour ordering $\{1,2,3,4,5,6\}$, which we will refer to as $\sigma_A$. In the NMRK region these cross ratios lie in the range $[0,1]$, and are studied further in appendix~\ref{sec:crossratio}. 
We choose to label the unphysical singularities of the BCFW representations by the cross ratio which tends to unity as this unphysical singularity surface is approached,
\begin{equation}
P_x^A=0 \quad\Rightarrow\quad x_A=1, \qquad x \in \{u,v,w\}.
\end{equation}
This notation will be helpful in section~\ref{sec:N4opp-hel} for demonstrating that the NMHV one-loop six-gluon amplitudes in $\cN=4$ are free from unphysical singularities.
We use the term $P_x^A$ to generically refer to the unphysical-singularity surfaces of the terms in the BCFW representation of the amplitudes. We list these unphysical-singularity surfaces in NMRK in terms of both light-cone coordinates and the minimal variables,
\begin{equation}
	\begin{array}{rrrrrrr}
		\langle 1 |2+3|4]=0 \quad& \ \leftrightarrow\quad & P^A_{v}=0 \quad& \ \leftrightarrow\quad & q_{2\perp}^*-p_{4\perp}^*\displaystyle{\frac{p_5^+}{p_4^+}}=0 \quad&\ \leftrightarrow\quad  & 1+X\bar{z}=0 \,,\\
		\langle 3 |4+5|6]=0 \quad& \leftrightarrow\quad & P^A_{u}=0 \quad& \leftrightarrow\quad & p_{4\perp}+p_{5\perp}=0\quad & \leftrightarrow \quad & w-z=0 \,,\\
		\langle 5 |6+1|2]=0 \quad& \leftrightarrow\quad & P^A_{w}=0 \quad& \leftrightarrow\quad & q_{2\perp}^*+p_{5\perp}^*\displaystyle{\frac{p_4^-}{p_5^-}}=0\quad & \leftrightarrow \quad & w+X|z|^2=0 \,,\\
	\end{array}
\label{eq:unphys}
\end{equation}
which provides intuition for the origin and nature of the unphysical singularities that appear in \eqns{kosc2}{kosc3} below.  
Equations~(\ref{eq:BCFWstringsA}) and (\ref{eq:BCFWstringsAmin}) list the full expressions for the NMRK limit of these spinor strings in lightcone and minimal coordinates respectively, from which the unphysical singularity surfaces of \eqn{eq:unphys} can be identified.

It is interesting to note that a given BCFW representation of a tree amplitude can give rise to different representations of the opposite-helicity two-gluon CEV. As an example, we can obtain a representation for this CEV from \eqn{eq:BCFWexample} by taking the central NMRK limit of this amplitude and dividing by impact factors and $t$-channel propagators according in \eqn{NLOfactorization2}. In this way we obtain
\begin{align}
	\begin{split}
		\left. A^{gg(0)}(q_1, p^\oplus_4, p^\ominus_5, q_3)\right|_2 = R^A_{uv}+R^A_{vw}+R^A_{wu}\, ,
	\end{split}
	\label{kosc2}
\end{align}
with
\begin{align}
	\begin{split}
		R^A_{uv}&=\frac{X^3|w-1|^2(\bar{z}-1)|z|^2}{(X+1)(\bar{w}+X \bar{z})(w-z) (1+X\bar{z}) }\,,\\
		R^A_{vw}&=\frac{zX(w-1)  (z-1) }{(1+X |z|^2)(1+X\bar{z})(w+X|z|^2)}\,, \\
		R^A_{wu}&=-\frac{z^3X^3(\bar{w}-1)  |z-1|^2  |z|^2}{(w+X z)(|w|^2+X |z|^2)(w+X|z|^2)(w-z)}\,.
	\end{split}
	\label{eq:RA}
\end{align}
Eq.~(\ref{kosc2}) is our second representation for the opposite helicity vertex, equivalent to eq.~(\ref{eq:koscmin}) above.
In addition to the physical singularities of \eqn{eq:physsing} and \eqn{eq:physsing2}, individual terms in eq.~(\ref{kosc2}) feature the unphysical singularities of \eqn{eq:unphys}. We denote the rational terms in (\ref{eq:RA}) by their singularities, where $R^A_{xy}$ features unphysical poles at $P_x^A=0$ and $P_y^A=0$. 

We can obtain another equivalent representation for the opposite-helicity vertex from \eqn{eq:BCFWexample} by taking the central NMRK limit of
\[
M^{(0)}_{(\ref{eq:BCFWexample})}(p_3^\oplus, p_4^\ominus,p_5^\oplus,p_6^\oplus,p_1^\ominus,p_2^\ominus)^*.
\]
This gives us our third representation of the CEV,
\begin{align}
	\begin{split}
		\left. A^{gg(0)}(q_1, p^\oplus_4, p^\ominus_5, q_3)\right|_3 = R^A_{\bar{u}\bar{v}}+R^A_{\bar{v}\bar{w}}+R^A_{\bar{w}\bar{u}}\, ,
	\end{split}
	\label{kosc3}
\end{align}
with
\begin{align}
	\begin{split}
	R^A_{\bar{u}\bar{v}}&=\frac{X |w-1|^2  (z-1) |z|^2}{(X+1)(w+X z) (\bar{w}-\bar{z}) (1+X z) }\,,\\
	R^A_{\bar{v}\bar{w}}&=\frac{X^3(\bar{w}-1)   (\bar{z}-1)z^4 \bar{z}}{ (1+X |z|^2)(1+Xz) (\bar{w}+X|z|^2)}\,,\\
	R^A_{\bar{w}\bar{u}}&=-\frac{Xz(w-1) \bar{w}^4  |z-1|^2  }{ (\bar{w}+X \bar{z})(|w|^2+X |z|^2) (\bar{w}+X|z|^2)(\bar{w}-\bar{z}) } \,.
\end{split}
\label{eq:RAb}
\end{align}
The notation $R^A_{\bar{x}\bar{y}}$ indicates these rational terms have unphysical singularities at $\bar{P}_x^A=0$ and $\bar{P}_y^A=0$, where as usual, we use bar to denote complex conjugation. 
Of course, these six rational terms are not independent due to the relation,
\begin{equation}
R^A_{uv}+R^A_{vw}+R^A_{wu}=R^A_{\bar{u}\bar{v}}+R^A_{\bar{v}\bar{w}}+R^A_{\bar{w}\bar{u}}.
\end{equation}
These rational terms, when divided by the appropriate MHV central-emission vertex, correspond to the NMRK limit of the dual-conformal-invariant $R$-invariants of $\mathcal{N}=4$ SYM~\cite{Drummond:2008vq}. 
In this paper however, we do not normalise by the MHV vertex as this will avoid the introduction of the MHV phase, and will lead to simpler expressions when we take further kinematic limits of these rational terms.

Having studied the physical and spurious poles of the BCFW representation of the opposite-helicity CEV, we briefly return to the CSW representation of the vertex, \eqn{eq:AggpmCSW}. 
We write this in terms of the minimal variables in~\eqn{eq:AggpmCSWmin}. We can now see that this representation has the expected physical singularities of~\eqns{eq:physsing}{eq:physsing2}, and also spurious poles at $P_{u}^A=0$ and $\bar{P}_{v}^A=0$. 
In addition to these representations, one can find a compact representation of the central-emission vertex, \eqn{eq:take5}, which only has a single spurious pole at $P_{u}^A=0$.
We have checked analytically that the five representations of the opposite-helicity vertex in
eqs.~(\ref{eq:koscmin}), (\ref{kosc2}), (\ref{kosc3}), (\ref{eq:AggpmCSWmin}) and (\ref{eq:take5}) are all equivalent.
The equivalence of these expressions shows that $P_x^A=0$ are not true singularities of the central-emission vertex, and we say that they are spurious, or removable poles of this expression. 

\subsection{Other colour orderings}
\label{sec:other}
\begin{figure}[htb]
\centering
    \subfigure[]{\label{fig:colour_connections_A}\includegraphics[scale=0.15]{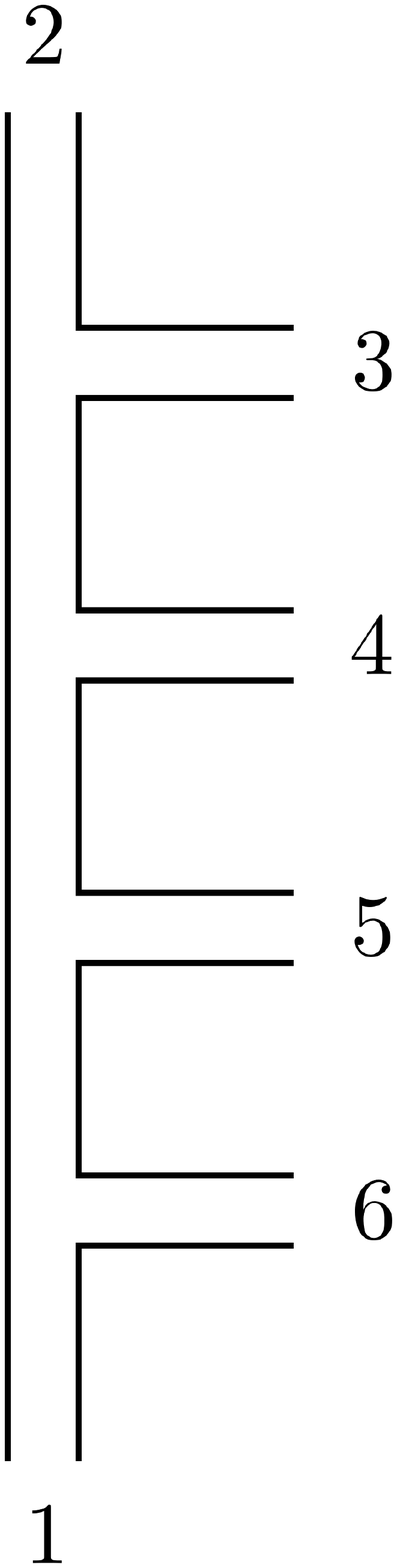}}
    \hspace*{20pt}
    \subfigure[]{\label{fig:colour_connections_Ax}\includegraphics[scale=0.15]{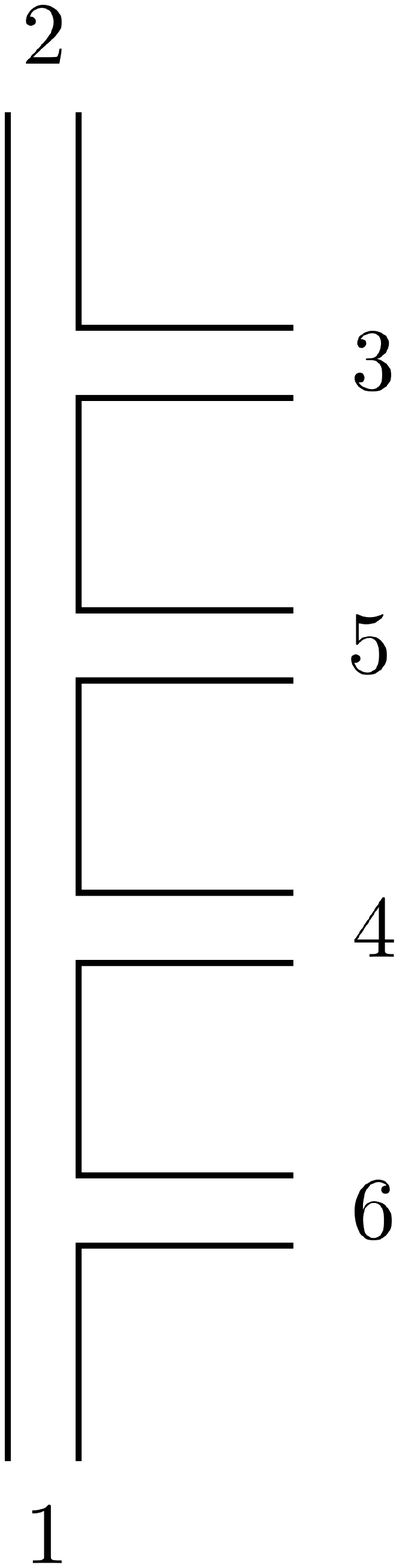}}
    \hspace*{20pt}
    \subfigure[]{\label{fig:colour_connections_B}\includegraphics[scale=0.15]{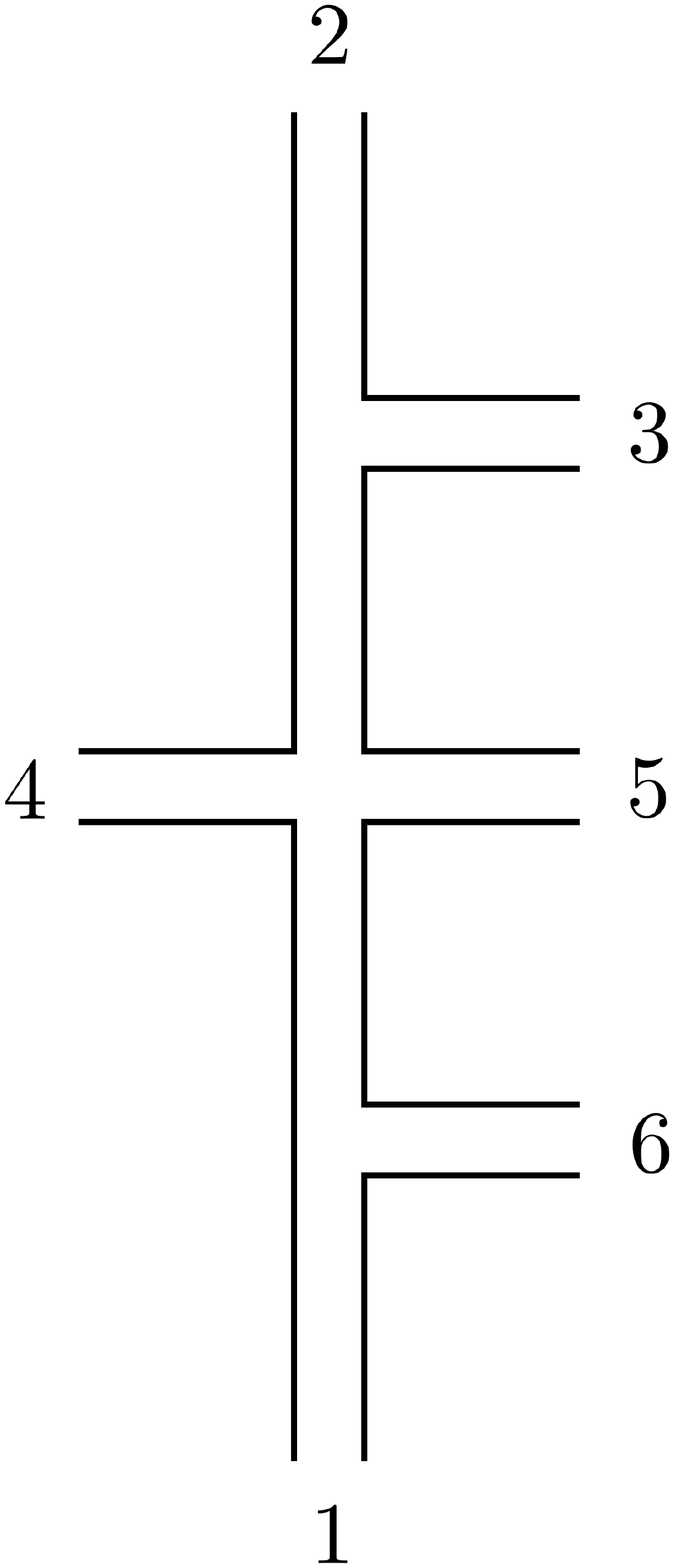}}
    \caption{The colour connections of the three representative colour orderings we discuss in this paper, displayed on a two-sided rapidity plot: (a) $\sigma_A$;  (b) $\sigma_{A'}$; (c) $\sigma_B$, which uses 't Hooft graphical notation, with untwisted colour lines. Modulo cyclic permutations of the colour ordering, the gluons which fall between incoming gluons 2 and 1 are placed on the ``front'' side of the plot, while the gluons which fall between 1 and 2 are placed on the ``back'' side.
    In MRK, the gluons are ordered in rapidity both on the front and on the back sides, however the relative ordering between the two sides is immaterial~\cite{DelDuca:1993pp,DelDuca:1995zy}. On the front side, colour and rapidity have the same ordering; on the back, the opposite ordering. In NMRK, the strict orderings may be relaxed on either side as appropriate~\cite{DelDuca:1995ki}.}
    \label{fig:colour_connections}
\end{figure}
So far we have only investigated a single colour ordering, which we termed $\sigma_A$. All colour orderings were studied in NMRK at tree-level in ref.~\cite{DelDuca:1995ki}. There the leading behaviour was found to be given by colour orderings that have $p_1$ adjacent to $p_6$ and $p_2$ adjacent to $p_3$. An interchange of $p_1\leftrightarrow p_6$ or $p_2\leftrightarrow p_3$ leads only to a change in the sign of the amplitude. In addition to $\sigma_A$ we should therefore consider the representative orderings $\sigma_{A'}=\{1,2,3,5,4,6\}$ and $\sigma_B=\{1,4,2,3,5,6\}$, which are displayed in fig.~\ref{fig:colour_connections} using the two-sided rapidity plot~\cite{DelDuca:1993pp,DelDuca:1995zy}\footnote{An earlier version of the two-sided rapidity plot, for which though colour was absent, was used in multiperipheral and dual models~\cite{Weis:1971yxx,Veneziano:1973nw}.}. The ordering $\sigma_{A'}$, is related to the former by a simple interchange of $p_{4}\leftrightarrow p_5$, which we may directly apply to the functions in~\eqn{kosc}. In terms of the minimal variables this operation is given by
\begin{align}
	\begin{split}
		z&\xrightarrow[p_{4}\leftrightarrow p_5]{} 1+w\frac{1- z}{z},\\
		w&\xrightarrow[p_{4}\leftrightarrow p_5]{} 1+z\frac{1-w}{w},\\
		X&\xrightarrow[p_{4}\leftrightarrow p_5]{} \frac{1}{X}.
	\end{split}
	\label{eq:AtoAx}
\end{align}
Throughout this paper we will use a prime to refer to an interchange of $p_4 \leftrightarrow p_5$, for example
\begin{equation}
q_2\equiv q_1-p_4=q_3+p_5, \qquad\text{so}\qquad    q_2'\equiv q_1-p_5=q_3+p_4,
\label{q2prime}
\end{equation}
which is the relevant $t$-channel momentum for the $\MRKx$ limit. For example, the equal-helicity tree-level CEV in the minimal variables is given by
\begin{equation}
		A^{gg(0)}(q_1, p^\oplus_5, p^\oplus_4, q_3) = \frac{q_{1\perp}^*}{q_{1\perp}}\frac{w(w-1)(z-1)}{(w+Xz)}\,.
		\label{eq:koscmin_same-hel_Aprime}
\end{equation}
For completeness, we list the full analysis of the BCFW representations of the opposite-helicity vertex for the ordering $\sigma_{A'}$ in appendix \ref{sec:BCFWAx}. 

In the central NMRK limit, the colour-ordered amplitude of the remaining ordering $\sigma_B$ similarly factorises as
\begin{equation}
\label{NLOfactorizationB}    
 M^{(0)}_{6g}(p_1^{\nu_1}, p_4^{\nu_4}, p^{\nu_2}_2, p^{\nu_3}_3, p_5^{\nu_5}, p_6^{\nu_6}) 
 = s\, C^{g(0)}(p_2^{\nu_2}, p_3^{\nu_3}) \frac1{t_1} B^{gg(0)}(q_1, p^{\nu_4}_4, p^{\nu_5}_5, q_3) \frac1{t_3} C^{g(0)}(p_1^{\nu_1}, p_6^{\nu_6}) \,,
\end{equation}
However, the tree-level CEV for the ordering $\sigma_B$ is not independent to that of the orderings $\sigma_A$ and $\sigma_{A'}$ due to the relationship~\cite{DelDuca:1995ki},
\begin{align}
	A^{gg(0)}(q_1, p^{\nu_4}_4, p^{\nu_5}_5, q_3)+A^{gg(0)}(q_1, p^{\nu_5}_5, p^{\nu_4}_4, q_3)+B^{gg(0)}(q_1, p^{\nu_4}_4, p^{\nu_5}_5, q_3)=0\,.
	\label{eq:ABrelation}
\end{align}
This relation follows~\cite{DelDuca:1999iql} from the U(1) decoupling equation~\cite{Berends:1988zn,Bern:1990ux}, or as the leading contribution in NMRK to a Kleiss-Kuijf relation~\cite{Kleiss:1988ne} 
which relates colour-ordered amplitudes at tree level~\cite{DelDuca:1999rs}.
Eq.~(\ref{eq:ABrelation}) is simple to verify in the equal-helicity case using \eqns{eq:koscmin_same-hel}{eq:koscmin_same-hel_Aprime}, and
\begin{equation}
	B^{gg(0)}(q_1, p^\oplus_4, p^\oplus_5, q_3) = \frac{q_{1\perp}^*q_{3\perp}}{p_{4\perp}p_{5\perp}}=-\frac{q_{1\perp}^*}{q_{1\perp}}(w-1)(z-1).
	\label{eq:B0pp}
\end{equation}

To demonstrate the relationship given in \eqn{eq:ABrelation} for the opposity-helicity case, it will be easier to first obtain a representation of $B^{gg(0)}$ in the minimal variables. 
At tree level, there are therefore only two independent colour orderings in NMRK, as in \eqn{NLOfactorization}. However, in the one-loop MHV amplitudes of $\cN=4$ SYM, the tree-level amplitudes for $\sigma_A$, $\sigma_{A'}$ and $\sigma_B$ will all be multiplied by distinct transcendental functions, so these colour orderings are independent beyond tree level. Furthermore, in the one-loop NMHV amplitudes for $\cN=4$ SYM, we encounter rational coefficients of transcendental functions which are not simply tree-level amplitudes. However, as mentioned in the outset of section~\ref{sec:tcev}, these rational coefficients consist of the individual terms in the BCFW representations of the tree amplitudes. For this reason, we investigate the BCFW-based representation of the tree-level amplitudes for the $\sigma_B$ ordering in this section, with an emphasis on the individual rational terms.

We now perform a study analogous to section~\ref{sec:bcfw} for the ordering $\sigma_B$. We first list the cross ratios for this colour ordering, 
\begin{alignat}{7}
& u_B 
&\,\,=\,\, 
& \frac{s_{14}s_{35}}{s_{142}s_{235}}
& \,\,=\,\,
& \frac{p_4^+p_5^-}{|q'_{2\perp}|^2+p_4^+p_5^-}
& \,\,=\,\,
& \frac{X|z|^2}{|w+z-wz|^2+X|z|^2}\nn
\\
& v_B 
& \,\,=\,\, 
& \frac{s_{42}s_{56}}{s_{423}s_{142}} 
& \,\,=\,\,
& \frac{p_4^-p_5^+}{|q_{2\perp}|^2+p_4^-p_5^+}
& \,\,=\,\,
& \frac{1}{1+X|z|^2}\label{eq:uvwB}
\\
& w_B
& \,\,=\,\,
& \frac{s_{23}s_{61}}{s_{235}s_{423}}
& \,\,=\,\,
& \frac{(-t_1)(-t_3)}{(|q_{2\perp}|^2+p_4^-p_5^+)(|q'_{2\perp}|^2+p_4^+p_5^-)}
& \,\,=\,\,
&  \frac{|w-1|^2|z-1|^2X|z|^2}{(1+X|z|^2)(|w+z-wz|^2+X|z|^2)}
\nn
\end{alignat}
which, like \eqn{eq:uvwA}, lie in the range $[0,1]$ for our physical kinematic region.
Returning to our example amplitude, the NMRK limit of $M^{(0)}_{\text{(\ref{eq:BCFWexample})}}(p_1^\oplus,p_4^\oplus,p_2^\ominus,p_3^\oplus,p_5^\ominus,p_6^\ominus)$ yields the representation,
\begin{align}
	\begin{split}
		B^{gg(0)}(q_1, p^\oplus_4,p^\ominus_5,  q_3) = R^{B}_{uv}+R^{B}_{vw}+R^{B}_{wu}\, ,
	\end{split}
	\label{eq:Bgg0min}
\end{align}
with
\begin{align}
\label{eq:RB}
	\begin{split}
	R^{B}_{uv}&=\frac{z|w-1|^2 |z-1|^2  }{\bar{z} (w+z-wz)}\,,\\
	R^{B}_{vw}&=-\frac{z(w-1) (z-1) }{\bar{z} (1+X|z|^2) (w+z-wz+X |z|^2)}\,,\\
	R^{B}_{wu}&=-\frac{(\bar{w}-1) X^2 z^3 (\bar{z}-1) \bar{z}}{(w+z-wz) (w+z-wz+X |z|^2)(|w+z-wz|^2+X|z|^2)}\,,
	\end{split}
\end{align}
while the NMRK limit of $M^{(0)}_{{\text{(\ref{eq:nmhvmmmppp})}}}(p_1^\ominus,p_4^\ominus,p_2^\ominus,p_3^\oplus,p_5^\oplus,p_6^\oplus)$ yields the representation,
\begin{align}
	\begin{split}
		B^{gg(0)}(q_1, p^\oplus_4,p^\ominus_5,  q_3) = R^{B}_{\bar{u}\bar{v}}+R^{B}_{\bar{v}\bar{w}}+R^{B}_{\bar{w}\bar{u}}\, ,
	\end{split}
	\label{eq:Bgg0min2}
\end{align}
with
\begin{align}
	\begin{split}
		R^{B}_{\bar{u}\bar{v}}&=0 \,,\\
		R^{B}_{\bar{v}\bar{w}}&=-\frac{(\bar{w}-1)(\bar{z}-1)|z|^2 z^2 X^2  }{(1+X|z|^2) (\bar{w}+\bar{z}-\bar{w} \bar{z}+X |z|^2)} \,,\\
		R^{B}_{\bar{w}\bar{u}}&=-\frac{z(w-1) (z-1)  (\bar{w}+\bar{z}- \bar{w}\bar{z})^3}{\bar{z} (\bar{w}+\bar{z}-\bar{w} \bar{z}+X |z|^2) (|w+z-wz|^2+X|z|^2)} \,.
	\end{split}
	\label{eq:RBb}
\end{align}
One can now verify the relation \eqn{eq:ABrelation} using the representations in \eqn{eq:Bgg0min} or \eqn{eq:Bgg0min2} along with those obtained in section~\ref{sec:bcfw}.

Recalling \eqn{eq:s3nmrk} and \eqn{eq:s3nmrkmin}, the physical-singularity surfaces of this ordering are 
\begin{equation}
	\begin{array}{rrrrrrr}
		s_{142} = 0  \quad& \phantom{m}\leftrightarrow \quad& p_3^+p_6^-=0 \,,
		\quad&  &  \\
		s_{234} = 0 \quad & \leftrightarrow \quad& |q_{2\perp}|^2+p_4^-p_5^+=0
		\quad& \phantom{m}\leftrightarrow\quad & 1+X |z|^2 = 0 \,,\\
		s_{235} = 0 \quad & \leftrightarrow \quad & |q_{2\perp}^\prime|^2+p_4^+p_5^-=0  \quad
		& \leftrightarrow \quad& |w+z-wz|^2+X|z|^2 = 0 \,,
	\end{array}
	\label{eq:physsingB}
\end{equation}
where we note that there is no singularity in the collinear limit, eq.~(\ref{eq:physsing}). In the NMRK limit $s_{142}\sim s_{12}$ which does not depend on the central degrees of freedom of the process. We therefore have not listed an entry for this Mandelstam invariant in terms of the minimal variables. 
With reference to \eqns{eq:BCFWstringsB}{eq:BCFWstringsBmin}, the unphysical-singularity surfaces are\footnote{In the minimal coordinates, the limit $q_{2\perp}^*\to0$ is equivalent to taking $\bar{w}$ and $\bar{z}$ to zero at the same rate. The ratio $\bar{w}/\bar{z}$ is well defined in this limit. }
\begin{equation}
	\begin{array}{rrrrrrr}
		\langle 1 |4+2|3]=0 & \,\,\,\leftrightarrow\,\,\, & P^{B}_{v}=0 & \,\,\, \leftrightarrow\,\,\, & q_{2\perp}^*=0\,, & \,\,\,\,\,\,  & \\
		\langle 2 |3+5|6]=0 & \,\,\,\leftrightarrow\,\,\, & P^{B}_{u}=0 & \,\,\,\leftrightarrow\,\,\, & q'_{2\perp}=0 & \,\,\,\leftrightarrow\,\,\, & w+z-wz=0 \,,\\
		\langle 5 |6+1|4]=0 & \,\,\,\leftrightarrow\,\,\, & P^{B}_{w}=0 & \,\,\,\leftrightarrow\,\,\, &
			{\displaystyle \frac{q_{2\perp}^*}{p_{4\perp}^*}}
			+	{\displaystyle\frac{q_{2\perp}'}{p_{5\perp}} \frac{p_5^+}{p_4^+}}=0
		& \,\,\,\leftrightarrow\,\,\, & w+z-wz+X |z|^2=0 \,.\\
	\end{array}
	\label{eq:unphysB}
\end{equation}
We note that under the transformation in eq.~(\ref{eq:AtoAx}) one obtains
\begin{align}
u_B{\underset{p_4\leftrightarrow p_5}\longleftrightarrow}v_B\,, \qquad w_B{\underset{p_4\leftrightarrow p_5}\longleftrightarrow}w_B \,, \label{eq:uvwB_swap45}
\end{align}
and
\begin{align}
&R^B_{uv}{\underset{p_4\leftrightarrow p_5}\longleftrightarrow}(R^B_{uv})^*,
&R^B_{vw}{\underset{p_4\leftrightarrow p_5}\longleftrightarrow}\left(R^{B}_{wu}\right)^*,
\nonumber
\\
&R^B_{\bar{u}\bar{v}}{\underset{p_4\leftrightarrow p_5}\longleftrightarrow}(R^B_{\bar{u}\bar{v}})^*=0,
&R^B_{\bar{v}\bar{w}}{\underset{p_4\leftrightarrow p_5}\longleftrightarrow}\left(R^{B}_{\bar{w}\bar{u}}\right)^*.
\label{eq:RBswap45}
\end{align}
These properties will be useful for organising the one-loop 
opposite-helicity amplitude in section \ref{sec:N4opp-hel}.
\subsection{The target-projectile exchange}
\label{sec:target-projectile}

The amplitude in \eqn{NLOfactorization} is manifestly symmetric under the target-projectile exchange.
The manifest symmetry is broken on the right-hand side of \eqn{NLOfactorization2} by the choice of impact factors in \eqn{centrc}, which is dictated by the momentum representation, \eqn{spro},
of the spinor products. However, in transverse-momentum space the impact factors in (\ref{centrc}) are just phases, and hence the central-emission vertex for the emission of two gluons is invariant under target-projectile exchange up to an overall phase. We may consider then the target-projectile map $F$, which interchanges
\beqa
	p_1\underset{F}{\leftrightarrow} p_2 \,, \qquad
	p_3\underset{F}{\leftrightarrow} p_6\,, \qquad
	p_4\underset{F}{\leftrightarrow} p_5 \,.\label{eq:target-projectile}
\eeqa
In NMRK the target-projectile map acts as
\beqa
	q_{1\perp}\underset{F}{\leftrightarrow} -q_{3\perp} \,, \qquad
	p_{4\perp}\underset{F}{\leftrightarrow} p_{5\perp}\,, \qquad
	p_{4}^\pm\underset{F}{\leftrightarrow} p_{5}^\mp \,.
\eeqa
In terms of the minimal set $\{w,z,X\}$ these flip relations can be expressed as
\beqa
	w\underset{F}{\leftrightarrow} z \,, \qquad
	X\underset{F}{\mapsto} \frac{|z|^2}{|w|^2}X \,.
	\label{fmapwzx}
\eeqa
Under these flips, the cross ratios for the $\sigma_A$ ordering, \eqn{eq:uvwA}, change as
\begin{equation}
    u_A\underset{F}{\leftrightarrow}u_A\,,
    \qquad v_A\underset{F}{\leftrightarrow}w_A\,,
\label{eq:uAF}
\end{equation}
while the spurious poles change as
\beqa
	w+X|z|^2 \xrightarrow[F]{} z (1+X\bar{z})\,, \qquad
	w-z \xrightarrow[F]{}  -(w-z)\,, \qquad
	1+X\bar{z} \xrightarrow[F]{} \frac{w+X|z|^2}{w} \,, 
\eeqa
i.e.~the first and third spurious-pole surfaces in \eqn{eq:unphys} are swapped. Furthermore, the rational coefficients~(\ref{eq:RA}) change as
\begin{align}
	R^A_{uv}&\xrightarrow[F]{}\frac{w\bar{z}}{\bar{w}z}R^A_{wu}\,, \qquad
	R^A_{\bar{u}\bar{v}}\xrightarrow[F]{}\frac{w\bar{z}}{\bar{w}z}R^A_{\bar{w}\bar{u}}\,, \nn\\
	R^A_{vw}&\xrightarrow[F]{}\frac{w\bar{z}}{\bar{w}z}R^A_{vw}\,, \qquad
	R^A_{\bar{v}\bar{w}}\xrightarrow[F]{}\frac{w\bar{z}}{\bar{w}z}R^A_{\bar{v}\bar{w}}\,, \label{eq:rflip}\\
	R^A_{wu}&\xrightarrow[F]{}\frac{w\bar{z}}{\bar{w}z}R^A_{uv}\,, \qquad
	R^A_{\bar{w}\bar{u}}\xrightarrow[F]{}\frac{w\bar{z}}{\bar{w}z}R^A_{\bar{u}\bar{v}}\,, \nn
\end{align}
from which we see that both eqs.~(\ref{kosc2}) and (\ref{kosc3}) are invariant under the target-projectile map $F$, up to an overall phase.
By contrast, for the $\sigma_B$ ordering we find that each cross ratio (\eqn{eq:uvwB}) is invariant under this map,
\begin{equation}
    u_B\underset{F}{\leftrightarrow}u_B\,,
    \qquad v_B\underset{F}{\leftrightarrow}v_B\,,
    \qquad w_B\underset{F}{\leftrightarrow}w_B\,,
    \label{eq:ubF}
\end{equation}
the spurious singularity surfaces~(\ref{eq:unphysB}) are likewise invariant, while each rational term in \eqns{eq:Bgg0min}{eq:Bgg0min2} is invariant up to an overall phase,
\begin{align}
	R^B_{xy}&\xrightarrow[F]{}\frac{w\bar{z}}{\bar{w}z}R^B_{xy}\,, \qquad
	R^B_{\bar{x}\bar{y}}\xrightarrow[F]{}\frac{w\bar{z}}{\bar{w}z}R^B_{\bar{x}\bar{y}}\,, \qquad x,y \in \{u,v,w\}, x \neq y\,. \nn
\end{align}

Having investigated the tree-level CEV, and in particular the rational terms $R_{xy}^I$ which comprise the BCFW representations of this CEV, we proceed to the study of this vertex at one loop.

\section{The two-gluon central-emission vertex at one loop}
\label{sec:6g1lamp}

In this section we will extract the two-gluon central-emission vertex at one loop in $\cN=4$ SYM. The plan is analogous to the extraction of the tree-level vertex in section~\ref{sec:tcev}. We will begin with the one-loop amplitude in general kinematics, and find the leading behaviour in the central NMRK limit defined in appendix~\ref{sec:appd}. We expect that in this limit, the amplitude will factorise into the known one-loop impact factors and Regge trajectories associated with the two large rapidity spans, and a new piece which depends only on the central degrees of freedom, i.e.~the gluons 4 and 5 and the transverse $t$-channel momenta. We will identify this new central piece with the one-loop CEV. 

The study of the two-gluon CEV in $\cN=4$ SYM is not only of interest in its own right; via the supersymmetric decomposition of QCD amplitudes at one loop \cite{Bern:1993mq,Bern:1994zx,Bern:1994cg}, it is also an ingredient for the two-gluon CEV in QCD. As such, we will begin by clarifying the connection between the two. 
 
The colour decomposition we use for the QCD one-loop six-gluon amplitude is~\cite{DelDuca:1999rs},
\beqa
	\mathcal{M}^{(1)}_{\QCD} &=& \gs^6 \sum_{\sigma \in S_{5}/\mathcal{R}}\Big[\tr\left(F^{a_{\sigma_1}}\cdots F^{a_{\sigma_6}}\right)M^{(1)[\mathbf{8}]}_{\QCD}\left(p_{\sigma_1}^{\nu_{\sigma_1}},\dots,p_{\sigma_6}^{\nu_{\sigma_6}}\right) \nn \\
	&&\qquad\qquad + N_f\tr\left(T^{a_{\sigma_1}}\cdots T^{a_{\sigma_6}}\right)M^{(1)[\mathbf{3}]}_{\QCD}\left(p_{\sigma_1}^{\nu_{\sigma_1}},\dots,p_{\sigma_6}^{\nu_{\sigma_6}}\right)\Big]\,,
	\label{eq:1loopDDM}
\eeqa
where $S_5=S_6/\mathbb{Z}_6$ is the group of non-cyclic permutations and $\mathcal{R}$ is the total reflection, e.g. $\mathcal{R}(1,2,3,4,5,6)=(6,5,4,3,2,1)$. The superscript $[\mathbf{R}]$ on colour-ordered amplitudes denotes the colour representation of the particle circulating in the loop. We remind the reader that $F^{a_i}$ and $T^{a_i}$ are generators in the adjoint and fundamental representations respectively. For convenience we define separate colour-dressed amplitudes for the gluon-loop and quark-loop contributions,
\begin{align}
	\mathcal{M}^{(1)[\mathbf{8}]}_{\QCD}& =
	\gs^6 \sum_{\sigma \in S_{5}/\mathcal{R}}\tr\left(F^{a_{\sigma_1}}\cdots F^{a_{\sigma_6}}\right)M^{(1)[\mathbf{8}]}_{\QCD}\left(p_{\sigma_1}^{\nu_{\sigma_1}},\dots,p_{\sigma_6}^{\nu_{\sigma_6}}\right) \,,\\
	\mathcal{M}^{(1)[\mathbf{3}]}_{\QCD}& =
	\gs^6 \sum_{\sigma \in S_{5}/\mathcal{R}}\tr\left(T^{a_{\sigma_1}}\cdots T^{a_{\sigma_6}}\right)M^{(1)[\mathbf{3}]}_{\QCD}\left(p_{\sigma_1}^{\nu_{\sigma_1}},\dots,p_{\sigma_6}^{\nu_{\sigma_6}}\right) \,,
	\label{eq:1loopDDMs}
\end{align}
such that the QCD amplitude of eq.~(\ref{eq:1loopDDM}), with $N_f$ Dirac fermions in the fundamental representation, is obtained by
\begin{align}
	\mathcal{M}^{(1)}_{\QCD}=\mathcal{M}^{(1)[\mathbf{8}]}_{\QCD}+N_f \mathcal{M}^{(1)[\mathbf{3}]}_{\QCD}\,.
\end{align}

Colour-ordered gluon amplitudes in QCD at one loop may be further decomposed into supersymmetric multiplets circulating in the loop,
\begin{subequations}
\begin{align}
	M^{(1)[\mathbf{8}]}_{\QCD}&=M^{(1)}_{\mathcal{N}=4}-4M^{(1)}_{\mathcal{N}=1 \chi}+M^{(1)}_{\mathrm{scalar}} \,,
	\label{MQCD8_decom}
	\\
	M^{(1)[\mathbf{3}]}_{\QCD}&=M^{(1)}_{\mathcal{N}=1 \chi}-M^{(1)}_{\mathrm{scalar}} \label{MQCD3_decom}\,.
\end{align}
\end{subequations}
To bring this supersymmetric decomposition to the level of colour-dressed QCD amplitudes we define the colour-dressed contribution from the $\mathcal{N}=4$ multiplet,
\beq
	\mathcal{M}^{(1)}_{\mathcal{N}=4} = \gs^6 \sum_{\sigma \in S_{5}/\mathcal{R}}\tr\left(F^{a_{\sigma_1}}\cdots F^{a_{\sigma_6}}\right)M^{(1)}_{\mathcal{N}=4}\left(p_{\sigma_1}^{\nu_{\sigma_1}},\dots,p_{\sigma_6}^{\nu_{\sigma_6}}\right) \,.
	\label{eq:N41loop}
\eeq
For the colour-dressed contributions from the $\mathcal{N}=1$ multiplet and the scalar circulating in the loop, we separately consider the cases where these are in the adjoint or fundamental representations,
\begin{subequations}
\begin{align}
	\mathcal{M}^{(1)[\mathbf{8}]}_{\mathcal{N}=1\chi}&= \gs^6\, \sum_{\sigma \in S_{5}/\mathcal{R}}\tr\left(F^{a_{\sigma_1}}\cdots F^{a_{\sigma_6}}\right)M^{(1)}_{\mathcal{N}=1\chi}
	\left(p_{\sigma_1}^{\nu_{\sigma_1}},\dots,p_{\sigma_6}^{\nu_{\sigma_6}}\right)
	\label{eq:N11loopAdj}\,,\\
	\mathcal{M}^{(1)[\mathbf{3}]}_{\mathcal{N}=1\chi}&= \gs^6\, \sum_{\sigma \in S_{5}/\mathcal{R}}\tr\left(T^{a_{\sigma_1}}\cdots T^{a_{\sigma_6}}\right)M^{(1)}_{\mathcal{N}=1\chi}
	\left(p_{\sigma_1}^{\nu_{\sigma_1}},\dots,p_{\sigma_6}^{\nu_{\sigma_6}}\right)
	\label{eq:N11loopFun}\,,\\
	\mathcal{M}^{(1)[\mathbf{8}]}_{\mathrm{scalar}}&= \gs^6\, \sum_{\sigma \in S_{5}/\mathcal{R}}\tr\left(F^{a_{\sigma_1}}\cdots F^{a_{\sigma_6}}\right)M^{(1)}_{\mathrm{scalar}}
	\left(p_{\sigma_1}^{\nu_{\sigma_1}},\dots,p_{\sigma_6}^{\nu_{\sigma_6}}\right)
	\label{eq:scalar1loopAdj}\,,\\
	\mathcal{M}^{(1)[\mathbf{3}]}_{\mathrm{scalar}}&= \gs^6\, \sum_{\sigma \in S_{5}/\mathcal{R}}\tr\left(T^{a_{\sigma_1}}\cdots T^{a_{\sigma_6}}\right)M^{(1)}_{\mathrm{scalar}}
	\left(p_{\sigma_1}^{\nu_{\sigma_1}},\dots,p_{\sigma_6}^{\nu_{\sigma_6}}\right)
	\label{eq:scalar1loopFun}\,,
\end{align}
\end{subequations}
such that we can write the QCD one-loop colour-dressed  amplitude as
\begin{align}
	\mathcal{M}^{(1)}_{\QCD}=\left(\mathcal{M}^{(1)}_{\mathcal{N}=4}-4\mathcal{M}^{(1)[\mathbf{8}]}_{\mathcal{N}=1 \chi}+\mathcal{M}^{(1)[\mathbf{8}]}_{\mathrm{scalar}}\right)+N_f \left(\mathcal{M}^{(1)[\mathbf{3}]}_{\mathcal{N}=1 \chi}-\mathcal{M}^{(1)[\mathbf{3}]}_{\mathrm{scalar}}\right).
	\label{calMQCD_decomposed}
\end{align}
At one loop, this decomposition holds also at the level of the Regge-factorised expressions. For example, we can write the one-loop coefficient of the impact factor in~\eqn{eq:if1} as 
\begin{align}
	c^{g(1)}_{\QCD}=\left(c^{g(1)}_{\mathcal{N}=4}-4c^{g(1)[\mathbf{8}]}_{\mathcal{N}=1 \chi}+c^{g(1)[\mathbf{8}]}_{\mathrm{scalar}}\right)+N_f \left(c^{g(1)[\mathbf{3}]}_{\mathcal{N}=1 \chi}-c^{g(1)[\mathbf{3}]}_{\mathrm{scalar}}\right).
	\label{eq:if41s}
\end{align}

In this paper we focus on the ${\cal N}=4$ component, $\mathcal{M}^{(1)}_{\mathcal{N}=4}$, of the one-loop amplitude.
Let us begin by discussing the relevant helicity configurations. As at tree level, there are two distinct helicity configurations for the one-loop two-gluon central-emission vertex. 
In order to derive this CEV we take the central NMRK limit, as defined in eq.~(\ref{nmrapp}), of $\mathcal{M}_{\mathcal{N}=4}^{(1)}(p_1^{\nu_1},p_2^{\nu_2}, p_3^{-\nu_2}, p^{\nu_4}_4, p^{\nu_5}_5, p_6^{-\nu_1})$.
Thus, to obtain the same-helicity vertex, i.e. $\nu_4= \nu_5$, the six-gluon amplitude we begin with must necessarily be an \hbox{(anti-)MHV} amplitude. In section~\ref{sec:N4same-hel} we take the central NMRK limit of \hbox{(anti-)MHV} one-loop six-gluon amplitudes in $\cN=4$ SYM to obtain the one-loop two-gluon CEV of this theory.
In turn, in order to extract the one-loop opposite-helicity vertex we must instead take the NMRK limit of a helicity configuration with $\nu_4\neq \nu_5$, which is necessarily an NMHV amplitude. This study is performed in section~\ref{sec:N4opp-hel}.

In both cases we can simplify the procedure by recalling some results from section~\ref{sec:tcev}. Although \eqn{eq:N41loop} consists of a sum of 60 colour orderings, only 12 of these are not power suppressed by the rational terms in the central NMRK limit. This is straightforward for the MHV amplitudes, which are proportional to the tree-level result, up to logarithmically varying functions, as we will see in \eqn{eq:N4MHV}. For the NMHV amplitudes, eqs.~(\ref{eq:N4mmmppp})--(\ref{eq:N4mpmpmp}), 
this is not the case. However, each of the pairs of individual rational terms in these amplitudes are power suppressed in the NMRK limit for colour orderings that do not have both $p_2$ and $p_3$ adjacent, and $p_6$ and $p_1$ adjacent. This means we need only study the 12 colour orderings that are leading at tree-level also for the one-loop NMHV amplitudes.
It will be useful to consider the representative colour orderings,
\begin{align}
\sigma_A=\left\{1,2,3,4,5,6\right\}, \quad
\sigma_{A'}=\left\{1,2,3,5,4,6\right\}, \quad
\sigma_B=\left\{1,4,2,3,5,6\right\},
\label{eq:sigmaAAxB}
\end{align}
that we introduced in section~\ref{sec:tcev}. 
We further introduce the notation $\overset{\leftrightarrow}{\sigma}$ to indicate a kinematic (momentum and spin) interchange of $2 \leftrightarrow 3$, and $\underset{\leftrightarrow}{\sigma}$ to indicate a kinematic interchange 
of~$1 \leftrightarrow 6$. As an example of this notation,
\begin{equation}
    \overset{\leftrightarrow}{\underset{\leftrightarrow}{\sigma_A}}=\left\{6,3,2,4,5,1\right\}.
\end{equation}
The amplitude can thus be written as a decomposition into symmetric ($+$) and antisymmetric ($-$) components, as done in \eqn{eq:signsymm},
\begin{equation}
\mathcal{M}^{(1)}=\mathcal{M}^{(1)(+,+)}+\mathcal{M}^{(1)(+,-)}+\mathcal{M}^{(1)(-,+)}+\mathcal{M}^{(1)(-,-)}.
\label{eq:Msig-sig}
\end{equation}
Owing to Bose symmetry for gluon amplitudes, these kinematic interchanges translate directly into interchanges of the corresponding colour indices, so for example $\mathcal{M}^{(1)(-,-)}$ corresponds to 
the component of the amplitude in which an antisymmetric colour representation is exchanged in both the $t_1$ and $t_3$ channels, $\mathcal{M}^{(1)(+,-)}$ corresponds to one in which a symmetric representation is exchanged in the $t_1$ channel, while an antisymmetric one in the $t_3$ channel, etc. In the central NMRK limit it will be useful to organise the colour-dressed amplitudes according to this symmetry. For example, when there is an adjoint representation circulating in the loop we obtain 
\begin{align}
    \begin{split}
    \mathcal{M}^{(1)(-,-)}&= \gs^6\, F^{a_3 a_2 c_1}F^{a_6 a_1 c_3}\\\times\frac{1}{4}\bigg\{&\tr\left(F^{c_1}F^{a_4}F^{a_5}F^{c_3}\right)\times \left(M^{(1)}(\sigma_A)-M^{(1)}(\overset{\leftrightarrow}{\sigma_A})-M^{(1)}(\underset{\leftrightarrow}{\sigma_A})+M^{(1)}(\overset{\leftrightarrow}{\underset{\leftrightarrow}{\sigma_A}})\right)\\
    +&\tr\left(F^{c_1}F^{a_5}F^{a_4}F^{c_3}\right)\times \left(M^{(1)}(\sigma_{A'})-M^{(1)}(\overset{\leftrightarrow}{\sigma_{A'}})-M^{(1)}(\underset{\leftrightarrow}{\sigma_{A'}})+M^{(1)}(\overset{\leftrightarrow}{\underset{\leftrightarrow}{\sigma_{A'}}})\right)\\
    +&\tr\left(F^{c_1}F^{a_4}F^{c_3}F^{a_5}\right)\times \left(M^{(1)}(\sigma_B)-M^{(1)}(\overset{\leftrightarrow}{\sigma_B})-M^{(1)}(\underset{\leftrightarrow}{\sigma_B})+M^{(1)}(\overset{\leftrightarrow}{\underset{\leftrightarrow}{\sigma_B}})\right)
    \bigg\} \,,
    \end{split}
    \label{eq:Modd-odd}
\end{align}
which in particular is valid for one-loop gluon amplitudes in $\cN=4$ SYM, \eqn{eq:N41loop}.
The analogous expressions for the other signatures are listed in eqs.~(\ref{eq:Meven-even})--(\ref{eq:Modd-even}).

Before extracting the same-helicity CEV in section~\ref{sec:N4same-hel} and the opposite-helicity CEV in section~\ref{sec:N4opp-hel}, we list here the known one-loop coefficients of the impact factor and Lipatov vertex in ${\mathcal{N}=4}$ SYM, which will be of use in our extraction of both helicity configurations of the two-gluon CEV. The one-loop coefficient, \eqn{eq:if1}, of the impact factor in ${\mathcal{N}=4}$ SYM is
\begin{align}
	c^{g(1)}_{\mathcal{N}=4}(t)=N_c\kappa_\Gamma 
	\left(\frac{\mu^2}{-t}\right)^\epsilon	\left( -\frac{2}{\epsilon^2}  + \frac1{\epsilon} \log\left(\frac{\tau}{-t}\right)
	+ \frac{\pi^2}{2} - \frac{\delta_R}{6} \right)\,,
	\label{eq:n4-1loopif}
\end{align}
where $\delta_R=1$ in conventional dimensional regularisation (CDR) and in 't-Hooft-Veltman (HV) schemes,
while $\delta_R=0$ in dimensional reduction. The QCD one-loop impact factor, from which \eqn{eq:n4-1loopif} may be understood through the decomposition of \eqn{eq:if41s}, was computed for $\delta_R=1$ in~\cite{Fadin:1992zt,Fadin:1993wh,DelDuca:1998cx,Bern:1998sc}, and for $\delta_R=0$ in~\cite{DelDuca:1998cx,Bern:1998sc}.
Secondly, the single-gluon CEV to one-loop accuracy, \eqn{eq:oneloopcev},
in ${\mathcal{N}=4}$ SYM is given by
\begin{align}
	\begin{split}
		v^{g(1)}_{\mathcal{N}=4}(t_1,|p_{4\perp}|^2,t_2)
		=&N_c \kappa_{\Gamma} \Bigg(- \frac{1}{\epsilon^2}\, \left(\frac{\mu^2}{ |p_{4\perp}|^2}\right)^\epsilon+ \frac{\pi^2}{3}
		\\&+ \frac{1}{\epsilon}\, \left[ \left(\frac{\mu^2}{-t_1}\right)^\epsilon + \left(\frac{\mu^2}{-t_2}\right)^\epsilon \right]
		\log\left( \frac{\tau}{|p_{4\perp}|^2} \right)- \frac{1}{2} \log^2\left( \frac{t_1}{t_2} \right) \Bigg) \,.
	\end{split}
	\label{eq:cev1n4}
\end{align}
Note that \eqn{eq:cev1n4} is independent of the helicity of the produced particle, and of the regularisation parameter $\delta_R$. The one-loop coefficient was computed in QCD 
 in~\cite{Fadin:1993wh,Fadin:1994fj,Fadin:1996yv,DelDuca:1998cx,Bern:1998sc}, from which the ${\mathcal{N}=4}$ part, \eqn{eq:cev1n4}, may be extracted. We note that since in ${\mathcal{N}=4}$ SYM the beta function vanishes, \eqns{eq:n4-1loopif}{eq:cev1n4} do not need to be renormalised. In this paper we use the dimensional reduction choice with $\delta_R=0$, 
 preserving the uniform transcendentality property of the $\cN=4$ SYM amplitude.

\subsection{Same-helicity vertex}
\label{sec:N4same-hel}

In this section we take the NMRK limit of the MHV amplitude,
\begin{equation}
\mathcal{M}_{\mathcal{N}=4}^{(1)}(p_1^{\nu_1},p_2^{\nu_2}, p_3^{-\nu_2}, p^{\oplus}_4, p^{\oplus}_5, p_6^{-\nu_1}).
\label{eq:cMsame-hel}
\end{equation}
After the CEV for $\nu_4=\nu_5=\oplus$ has been obtained, the CEV for $\nu_4=\nu_5=\ominus$ may be obtained by complex conjugation, just as for the tree-level vertex, \eqn{eq:compl}. 
The colour-ordered amplitudes for one-loop six-gluon MHV amplitudes in $\mathcal{N}=4$ SYM are \cite{Bern:1994zx}
\begin{align}
	M^{(1)}_{\mathcal{N}=4}(p_{\sigma_1}^{\nu_{\sigma_1}},\cdots,p_{\sigma_6}^{\nu_{\sigma_6}})=\frac{\cg}{(4\pi)^2} M^{(0)}(p_{\sigma_1}^{\nu_{\sigma_1}},\cdots,p_{\sigma_6}^{\nu_{\sigma_6}})V_{6}(\sigma_1,\cdots,\sigma_{6}) \,,
	\label{eq:N4MHV}
\end{align}
where the transcendental terms are given by the function,
\begin{align}
	\begin{split}
		V_{6}(\sigma_1,\cdots,\sigma_{6})&=
		\sum_{i=1}^{6}-\frac{1}{\epsilon^2}\left(\frac{\mu^2}{-t_i^{[2]}}\right)^\epsilon
		-\sum_{i=1}^{6}\log\left(\frac{-t_i^{[2]}}{-t_i^{[3]}}\right)\log\left(\frac{-t_{i+1}^{[2]}}{-t_i^{[3]}}\right)\\
		&+\frac{1}{2}\sum_{i=1}^{3}\log^2\left(\frac{-t_i^{[3]}}{-t_{i+1}^{[3]}}\right)-\sum_{i=1}^{3}\Li\left[1-\frac{(-t_i^{[2]})(-t_{i+3}^{[2]})}{(-t_i^{[3]})(-t_{i-1}^{[3]})}\right]+\pi^2,
	\end{split}
	\label{eq:V6g}
\end{align}
which does not depend on the helicities of the gluons.
Here we have used the notation,
\begin{equation}
	t_i^{[r]}=(p_{\sigma_i}+\cdots+p_{\sigma_{i+r-1}})^2.
	\label{eq:ti}
\end{equation}
When analytically continuing to the physical region, as discussed in appendix~\ref{sec:appa}, we give all Mandelstam invariants a small positive imaginary part, $t_i^{[m]}\to t_i^{[m]}+i 0$. For logarithms of ratios of invariants, this prescription yields
\begin{equation}
    \log\left(\frac{-t_i^{[m]}}{-t_j^{[n]}}\right)=\log\left(\left|\frac{t_i^{[m]}}{t_j^{[n]}}\right|\right)-i\pi \left(\Theta(t_i^{[m]})-\Theta(t_j^{[n]}) \right),
    \label{eq:log_prescription}
\end{equation}
where the Heaviside function $\Theta(x)$ is defined as $\Theta(x)=1$ for $x>0$ and $\Theta(x)=0$ otherwise. 
Care must be taken when performing the analytic continuation of the arguments of the dilogarithms in \eqn{eq:V6g} to the physical region, as discussed in appendix~\ref{sec:absorp}.

From \eqn{eq:N4MHV} we see that the one-loop amplitudes in $\mathcal{N}=4$ SYM in the same-helicity configuration
are given by the tree-level gluon amplitude times the transcendental function $V_6$. As a step towards taking the full NMRK limit of \eqn{eq:cMsame-hel}, let us keep only the leading power terms of the rational functions. This means we neglect all but the 12 leading colour orderings in \eqn{eq:cMsame-hel}, and for these 12 leading colour orderings, we take the NMRK limit of the tree-level colour-ordered amplitudes that appear in \eqn{eq:N4MHV}. We recall that in the NMRK limit, an interchange of $p_2^{\nu_2}\leftrightarrow p_3^{\nu_3}$ or $p_6^{\nu_6}\leftrightarrow p_1^{\nu_1}$ leads only to a sign change of the leading tree-level amplitude. This lets us write, for example, the $(-,-)$ component of the colour-dressed amplitude in the NMRK limit as
\begin{align}
    \begin{split}
    \mathcal{M}^{(1)(-,-)}_{\mathcal{N}=4}&(p_4^{\nu},p_5^{\nu})\toNMRK \gs^6\frac{\cg}{(4\pi)^2}\, F^{a_3 a_2 c_1}F^{a_6 a_1 c_3}\\\times\frac{1}{4}\bigg\{&\tr\left(F^{c_1}F^{a_4}F^{a_5}F^{c_3}\right) \left.M^{(0)}(\sigma_A)\right|_{\mathrm{NMRK}}\left(V_6(\sigma_A)+V_6(\overset{\leftrightarrow}{\sigma_A})+V_6(\underset{\leftrightarrow}{\sigma_A})+V_6(\overset{\leftrightarrow}{\underset{\leftrightarrow}{\sigma_A}})\right)\\
    +&\tr\left(F^{c_1}F^{a_5}F^{a_4}F^{c_3}\right) \left.M^{(0)}(\sigma_{A'})\right|_{\mathrm{NMRK}}\left(V_6(\sigma_{A'})+V_6(\overset{\leftrightarrow}{\sigma_{A'}})+V_6(\underset{\leftrightarrow}{\sigma_{A'}})+V_6(\overset{\leftrightarrow}{\underset{\leftrightarrow}{\sigma_{A'}}})\right)\\
    +&\tr\left(F^{c_1}F^{a_4}F^{c_3}F^{a_5}\right) \left.M^{(0)}(\sigma_B)\right|_{\mathrm{NMRK}}\left(V_6(\sigma_B)+V_6(\overset{\leftrightarrow}{\sigma_B})+V_6(\underset{\leftrightarrow}{\sigma_B})+V_6(\overset{\leftrightarrow}{\underset{\leftrightarrow}{\sigma_B}})\right)
    \bigg\},
    \end{split}
    \label{eq:M--}
\end{align}
where we factorised the tree amplitude for each colour ordering. The positive relative signs in this equation, in contrast to \eqn{eq:Modd-odd}, are due to the sign flips in the tree amplitude $M^{(0)}(\sigma)$.
Here, and in the following, we suppress the momentum and helicity arguments of $\cM_{\cN=4}^{(1)}$, apart from the two central gluons 4 and 5.
Next, we write the tree-level colour-ordered amplitudes in the central NMRK in a factorized form according to eq.~(\ref{NLOfactorization2}), e.g.~for the $A$ colour ordering we use 
\begin{equation}
\left.M^{(0)}(\sigma_A)\right|_{\mathrm{NMRK}}=C^{g(0)}\left(p_2^{\nu_2}, p_3^{-\nu_2}\right) 
\frac{1}{t_1}
A^{gg(0)}(q_1, p^\oplus_4, p^\oplus_5, q_3)
\frac{1}{t_3}C^{g(0)}\left( p_1^{\nu_1},p_6^{-\nu_1}\right)
\end{equation}
and similarly for the $A^\prime$ and $B$ orderings.

Turning our attention to the transcendental functions, we 
observe a major simplification in the central NMRK limit upon considering \emph{the dispersive part of the amplitude}, defined by taking the real part of the transcendental functions.
We find that the latter
is given solely by the $(-,-)$ component, eq.~(\ref{eq:M--}), as the dispersive parts of the remaining components of eq.~(\ref{eq:Msig-sig}), quoted in eqs.~(\ref{eq:Meven-even})--(\ref{eq:Modd-even}), vanish.
This simplification is due to the fact that, for the 12 leading configurations (shown in e.g. \eqn{eq:M--}),  the real part of the transcendental functions $V_6$ in the central NMRK limit is unchanged by an interchange of $p_2\leftrightarrow p_3$ or $p_6\leftrightarrow p_1$. In conclusion, the dispersive part of the colour-dressed amplitude in the central NMRK limit is given by
\begin{align}
\begin{split}
	&\disp{\mathcal{M}^{(1)}_{\mathcal{N}=4}(p_4^\oplus, p_5^\oplus)}\,\,\toNMRK \\
	&\hspace*{30pt}\,\gs^6\, \frac{\kappa_{\Gamma}}{(4\pi)^2}s_{12}F^{a_3 a_2 c_1} C^{g(0)}\left(p_2^{\nu_2}, p_3^{-\nu_2}\right)\frac{1}{t_1}\frac{1}{t_3}F^{a_6 a_1 c_3}C^{g(0)}\left( p_1^{\nu_1},p_6^{-\nu_1}\right)\,\, \times\\
	&\hspace*{30pt} \bigg\{  
	\tr \left(F^{c_1}F^{a_4}F^{a_5}F^{c_3}\right)A^{gg(0)}(q_1, p^\oplus_4, p^\oplus_5, q_3)\real{\left.V_6(\sigma_{A})\right|_{\rm NMRK}}\\
	&\hspace*{30pt}\hphantom{\Big\{}+\tr \left(F^{c_1}F^{a_5}F^{a_4}F^{c_3}\right)A^{gg(0)}(q_1, p^\oplus_5, p^\oplus_4, q_3)\real{\left.V_6(\sigma_{A'})\right|_{\rm NMRK}}\\
	&\hspace*{30pt}\hphantom{\Big\{}+\tr \left(F^{c_1}F^{a_5}F^{c_3}F^{a_4}\right)B^{gg(0)}(q_1, p^\oplus_4, p^\oplus_5, q_3)\real{\left.V_6(\sigma_{B})\right|_{\rm NMRK}}
	\bigg\}.
	\label{eq:disp_same-hel}
\end{split}
\end{align}
By contrast, the absorptive part of the amplitude, defined by taking the imaginary part of the transcendental functions, in NMRK receives contribution from all terms in the decomposition of~\eqn{eq:Msig-sig}. The absorptive part of the amplitude is discussed further in appendix~\ref{sec:absorp}, and the four signature components are listed in eqs.~(\ref{eq:abN41loop})--(\ref{eq:absorpM++}). In section~\ref{sec:NLO_xs} we show that this absorptive part does not contribute to the colour-summed NLO $gg\to gggg$ squared matrix element in the central NMRK limit. This means that, in particular, only the dispersive part of the one-loop amplitude contributes to the dijet cross section at NNLL accuracy. We now move to the study of the real part of $V_6$ for the relevant colour orderings, leaving the analogous discussion of the imaginary parts of $V_6$ to appendix \ref{sec:absorp}.

As noted in ref.~\cite{Bartels:2008ce}, the real part of $V_6$ can be written, \emph{without any approximation}, as a combination of one-loop functions that describe amplitudes in MRK -- the one-loop Regge trajectory of eq.~(\ref{alph1}), the impact factor of eq.~(\ref{eq:n4-1loopif}) and the Lipatov vertex, eq.~(\ref{eq:cev1n4}) -- plus a remainder term that vanishes in the MRK limit. We will refer to this term as the NMRK remainder function ($\Delta V$ below). The arguments of the one-loop MRK functions can be chosen such that the remainder function is free from infrared (IR) divergences, and is a function of conformally-invariant cross ratios alone. For future convenience we choose to group these one-loop MRK functions into two components. One component is
a function which collects the `central' behaviour that will be relevant for our central NMRK limit,
\begin{equation}
U(\tilde{t}_1,\tilde{\eta}_{12},\tilde{t}_2,\tilde{s}_2,\tilde{\eta}_{23},\tilde{t}_3) = v^{g(1)}_{\mathcal{N}=4}\left(\tilde{t}_1,\tilde{\eta}_{12},\tilde{t}_2\right)
    +\alpha^{(1)}(\tilde{t}_2)\log\left(\frac{\tilde{s}_2}{\tau}\right)
    +v^{g(1)}_{\mathcal{N}=4}\left(\tilde{t}_{2},\tilde{\eta}_{23},\tilde{t}_3\right) \,.
    \label{eq:U}
\end{equation}
The other component collects the remaining `non-central' behaviour,
\begin{equation}
		E(\tilde{t}_1,\tilde{s}_1;\tilde{t}_3,\tilde{s}_3)=c^{g (1)}_{\mathcal{N}=4}(\tilde{t}_1;\tau)
		+\alpha^{(1)}(\tilde{t}_1) \log\left(\frac{\tilde{s}_1}{\tau}\right) +\alpha^{(1)}(\tilde{t}_3) \log\left(\frac{\tilde{s}_3}{\tau}\right) 
		+ c^{g (1)}_{\mathcal{N}=4}(\tilde{t}_3;\tau) \,.
\label{eq:E}
\end{equation}
The arguments decorated with tildes in \eqn{eq:U} and \eqn{eq:E} are placeholders: the arguments assigned  will differ for each colour ordering in \eqn{eq:disp_same-hel}. However, we will always make the natural choice of assigning $\tilde{t}_1=t_1=s_{23}$ and $\tilde{t}_3=t_3=s_{61}$, and we will assign $\tilde{t}_2=t_2=s_{234}$ when considering the MRK limit, and $\tilde{t}_2=t'_2=s_{235}$ when considering the $\MRKx$ limit. Based on the structure of $V_6$ in \eqn{eq:V6g}, for a given colour ordering we will also choose the remaining  arguments to be constructed from colour-ordered two- and three-particle Mandelstam invariants. This choice will be informed by our knowledge of the MRK and $\MRKx$ limit of this amplitude. However, these criteria do not lead to a unique assignment of the arguments of \eqns{eq:U}{eq:E}: 
the remaining freedom will be discussed separately for each colour ordering.

As a first example, let us see how this re-writing can be performed for the real part of $V_6(\sigma_A)$. We first separate the non-central function $E$, making the natural choice of Mandelstam variables for its arguments,
\begin{equation}
	\begin{split}
		N_c\kappa_\Gamma \real{V_6(\sigma_{A}) }
		&=E(s_{23},s_{34};s_{61},s_{56})+V_A \,.
	\end{split}
	\label{eq:V6Aexact}
\end{equation}
This choice of arguments for $E$ is the simplest choice which is compatible with the known MRK limit\footnote{As this function does not depend on the central degrees of freedom in the NMRK, no further simplification is obtained by taking the further MRK limit.},
\begin{equation}
E(s_{23},s_{34};s_{61},s_{56})\xrightarrow[\mathrm{(N)MRK}]{} E(-|q_{1\perp}|^2, p_3^+p_4^-; -|q_{3\perp}|^2,p_5^+p_6^-)\,.
\end{equation}

The function $V_A$ in eq.~(\ref{eq:V6Aexact}) collects the remaining `central' behaviour and it can be identified as the transcendental function associated with the two-gluon CEV for the $A$ colour ordering. 
As anticipated, this function itself can be neatly written as
\begin{equation}
	\begin{split}
		V_A=U_A +\Delta V_A(u_A,v_A,w_A) \,,
	\end{split}
	\label{eq:VA}
\end{equation}
where, again guided by the MRK limit, we fix  $U_A$ to equal the function $U$, \eqn{eq:U}, with arguments assigned as follows
\begin{equation}
U_A = U\left(s_{23},\frac{s_{34}s_{45}}{s_{345}\sqrt{u_A}},s_{234},s_{45},\frac{s_{45}s_{56}}{s_{456}\sqrt{u_A}},s_{61}\right)\,.
    \label{eq:UA}
\end{equation}
This choice of kinematic arguments leads to a simple NMRK remainder function $\Delta V_A$, which is IR-finite, vanishes in the MRK limit and depends only on the cross ratios, \eqn{eq:uvwA},
\begin{align}
\begin{split}
\label{eq:deltaVA}
  \Delta V_A(u_A,v_A,w_A)=N_c\kappa_\Gamma \bigg( &\frac{\pi^2}{3}-\frac{1}{4}\log^2(u_A)-\frac{1}{2}\log(u_A)\log(v_Aw_A)\\&-\Li\left(1-u_A\right)-\Li\left(1-v_A\right)-\Li\left(1-w_A\right)\bigg) \,.
\end{split}
\end{align}
The MRK limit of the cross ratios is $u_A\to1$, $v_A\to0$, $w_A\to0$ (see \eqn{eq:uvwMRK}).  It is easy to see that \eqn{eq:deltaVA} vanishes in this limit.

As mentioned above, \eqn{eq:UA} is not the unique choice which leads to an IR-finite NMRK remainder. In \eqn{eq:UA} we made the further choice of requiring $U_A$ to be invariant under target-projectile exchange, \eqn{eq:target-projectile}. Yet even this physically-motivated requirement does not uniquely specify arguments for $U_A$: a rescaling of the $\tilde{t}_2$ variable in \eqn{eq:UA} by an arbitrary power of $u_A$, i.e. $s_{234}\to s_{234} \times(u_A)^p$ for real $p$, yields a NMRK remainder function which is an IR-finite function of cross ratios and is invariant under target-projectile exchange. 
Our ultimate choice of arguments in \eqn{eq:UA} was guided by the intuition that the trajectory should be evaluated exactly at the $t$-channel invariant $t_2=s_{234}$ also away from the strict MRK limit. It would be interesting to explore this same procedure at higher orders in $\epsilon$, or at two loops, where, if such an organisation proves possible, there may be additional constraints on the kinematic arguments.

\begin{figure}[hbt]
\centering
     \includegraphics[width=1.0 \linewidth]{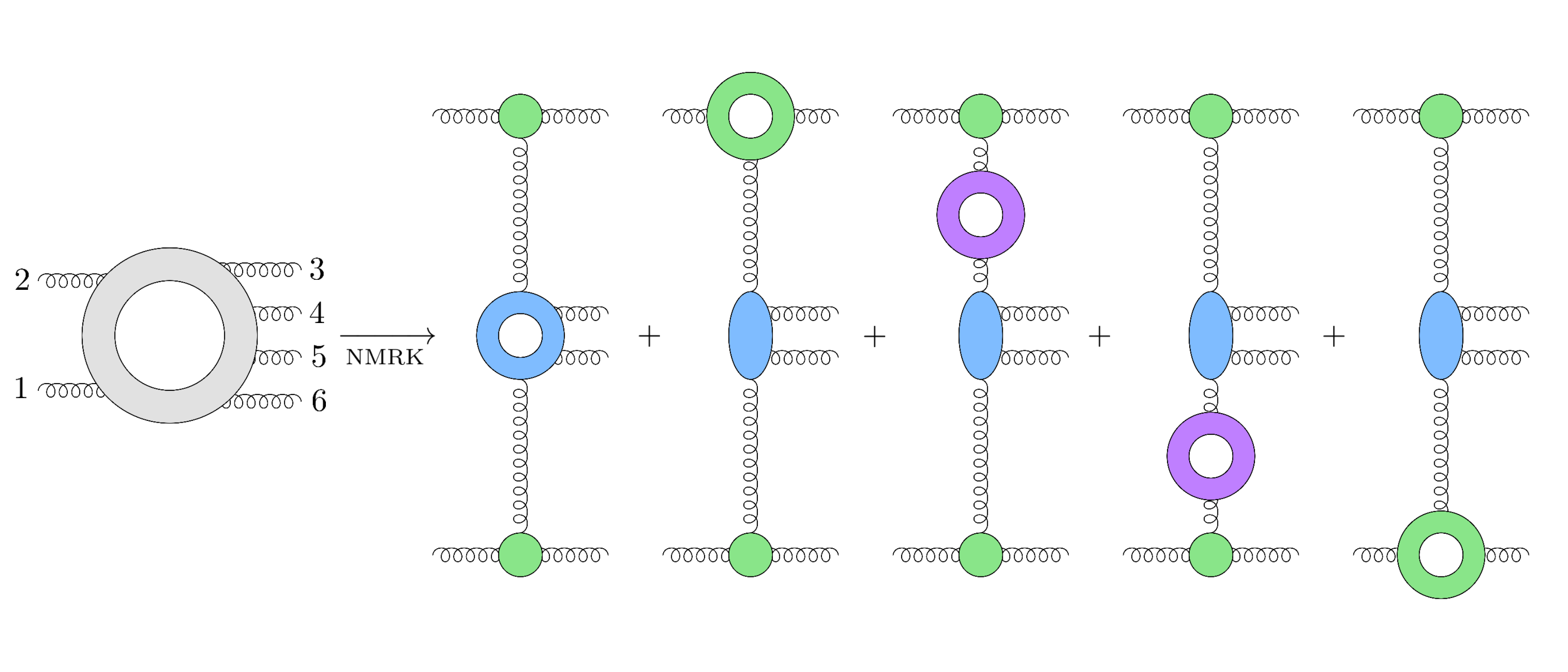}
    \caption{Schematic illustration of the behaviour of the dispersive part of the one-loop partial amplitude $M^{(1)}_{\cN=4}(\sigma_A)$ in the central NMRK limit. The first diagram on the right-hand side represents the transcendental functions that depend on the central degrees of freedom, $V_A$, which we identify as the one-loop correction to the two-gluon CEV for this colour ordering. The remaining four diagrams on the right-hand side are given by the known one-loop impact factors and gluon Regge trajectory contributions, and are collected in the function $E$ of eq.~(\ref{eq:E}).}
    \label{fig:Afactorisation}
\end{figure}

We stress that \eqn{eq:V6Aexact} holds in general kinematics if no approximation is made to the kinematic invariants.
However, we will proceed to make the kinematic approximations valid in the central NMRK limit, in order to exploit the invariance of the real part under the interchanges $p_2\leftrightarrow p_3$ or $p_6\leftrightarrow p_1$ in this limit, as in eq.~(\ref{eq:disp_same-hel}). Thus, we specify \eqn{eq:VA} to the
NMRK limit, by making the approximation,
\begin{equation}
	\begin{split}
		U_A\toNMRK U\left(-|q_{1\perp}|^2,\frac{p_4^- \ s_{45}}{p_4^-+p_5^-},-|q_{2\perp}|^2-p_5^+p_4^-,s_{45},\frac{s_{45} \ p_5^+}{p_4^++p_5^+},-|q_{3\perp}|^2\right)\,,
	\end{split}
	\label{eq:UAnmrk}
\end{equation}
and also taking the NMRK approximation of the cross ratios, as listed in \eqn{eq:uvwA}, for the arguments of \eqn{eq:deltaVA}. This NMRK approximation of $V_A$ will later form one of the building blocks of the colour-dressed two-gluon CEV. The structure of the NMRK limit of the dispersive part of $M^{(1)}_{\cN=4}$ is depicted in fig.~\ref{fig:Afactorisation}.

The function $V_6(\sigma_{A'})$ is simply related to $V_6(\sigma_{A})$ by an interchange of $p_4\leftrightarrow p_5$. We now use the known $\MRKx$ limit of the amplitude as our guide, writing
\begin{align}
		N_c\kappa_\Gamma \real{\left.V_6(\sigma_{A'})\right|_{\rm NMRK}}&=E(s_{23},s_{35};s_{61},s_{46})+V_{A'} \,,\label{eq:ReVAx}\\
V_{A'}&=U_{A'}+\Delta V_{A}(u_{A'},w_{A'},v_{A'}) \,,
	\label{eq:VAx}
\end{align}
where $U_{A'}$ is related to \eqn{eq:UA} by the interchange of $p_4\leftrightarrow p_5$. As was true for the $A$ ordering, in order to use the compact form of \eqn{eq:disp_same-hel}, we will need to make the NMRK approximations to the arguments of $\Delta V_A$ in \eqn{eq:VAx}, and similarly we make the approximation,
\begin{equation}
	\begin{split}
		U_{A'}\toNMRK U\left(-|q_{1\perp}|^2,\frac{p_5^- \ s_{45}}{p_4^-+p_5^-},-|q'_{2\perp}|^2-p_4^+p_5^-,s_{45},\frac{s_{45} \ p_4^+}{p_4^++p_5^+},-|q_{3\perp}|^2\right)\,.
	\end{split}
	\label{eq:UAxnmrk}
\end{equation}
Note that the NMHV remainder function for the $A'$ ordering is the same as for the $A$ ordering, because $\Delta V_A(u,w,v) = \Delta V_A(u,v,w)$ based on~\eqn{eq:deltaVA}.

We now turn to study of the real part of $V_6(\sigma_B)$. The rational coefficient of this transcendental function, $M^{(0)}(\sigma_B)$, has a leading contribution in both the MRK and $\MRKx$ limits. It will be useful therefore to re-write this transcendental function in two ways: one guided by the known MRK limit of the amplitude, and the other by the known $\MRKx$ limit. As usual, we first separate the central and non-central functions,
\beqa
    N_c\kappa_\Gamma \real{V_6(\sigma_{B})} &=&E(s_{23},-s_{42};s_{61},s_{56})+V_B = E(s_{23},s_{35};s_{61},-s_{14})+V_{B'}\,.
    \label{eq:V6Bexact}
\eeqa
The forms of the central functions $V_B$ and $V_{B'}$ are guided, respectively, by the known MRK and $\MRKx$ limits of this function -- note that both refer to the same colour ordering. 
These central functions are given by
\begin{equation}
		V_B=U_B +\Delta V_B(u_B,v_B,w_B), \qquad \qquad V_{B'}=U_{B'} +\Delta V_B(v_B,u_B,w_B) \,,
	\label{eq:VB}
\end{equation}
where we have chosen to use the kinematic arguments,
\begin{align}
\begin{split}
    U_B&=U\left(s_{23},\frac{s_{14}s_{42}}{s_{142}u_B},s_{234},-s_{235}, \frac{s_{35}s_{56}}{s_{356}u_B},s_{61}\right)\,,\\
    U_{B'}&=U\left(s_{23},\frac{s_{35}s_{56}}{s_{356}v_B},s_{235},-s_{234}, \frac{s_{14}s_{42}}{s_{142}v_B},s_{61}\right)\,,
    \label{eq:UB}
\end{split}
\end{align}
for the one-loop MRK functions. 

Unlike the $A$ colour ordering, target-projectile symmetry does not constrain the arguments of \eqn{eq:UB}: from \eqn{eq:ubF} we see that in NMRK, each cross ratio is invariant under target-projectile exchange.
Our choice of kinematic arguments in \eqn{eq:UB} were instead motivated by the requirement that $U_B$ and $U_{B'}$ are related by the map which interchanges~$\{p_2 \leftrightarrow p_3, \ p_4 \leftrightarrow p_5, \ p_1 \leftrightarrow p_6 \}$, corresponding to a symmetry of the real part of $V_6(\sigma_B)$.
We will refer to this map as $G$. For the natural assignment of variables in~\eqn{eq:V6Bexact}, the real parts of $E$ are related by this map.\footnote{Note that we always choose the signs of the arguments of $E$ such that it is in fact a real function.} 
Choosing $U_B$ and $U_{B'}$ to be related by $G$ therefore means the respective NMRK remainder functions are also related by this map. For the cross ratios, \eqn{eq:uvwB}, the map $G$ leads to $u_B \,\,{\underset{G}\leftrightarrow}\,\, v_B$, while $w_B$ is invariant under $G$. The NMRK remainder function that follows from the choices made in \eqn{eq:UB} is
\beqa
\label{eq:deltaVB}
  \lefteqn{ \Delta V_B(u_B,v_B,w_B) = \Delta V_B(v_B,u_B,w_B)  } \\
  &=& N_c\kappa_\Gamma \bigg( \frac{\pi^2}{3}-\log(u_B)\log(v_B)-\Li\left(1-u_B\right)-\Li\left(1-v_B\right)-\Li\left(1-w_B\right)\bigg) \,,\nn
\eeqa
which is indeed invariant under $G$. We note, however, that by considering suitable rescaling of the arguments of $U_B$ and $U_{B'}$ by powers of $u_B$ or $v_B$ respectively (the quantities which tend to unity in the MRK or $\MRKx$ limits respectively), there is one alternative set of kinematic assignments which lead to an NMRK remainder function that possesses the same desirable properties as \eqn{eq:deltaVB}: as for the $\sigma_A$ ordering, our final choice of arguments in \eqn{eq:UB} was determined by requiring that $\tilde{t}_2$ is identified with the relevant $t$-channel momentum even in general kinematics, namely $t_2$ and $t_2^\prime$ in the MRK and $\MRKx$ limits, respectively.

In order to use \eqn{eq:V6Bexact} within the compact form of \eqn{eq:disp_same-hel}, we must make the NMRK approximations
\beqa
    U_B&\toNMRK & U\bigg(
    -|q_{1\perp}|^2,
    -\frac{p_4^-}{p_5^-} t_2^{\prime},
    t_2,-t_2^\prime, 
   - \frac{p_5^+}{p_4^+}t_2^\prime,
    -|q_{3\perp}|^2
    \bigg)\,,\nn\\
    U_{B'}&\toNMRK & U\bigg(
    -|q_{1\perp}|^2,
    -\frac{p_5^-}{p_4^-}t_2,
    t_2^\prime,-t_2, 
    -\frac{p_4^+}{p_5^+}t_2,
    -|q_{3\perp}|^2
    \bigg)\,,
    \label{eq:UBnmrk}
\eeqa
where it is understood that on the right hand side we use the NMRK limits of
\begin{equation}
t_2=s_{234}\toNMRK-\left(|q_{2\perp}|^2+p_5^+p_4^-\right),\,\qquad
t_2^\prime=s_{235}\toNMRK-\left(|q'_{2\perp}|^2+p_4^+p_5^-\right)\,.
\end{equation}
Similarly, we apply the NMRK approximations listed in \eqn{eq:uvwB} for the cross-ratio arguments of \eqn{eq:deltaVB}. We see that in  the NMRK limit, $U_B$ and $U_{B'}$ (and therefore $V_B$ and $V_{B'}$) are simply related by an interchange of $p_4 \leftrightarrow p_5$.

Having studied the three colour orderings of \eqn{eq:disp_same-hel}, we will now move to a basis of colour structures that more closely resembles the colour structure of the tree-level amplitude, \eqn{NLOfactorization}. This basis will let us identify one-loop corrections associated with each of the tree-level colour structures, as well as highlight the new colour structure that starts contributing at one-loop level. 
Denoting the three traces by
\begin{align}
T_A=\tr \left(F^{c_1}F^{a_4}F^{a_5}F^{c_3}\right),\qquad
T_{A^\prime}=\tr \left(F^{c_1}F^{a_5}F^{a_4}F^{c_3}\right),\qquad
T_B=\tr \left(F^{c_1}F^{a_5}F^{c_3}F^{a_4}\right),
\end{align}
we consider the following rotation of the sum inside the curly brackets in~\eqn{eq:disp_same-hel}
\begin{align}
\label{colour-rotation}
\begin{split}
T_AM_A+T_{A^\prime}M_{A^\prime}+T_BM_B&=
 (T_A-T_B) \times \frac13 (2M_{A}-M_{A^\prime}-M_B)
\\&+ (T_{A^\prime}-T_B) \times \frac13 (2M_{A^\prime}-M_A-M_B)
\\&+\frac13 (T_A+T_{A^\prime}+T_B) \times (M_A+M_{A^\prime}+M_B)\,.
\end{split}
\end{align}
Using the identities,
\begin{align}
	\tr\left(F^aF^bF^c\right)=N_c F^{abc} \,, \qquad
	\tr\left(F^aF^b\left[F^c,F^d\right]\right)=N_c F^{abe}F^{ecd} \,,
\end{align}
we relate the first two combinations of colour structure on the right-hand side of eq.~(\ref{colour-rotation}) to the tree-level structures $(F^{a_{\sigma_4}}F^{a_{\sigma_5}})_{c_1c_3}$ of eq.~(\ref{NLOfactorization}), and define the third colour factor in terms of the fully-symmetric trace:
\begin{align}
	d^{a_1 a_2 a_3 a_4}_{A}=\frac{1}{4!}\sum_{\sigma \in S_4}\tr\left(F^{a_{\sigma_1}}F^{a_{\sigma_2}}F^{a_{\sigma_3}}F^{a_{\sigma_4}}\right)\,.
	\label{eq:dA}
\end{align}
Making use of the additional relations that exist within NMRK, namely
\begin{align}
\begin{split}
\left.E(t_1,-s_{42};t_3,s_{56})\right|_{\NMRK}&=\left.E(t_1,s_{34};t_3,s_{56})\right|_{\NMRK}\,,\\
\left.E(t_1,s_{35};t_3,-s_{14})\right|_{\NMRK}&=\left.E(t_1,s_{35};t_3,s_{46})\right|_{\NMRK}\,,
\end{split}
\end{align}
we now express \eqn{eq:disp_same-hel} as
\begin{align}
\begin{split}
&\mathrm{Disp}\left[
\mathcal{M}^{(1)}_{\cN=4}(p_4^\oplus, p_5^\oplus)\right]\toNMRK
\frac{\gs^6}{(4\pi)^2}\, 
s_{12}F^{a_3 a_2 c_1}
C^{g(0)}\left(p_2^{\nu_2}, p_3^{-\nu_2}\right)
\frac{1}{t_1}  \times  \\
&\sum_{\sigma \in S_2}\Bigg\{  
		(F^{a_{\sigma_4}}F^{a_{\sigma_5}})_{c_1c_3}\left(A^{gg(1)}_{\vDash}(q_1, p^\oplus_{\sigma_4}, p^\oplus_{\sigma_5}, q_3)+A^{gg(0)}(q_1, p^\oplus_{\sigma_4}, p^\oplus_{\sigma_5}, q_3)E(t_1,s_{3\sigma_4};t_3,s_{\sigma_56})\right)
		\\
		&
		\qquad \qquad
		+\frac{1}{N_c}d_{A}^{c_1 a_{\sigma_4} a_{\sigma_5} c_3}A^{gg(1)}_{d}(q_1, p^\oplus_{\sigma_4}, p^\oplus_{\sigma_5}, q_3) \Bigg\}
\times\frac{1}{t_3}
F^{a_6 a_1 c_3}
C^{g(0)}\left(p_1^{\nu_1},p_6^{-\nu_1}\right)\,,
\label{eq:DispFFd}
\end{split}
\end{align}
where the totally symmetric colour structure $d_A$ starts contributing at one-loop level, with
\begin{align}
	A^{gg(1)}_{d}(q_1, p^\oplus_4, p^\oplus_5, q_3)
	=
	A^{gg(0)}(q_1, p^\oplus_4, p^\oplus_5, q_3)\left(V_{A}-V_B\right)
	\,.
	\label{eq:Agg1_d}
\end{align}
The symmetry of the $d_A$ colour structure under the permutation $S_2$ of \eqn{eq:DispFFd} means that after the sum over the $\sigma_4, \sigma_5$ permutations is performed,  there are only three independent colour structures in this equation, as in \eqn{eq:disp_same-hel}. 

For the kinematic coefficient of the tree-level colour structures we find, using \eqns{eq:ABrelation}{colour-rotation},
\begin{align}
\begin{split}
	&A^{gg(1)}_{\vDash}(q_1, p^\oplus_4, p^\oplus_5, q_3) \\
	&=\frac{1}{3}\left(
	A^{gg(0)}(q_1, p^\oplus_4, p^\oplus_5, q_3)\left(2V_A+V_B\right)
	-A^{gg(0)}(q_1, p^\oplus_5, p^\oplus_4, q_3)\left(V_{A'}-V_{B'}\right)\right)\,.
	\label{eq:Agg1_F}
\end{split}
\end{align}
We interpret this term as the one-loop coefficient of the all-orders expansion of this colour structure,
\begin{align}
\begin{split}
&A^{gg}_{\vDash}(q_1, p^{\nu_4}_4, p^{\nu_5}_5, q_3) = 
A^{gg(0)}_{\vDash}(q_1, p^{\nu_4}_4, p^{\nu_5}_5, q_3) + \frac{\as}{4\pi} A^{gg(1)}_{\vDash}(q_1, p^{\nu_4}_4, p^{\nu_5}_5, q_3) + \ord(\as^2)\, ,
\end{split}
\label{eq:Agg_F}
\end{align}
with 
\begin{align}
	&A^{gg(0)}_{\vDash}(q_1, p^{\nu_4}_4, p^{\nu_5}_5, q_3) =A^{gg(0)}(q_1, p^{\nu_4}_4, p^{\nu_5}_5, q_3)\,.
\end{align}
As the structure $d_A$ first appears at one-loop level we write
\begin{align}   
\begin{split}   
	A^{gg}_{d}(q_1, p^{\nu_4}_4, p^{\nu_5}_5, q_3) &=  \frac{\as}{4\pi} A^{gg(1)}_{d}(q_1, p^{\nu_4}_4, p^{\nu_5}_5, q_3) + \ord(\as^2)\, ,
\end{split} 
\label{eq:Agg_d}
\end{align}
in anticipation of higher-order corrections to this colour structure. We note that $A^{gg(1)}_{d}$ and $A^{gg(1)}_{\vDash}$ both depend on $\mu^2$ and $\tau$, but as for the one-loop coefficients of the impact factor and Lipatov vertex, we omit this dependence for brevity.

\begin{figure}[htb]
\centering
     \includegraphics[scale=0.35]{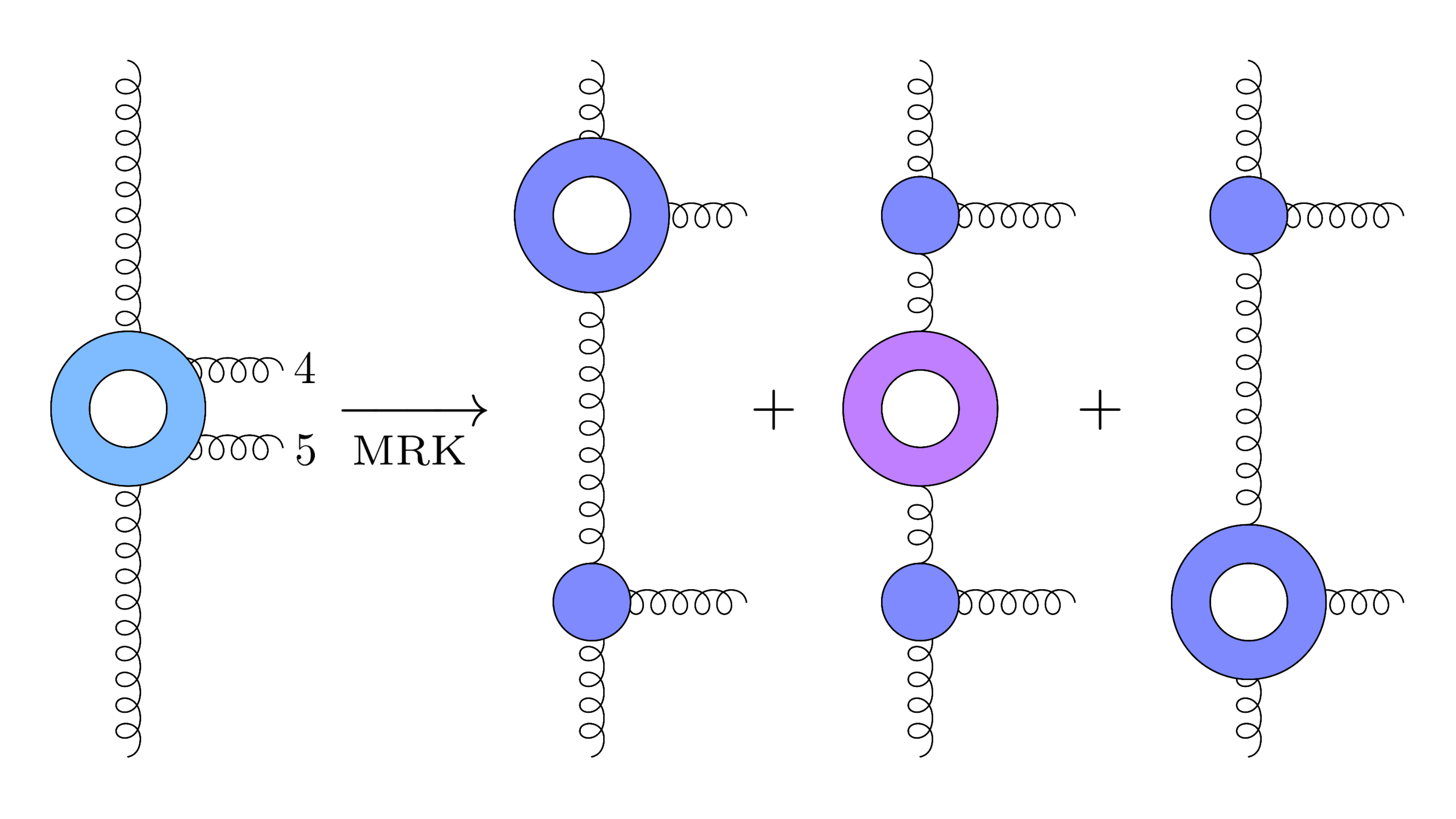}
    \caption{Schematic illustration of the further factorisation of the one-loop two-gluon CEV for the ordering $\sigma_A$ in the MRK limit.}
    \label{fig:MRK_factorisation}
\end{figure}
The notation introduced in this section makes it straightforward to verify that the \eqns{eq:Agg1_F}{eq:Agg1_d} yield the expected MRK limit. We first note that by construction, \eqns{eq:VA}{eq:VB} both tend to the same MRK limit,
\begin{align}
	\begin{split}
		\left.U_A\right|_{\mathrm{MRK}}=\left.U_B\right|_{\mathrm{MRK}}=U_{\mathrm{MRK}}=&
	U(-|q_{1\perp}|^2,|p_{4\perp}|^2,-|q_{2\perp}|^2,p_4^+p_5^-,|p_{5\perp}|^2,-|q_{3\perp}|^2).
	\end{split}
	\label{eq:UMRK}
\end{align}
Using the MRK limits of the relevant cross ratios listed in \eqn{eq:uvwMRK} together with the values of the dilogarithm,
\begin{equation}
    \Li \left(0\right)=0, \qquad \Li \left(1\right)=\frac{\pi^2}{6},
\end{equation}
we see that both $\Delta V_A(u_A,v_A,w_A)$ and $\Delta V_B(u_B,v_B,w_B)$ vanish in the MRK limit as desired. This leads to the simple results,
\begin{align}
	A^{gg(1)}_{\vDash}(q_1, p^\oplus_4, p^\oplus_5, q_3)
	&\toMRK
	V^{g(0)}(q_1,p_4^\oplus,q_2)\frac{1}{t_2}V^{g(0)}(q_2,p_5^\oplus,q_3) \, U_{\mathrm{MRK}}\, \label{eq:AggF1_MRK},\\
	A^{gg(1)}_{d}(q_1, p^\oplus_4, p^\oplus_5, q_3)
	&\toMRK
	0\, .
	\label{eq:Aggd1_MRK}
\end{align}
Equation~(\ref{eq:AggF1_MRK}) is depicted schematically in fig.~\ref{fig:MRK_factorisation}. The function $U_{\mathrm{MRK}}$ generates all three of the loop corrections on the right-hand side of the figure, thanks to the definition~(\ref{eq:U}).
A similar analysis shows that
\begin{align}
	A^{gg(1)}_{\vDash}(q_1, p^\oplus_5, p^\oplus_4, q_3)
	&\toMRKx
	V^{g(0)}(q_1,p_5^\oplus,q_2^\prime)\frac{1}{t_2^\prime}V^{g(0)}(q_2^\prime,p_4^\oplus,q_3)
	U_{\MRKx}\label{eq:AggF1_MRKx}\, ,\\
	A^{gg(1)}_{d}(q_1, p^\oplus_5, p^\oplus_4, q_3)
	&\toMRKx
	0\, ,
	\label{eq:Aggd1_MRKx}
\end{align}
where 
\begin{equation}
	\begin{split}
		\left.U_{A'}\right|_{\MRKx}=\left.U_{B'}\right|_{\MRKx}=U_{\mathrm{MRK'}}=
	U(-|q_{1\perp}|^2,|p_{5\perp}|^2,-|q'_{2\perp}|^2,p_5^+p_4^-,|p_{4\perp}|^2,-|q_{3\perp}|^2)\,.
	\end{split}
	\label{eq:UxMRKx}
\end{equation}

In appendix \ref{sec:lim} we investigate the soft and collinear limits of \eqn{eq:DispFFd}.
In section \ref{sec:all-orders} we conjecture how the one-loop results of this section may be extended to all orders. In the next subsection we obtain expressions analogous to \eqns{eq:Agg1_F}{eq:Agg1_d} 
but for the case where gluons 4 and 5 have opposite helicities.

\subsection{Opposite-helicity vertex}
\label{sec:N4opp-hel}
We now perform an analogous study to section~\ref{sec:N4same-hel} for the opposite-helicity case, where we must start from an NMHV helicity configuration.
We present the procedure for a configuration of the form $\mathcal{M}_{\mathcal{N}=4}^{(1)}(p_1^{\nu_1},p_2^{\nu_2}, p_3^{-\nu_2}, p^\oplus_4, p^\ominus_5, p_6^{-\nu_1})$. As for the tree-level vertex, after the CEV for $\nu_4=-\nu_5=\oplus$ has been obtained, the CEV for $\nu_4=-\nu_5=\ominus$ may be obtained by complex conjugation.
The colour-ordered one-loop six-gluon NMHV amplitudes in $\mathcal{N}=4$ SYM are given in eqs.~(\ref{eq:N4mmmppp})--(\ref{eq:N4mpmpmp}). Unlike \eqn{eq:N4MHV}, these NMHV amplitudes are not proportional to the tree-level amplitude. Instead, the transcendental functions of the NMHV amplitude, $W_i$ are multiplied by pairs of the rational functions that appear in the BCFW representations of the tree-level amplitude listed in appendix~\ref{sec:treeNMHV}. In the NMRK, after dividing by the appropriate impact factors and $t$-channel propagators, these rational coefficients are simply given by sums of pairs of the rational functions which we have studied in sections~\ref{sec:bcfw} and~\ref{sec:other},
\begin{equation}
S_{xy}^I=R_{xy}^I+R_{\bar{x}\bar{y}}^I \qquad x,y \in \{u,v,w\}, \quad I \in \{A,A',B\}.
\label{eq:Sxy}
\end{equation}
Like the tree-level amplitudes, these rational coefficients of the transcendental functions $W_i$ change only by a sign under exchange of $p_2^{\nu_2}\leftrightarrow p_3^{\nu_3}$ or $p_6^{\nu_6}\leftrightarrow p_1^{\nu_1}$. For the 12 leading colour orderings, we further find that an exchange of $p_2\leftrightarrow p_3$ or $p_6\leftrightarrow p_1$ does not alter the real part of the functions $W_i$. These properties collectively allow the dispersive part of the amplitude to be written as
\begin{align}
	\begin{split}
		&\mathrm{Disp}\left[\cM^{(1)}_{\mathcal{N}=4}(p_4^\oplus,p_5^\ominus)\right]\toNMRK  \\
	&	\hspace*{10pt} \gs^6\frac{\cg}{(4\pi)^2}\,s_{12}\, F^{a_3 a_2 c_1}\, C^{g(0)}(p_2^{\nu_2},p_3^{-\nu_2})\, \frac1{t_1}\, \frac1{t_3}\, F^{a_6 a_1 c_3}C^{g(0)}(p_1^{\nu_1},p_6^{-\nu_1})\,\, \\
		&\hspace*{10pt}\times  \bigg\{  
		\tr \left(F^{c_1}F^{a_4}F^{a_5}F^{c_3}\right)\Big[ S^A_{uv}\real{W_1(\sigma_A)}+S^A_{vw}\real{W_2(\sigma_A)}+S^A_{wu}\real{W_3(\sigma_A)} \Big]\\
		&\hspace*{10pt}\quad +\tr \left(F^{c_1}F^{a_5}F^{a_4}F^{c_3}\right)\Big[ S^{A'}_{uv}\real{W_1(\sigma_{A'})}+S^{A'}_{vw}\real{W_2(\sigma_{A'})}+S^{A'}_{wu}\real{W_3(\sigma_{A'})} \Big]\\
		&\hspace*{10pt}\quad +\tr \left(F^{c_1}F^{a_5}F^{c_3}F^{a_4}\right)\Big[ S^B_{uv}\real{W_1(\sigma_{B})}+S^B_{vw} \real{W_2(\sigma_{B})}+S^B_{wu}\real{W_3(\sigma_{B})} \Big]
		\bigg\} \,,
	\end{split}
	\label{eq:N4NMHVReW}
\end{align}
analogously to \eqn{eq:disp_same-hel}. On the right-hand side of \eqn{eq:N4NMHVReW}, the transcendental functions are understood to be taken in the NMRK limit, but for brevity we do not indicate this explicitly.
As in section~\ref{sec:N4same-hel}, we find that the dispersive part of the amplitude is given solely by $(-,-)$ exchange. This is not true for the absorptive part, which is studied in appendix~\ref{sec:absorp}. We will see in section~\ref{sec:NLO_xs} that the absorptive part does not contribute to the helicity- and colour-summed NLO squared matrix element for $gg \to gggg$, or to the dijet cross section at NNLL. The absorptive part will only contribute at N$^3$LL. For the remainder of this section we focus on the dispersive part of the amplitude, \eqn{eq:N4NMHVReW}.

Unlike the tree-level expressions (e.g.~\eqn{kosc2}), the sums of two rational terms, \eqn{eq:Sxy}, do contain genuine unphysical poles. To demonstrate that the dispersive part of the NMRK amplitude, \eqn{eq:N4NMHVReW}, is nonetheless free from unphysical singularities it is useful to use the fact that the sum of the three $S_{xy}^I$ is twice the tree-level CEV to rewrite  \eqn{eq:N4NMHVReW} as 
\begin{align}
	\begin{split}
		&\mathrm{Disp}\left[\mathcal{M}^{(1)}_{\mathcal{N}=4}(p_4^\oplus,p_5^\ominus)\right]
		\toNMRK\gs^6\frac{\cg}{(4\pi)^2}\, s_{12} \, F^{a_3 a_2 c_1}\, C^{g(0)}(p_2^{\nu_2},p_3^{-\nu_2})\, \frac1{t_1}\, \\
		&\times \bigg\{  
		\tr\left( F^{c_1}F^{a_4}F^{a_5}F^{c_3}\right)
		\Big[ 2A^{gg(0)}(q_{1},p_4^{\oplus},p_5^\ominus,q_{3})\real{W_2(\sigma_A)}\\&\hspace{12 em}+S^A_{uv}\real{W_1(\sigma_A)-W_2(\sigma_A)} +S^A_{wu}\real{W_3(\sigma_A)-W_2(\sigma_A)}  \Big]\\
		&\,
		+\tr\left(F^{c_1}F^{a_5}F^{a_4}F^{c_3}\right)
		\Big[ 2A^{gg(0)}(q_{1},p_5^\ominus,p_4^{\oplus},q_{3})\real{W_2(\sigma_{A'})}\\&\hspace{12 em}
		+S^{A'}_{uv}\real{W_1(\sigma_{A'})-W_2(\sigma_{A'})} +S^{A'}_{wu}\real{W_3(\sigma_{A'})-W_2(\sigma_{A'})} \Big]\\
		&\, +\tr\left(F^{c_1}F^{a_5}F^{c_3}F^{a_4}\right)
		\Big[ 2B^{gg(0)}(q_{1},p_4^{\oplus},p_5^\ominus,q_{3})\real{W_1(\sigma_B)}
		\\&\hspace{12 em}
		+S^B_{vw}\real{W_2(\sigma_B)-W_1(\sigma_B)} +S^B_{wu}\real{W_3(\sigma_B)-W_1(\sigma_B)}  \Big]
		\bigg\}\\
		&\times\frac1{t_3}\, F^{a_6 a_1 c_3}C^{g(0)}(p_1^{\nu_1},p_6^{-\nu_1}) \,.
	\end{split}
	\label{eq:N4NMHVReWA+R}
\end{align}
This rewriting simplifies our task for two reasons:
\begin{itemize}
\item From section~\ref{sec:tcev} we know the functions $A^{gg(0)}$ and $B^{gg(0)}$ are free from unphysical singularities.
\item The difference between two cyclic permutations of $W$ is much simpler than a given permutation of $W$.
\end{itemize}
To expand on the second point, the difference between two cyclic permutations of $W$ can be expressed as a single product of logarithms whose arguments are simple functions of the cross ratios. This property holds even in general kinematics, although in the following we implicitly consider the cross ratios to be taken in the NMRK limit.
Let us first consider these terms in the $\sigma_A$ ordering,
\begin{align}
\begin{split}
	W_1(\sigma_A)-W_2(\sigma_A)&=\log\left(v_A\right)\log\left(\frac{w_A}{u_A}\right) \,,\\
	W_3(\sigma_A)-W_2(\sigma_A)&=\log\left(w_A\right)\log\left(\frac{v_A}{u_A}\right) \,.
	\label{eq:WA12_WA32}
\end{split}
\end{align}
We note in passing that these functions are related by the interchange of $u_A\leftrightarrow v_A$, which has the physical interpretation of target-projectile exchange, \eqn{eq:uAF}. This provides our motivation for arranging \eqn{eq:N4NMHVReWA+R} as we have done, where 
 $W_2(\sigma_A)$ multiplies the corresponding tree-level CEV, while the remaining terms are related by
\begin{align}
\begin{split}
    S^A_{uv}\real{W_1(\sigma_A)-W_2(\sigma_A)}\xrightarrow[F]{} \frac{w \bar{z}}{\bar{w}z}S^A_{wu}\real{W_3(\sigma_A)-W_2(\sigma_A)}.
\end{split}
\end{align}

Table \ref{table:spurA} lists the behaviour of the cross ratios at each unphysical-singularity surface. From this table we see that at $P^A_{v}=0$ or $P^A_{w}=0$, the transcendental terms which multiply the divergent rational terms in \eqn{eq:N4NMHVReWA+R} vanish. For the remaining unphysical singularity at $P^A_{u}=0$, we find
\begin{align}
	\begin{split}
S^A_{uv}\left(W_1(\sigma_A)-W_2(\sigma_A)\right) +&S^A_{wu}\left( W_3(\sigma_A)-W_2(\sigma_A)\right)\\ &\xrightarrow[P^A_u\to0]{}\left(S^A_{uv}+S^A_{wu}\right)\log^2\left(\frac{|z-1|^2X}{(1+X)(1+|z|^2X)}\right),
	\end{split}
\end{align}
which is finite at $P^A_{u}=0$, because in the sum $S^A_{uv}+S^A_{wu} = 2 A^{gg(0)} - S^A_{vw}$, the apparent pole at $P^A_{u}=0$ is a removable singularity.
\begin{table}[hb]
	\centering
	\begin{tabular}{c|ccc}
		& $P^A_{u}=0$ & $P^A_{v}=0$ & $P^A_{w}=0$ \\ \hline
		$u_A$                  & 1           & $\frac{|w-1|^2X}{(1+X)(1+|w|^2X)}$           & $\frac{|z-1|^2X}{(1+X)(1+|z|^2X)}$           \\
		$v_A$                  & $\frac{|z-1|^2X}{(1+X)(1+|z|^2X)}$           & 1           & $\frac{|z-1|^2X}{(1+X)(1+|z|^2X)}$           \\
		$w_A$                  & $\frac{|z-1|^2X}{(1+X)(1+|z|^2X)}$           & $\frac{|w-1|^2X}{(1+X)(1+|w|^2X)}$           & 1          
	\end{tabular}
\caption{Behaviour of the cross ratios in the NMRK limit at the three unphysical-singularity surfaces of the rational terms of the ordering $\sigma_A$, listed in \eqn{eq:unphys}.}
\label{table:spurA}
\end{table}

The $\sigma_{A'}$ ordering is exactly analogous. For the $\sigma_{B}$ ordering we have singled out in eq.~(\ref{eq:N4NMHVReWA+R}) the transcendental function $W_1(\sigma_B)$ to multiply the tree-level amplitude. The other transcendental functions that appear in this colour structure are then
\begin{align}
\begin{split}
	W_2(\sigma_B)-W_1(\sigma_B)&=\log\left(v_B\right)\log\left(\frac{u_B}{w_B}\right) \,,\\
	W_3(\sigma_B)-W_1(\sigma_B)&=\log\left(u_B\right)\log\left(\frac{v_B}{w_B}\right)\,,
	\label{eq:WB21_WB31}
\end{split}
\end{align}
which are related to each other by the map $G$ introduced in section~\ref{sec:N4same-hel}. For the NMHV tree-level rational functions in the NMRK limit, this map is equivalent to the composition of interchanging  $p_4\leftrightarrow p_5$ with complex conjugation, i.e.~the remaining terms in \eqn{eq:N4NMHVReWA+R} have the property,
\begin{align}
S^B_{vw} \left(W_2(\sigma_B)-W_1(\sigma_B)\right)\,\,{\underset{p_4 \leftrightarrow p_5}\longleftrightarrow}\,\, \left(S^B_{wu}\right)^*\left(W_3(\sigma_B)-W_1(\sigma_B)\right) \,,
\label{eq:SBflip}
\end{align}
where we have used \eqn{eq:RBswap45}.
Using the properties summarised in table~\ref{table:spurB}, we see that this ordering is also free from unphysical singularities. At $P_{u}^B=0$ or $P_{v}^B=0$, the rational terms in \eqn{eq:N4NMHVReWA+R} which diverge are multiplied by transcendental terms (\eqn{eq:WB21_WB31}) that vanish. At $P_w^B=0$ we see that the transcendental terms in \eqn{eq:WB21_WB31} become equal, allowing us to write
\begin{align}
	\begin{split}
S^B_{vw}(W_2(\sigma_B)-W_1(\sigma_B)) +S^B_{wu}(W_3(\sigma_B)-W_1(\sigma_B)) \xrightarrow[P^B_w\to0]{}(S^B_{vw}+S^B_{wu})\log^2\left(\frac{1}{1+|z|^2X}\right).
	\end{split}
\end{align}
The sum of rational terms $S^B_{vw}+S^B_{wu}$ is finite at $P_w^B=0$, so we have shown that the dispersive part of the amplitude \eqn{eq:N4NMHVReWA+R} is free from unphysical singularities, as expected.
\begin{table}[ht]
	\centering
	\begin{tabular}{c|ccc}
		& $P_{u}^B=0$ & $P_{v}^B=0$ & $P_{w}^B=0$ \\ \hline
		$u_B$                  & 1           & $\frac{1}{1+|1+\frac{w}{z}|^2\frac{1}{X}}$           & $\frac{1}{1+|z|^2X}$           \\
		$v_B$                  & $\frac{1}{1+|z|^2X}$           & 1           & $\frac{1}{1+|z|^2X}$           \\
		$w_B$                  & $\frac{1}{1+|z|^2X}$           & $\frac{1}{1+|1+\frac{w}{z}|^2\frac{1}{X}}$           & 1 
	\end{tabular}
\caption{Behaviour of the cross ratios in the NMRK at the three unphysical-singularity surfaces of the rational terms of the ordering $\sigma_B$, listed in \eqn{eq:unphysB}.}
\label{table:spurB}
\end{table}
\par
As in section~\ref{sec:N4same-hel}, we can write the real part of the transcendental functions exactly in terms of the one-loop trajectory, impact factors and Lipatov vertex, plus a NMRK remainder term. We first separate the transcendental function into a `central' and `non-central' pieces,
\begin{align}
\begin{split}
    2 N_c\kappa_\Gamma \real{W_2(\sigma_A)}=&E(t_1,s_{34};t_3,s_{56})+2 W_2^A \,,
    \end{split}
    \label{eq:W2Aexact}
\end{align}
with $E$ given in \eqn{eq:E}, and we further separate the central function into
\begin{equation}
    2W_2^A=U_A+\Delta W_A(u_A,v_A,w_A) \,.
    \label{eq:W2A}
\end{equation}
The function $U_A$ is given by \eqn{eq:UA}. In other words, this is a helicity-independent term. In contrast, for the opposite-helicity case the remainder function is given by
\begin{equation}
\label{eq:deltaWA}
  \Delta W_A(u_A,v_A,w_A)=N_c\kappa_\Gamma\left(
  -\frac{1}{4}\log^2\left(u_A\right)
  +\frac{1}{2}\log(u_A)\log(v_A w_A)-\log(v_A)\log(w_A)\right)\,.
\end{equation}
Note that, as for the same-helicity case, \eqn{eq:deltaVA}, the variable assignment in \eqn{eq:UA} leads to a NMRK remainder function which is IR-finite and symmetric under $v_A \leftrightarrow w_A$. However, we note that unlike \eqn{eq:deltaVA}, \eqn{eq:deltaWA} does not vanish in the MRK limit. We will return to this point later. When working in NMRK, in order to exploit the equality of the real parts of the transcendental functions under an exchange of $p_1 \leftrightarrow p_6$ or $p_2 \leftrightarrow p_3$, we make the approximation \eqn{eq:UAnmrk}, and we make the NMRK approximation of the cross ratios in \eqn{eq:deltaWA}.
For the $A'$ ordering we similarly write
\begin{align}
\begin{split}
 2 N_c\kappa_\Gamma \real{W_2(\sigma_{A'})}&=E(t_1,s_{35};t_3,s_{46})+ 2W_2^{A'}\,,\\
 2W_2^{A'}&=U_{A'}+\Delta W_{A}(u_{A'},w_{A'},v_{A'}).
\end{split}
\end{align}
where $U_{A'}$ is related to $U_{A}$ by exchange of $p_4\leftrightarrow p_5$. 

As in section~\ref{sec:N4same-hel}, we can write the transcendental function of the $\sigma_B$ ordering in terms of the one-loop factorised pieces in two different ways,
\begin{align}
\begin{split}
    2 N_c\kappa_\Gamma \real{W_1(\sigma_B)}=E(t_1,-s_{24};t_3,s_{56})+2W_1^B=E(t_1,s_{35};t_3,-s_{14})+2W_1^{B'}\,,
    \end{split}
    \label{eq:W1Bexact}
\end{align}
with
\begin{align}
\begin{split}
    2W_1^{B}&=U_{B}+\Delta W_{B}(u_B,v_B,w_B), \qquad \qquad
    2W_1^{B'}=U_{B'}+\Delta W_{B}(v_B,u_B,w_B)\,.
\end{split}
\end{align}
The helicity-independent functions $U_B$ and $U_{B'}$ are given in \eqn{eq:UB}, and the NMRK remainder is
\begin{align}
\label{eq:deltaWB}
    \Delta W_{B}(u_B,v_B,w_B)=N_c\kappa_\Gamma\left(\log(v_B)\log\left( \frac{w_B}{u_B}\right)+\log(u_B)\log\left( \frac{w_B}{v_B}\right)\right)\,.
\end{align}
We note that, similar to \eqn{eq:deltaVB}, \eqn{eq:deltaWB} is IR finite, symmetric under $u_B \leftrightarrow v_B$, but does not vanish in either MRK or $\MRKx$ limits. We emphasise that \eqn{eq:W1Bexact} is exact, but in the following we make the NMRK approximation~(\ref{eq:UBnmrk}) for $U_B$ or $U_{B'}$, after which the functions $U_B$ and $U_{B'}$ are related by $p_4\leftrightarrow p_5$. We also make the NMRK approximation to the cross ratios in \eqn{eq:deltaWB}, after which this function is symmetric under $p_4\leftrightarrow p_5$. 

Following the procedure of section~\ref{sec:N4same-hel}, we now wish to express the opposite-helicity amplitude 
using the colour-structure basis of \eqn{eq:DispFFd}. To this end, we apply the colour rotation of eq.~(\ref{colour-rotation}) to the sum inside the curly brackets in eq.~(\ref{eq:N4NMHVReWA+R}). We find that the structure of the kinematic coefficients is similar to the same-helicity case, motivating the following expression for general helicities, generalising \eqn{eq:DispFFd},
\beqa
	&& \disp{\mathcal{M}^{(1)}_{\mathcal{N}=4}(p_4^{\nu_4}, p_5^{\nu_5})}  \xrightarrow[\rm NMRK] 
	\nn\\&& 
		\frac{\gs^6}{(4\pi)^2}\, 
		s_{12}F^{a_3 a_2 c_1}C^{g(0)}\left(p_2^{\nu_2}, p_3^{-\nu_2}\right)\frac{1}{t_1}\,  \times \nn\\
		&& \sum_{\sigma \in S_2}\Bigg\{  
		(F^{a_{\sigma_4}}F^{a_{\sigma_5}})_{c_1c_3}\left(A^{gg(1)}_{\vDash}(q_1, p^{\nu_{\sigma_4}}_{\sigma_4}, p^{\nu_{\sigma_5}}_{\sigma_5}, q_3)+A^{gg(0)}(q_1, p^{\nu_{\sigma_4}}_{\sigma_4}, p^{\nu_{\sigma_5}}_{\sigma_5}, q_3)E(t_1,s_{3\sigma_4};t_3,s_{\sigma_5 6})\right) \nn\\
		&& \qquad\qquad\qquad +\frac{1}{N_c}d_{A}^{c_1 a_{\sigma_4} a_{\sigma_5} c_3}A^{gg(1)}_{d}(q_1, p^{\nu_{\sigma_4}}_{\sigma_4}, p^{\nu_{\sigma_5}}_{\sigma_5}, q_3)
		\Bigg\}
		\times
		\frac{1}{t_3}F^{a_6 a_1 c_3}C^{g(0)}\left( p_1^{\nu_1},p_6^{-\nu_1}\right) \,.\label{eq:DispFFdgen}
\eeqa
For the opposite-helicity case we find that the kinematic coefficient of the tree-level colour structure is
\beqa
	\lefteqn{ A^{gg(1)}_{\vDash}(q_1, p^\oplus_4, p^\ominus_5, q_3) } \nn\\ &=&\frac{1}{3}
	\Bigg[
	2A^{gg(0)}(q_1, p^\oplus_4, p^\ominus_5, q_3)
	\Big(2W^A_2+W^B_1\Big)
	-2A^{gg(0)}(q_1, p^\ominus_5, p^\oplus_4, q_3)\Big(W^{A'}_2-W^{B'}_1\Big)
	\nn \\&&\quad +2\Big(S^A_{uv}(W^A_1-W^A_2) +S^A_{wu}( W^A_3-W^A_2)\Big)
	-\Big( S^{A'}_{uv}(W^{A'}_1-W^{A'}_2) +S^{A'}_{wu}( W^{A'}_3-W^{A'}_2)\Big)\nn\\
	&& \quad -\Big(S^B_{vw} (W^B_2-W^B_1) +S^B_{wu}( W^B_3-W^B_1)\Big)
	\Bigg] \,.
	\label{eq:Agg1pm_F}
\eeqa
There is some freedom in how to partition the kinematic coefficient of $d_A$ as a sum over colour orderings. Before discussing our choice of how we organise this term, let us investigate the MRK limit of \eqn{eq:Agg1pm_F}. We first drop the rational terms that are power suppressed in this limit,
\beqa
	\lefteqn{ A^{gg(1)}_{\vDash}(q_1, p^\oplus_4, p^\ominus_5, q_3) } \nn\\ &\toMRK& \frac{1}{3}\Bigg[
	A^{gg(0)}(q_1, p^\oplus_4, p^\ominus_5, q_3)
	\Big(3U_{\mathrm{MRK}}+2\Delta W_A(u_A,v_A,w_A)+\Delta W_B(u_B,v_B,w_B) \Big)
	\nn\\ &&\quad +2\Big(S^A_{uv}(W^A_1-W^A_2) +S^A_{wu}( W^A_3-W^A_2)\Big)
	-S^B_{vw} \big(W^B_2-W^B_1\big)
	\Bigg] \,,
	\label{eq:Agg1pm_F_MRK}
\eeqa
where $U_{\mathrm{MRK}}$ is the helicity-independent MRK limit of the central transcendental functions, introduced in \eqn{eq:UMRK}. We now note that the nonvanishing terms of $\Delta W_A$ and $\Delta W_B$ cancel against the MRK limits of the terms proportional to $S^A$ and $S^B$ respectively, i.e.
\begin{align}
\begin{split}
	A^{gg(0)}(q_1, p^\oplus_4, p^\ominus_5, q_3)\Delta W_A(u_A,v_A,w_A) &\underset{\mathrm{MRK}}{=} -S^A_{uv}(W^A_1-W^A_2) -S^A_{wu}(W^A_3-W^A_2) \,,\\
	A^{gg(0)}(q_1, p^\oplus_4, p^\ominus_5, q_3)\Delta W_B(u_B,v_B,w_B) &\underset{\mathrm{MRK}}{=}
	S^B_{vw} (W^B_2-W^B_1) \,.
	\label{eq:NMHV_MRK_cancel}
	\end{split}
\end{align}
With this we see that the MRK limit of \eqn{eq:Agg1pm_F_MRK}, 
\begin{align}
\begin{split}
	A^{gg(1)}_{\vDash}(q_1, p^\oplus_4, p^\ominus_5, q_3)
	&\toMRK
	V^{g(0)}(q_1,p_4^\oplus,q_2)\frac{1}{t_2}V^{g(0)}(q_2,p_5^\ominus,q_3)
	\, U_{\mathrm{MRK}}\, ,
	\label{eq:Agg1_F_MRK}
\end{split}
\end{align}
coincides with the MRK limit of the same-helicity amplitude, \eqn{eq:AggF1_MRK}, manifesting the helicity independence of the MRK limit.
A similar analysis shows that the $\MRKx$ limit of \eqn{eq:Agg1pm_F_MRK},
\begin{align}
\begin{split}
	A^{gg(1)}_{\vDash}(q_1, p^\ominus_5, p^\oplus_4, q_3)
	&\toMRKx
	V^{g(0)}(q_1,p_5^\ominus,q_2^\prime)\frac{1}{t_2^\prime}V^{g(0)}(q_2^\prime,p_4^\oplus,q_3)
	\, U_{\mathrm{MRK'}}\, ,
	\label{eq:Agg1_F_MRKx}
\end{split}
\end{align}
coincides with the $\MRKx$ limit of the same-helicity vertex, \eqn{eq:AggF1_MRKx}.

Equation~(\ref{eq:NMHV_MRK_cancel}) also provides our motivation for writing the kinematic coefficient of the totally symmetric colour structure as
\begin{align}
	\begin{split}
	A^{gg(1)}_{d}(q_1, p^\oplus_4, p^\ominus_5, q_3)&=
	2A^{gg(0)}(q_1, p^\oplus_4, p^\ominus_5, q_3)\left(W_2^A-W_1^B\right)\\
	&+S^A_{uv}(W^A_1-W^A_2) +S^A_{wu}( W^A_3-W^A_2)+S^B_{vw}(W^B_2-W^B_1),
	\end{split}
	\label{eq:Agg1pm_d}
\end{align}
with the aforementioned property \eqn{eq:SBflip} leading to
\begin{align}
	\begin{split}
	A^{gg(1)}_{d}(q_1,  p^\ominus_5,p^\oplus_4, q_3) &=
	2A^{gg(0)}(q_1, p^\ominus_5, p^\oplus_4, q_3)\left(W^{A'}_2-W_1^{B'}\right)\\
	&+S^{A'}_{uv}(W^{A'}_1-W^{A'}_2) +S^{A'}_{wu}(W^{A'}_3-W^{A'}_2)
	+S^B_{wu}( W^B_3-W^B_1).
	\end{split}
	\label{eq:Agg1mp_d}
\end{align}
This choice gives us the simple MRK and MRK$'$ limits,
\begin{align}
\begin{split}
A^{gg(1)}_{d}(q_1, p^\oplus_4, p^\ominus_5, q_3)
	&\toMRK
	0\, ,\\
	A^{gg(1)}_{d}(q_1, p^\ominus_5, p^\oplus_4,  q_3)
	&\toMRKx
	0\, .
	\label{eq:Agg1pm_d_MRK}
\end{split}
\end{align}
similar to eqs.~(\ref{eq:Aggd1_MRK}) and (\ref{eq:Aggd1_MRKx}).

\subsection{NLO \texorpdfstring{$gg \to gggg$}{\textit{gg} to \textit{gggg}} squared matrix element in the central NMRK limit}

\label{sec:NLO_xs}
Having obtained the central NMRK limit of the six-gluon one-loop amplitude in ${\cN=4}$~SYM for all leading helicity configurations, we now consider the contribution of this amplitude to the squared matrix element for
 $gg\to gggg$ at NLO, i.e. the interference of this amplitude with the tree-level amplitude given in eqs.~(\ref{NLOfactorization}) and (\ref{NLOfactorization2}), summed over all colour indices:
\begin{equation}
    2\real{\cM^{(0)*}_{6g}\cM^{(1)}_{6g}}=
    2\real{\cM^{(0)*}_{6g}\mathrm{Disp}\left[\cM^{(1)}_{6g}\right]}-2\imag{\cM^{(0)*}_{6g}\mathrm{Absorp}\left[\cM^{(1)}_{6g}\right]}\,.
    \label{eq:1loop_interference}
\end{equation}
We consider these two contributions in turn.

\subsubsection*{Dispersive part} 
\begin{figure}[bht]
\centering
    \subfigure[]{\label{fig:colour_graphs_FF}\includegraphics[scale=0.16]{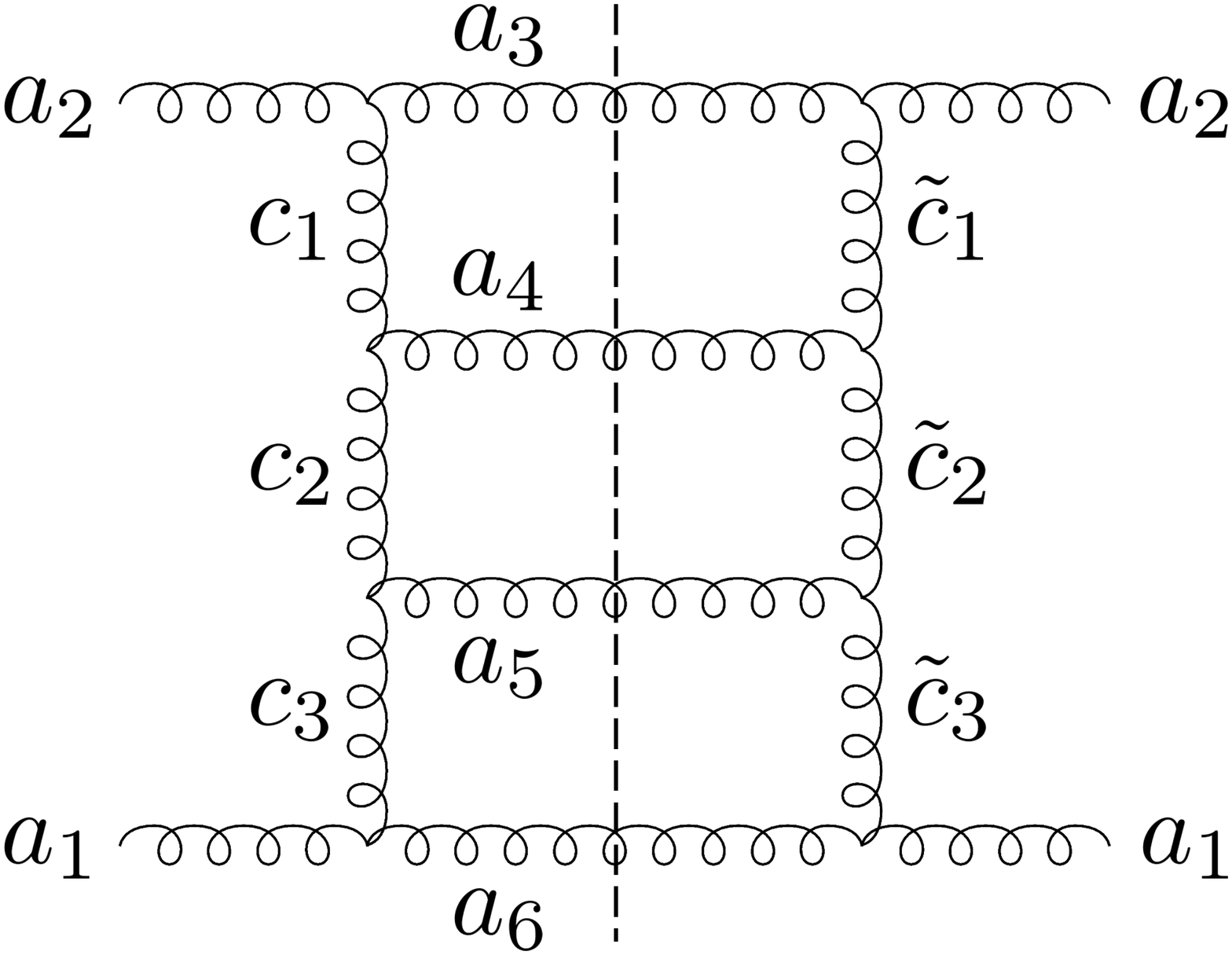}}
    \subfigure[]{\label{fig:colour_graphs_FFx}\includegraphics[scale=0.16]{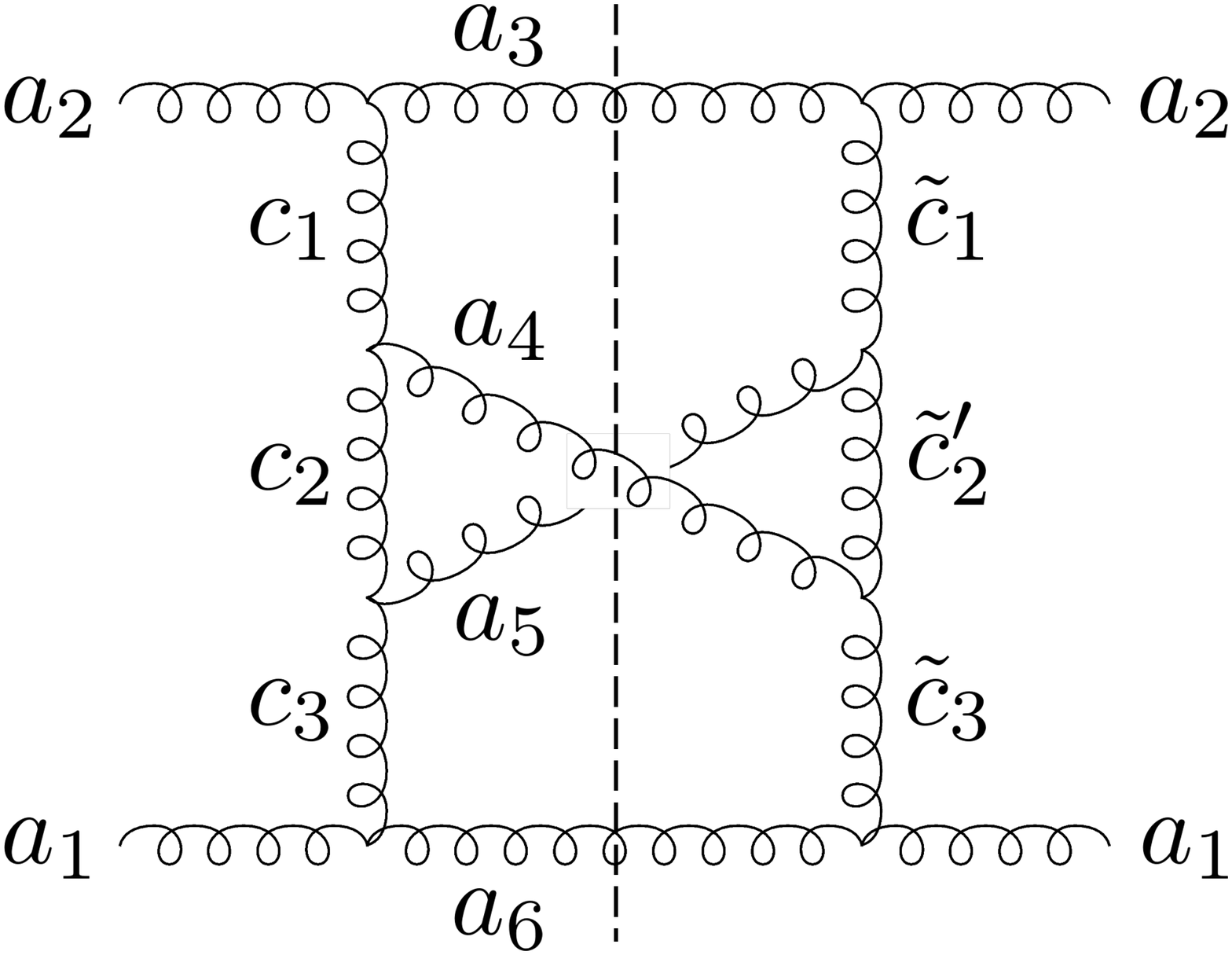}}
    \subfigure[]{\label{fig:colour_graphs_d}\includegraphics[scale=0.16]{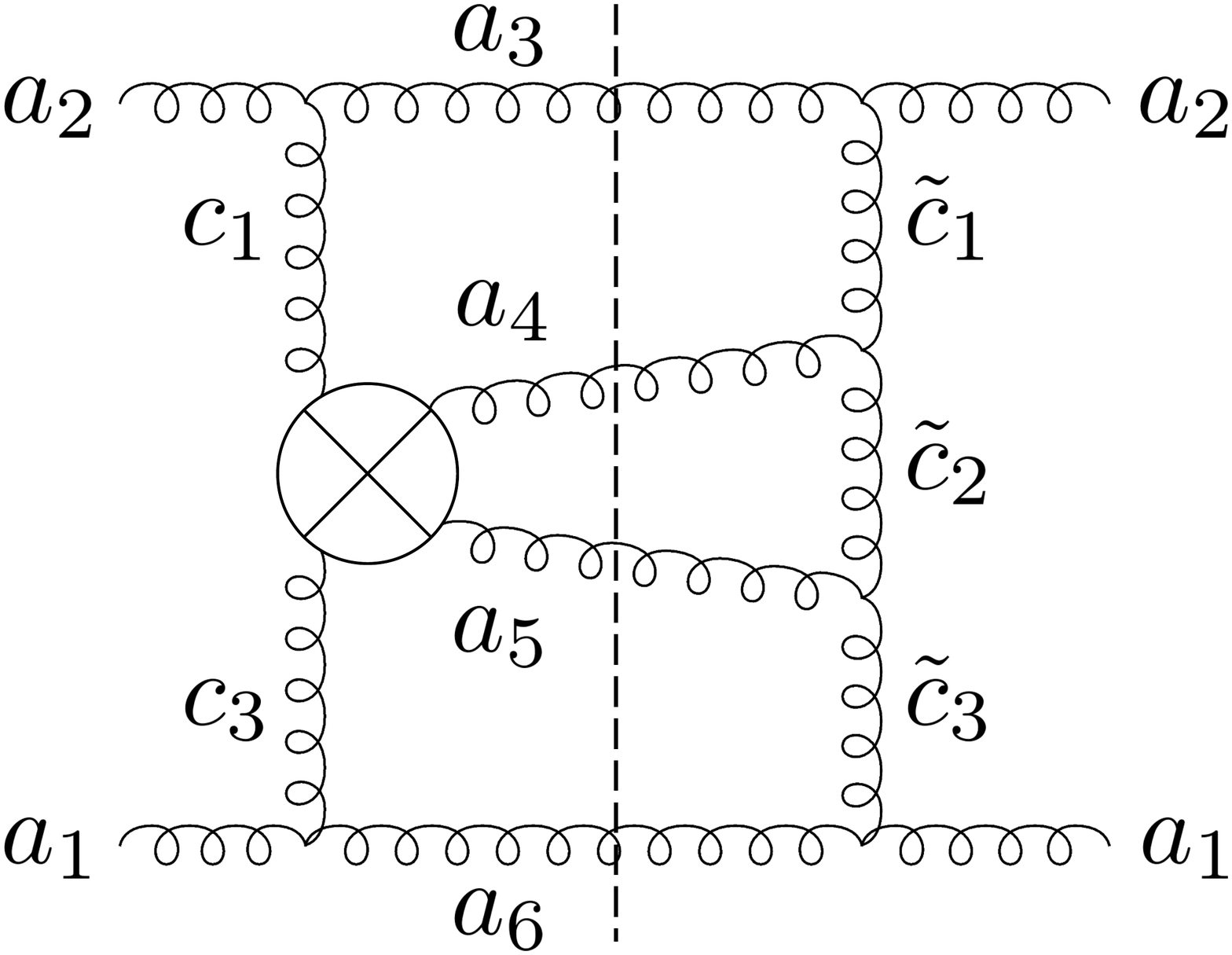}}
    \caption{Three of the colour structures contributing to  $\cM^{(0)*}_{6g}\mathrm{Disp}\big[\cM^{(1)}_{6g}\big]$ using the basis of one-loop colour structures in \eqn{eq:DispFFdgen}. The remaining three colour structures can be obtained by interchanging $a_4 \leftrightarrow a_5$. A three-gluon vertex represents the adjoint generator $F$, while the $\otimes$~symbol represents the fully symmetric trace $d_A$ of eq.~(\ref{eq:dA}).}
    \label{fig:interference}
\end{figure}

We first consider the contribution of the dispersive part of the one-loop amplitude, \eqn{eq:DispFFdgen},  to \eqn{eq:1loop_interference}. 
The colour structures contributing to  $\cM^{(0)*}_{6g}\mathrm{Disp}\big[\cM^{(1)}_{6g}\big]$ are depicted as interference diagrams in figure~\ref{fig:interference}. For each of them, after summing over the colour indices $a_2$, $a_3$ and $a_1$, $a_6$, the factors associated with the impact factors are proportional to the identity in the adjoint representation, i.e.
\begin{equation}
\tr(F^{c_1}F^{\tilde{c}_1})=2N_c\delta^{c_1\tilde{c}_1}\,,
\qquad \qquad\tr(F^{c_3}F^{\tilde{c}_3})=2N_c\delta^{c_3\tilde{c}_3}\,.
\label{eq:squared_IFs}
\end{equation}

We then see that the colour structures involving the totally symmetric tensor $d_A$ vanish after contraction with the partially antisymmetric tree-level colour-structure:
\begin{align}
\label{d_A_demise}
4 N_c^2\delta^{c_1\tilde{c}_1} \delta^{c_3\tilde{c}_3} d_A^{c_1a_4 a_5 c_3}(F^{a_4}F^{a_5})_{\tilde{c}_1 \tilde{c}_3}
=
-8 N_c^2 d_A^{c_1a_4 a_5 c_3}f^{c_1 a_4\tilde{c}_2} f^{\tilde{c}_2 a_5c_3}
=0.
\end{align}
The coefficient $A_d^{gg(1)}$ of the two-gluon CEV in \eqn{eq:DispFFdgen} therefore does not contribute to the NLO $gg \to 4\, \text{jet}$ cross section, or by extension, to the N$^3$LO dijet cross section after phase-space integration. We emphasise that this conclusion applies to any helicity configuration of the one-loop two-gluon CEV. 

\begin{figure}[hbt]
\centering
     \includegraphics[scale=0.2]{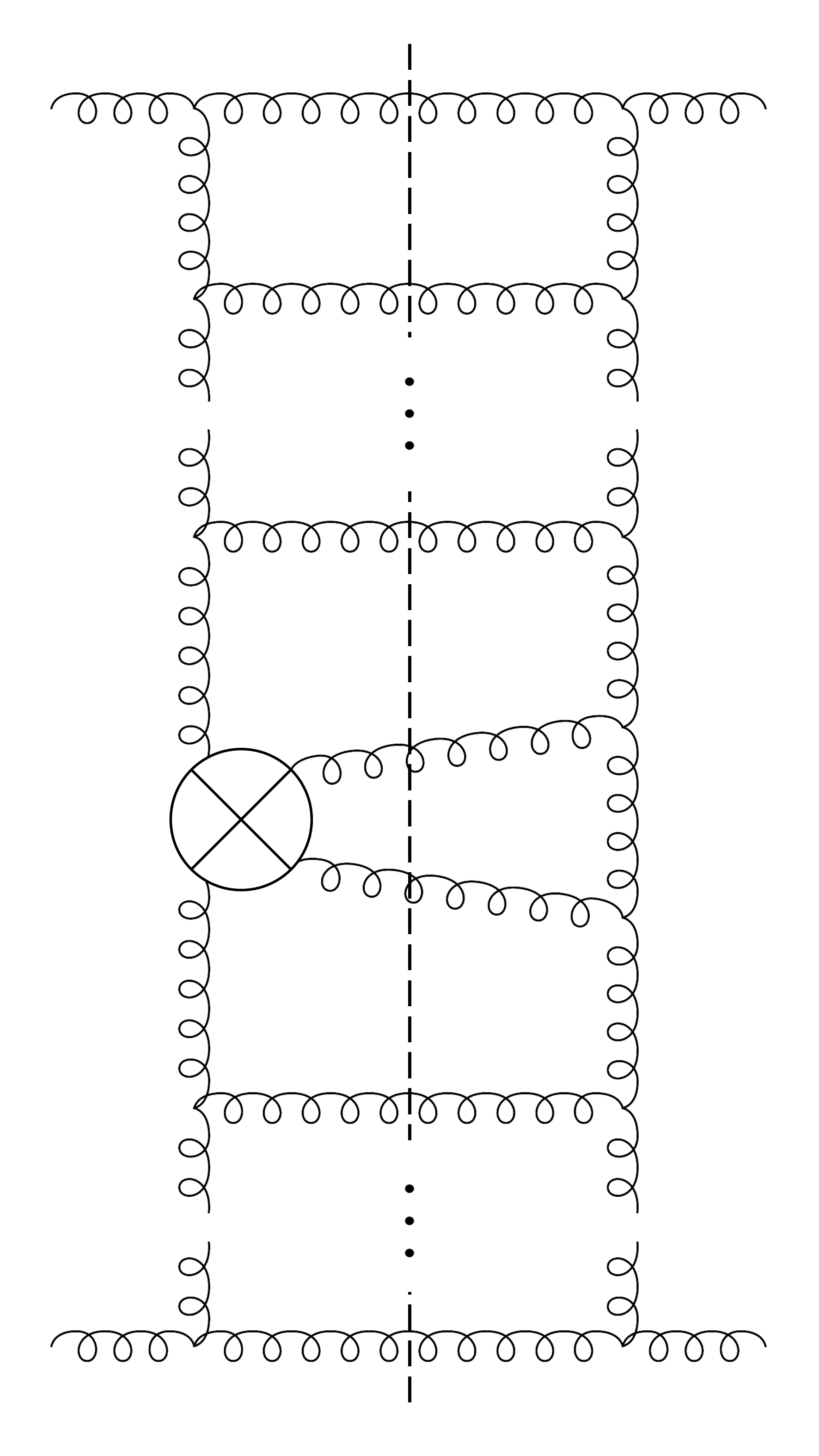}
    \caption{A colour structure of an interference, or `ladder' diagram at NNLL accuracy, containing one instance of the one-loop two-gluon CEV, and all other real emissions given by tree-level one-gluon CEVs. The $\otimes$ symbol represents the fully symmetric trace $d_A$ of the two-gluon CEV. This colour structure vanishes upon summing over all colour indices, as discussed in the main text. 
    }
    \label{fig:ladder_dA}
\end{figure}

By a similar argument, the symmetric colour structure of the two-gluon CEV does not contribute to the NNLL BFKL kernel. 
To see this we note that at this logarithmic accuracy, the relevant interference, or `ladder' diagrams describing an arbitrary number of real emissions along the ladder, including precisely one instance of the one-loop two-gluon CEV, have all other `rungs' generated by the interference of tree-level impact factors and one-gluon CEVs (those appearing in fig.~\ref{fig:BFKL_LL_Building_blocks} (b)). The colour structure containing $d_A$ of such a ladder diagram is depicted in figure~\ref{fig:ladder_dA}. Given that the colour structure of the tree-level impact factors are either a fundamental generator or adjoint structure constant, and that the colour-structure of any one-gluon CEV is simply a structure constant, it follows that upon summing over the colour indices, the entire ladder of rungs above and below the one-loop two-gluon CEV collapse with the result proportional the identity in the adjoint representation (with each rung yielding an extra factor of $2N_c$). At the final stage, the contraction between $d_A$ and the structure constant yields a vanishing result, just as in eq.~(\ref{d_A_demise}).

We further note that the symmetric colour structure of the two-gluon CEV is expected to contribute to the cross section and to the BFKL kernel at N$^3$LL accuracy: at this order one must consider the square of the one-loop two-gluon CEV, where the interference involving two $d_A$ terms survives, and yields a quartic Casimir invariant.

Having seen that the symmetric trace of four generators, $d_A$, which appears in the two-gluon CEV, does not contribute to the cross section or the BFKL kernel at NNLL, it is interesting to ask whether such symmetric colour structures contribute at this logarithmic accuracy to \emph{any} of the interference diagrams making up the BFKL kernel in fig.~\ref{fig:BFKL_NNLL_Building_blocks}. The answer is negative: upon summing over colour indices, the impact factors in any of these interference diagrams would be proportional to the identity in the adjoint representation. This argument holds for both gluon or quark impact factors. Having contracted the impact factors, the virtual loops and the central gluons only involve six colour vertices in total, which cannot form a quartic Casimir in any of the contributions in fig.~\ref{fig:BFKL_NNLL_Building_blocks}. 
This may be contrasted with contributions to the $2 \to 2$ amplitude from triple Reggeon exchange, which do involve quartic Casimirs at NNLL accuracy~\cite{Falcioni:2020lvv,Falcioni:2021buo,Falcioni:2021dgr}. 

\subsubsection*{Absorptive part} 
We now consider the contribution of the absorptive part of the one-loop amplitude to \eqn{eq:1loop_interference}. As mentioned above, for the absorptive part of the one-loop amplitude, all signature components contribute at leading power in the NMRK limit, although they do not have logarithmic dependence on the large invariants $s_{34}$ or $s_{56}$. The four signature components are listed in eqs.~(\ref{eq:abN41loop})--(\ref{eq:absorpM++}) for $\nu_4=\nu_5$ and in eqs.~(\ref{eq:absorpM--nmhv})--(\ref{eq:absorpM++nmhv}) for \hbox{$\nu_4=-\nu_5$}. However, the leading-power behaviour of the tree-level amplitude is given by the $(-,-)$ component alone, and this projects out the $(-,-)$ component of the absorptive one-loop amplitude
in $\cM^{(0)*}_{6g}\mathrm{Absorp}\big[\cM^{(1)}_{6g}\big]$ upon summing over the external colour indices. For the $(-,-)$ component of the absorptive part of the amplitude (\eqn{eq:abN41loop} for the $\nu_4=\nu_5$ case and \eqn{eq:absorpM--nmhv} for the $\nu_4=-\nu_5$ case), we can factor out an adjoint generator associated to the tree-level impact factors.  Upon considering the interference between these absorptive parts and the tree-level amplitude, we therefore find that the colour factors associated with the impact factors are identical to the dispersive case above, \eqn{eq:squared_IFs}.

To proceed and evaluate the imaginary part of $\cM^{(0)*}_{6g}\mathrm{Absorp}\big[\cM^{(1)}_{6g}\big]$ we now distinguish between the same-helicity case, where the relevant absorptive part is given in \eqn{eq:abN41loop}, and the opposite-helicity case, where the relevant absorptive part is given in \eqn{eq:absorpM--nmhv}. For the former case, the colour factors which contain $\tr(F^{c_1} F^{a_4} F^{a_5} F^{c_3})$ or $\tr(F^{c_1} F^{a_5} F^{a_4} F^{c_3})$ do not contribute to the absorptive interference term, due to either the vanishing of the colour factor or the reality of its kinematic coefficient. The remaining contribution is proportional to
\begin{align}
\begin{split}
&\tr(F^{c_1} F^{a_5} F^{c_3} F^{a_4})(F^{a_4}F^{a_5})_{c_1 c_3}\mathrm{Im}\left[B^{gg(0)}(q_1, p^\oplus_4, p^\oplus_5, q_3)A^{gg(0)*}(q_1, p^\oplus_4, p^\oplus_5, q_3)\right]\\
+\,&\tr(F^{c_1} F^{a_5} F^{c_3} F^{a_4})(F^{a_5}F^{a_4})_{c_1 c_3}\mathrm{Im}\left[B^{gg(0)}(q_1, p^\oplus_4, p^\oplus_5, q_3)A^{gg(0)*}(q_1, p^\oplus_5, p^\oplus_4, q_3)\right]=0.
\end{split}
\end{align}
To show that this quantity is zero we first note that the two colour factors are equal, which can be shown using the reversal-invariance property of the trace of four adjoint generators, along with a relabelling of the indices $a_4$ and $a_5$. After factoring out the common colour factor one is left with the imaginary part of a kinematic factor which can be shown to be real using \eqn{eq:ABrelation}. We conclude that the absorptive part of the amplitude does not contribute to the squared matrix element for $gg\to gggg$ at NLO in the central NMRK limit for the helicity configurations with $\nu_4=\nu_5$.

Repeating the above analysis for the opposite-helicity case we find a non-vanishing result. Specifically, for $\nu_4=\oplus$ and $\nu_5=\ominus$ we find that
\begin{align}
\label{eq:NLO_Absorp}
\begin{split}
2\mathrm{Im}
\Big[\cM^{(0)*}_{6g}(p^\oplus_4, p^\ominus_5)&
\, \mathrm{Absorp}
\left[\cM^{(1)}_{6g}(p^\oplus_4, p^\ominus_5)\right]
\Big]
\toNMRK 2 \, \gs^8\frac{\as}{4\pi}\frac{\pi}{2}\frac{s^2}{t_1^2 t_3^2}\left|C^{g(0)}(p_2^{\nu_2},p_3^{-\nu_2}) \right|^2
\\
\times 8 \, N_c^5(N_c^2-1)\bigg\{ 
&\mathrm{Im}\left[\Big(S_{uv}^A\log(v_A)+S_{wu}^A\log(w_A)\Big)A^{gg(0)*}(q_1, p^\oplus_4, p^\ominus_5, q_3)\right]\\
+&\mathrm{Im}\left[\left(S_{uv}^{A'}\log(v_{A'})+S_{wu}^{A'}\log(w_{A'})\right)A^{gg(0)*}(q_1, p^\ominus_5, p^\oplus_4, q_3)\right]\\
-&\mathrm{Im}\left[S_{uv}^B
    B^{gg(0)*}(q_1, p^\oplus_4, p^\ominus_5, q_3)\right]\log\left(\frac{w_B}{u_B v_B}\right)\bigg\}\times \left|C^{g(0)}(p_1^{\nu_1},p_6^{-\nu_1}) \right|^2\,,
    \end{split}
\end{align}
which is generically non-zero. 

Next we observe that the absorptive interference term for $\nu_4=\ominus$ and $\nu_5=\oplus$ is equal to the complex conjugate of \eqn{eq:NLO_Absorp}. Therefore, the absorptive part of the one-loop amplitude in NMRK does not contribute to the helicity-summed squared matrix element for $gg\to gggg$ at NLO. Thus, the contribution of the one-loop six-gluon amplitude in the central NMRK limit to the cross section at this logarithmic accuracy is entirely driven by the dispersive part.

\section{All-orders conjectures for amplitudes in a central NMRK limit}
\label{sec:all-orders}

In section \ref{sec:6g1lamp} we discussed the six-gluon ${\cal N}=4$ SYM amplitude to one-loop accuracy. However, the results of section \ref{sec:N4same-hel} and \ref{sec:N4opp-hel} are suggestive of the factorised form that amplitudes may take in the central NMRK limit at higher orders, and potentially at higher multiplicity. In this section we therefore make conjectures for the form that the dispersive part of amplitudes may take beyond one loop and beyond the six-gluon case.

\begin{figure}[hbt]
\centering
     \subfigure[]{\label{fig:6g_NMRK}\includegraphics[scale=0.3]{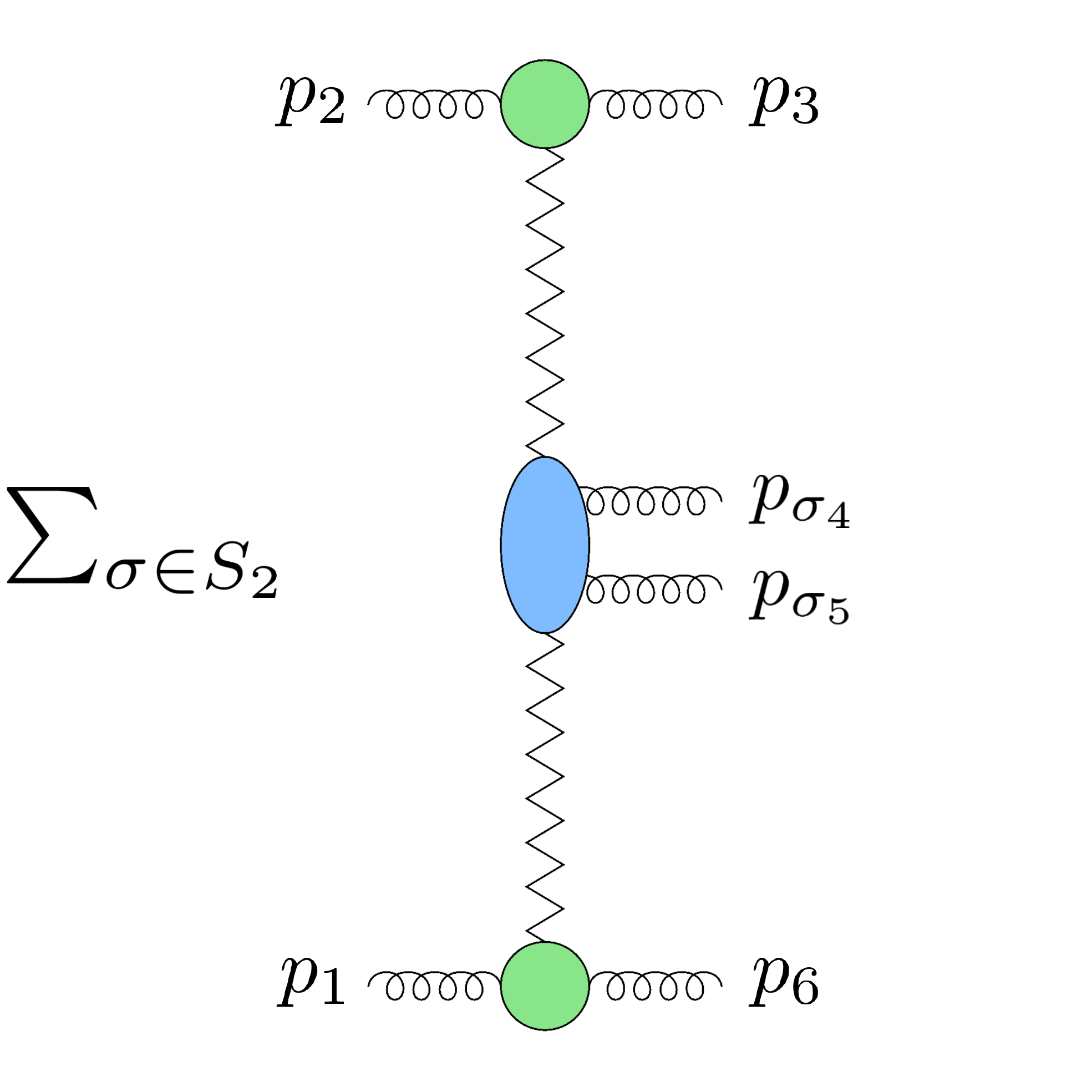}}
    \hspace*{40pt}
    \subfigure[]{\label{fig:ng_NMRK}\includegraphics[scale=0.3]{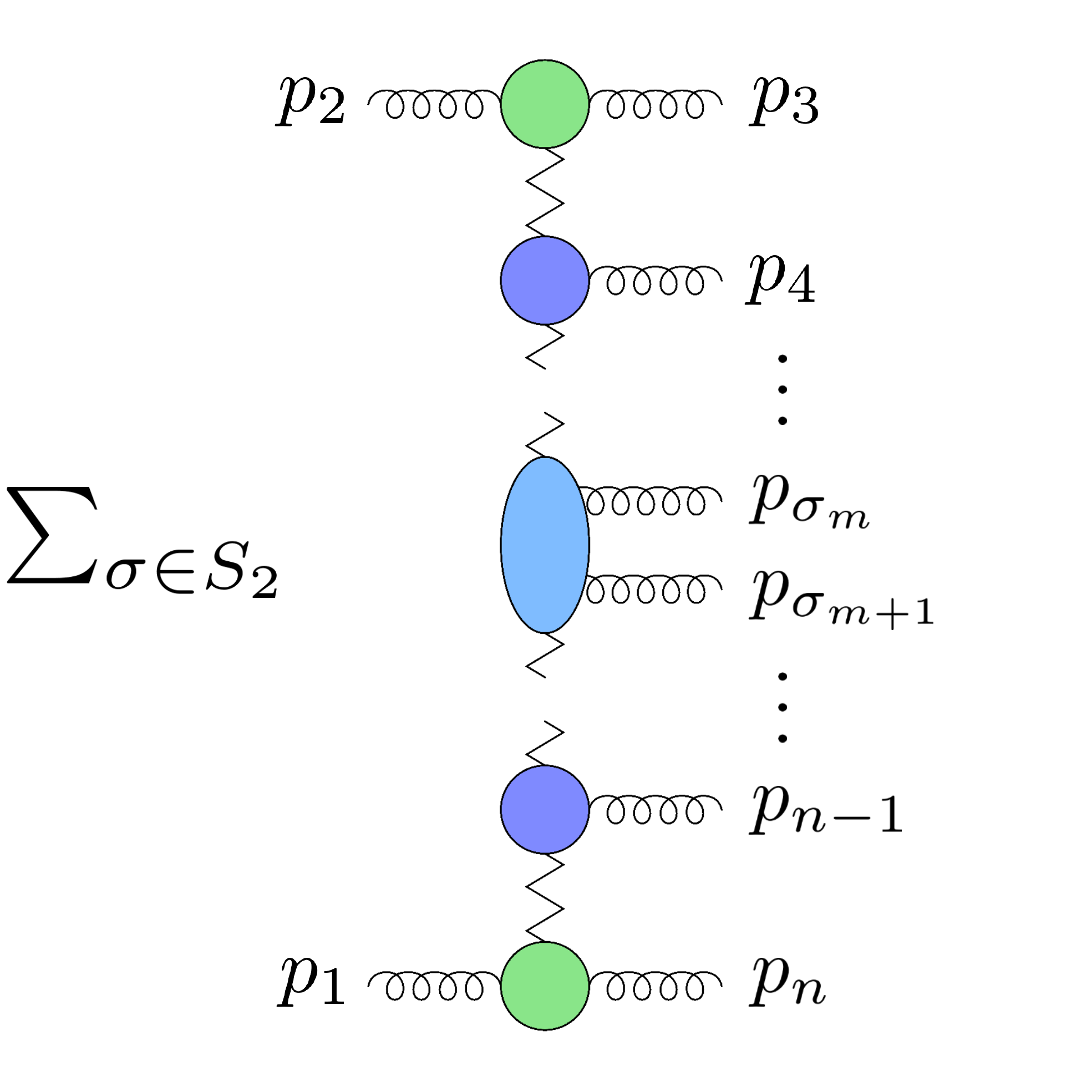}}
    \caption{(a) The expected factorised picture of the dispersive part of a six-gluon amplitude in the central NMRK limit. (b) The expected factorization of an $n$-gluon amplitude in a central NMRK limit. The coloured blobs represent impact factors, and one- and two gluon CEVs, defined to all orders in perturbation theory, similarly to figure~\ref{fig:ng_MRK}.  
}
    \label{fig:NMRK_factorisation}
\end{figure}

The separation of the non-central $E$ functions in \eqn{eq:DispFFdgen}, multiplying the tree-level two-gluon CEV, strongly suggests a factorised picture, analogous to \eqn{threeall}, for the production of two gluons in the central NMRK limit. The one-loop two-gluon CEV extracted this way is a component of BFKL at NNLL accuracy.
As noted in ref.~\cite{Canay:2021yvm}, one can place limits on the pattern of Reggeization by requiring that the MRK limit of said CEV agrees with the known behaviour of amplitudes in MRK. 
The following Reggeization ansatz for the production of two gluons in the central NMRK limit is compatible with both the known MRK limit and with the structure of the dispersive part of the one-loop amplitude~\eqn{eq:DispFFdgen},
\begin{align}
\begin{split}
&\disp{\cM^{(-,-)}_{6g}} \toNMRK 
s \left[\gs (F^{a_3})_{a_2c_1}\, C^{g}(p_2^{\nu_2}, p_3^{\nu_3}) \right]\\
& \hspace{1cm}\times\sum_{\sigma \in S_2}\Bigg\{
\frac{1}{t_1} \, \frac{1}{2}
\left[ \left(\frac{s_{3\sigma_4}}{\tau}\right)^{\alpha(t_1)} + \left(\frac{-s_{3\sigma_4}}{\tau}\right)^{\alpha(t_1)} \right]\\
&\hspace{1cm}\times 
\left[\gs^2 (F^{a_{\sigma_4}}F^{a_{\sigma_5}})_{c_1c_3}\, 
A^{gg}_{\vDash}(q_1,p^{\nu_{\sigma_4}}_{\sigma_4},p^{\nu_{\sigma_5}}_{\sigma_5},q_3) +\frac{1}{N_c}d^{c_1 a_{\sigma_4} a_{\sigma_5} c_3}_{A}A^{gg}_{d}(q_1, p^{\nu_{\sigma_4}}_{\sigma_4}, p^{\nu_{\sigma_5}}_{\sigma_5}, q_3) \right]\, 
\\
&\hspace{1cm}\times \frac{1}{t_3} \, \frac{1}{2}
\left[ \left(\frac{s_{{\sigma_5}6}}{\tau}\right)^{\alpha(t_3)} + \left(\frac{-s_{{\sigma_5}6}}{\tau}\right)^{\alpha(t_3)} \right]
\Bigg\}\\
&\hspace{1cm}\times
\left[\gs (F^{a_6})_{a_1c_3}\, C^{g}(p_1^{\nu_1}, p_6^{\nu_6}) \right] \,. 
\label{eq:DispM6g_ansatz}
\end{split}
\end{align}
This factorization is illustrated schematically in figure \ref{fig:NMRK_factorisation}~(a). The first thing to note is that, unlike amplitudes at LL, or signature-odd amplitudes at NLL, the amplitude is not fully factorised. Rather, \eqn{eq:DispM6g_ansatz} is rather a sum over two colour orderings, each of which is factorised in the familiar form of impact factors, Reggeized gluons and a two-gluon CEV.

Further to this point are the assumptions we have made regarding the colour structure $d_A$ beyond one-loop accuracy. In \eqn{eq:DispM6g_ansatz} we have assumed that this 
totally-symmetric colour factor couples to Reggeized gluons at higher orders in $\as$. Specifically, 
\eqn{eq:DispM6g_ansatz} assumes a particular form for this Reggeization, based on the partitioning of the kinematic coefficient of $d_A$, sections~\ref{sec:N4same-hel} and~\ref{sec:N4opp-hel}. While this partitioning seems natural from the point of view of requiring compatibility with the known MRK limit, the nature of Reggeization of this term of course cannot be determined from the one-loop amplitude studied in this paper. 
It may be constrained by taking the central NMRK limit of the two-loop six-gluon amplitude. A similar structure is accessible through the non-central NMRK limit of the two-loop five-gluon amplitude. 
\par
With these assumptions in mind, we may consider how \eqn{eq:DispM6g_ansatz} may be generalised to the case of $2\to n$ scattering in any of the central NMRK limits where the rapidity separation between a pair of consecutive central gluons denoted by $m$ and $m+1$ (with $4\leq m\leq n-2$) is not necessarily large, while all other rapidity separations are large.  
We propose the following ansatz for an arbitrary number of legs:
\begin{align}
\label{many_gluons_NMRK}
\begin{split}
&\disp{\cM^{[\mathbf{8}_a]}_{ng}} \toNMRK 
s 
\left[\gs (F^{a_3})_{a_2c_1}\, C^{g}(p_2^{\nu_2}, p_3^{\nu_3}) \right]\\
&\times\prod_{i=4}^{m-1} \left( 
\frac{1}{t_{i-3}} \, \frac{1}{2}
\left[ \left(\frac{s_{i-1,i}}{\tau}\right)^{\alpha(t_{i-3})} + \left(\frac{-s_{i-1,i}}{\tau}\right)^{\alpha(t_{i-3})} \right]
\left[\gs(F^{a_{i}})_{c_{i-3} c_{i-2}}V^g(q_{i-3},p_{i}^{\nu_{i}},q_{i-2})\right]
\right)\\
&\times\sum_{\sigma \in S_2}\Bigg\{
\frac{1}{t_{m-3}} \, \frac{1}{2}
\left[ \left(\frac{s_{m-1,\sigma_m}}{\tau}\right)^{\alpha(t_{m-3})} + \left(\frac{-s_{m-1,\sigma_m}}{\tau}\right)^{\alpha(t_{m-3})} \right]\\
&\times 
\bigg[
\gs^2 (F^{a_{\sigma_m}}F^{a_{\sigma_{m+1}}})_{c_{m-3}c_{m-1}}\, 
A^{gg}_{\vDash}(q_{m-3},p^{\nu_{\sigma_m}}_{\sigma_{m}},p^{\nu_{\sigma_{m+1}}}_{\sigma_{m+1}},q_{m-1})
\\& \hspace*{150pt}
+\frac{1}{N_c}d^{c_{m-3} a_{\sigma_m} a_{\sigma_{m+1}} c_{m-1}}_{A}A^{gg}_{d}(q_{m-3}, p^{\nu_{\sigma_{m}}}_{\sigma_m}, p^{\nu_{\sigma_{m+1}}}_{\sigma_{m+1}}, q_{m-1}) 
\bigg]\, 
\\
&\times \frac{1}{t_{m-1}} \, \frac{1}{2}
\left[
\left(\frac{s_{\sigma_{m+1},m+2}}{\tau}\right)^{\alpha(t_{m-1})} + \left(\frac{-s_{\sigma_{m+1},m+2}}{ \tau}\right)^{\alpha(t_{m-1})} \right]
\Bigg\}\\
&\times\prod_{i=m+2}^{n-1} \left( 
\left[\gs(F^{a_{i}})_{c_{i-3} c_{i-2}}V^g(q_{i-3},p_{i}^{\nu_{i}},q_{i-2})\right]
\frac{1}{t_{i-2}} \, \frac{1}{2}
\left[ \left(\frac{s_{i,i+1}}{\tau}\right)^{\alpha(t_{i-2})} + \left(\frac{-s_{i,i+1}}{\tau}\right)^{\alpha(t_{i-2})} \right]
\right)\\
&\times
\left[\gs (F^{a_n})_{a_1c_{n-3}}\, C^{g}(p_1^{\nu_1}, p_n^{\nu_n}) \right]\,,
\end{split}
\end{align}
which is written entirely in terms of the components extracted from  $n\leq 6$ amplitudes.  This factorised structure is depicted schematically in figure \ref{fig:NMRK_factorisation} (b), where real emissions that are strongly ordered in rapidity are described in a similar way to the MRK structure depicted in figure~\ref{fig:ng_MRK}.

\section{Conclusions}
\label{sec:conclusions}

Taking the one-loop six-gluon amplitudes of $\mathcal{N}=4$ SYM~\cite{Bern:1994zx,Bern:1994cg} into the central NMRK limit, in this paper we have obtained the one-loop corrections to the two-gluon central-emission vertex (CEV) in this theory, in both the same-helicity and opposite-helicity configurations. This is a first step toward the determination of the same quantity in QCD. 

In contrast with the one-gluon CEV, 
we find that the two-gluon CEV, \eqn{eq:Modd-odd}, receives one-loop corrections involving a new colour structure consisting of a fully symmetric trace of four adjoint generators, 
which is not present at tree level. We find, however, that this new structure does not contribute to the squared matrix element for $gg\to gggg$ at NLO, so it does not enter the NNLL BFKL kernel. We note, however that this symmetric colour structure does contribute to the squared matrix element at N$^3$LL accuracy, where the interference between two one-loop two-gluon CEVs gives rise to a quartic Casimir invariant.

As discussed in section \ref{sec:NLO_xs}, none of the NNLL contributions to the BFKL kernel, which are described by the interference diagrams in fig.~\ref{fig:BFKL_NNLL_Building_blocks}, gives rise to a quartic Casimir. This may be contrasted with contributions to the $2 \to 2$ amplitude from triple-Reggeon exchange, which do involve quartic Casimirs at NNLL accuracy~\cite{Falcioni:2020lvv,Falcioni:2021buo,Falcioni:2021dgr}. 

In addition to the new colour structure, also the kinematic dependence of the two-gluon CEV is rather complex. In the same-helicity two-gluon CEV, which is extracted from an MHV amplitude, dilogarithms survive in the NMRK limit, \eqns{eq:deltaVA}{eq:deltaVB}. The dilogarithms depend on conformally-invariant cross ratios (\ref{eq:uvwA}) and (\ref{eq:uvwB}), the same variables which characterize six-gluon amplitudes in planar $\mathcal{N}=4$ SYM~\cite{Drummond:2007au}.

Yet another layer of complexity arises in the case of the one-loop NMHV amplitude, from which the opposite-helicity two-gluon CEV is extracted.
Considering this amplitude in central NMRK, \eqn{eq:DispFFdgen}, \eqn{eq:Agg1pm_F} and \eqn{eq:Agg1pm_d},  we find rational coefficients which are not proportional to the tree amplitude, accompanied by different transcendental functions, \eqn{eq:N4NMHVReW}. 
This structure involves an interplay between spurious poles of these rational coefficients and zeroes of the transcendental functions multiplying them, see eqs.~(\ref{eq:N4NMHVReWA+R})--(\ref{eq:WB21_WB31}) and tables~\ref{table:spurA} and \ref{table:spurB}. 
These features propagate into the one-loop opposite-helicity CEV.

Another consequence of the more complicated structure of the one-loop NMHV amplitude is that, in contrast with the MHV case, an absorptive part survives in the squared matrix element for $gg\to gggg$ at NLO, after summing over colour indices, as shown in section~\ref{sec:NLO_xs}.
However, upon summing over gluon helicities, this absorptive contribution vanishes, and thus only the dispersive part of the one-loop six-gluon amplitude contributes to unpolarised cross sections at NNLL. 

In ref.~\cite{Bartels:2008ce}, it was noted that the transcendental functions of the one-loop six-gluon MHV amplitude
can be written exactly in terms of the known one-loop Regge components, \eqn{eq:V6Aexact}, consisting of impact factors, Reggeized gluons and Lipatov vertices, plus a remainder term (\ref{eq:deltaVA}) written in terms of the conformally-invariant cross ratios (\ref{eq:uvwA}). 
In \eqns{eq:W2Aexact}{eq:deltaWA}, we have shown that this holds also for the NMHV amplitude, and thus for the whole one-loop six-gluon amplitude in $\mathcal{N}=4$ SYM. 
This parallels the set-up of the 
two-loop six-gluon MHV amplitude in planar $\mathcal{N}=4$ SYM, which can be expressed in terms of two-loop Regge components~\cite{DelDuca:2008jg} and remainder functions depending on conformally-invariant cross ratios (\ref{eq:uvwA}), and which can be evaluated in the central NMRK limit without any loss of generality~\cite{DelDuca:2009au,DelDuca:2010zg}. It would be interesting to extend this analysis to the two-loop six-gluon amplitude in full $\mathcal{N}=4$ SYM.

It follows in particular that the transcendental functions making up the 
one-loop two-gluon CEV can be expressed in terms of one-loop MRK components, evaluated at NMRK kinematics, plus a remainder function.
The former is independent of the helicity configuration, eq.~(\ref{eq:U}), while the latter has a distinct functional form for each of the two helicity configurations.  

Our one-loop analysis of the two-gluon CEV provides first hints with regards to the factorization and exponentiation properties of the six-gluon amplitude.
In \eqn{eq:DispM6g_ansatz} we have proposed an exponentiation pattern of the six-gluon amplitude in the central NMRK limit in which rapidity ordering is partially linked with colour ordering. The guiding principles have been the locality of the two-gluon CEV in rapidity, and its factorisation in the MRK limit into two one-gluon CEVs connected by a Regge trajectory.
However, the link between rapidity and colour ordering in (\ref{eq:DispM6g_ansatz}) is only partial, because of the presence of a fully-symmetric colour structure, which occurs first at the one-loop level, \eqn{eq:DispFFdgen}: an evaluation of the two-loop amplitude in the NMRK limit is necessary to confirm or disprove this factorized structure. We leave the investigation of two-loop amplitudes in the NMRK limit to future work.

In the NLL corrections to the BFKL kernel (fig.~\ref{fig:BFKL_NLL_Building_blocks} b), the one-loop one-gluon CEV is multiplied by the corresponding tree-level CEV, with the gluon $p_4$ emitted centrally along the ladder which is
integrated over its phase space. No collinear divergences are allowed in MRK, \eqn{eq:mrk}, however in the phase-space integration, the gluon $p_4$ may become soft and yield a pole of $\ord(1/\epsilon)$. In order to generate all the finite terms of the squared amplitude, the one-loop one-gluon CEV must be evaluated to $\ord(\epsilon)$ in the limit that the gluon $p_4$ is soft~\cite{Fadin:1996yv,DelDuca:1998cx}, which requires the use of one-loop soft functions to $\ord(\epsilon)$~\cite{Bern:1998sc}.
In the NNLL corrections to the BFKL kernel, similar features are present: the one-loop two-gluon CEV is multiplied by the corresponding tree-level CEV with the gluons $p_4$ and $p_5$ emitted along the ladder, which are integrated over their phase space. In the phase-space integration, both gluons may become soft, as well as collinear to each other, yielding poles of up to $\ord(1/\epsilon^3)$. Accordingly, the one-loop two-gluon CEV must be evaluated to $\ord(\epsilon^3)$ in the limit that gluons $p_4$ and $p_5$ are soft, and/or collinear to each other. To higher order in $\epsilon$, the one-loop splitting  functions are known~\cite{Kosower:1999rx,Bern:1999ry}, but the one-loop double-soft functions are yet to be determined. We leave the analysis of the 
higher-order terms of the one-loop two-gluon CEV to future work.

For generic gluon momenta, in order to obtain the two-gluon CEV in QCD using the supersymmetric decomposition of one-loop amplitudes, the next step is to perform a similar NMRK analysis of the contributions of a $\mathcal{N}=1$ multiplet and of a scalar circulating in the loop.  We also leave this analysis for the future.

\section*{Acknowledgements}

We thank Claude Duhr for useful discussions.
EG is supported by the STFC Consolidated Grant ‘Particle Physics at the Higgs Centre’. EB and JMS are supported by the ERC Starting Grant 715049 ‘QCDforfuture’.  JMS is also supported by a Royal Society University Research Fellowship. This research was supported by the US Department of Energy under contract DE--AC02--76SF00515, and in part by the National Science Foundation under Grant No.~NSF PHY--1748958.  LD thanks the Kavli Institute for Theoretical Physics for hospitality during the completion of this work. For the purpose of open access, the authors have applied a Creative Commons Attribution (CC BY) licence to any Author Accepted Manuscript version arising from this submission.

\appendix
\section{Kinematic regions}
\label{sec:appkin}

\subsection{General multi-parton kinematics}
\label{sec:appa}

\begin{figure}[hbt]
\centering
     \includegraphics[scale=0.2]{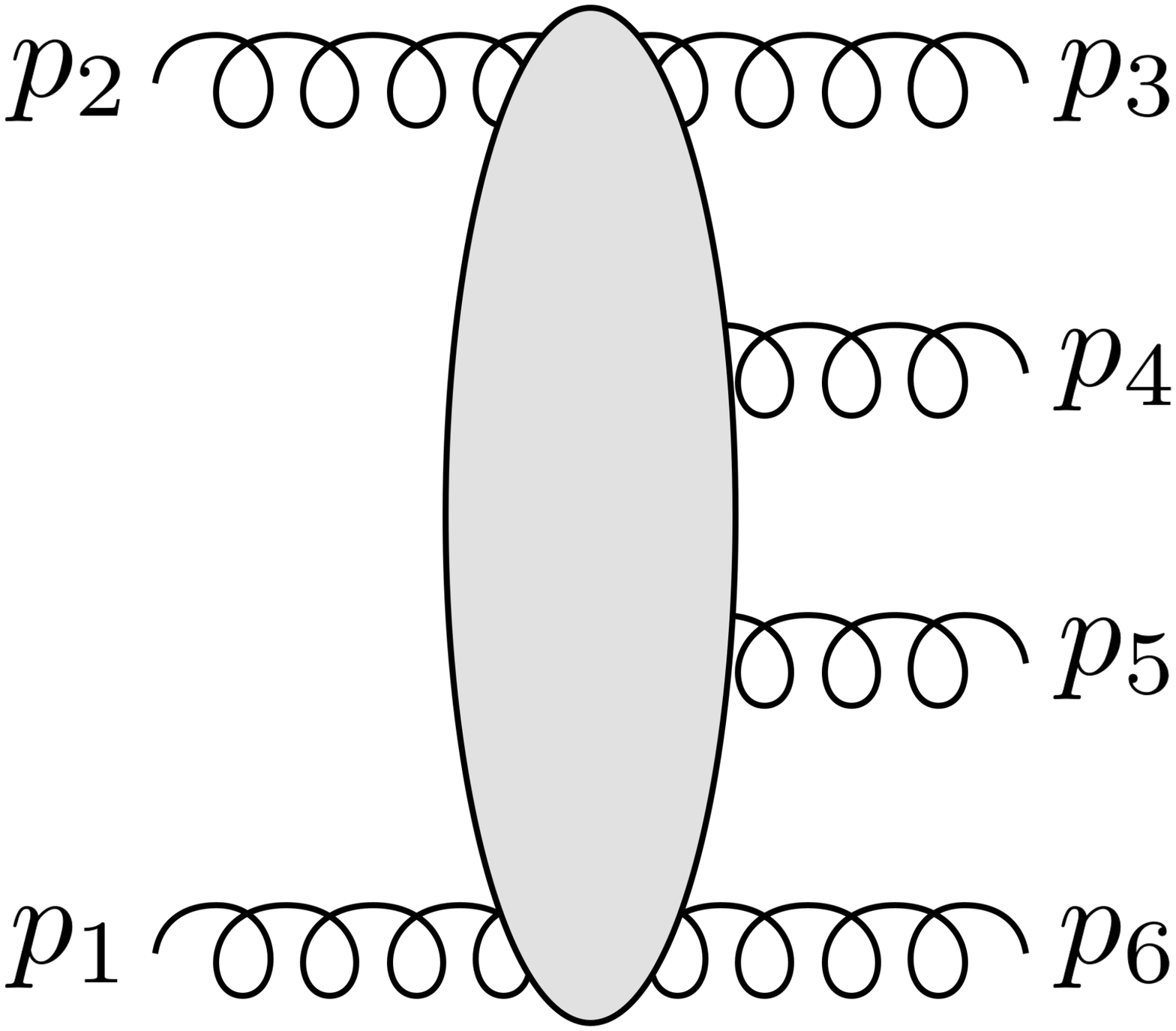}
    \caption{Depiction of the production of four gluons of momenta $p_i$, $i=3,4,5,6$, 
in the scattering between two partons of  
momenta $p_1$ and $p_2$. We take all momenta to be outgoing.}
    \label{fig:momenta}
\end{figure}

We consider the production of four partons of momenta $p_i$, $i=3,4,5,6$, 
in the scattering between two partons of  
momenta $p_1$ and $p_2$, as depicted in figure~\ref{fig:momenta}. By convention,
we consider the scattering in the unphysical region where all momenta 
are taken as outgoing, and then we analytically continue to the
physical region where $p_1^0<0$ and $p_2^0<0$. Thus
partons are incoming or outgoing depending on the sign
of their energy. Since the helicity of a positive-energy 
(negative-energy) massless spinor has the same (opposite) sign as its
chirality, the helicities assigned to the partons 
depend on whether they are incoming or outgoing.
We label outgoing (positive-energy) particles 
with their helicity; if they are incoming the 
actual helicity and charge are reversed, e.g.~an incoming left-handed
parton is labelled as an outgoing right-handed anti-parton.

Using light-cone coordinates $p^{\pm}= p^0\pm p^z $, and
complex transverse coordinates $p_{\perp} = p^x + i p^y$, with scalar
product,
\begin{equation}
2 p\cdot q = p^+q^- + p^-q^+ - p_{\perp} q^*_{\perp} - p^*_{\perp} q_{\perp}\,, 
\label{eq:scalprod}
\end{equation} 
the four-momenta are,
\begin{eqnarray}
p_1 &=& \left(\frac{p_1^-}{2}, 0, 0,\frac{-p_1^-}{2} \right) 
     \equiv  \left(0, p_1^-; 0, 0\right)\, ,\nonumber \\
p_2 &=& \left(\frac{p_2^+}{2}, 0, 0,\frac{p_2^+}{2} \right) 
     \equiv  \left(p_2^+ , 0; 0, 0 \right)\,  ,\label{in}\\
p_i &=& \left( \frac{p_i^+ + p_i^- }{2}, 
                {\rm Re}[p_{i\perp}],
                {\rm Im}[p_{i\perp}], 
                \frac{p_i^+ - p_i^- }{2} \right)
                \equiv
   |p_{i\perp}|\left( e^{y_i},  e^{-y_i}; 
\cos{\phi_i}, \sin{\phi_i}\right)\, \,,\nonumber
\end{eqnarray}
where $y$ is the rapidity, and $3\le i \le 6$. The first notation in \eqn{in} is the 
standard representation 
$p^\mu =(p^0,p^x,p^y,p^z)$, while in the second we have the $+$ and $-$
light-cone components on the left of the semicolon, 
and the transverse components on the right.

From momentum conservation,
\begin{eqnarray}
0 &=& \sum_{i=3}^6 p_{i\perp}\, ,\nonumber \\
p_2^+ &=& -\sum_{i=3}^6 p_i^+\, ,\label{nkin}\\ 
p_1^- &=& -\sum_{i=3}^6 p_i^-\, ,\nonumber
\end{eqnarray}
and using the scalar product \eqn{eq:scalprod},
the Mandelstam invariants may be written as,
\begin{eqnarray}
s &=& 2 p_1\cdot p_2 = \sum_{i,j=3}^6 p_i^+ p_j^- \nonumber\\ 
s_{2i} &=& 2 p_2\cdot p_i = -\sum_{j=3}^6 p_i^- p_j^+ \label{inv}\\ 
s_{1i} &=& 2 p_1\cdot p_i = -\sum_{j=3}^6 p_i^+ p_j^- \nonumber\\
s_{ik} &=& 2 p_i\cdot p_k = p_i^+p_k^- + p_i^-p_k^+ - p_{i\perp} p^\ast_{k\perp} - p^\ast_{i\perp} p_{k\perp}\,, \nonumber
\end{eqnarray}
with $3\le i, k \le 6$.

For the momenta in \eqn{in}, we find the following right-handed spinor products,
$\langle k \, p \rangle = \overline{u}_{-}(k) u_{+}(p)$,
in the notation of Ref.~\cite{DelDuca:1999iql},
\begin{eqnarray}
\langle p_i \, p_j\rangle &=& p_{i\perp}\sqrt{\frac{p_j^+}{p_i^+}} - p_{j\perp}
\sqrt{\frac{p_i^+}{p_j^+}}\, , \nonumber\\ 
\langle p_2 \, p_i\rangle &=& - i \sqrt{\frac{-p_2^+}{p_i^+}}\, p_{i\perp}\, ,\label{spro}\\ 
\langle p_i \, p_1\rangle &=&
i \sqrt{-p_1^- p_i^+}\, ,\nonumber\\ 
\langle p_2 \, p_1\rangle 
&=& -\sqrt{p_2^+p_1^-}\, ,\nonumber
\end{eqnarray}
where we have used the mass-shell condition,
\begin{equation}
|p_{i\perp}|^2 = p_i^+ p_i^-\,.
\label{eq:masssh}
\end{equation}
Left-handed spinor products, $[k \, p] = \overline{u}_{+}(k) u_{-}(p)$, are given by complex conjugation,
\beq
[k \, p] = {\rm sign}(k^0 p^0) \langle p \, k\rangle^\ast\,.
\eeq
Spinor products are antisymmetric,
\beq
\langle k\, p\rangle = - \langle p \, k\rangle\,,\qquad [k \, p] = -[p \, k]\,.
\eeq
We also use the currents, $[p| \gamma^{\mu} |k\rangle$ and $\langle p|\gamma^{\mu}|k]$,
which are related by charge conjugation,
\begin{equation}
[p| \gamma^{\mu} |k\rangle=\langle k|\gamma^{\mu}|p]\,,
\end{equation}
and complex conjugation,
\beq
[p| \gamma^{\mu} |k\rangle^{*}  = {\rm sign}(k^0 p^0) [k| \gamma^{\mu} |p\rangle \,.
\eeq
Through the Fierz rearrangement,
\begin{equation}
 \langle k|\gamma^{\mu}|p]\langle v|\gamma^{\mu}|q] = 2 \langle k \, v\rangle [q \, p]\,,
\end{equation}
and the Gordon identity,
\begin{equation}
 [p| \gamma^{\mu} |p\rangle= \langle p|\gamma^{\mu}|p]= 2 p^{\mu}\,,
\end{equation}
we obtain that
\begin{eqnarray}
&& \langle k|\slashed{q}|p] = \langle k \, q\rangle [q \, p]\,, \nonumber\\ 
&&  [p| \slashed{q} |k\rangle = [p \, q] \langle q \, k\rangle \,.
\end{eqnarray}

\subsection{Multi-Regge kinematics}
\label{sec:appb}

In multi-Regge kinematics, we require that the light-cone momenta of the gluons are strongly ordered and have comparable transverse momentum.  In the six-gluon case, we take
\begin{align}
\begin{split}
&p^+_3 \gg p^+_4 \gg p^+_5 \gg p^+_6\,, \quad\qquad p^-_3 \ll p^-_4 \ll p^-_5 \ll p^-_6\,,\\
&|p_{3\perp}| \simeq |p_{4\perp}| \simeq|p_{5\perp}| \simeq|p_{6\perp}|\,.
\label{eq:mrkapp}
\end{split}
\end{align}
The strong ordering of light-cone momenta in \eqn{eq:mrkapp} is equivalent to requiring a strong-ordering on the rapidities,
\begin{equation}
y_3 \gg y_4 \gg y_5 \gg y_6\,.
\label{MRK_hierarchy}
\end{equation}
Momentum conservation in \eqn{nkin} then becomes
\begin{eqnarray}
0 &=& \sum_{i=3}^6 p_{i\perp}\, ,\nonumber \\
p_2^+ &\simeq& -p_3^+\, ,\label{mrkin}\\ 
p_1^- &\simeq& -p_6^-\, .\nonumber
\end{eqnarray}
To leading accuracy, the Mandelstam invariants in \eqn{inv} are reduced to
\begin{eqnarray}
s &=& 2 p_1\cdot p_2 \simeq p_3^+ p_6^-\,, \nonumber\\ 
s_{2i} &=& 2 p_2\cdot p_i \simeq - p_3^+ p_i^- \,,\label{mrinv}\\ 
s_{1i} &=& 2 p_1\cdot p_i \simeq - p_i^+ p_6^-\,, \nonumber\\ 
s_{ij} &=& 2 p_i\cdot p_j \simeq p_i^+ p_j^-\,, \nonumber
\end{eqnarray}
with $3\le i, j \le 6$, and with $p^+_i > p^+_j$, for $i<j$.
Note that the mass-shell condition in \eqn{eq:masssh} implies that
\beq
s\, |p_{4\perp}|^2 |p_{5\perp}|^2 \simeq s_{34} s_{45} s_{56}\,,
\eeq
which is an example of the general multi-Regge constraint,
\beq
s_{ij}\, \prod_{k=i+1}^{j-1} |p_{k\perp}|^2 = \prod_{k=i}^{j-1} s_{k,k+1}\,.
\eeq
The spinor products in \eqn{spro} become,
\begin{eqnarray}
\langle p_i \, p_j\rangle &\simeq& -\sqrt{\frac{p_i^+}{p_j^+}}\,
p_{j\perp}\, \qquad {\rm for}\, y_i>y_j \;, \nonumber\\
\langle p_2 \,  p_i\rangle &\simeq& - i\sqrt{\frac{p_3^+}{p_i^+}}\,
p_{i\perp}\, ,\label{mrpro}\\ 
\langle p_i \, p_1\rangle 
&\simeq& i\sqrt{p_i^+ p_6^-}\, ,\nonumber\\ 
\langle p_2 \,  p_1\rangle &\simeq& -\sqrt{p_3^+ p_6^-}\, ,\nonumber
\end{eqnarray}
with $3\le i \le 6$. In \eqns{mrinv}{mrpro} only the leading terms are displayed.
Here and henceforth, it is understood that, if so required by the computation,
sub-leading terms may need to be retained.

In the Euclidean region, where the Mandelstam invariants are taken as all negative, the ordering in \eqn{eq:mrkapp} becomes
\beq
-s_{12} \gg -s_{123}, -s_{345} \gg -s_{34}, -s_{45}, -s_{56} \gg -s_{23},  -s_{61}, -s_{234} \,.
\eeq
Introducing a parameter $\lambda \ll 1$, the hierarchy in eq.~(\ref{MRK_hierarchy}) above is equivalent to the rescaling,
\beq
\{s_{123}, s_{345} \} = \ord(\lambda)\,, \qquad 
\{s_{34}, s_{45}, s_{56}\} = \ord(\lambda^2)\,, \qquad 
\{ s_{23}, s_{61}, s_{234} \} = \ord(\lambda^3)\,.
\eeq

\subsection{Central next-to-multi-Regge kinematics}
\label{sec:appd}

We now consider the production of four partons of momenta $p_i$, with $3\le i \le 6$,
with the new constraint that partons 4 and 5 are in the central region along the gluon ladder,
\begin{align}
\label{nmrapp}
\begin{split}
& p^+_3 \gg p^+_4 \simeq p^+_5 \gg p^+_6\,, \quad\qquad p^-_3 \ll p^-_4 \simeq p^-_5 \ll p^-_6\,
\,, \\ 
&|p_{3\perp}|
\simeq |p_{4\perp}| \simeq |p_{5\perp}| \simeq |p_{6\perp}|\, ,
\end{split}
\end{align}
and hence,
\begin{equation}
y_3 \gg y_4 \simeq y_5\gg y_6.
\label{NMRK_hierarchy}
\end{equation}

The leading contributions to momentum conservation are the same as in \eqn{mrkin}.
The Mandelstam invariants of \eqn{inv} become
\begin{eqnarray}
s &\simeq& p_3^+ p_6^-\,, \nonumber\\ 
s_{2i} &\simeq&  - p_3^+ p_i^-\,, \nn\\ 
s_{1i} &\simeq& - p_i^+ p_6^-\,, \label{nmrkinv} \\ 
s_{jk} &\simeq& p_j^+ p_k^-\,, \qquad j\ne4\,\,\, {\rm and} \,\,\, k\ne 5\,,\nonumber \\ 
s_{45} &=& p_4^+ p_5^- + p_4^- p_5^+ - p_{4\perp}^\ast p_{5\perp} - p_{5\perp}^\ast p_{4\perp} \nonumber\,, 
\end{eqnarray}
with $3\le i, j, k \le 6$, with $p^+_j > p^+_k$ for $j<k$. 
Note that the mass-shell condition in \eqn{eq:masssh} and the Mandelstam invariants in \eqn{nmrkinv} imply that
\begin{align}
\begin{split}
s &\simeq \frac{s_{34}\, s_{56} }{p_4^- p_5^+ } \\ &= \frac{s_{34}\, s_{56}\, p_4^+ p_5^- }{|p_{4\perp}|^2 |p_{5\perp}|^2}\, . \\
\end{split}
\end{align}
All three-particle Mandelstam invariants either simplify to a two-particle Mandelstam invariant, or are equal to one of the following, up to an overall sign,
\begin{align}
\begin{split}
s_{123} &\simeq (p_4^+ + p_5^+) p_6^- \,, \\
s_{234} &\simeq - |p_{3\perp} + p_{4\perp}|^2 - p_4^- p_5^+ \,, \\ 
s_{235} &\simeq - |p_{3\perp} + p_{5\perp}|^2 - p_4^+ p_5^- \,, \\ 
s_{345} &\simeq p_3^+ (p_4^- + p_5^-) \,. 
\label{eq:s3nmrk}
\end{split}
\end{align}
In the Euclidean region, where the Mandelstam invariants are taken as all negative, the ordering in \eqn{nmrapp} becomes
\beq
-s_{12} \gg -s_{34}, -s_{56}, -s_{123}, -s_{345} \gg -s_{23}, -s_{45}, -s_{61}, -s_{234}, -s_{235} \,.
\eeq
Introducing a parameter $\lambda \ll 1$, the hierarchy in eq.~(\ref{nmrapp}) above is equivalent to the rescaling,
\beq
\{s_{34}, s_{56}, s_{123}, s_{345} \} = \ord(\lambda)\,, \qquad \{ s_{23}, s_{45}, s_{61}, s_{234}, s_{235} \} = \ord(\lambda^2)\,.
\eeq
The spinor products in \eqn{spro} become
\begin{eqnarray}
\langle p_2 \,  p_1\rangle &\simeq& - \sqrt{p_3^+ p_6^-}\,,\nonumber\\ 
\langle p_2 \, p_k\rangle &=& -i \sqrt{\frac{-p_2^+}{p_k^+}}\, p_{k\perp}
\simeq -i \sqrt{\frac{p_3^+}{p_k^+}} p_{k\perp}\, , \nonumber\\
\langle p_k \, p_1\rangle &=& i \sqrt{-p_1^- p_k^+}\, 
\simeq i \sqrt{p_k^+ p_6^-}\, ,\label{cnrpro}\\
\langle p_j \, p_k\rangle &=& p_{j\perp}\sqrt{\frac{p_k^+}{p_j^+}} - p_{k\perp}
\sqrt{\frac{p_j^+}{p_k^+}} \simeq - p_{k\perp}\, \sqrt{\frac{p_j^+}{p_k^+}}\, , \qquad j\ne4\,\,\, {\rm and} \,\,\, k\ne 5 \nonumber \\
\langle p_4 \, p_5\rangle &=& p_{4\perp}\sqrt{\frac{p_5^+}{p_4^+}} - 
p_{5\perp}\sqrt{\frac{p_4^+}{p_5^+}}\, ,\nonumber
\end{eqnarray}
with $3\le j, k \le 6$, with $p^+_j > p^+_k$. 

The following spinor strings occur in the BCFW representation of NMHV tree-level amplitudes. We list the spinor strings for two representative colour orderings. For the colour ordering $\{1,2,3,4,5,6\}$ the relevant quantities are
\begin{align}
\begin{split}
	\langle p_1 | \slashed{p}_2+\slashed{p}_3|p_4 ] 
	&\simeq
	-i|q_{3\perp}|\sqrt{\frac{p_4^+}{p_6^+}}(q_{2\perp}^*+p_{4\perp}^*\frac{p_5^+}{p_4^+})\, ,\\
	\langle p_3 | \slashed{p}_4+\slashed{p}_5|p_6 ] 
	&\simeq
	-q_{3\perp}^*\sqrt{\frac{p_3^+}{p_6^+}}\left(p_{4\perp}+p_{5\perp}\right)\, ,\\
	\langle p_5 | \slashed{p}_6+\slashed{p}_1|p_2 ] 
	&\simeq
	-ip_{5\perp}\sqrt{\frac{p_3^+}{p_5^+}}(q_{2\perp}^*+p_{5\perp}^*\frac{p_4^-}{p_5^-})\, ,
\end{split}
\label{eq:BCFWstringsA}
\end{align}
while for the colour ordering $\{1,4,2,3,5,6\}$ the relevant quantities are
\begin{align}
\begin{split}
	\langle p_1 | \slashed{p}_4+\slashed{p}_2|p_3 ] 
	&\simeq
	-i|q_{3\perp}|\sqrt{\frac{p_3^+}{p_6^+}}(p_{5\perp}^*+q_{3\perp}^*)\, ,\\
	\langle p_2 |\slashed{p}_3+\slashed{p}_5|p_6 ] 
	&\simeq
	iq_{3\perp}^*\sqrt{\frac{p_3^+}{p_6^+}}(p_{4\perp}+q_{3\perp})\, ,\\
	\langle p_5 | \slashed{p}_6+\slashed{p}_1|p_4 ] 
	&\simeq
	-p_{5\perp}q_{2\perp}^*\sqrt{\frac{p_4^+}{p_5^+}}-p_{4\perp}^*q'_{2\perp}\sqrt{\frac{p_5^+}{p_4^+}} \, .\\
\end{split}
\label{eq:BCFWstringsB}
\end{align}

\section{Lorentz invariant quantities in minimal coordinates}
\label{sec:min-invts}

In this section we collect useful expressions for Lorentz-invariant quantities, in the central NMRK region of appendix~\ref{sec:appd}, written in terms of the minimal set of coordinates, $\{w,z,X\}$, introduced in \eqns{eq:wzparam2}{eq:xparam}. We use the shorthand notation for spinors,
\begin{equation}
    \langle i \, j \rangle=\langle p_{i} \, p_{j}\rangle, \qquad [i \, j]=[p_i \, p_j], \qquad \langle i|k+l| j ]=\langle p_{i}|\slash{p}_{l}+\slash{p}_{k}|p_{j}].
    \label{eq:short_spinor}
\end{equation}
Of particular importance to the central NMRK limit is the spinor product,
\beqa
\langle 4 \, 5 \rangle= -q_{1\perp}\, \frac{ w+X z}{w \sqrt{X} (z-1)} \,,\label{eq:Spa45min}
\eeqa
which is treated exactly in this limit. The three-particle invariants of \eqn{eq:s3nmrk} become
\begin{align}
\begin{split}
	s_{123} 
	&=
	|q_{1\perp}|^2\frac{p_4^+}{p_6^+}\frac{(1+X)|w-1|^2|z|^2}{X|z-1|^2|w|^2}\, ,\\
	s_{234} 
	&= 
	-|q_{1\perp}|^2\, \frac{(1+ X |z|^2)}{X |z-1|^2}\, ,\\
	s_{235}
	&=
	-|q_{1\perp}|^2\frac{(|w+z-wz|^2+X|z|^2)}{|z-1|^2|w|^2}\, ,\\
	s_{345} 
	&=
	|q_{1\perp}|^2\frac{p_3^+}{p_4^+}\frac{(|w|^2+X|z|^2)}{|z-1|^2|w|^2}\, .\\
\end{split}
\label{eq:s3nmrkmin}
\end{align}
The lightcone-momentum dependence cancels when considering physical quantities such as amplitudes.
The spinor strings of \eqn{eq:BCFWstringsA} become
\begin{align}
\begin{split}
	\langle 1 | 2+3|4 ] 
	&=
	-iq_{1\perp}^*|q_{1\perp}|\sqrt{\frac{p_4^+}{p_6^+}}\frac{|w-1||z|}{|z-1||w|}\frac{(1+X\bar{z})}{X(\bar{z}-1)}\, ,\\
	\langle 3 | 4+5|6 ] 
	&=
	|q_{1\perp}|^2\sqrt{\frac{p_3^+}{p_6^+}}\,\frac{\bar{z}(\bar{w}-1)(w-z)}{|w|^2|z-1|^2}\, ,\\
	\langle 5 | 6+1|2 ] 
	&=
	-i|q_{1\perp}|^2\sqrt{\frac{p_3^+}{p_5^+}}\frac{(w+X|z|^2)}{wX|z-1|^2}\, ,
\end{split}
\label{eq:BCFWstringsAmin}
\end{align}
while the spinor strings of \eqn{eq:BCFWstringsB} become
\begin{align}
\begin{split}
	\langle 1 | 4+2|3 ] 
	&=
	-iq_{1\perp}^*|q_{1\perp}|\sqrt{\frac{p_3^+}{p_6^+}}\frac{|w-1||z|}{|z-1||w|}\frac{\bar{z}}{(\bar{z}-1)}\, ,\\
	\langle 2 | 3+5|6 ] 
	&=
	-i|q_{1\perp}|^2\sqrt{\frac{p_3^+}{p_6^+}}\frac{\bar{z}(\bar{w}-1)(w+z-w z)}{|w|^2|z-1|^2}\, ,\\
	\langle 5 | 6+1|4 ] 
	&=
	-|q_{1\perp}|^2\frac{(w+z-wz+X|z|^2)}{w\sqrt{X}|z-1|^2}	\, .
\end{split}
\label{eq:BCFWstringsBmin}
\end{align}
The spurious surfaces listed in \eqns{eq:unphys}{eq:unphysB} can be recognised from \eqns{eq:BCFWstringsAmin}{eq:BCFWstringsBmin} respectively.
\subsection{Cross ratios in the NMRK limit}
\label{sec:crossratio}
In general kinematics, the finite, dual conformally invariant part of the six-gluon amplitude in planar $\cN=4$ SYM can be written in terms of multiple polylogarithms with
the following 9-letter alphabet,
\begin{equation}
\{u,v,w,1-u,1-v,1-w,y_u,y_v,y_w\}.
\end{equation}
The cross ratios $u$, $v$ and $w$ for the canonical ordering are given in \eqn{eq:uvwA}, although we appended a further $A$ subscript to distinguish this from the other colour orderings that are relevant away from the planar limit. The remaining letters can be constructed from the cross ratios. For $x \in \{u,v,w\}$, $y_x$ is defined by
\begin{align}
    \begin{split}
    y_x\equiv \frac{x-z_+}{x-z_-} \,,
    \end{split}
\end{align}
with 
\begin{equation}
    z_{\pm}\equiv\frac{1}{2}(u+v+w-1\pm \sqrt{\Delta}), \quad\qquad \Delta\equiv (1-u-v-w)^2-4 u v w.
\end{equation}
In this section we investigate these quantities for the two representative colour orderings used throughout this paper, namely $\sigma_A$ and  $\sigma_B$ of \eqn{eq:sigmaAAxB}. The corresponding quantities for $\sigma_{A'}$ can be obtained using the relations given in \eqn{eq:AtoAx}.

\subsection*{Permutation $\sigma_A$}
The cross ratios for this colour ordering, $u_A$, $v_A$ and $w_A$, are given in \eqn{eq:uvwA}. In NMRK they lie in the range $[0,1]$. In the central NMRK region we list the other six letters,
\begin{align}
	\begin{split}
		1-u_A&=\frac{X|w-z|^2}{(1+X)(|w|^2+X|z|^2)}\, ,\\
		1-v_A&=\frac{|1+Xz|^2}{(1+X|z|^2)(1+X)}\, ,\\
		1-w_A&=\frac{|w+X|z|^2|^2}{(|w|^2+X|z|^2)(1+X|z|^2)}\, ,
	\end{split}
\end{align}
\begin{align}
\begin{split}
	y_{u_A}&=\frac{(w+Xz)(1+X \bar{z})(\bar{w}+X|z|^2)}
	{(\bar{w}+X\bar{z})(1+X z)(w+X|z|^2)}\, ,\\
	y_{v_A}&=\frac{(\bar{w}-\bar{z})(z-1)(w+X|z|^2)}{(w-z)(\bar{z}-1)(\bar{w}+X|z|^2)}\, ,\\
	y_{w_A}&=\frac{\bar{z}(\bar{w}-1)(w-z)(1+X z)}{z(w-1)(\bar{w}-\bar{z})(1+X \bar{z})}\, .
	\end{split}
	\label{eq:yA}
\end{align}
We note that \eqn{eq:yA} are complex phases, which results from the quantity,
\begin{equation}
    \sqrt{\Delta_A}=\pm\frac{X
    \Big(
    (|w|^2-X|z|^2)(z-\bar{z})-|z|^2(1-X)(w-\bar{w})+ (1-X|z|^2)(w\bar{z}-\bar{w}z)
    \Big) 
    }{(1+X)(1+X|z|^2)(|w|^2+X|z|^2)} \,,
    \label{eq:rDeltaA}
\end{equation}
being purely imaginary. Thus its square is negative, $\Delta_A < 0$. One can get from any $2\to4$ scattering kinematics to NMRK by a dual conformal transformation~\cite{DelDuca:2009au,DelDuca:2010zg}.  Since $\Delta_A$ is dual conformally invariant, this means that $\Delta_A < 0$ for any nonsingular $2\to4$ scattering configuration. We further note that the minimal variables, introduced in section~\ref{sec:minset}, rationalise $\sqrt{\Delta_A}$.

\subsection*{Permutation $\sigma_B$}
The cross ratios for this colour ordering, $u_B$, $v_B$ and $w_B$, are given in \eqn{eq:uvwB}, which in NMRK lie in the range $[0,1]$. In the central NMRK region we list the other six letters,
\begin{align}
	\begin{split}
		1-u_B&=\frac{|w+z-w z|^2}{(|w+z-wz|^2+X|z|^2)}\, ,\\
		1-v_B&=\frac{X|z|^2}{(1+X|z|^2)}\, ,\\
		1-w_B&=\frac{|w+z-w z+X |z|^2|^2}{(1+X|z|^2)(|w+z-wz|^2+X|z|^2)}\, ,
	\end{split}
\end{align}
\begin{align}
\begin{split}
	y_{u_B}&=\frac{w+z-w z+X |z|^2}{\bar{w}+\bar{z}-\bar{w}\bar{z}+X |z|^2}\, ,\\
	y_{v_B}&=\frac{(w+z-wz) (\bar{w}+\bar{z} -\bar{w}\bar{z}+X |z|^2)}{(\bar{w}+\bar{z}-\bar{w} \bar{z}) (w+z-wz+X |z|^2)}\, ,\\
	y_{w_B}&=\frac{(w-1) (z-1) (\bar{w}+\bar{z}-\bar{w} \bar{z})}{(\bar{w}-1) (\bar{z}-1)(w+z-wz)}\,.
	\label{eq:yB}
	\end{split}
\end{align}
Again we find that \eqn{eq:yB} are phases which is a result of
\begin{equation}
    \sqrt{\Delta_B}=\pm\frac{X |z|^2\Big((\bar{w}+\bar{z}-\bar{w}\bar{z})-(w+z-w z)\Big)}{(1+X|z|^2)(|w+z-wz|^2+X|z|^2)} \,,
\end{equation}
being purely imaginary. We note that as in \eqn{eq:rDeltaA}, the minimal variables rationalise this quantity.

\section{The power-suppressed helicity configurations}
\label{sec:power-suppressed}

\begin{figure}[!htb]
\centering
    \subfigure[]{\label{fig:HelicityFlipNMHVc}\includegraphics[scale=0.15]{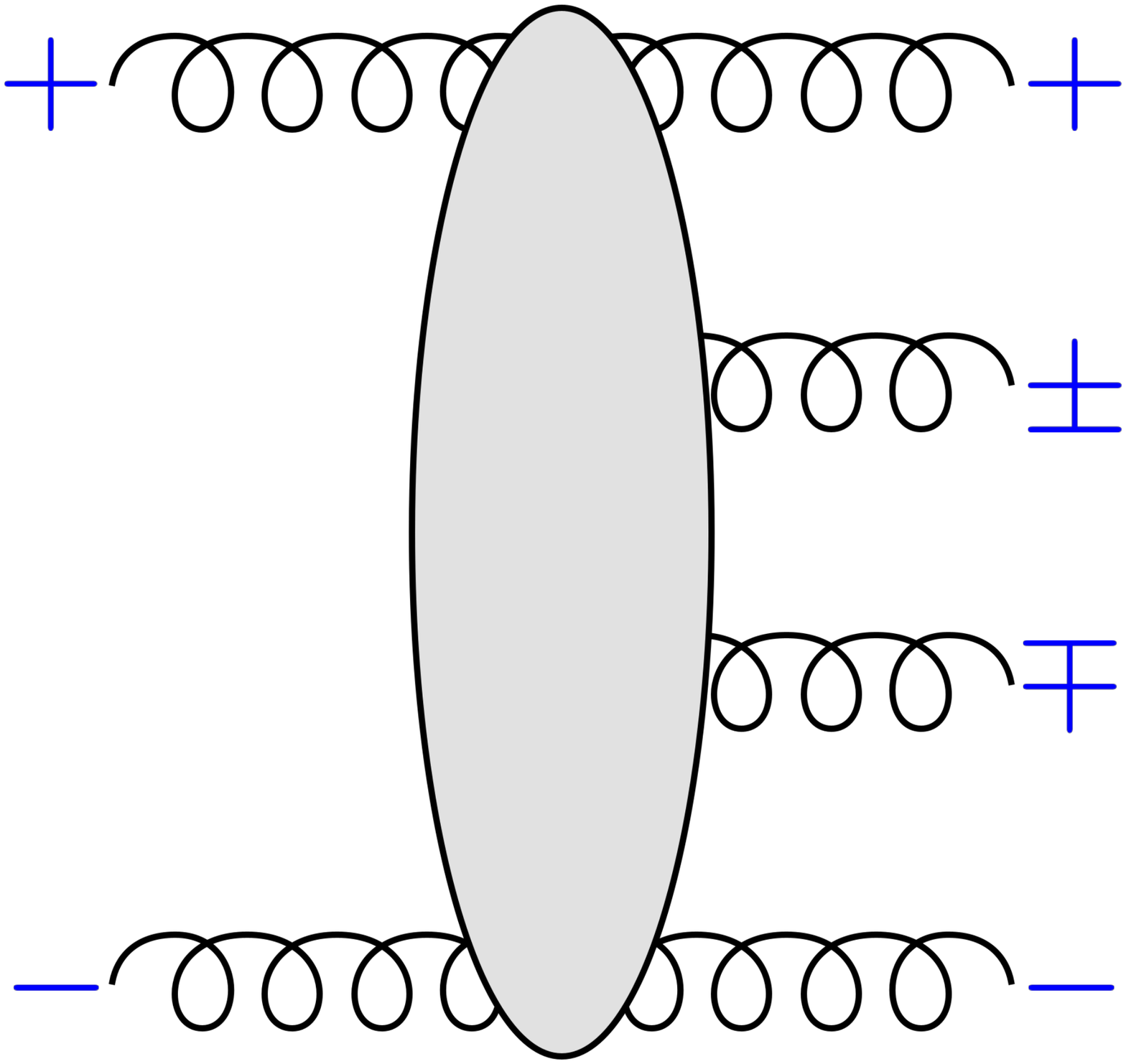}}
    \hspace*{20pt}
    \subfigure[]{\label{fig:HelicityFlipNMHVb}\includegraphics[scale=0.15]{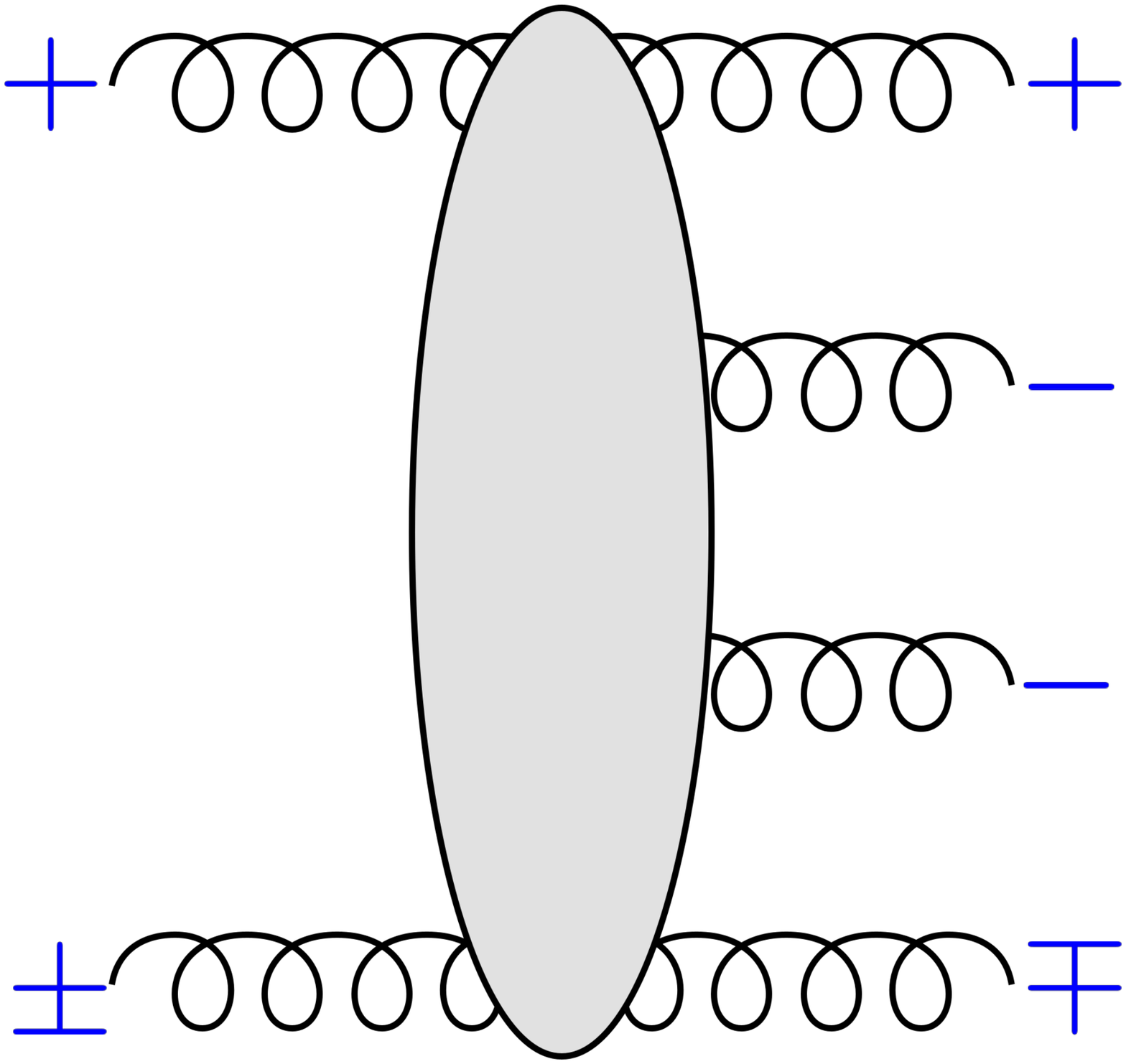}}
    \hspace*{20pt}
    \subfigure[]{\label{fig:HelicityFlipNMHVa}\includegraphics[scale=0.15]{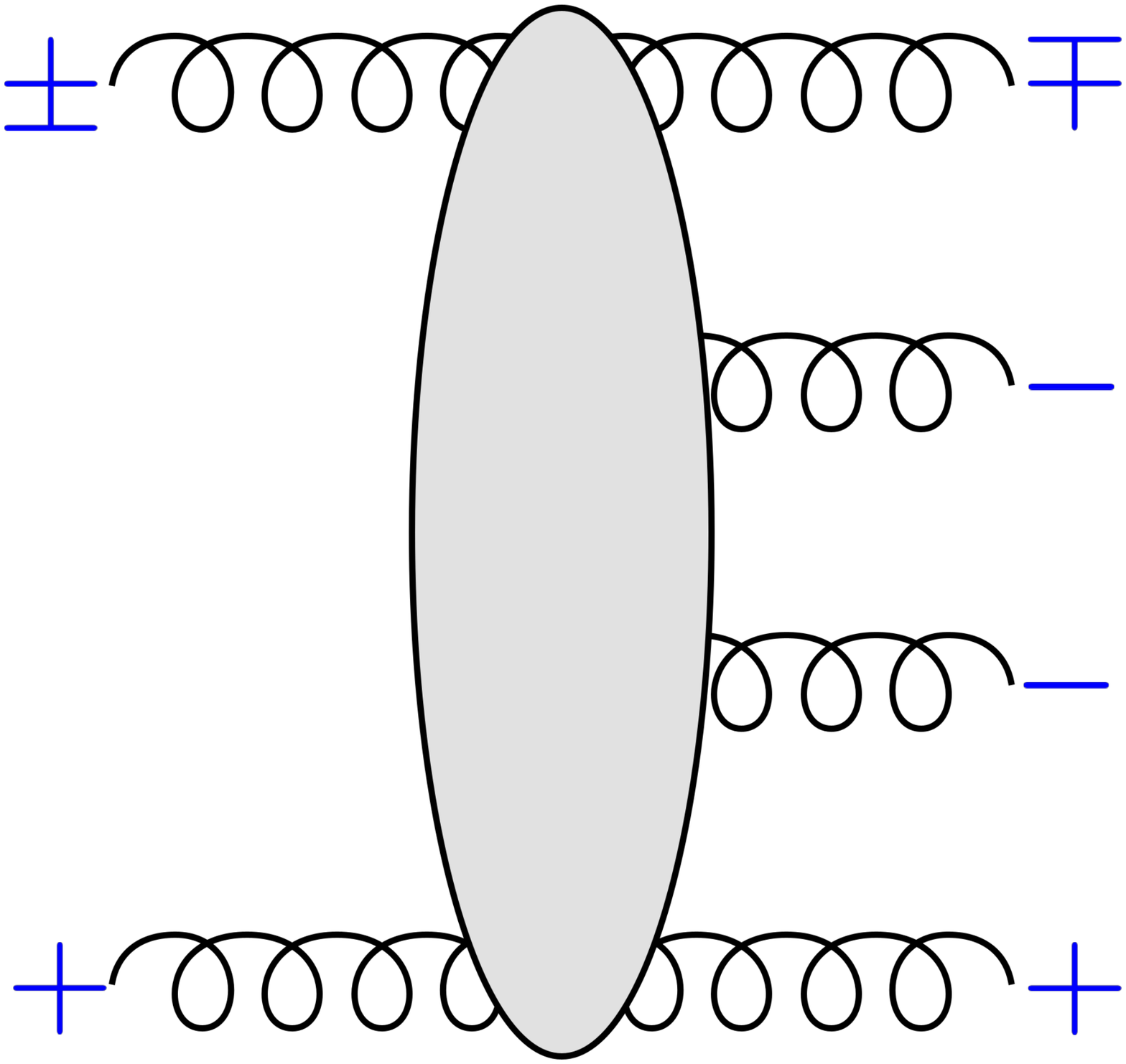}}
    \caption{Six of the NMHV helicity configurations that are power suppressed in the central NMRK limit. The remaining six 
    power-suppressed NMHV configurations can be obtained by flipping all helicities. The labelling of the gluons follows figure~\ref{fig:momenta}.}
    \label{fig:HelFlipNMHV}
\end{figure}

\begin{figure}[!htb]
\centering
    \subfigure[]{\label{fig:HelicityFlipMHVa}\includegraphics[scale=0.15]{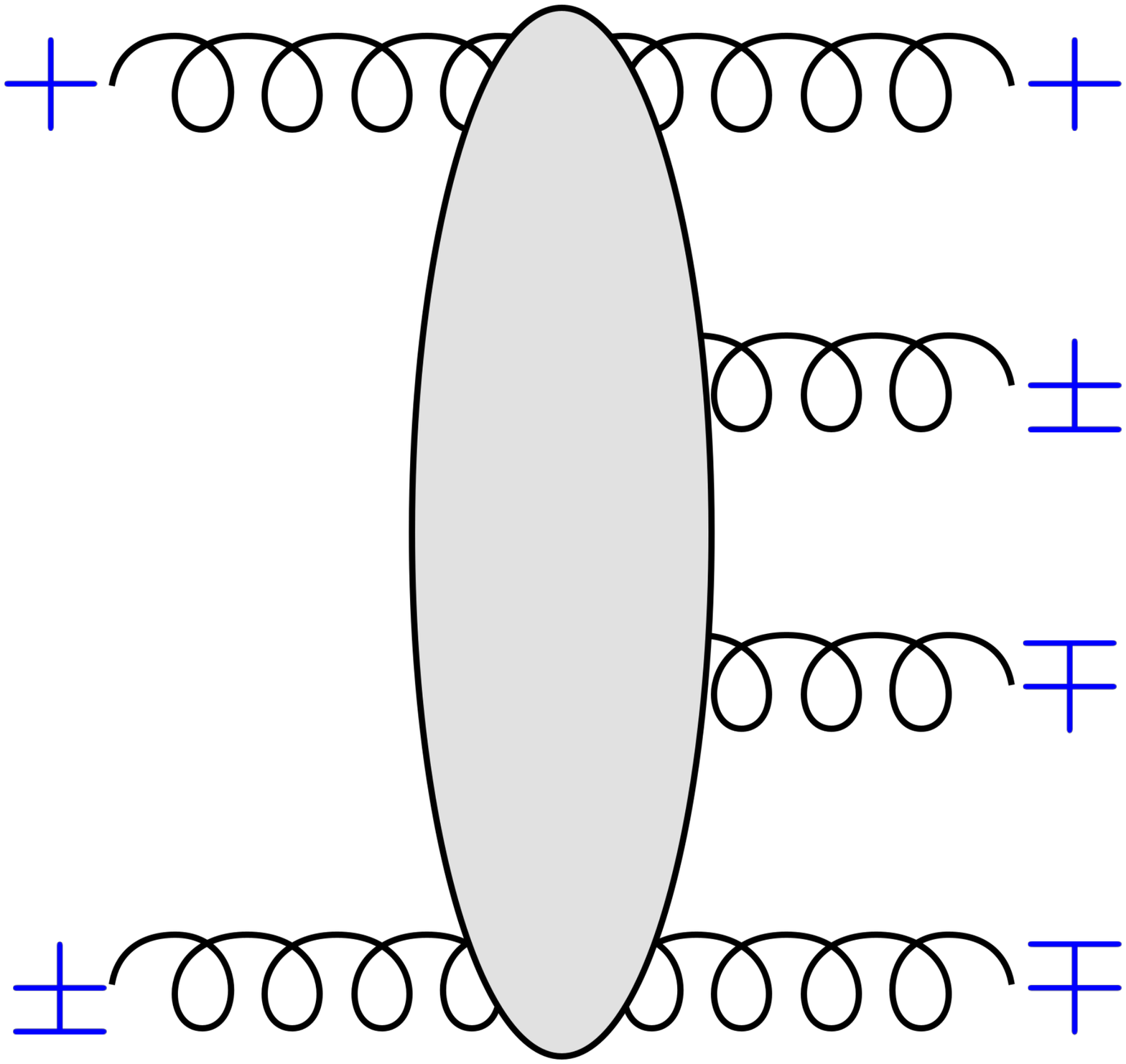}}
    \hspace*{20pt}
    \subfigure[]{\label{fig:HelicityFlipMHVb}\includegraphics[scale=0.15]{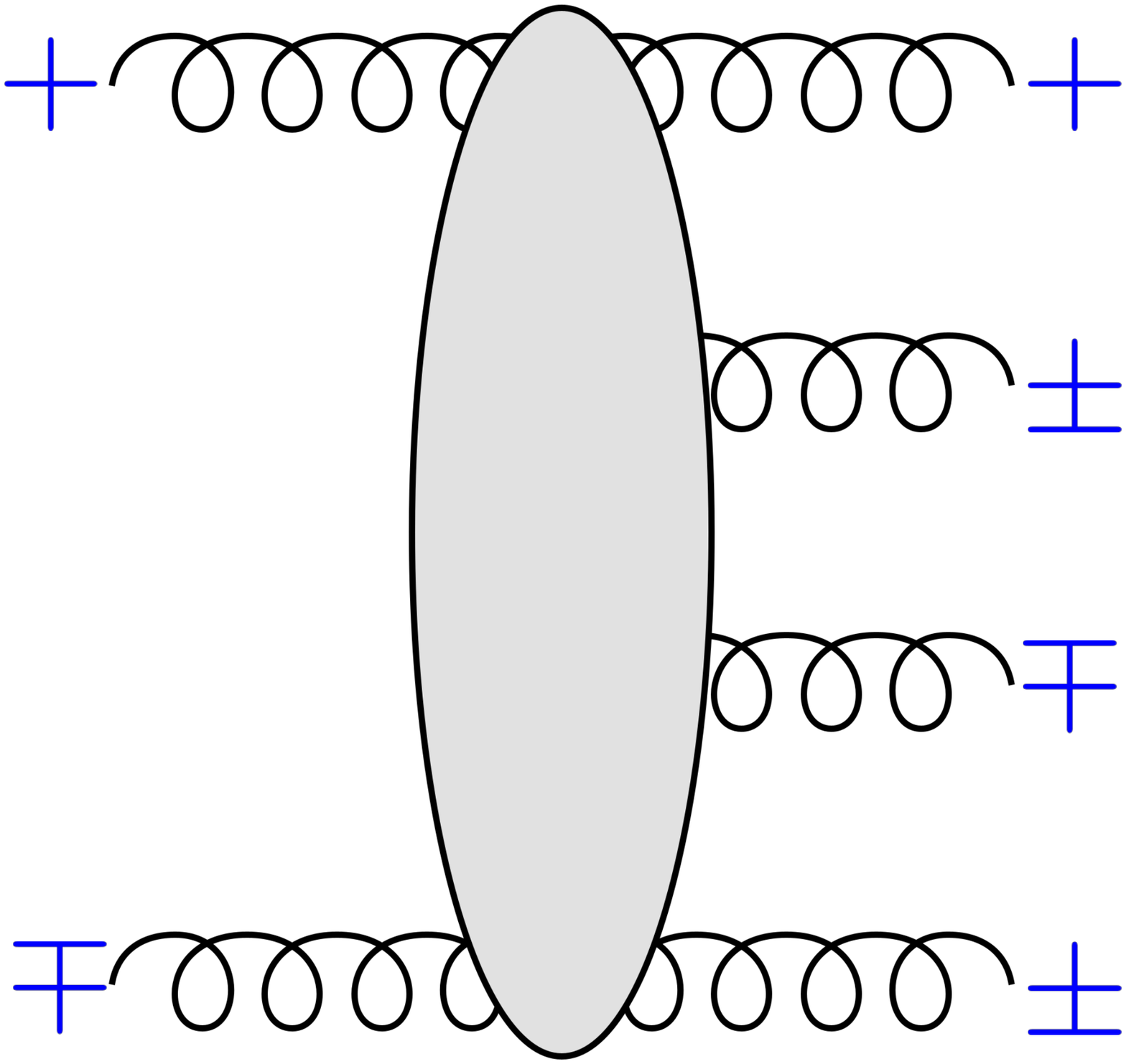}}
    \hspace*{20pt}
    \subfigure[]{\label{fig:HelicityFlipMHVc}\includegraphics[scale=0.15]{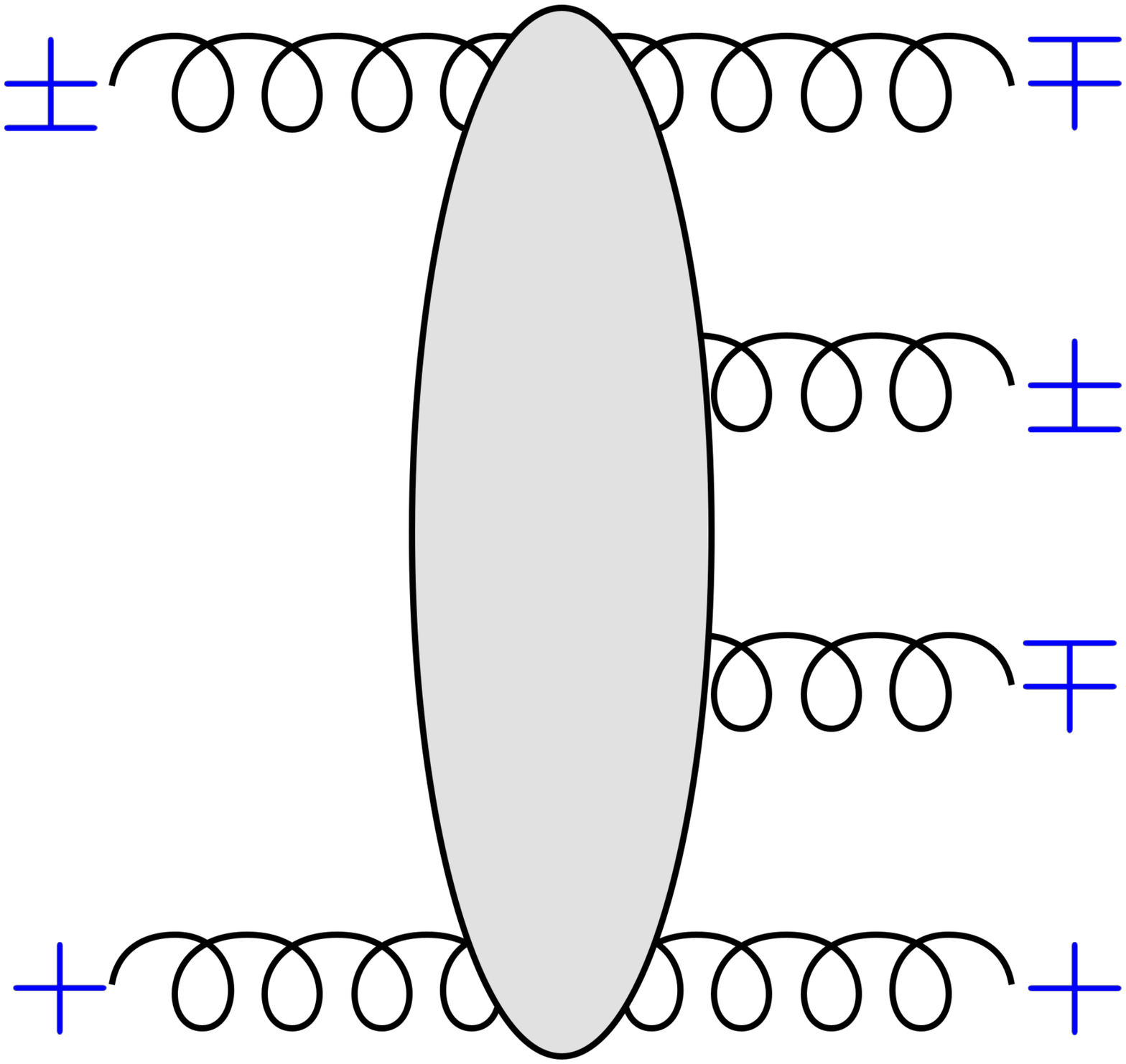}}\\
    \subfigure[]{\label{fig:HelicityFlipMHVd}\includegraphics[scale=0.15]{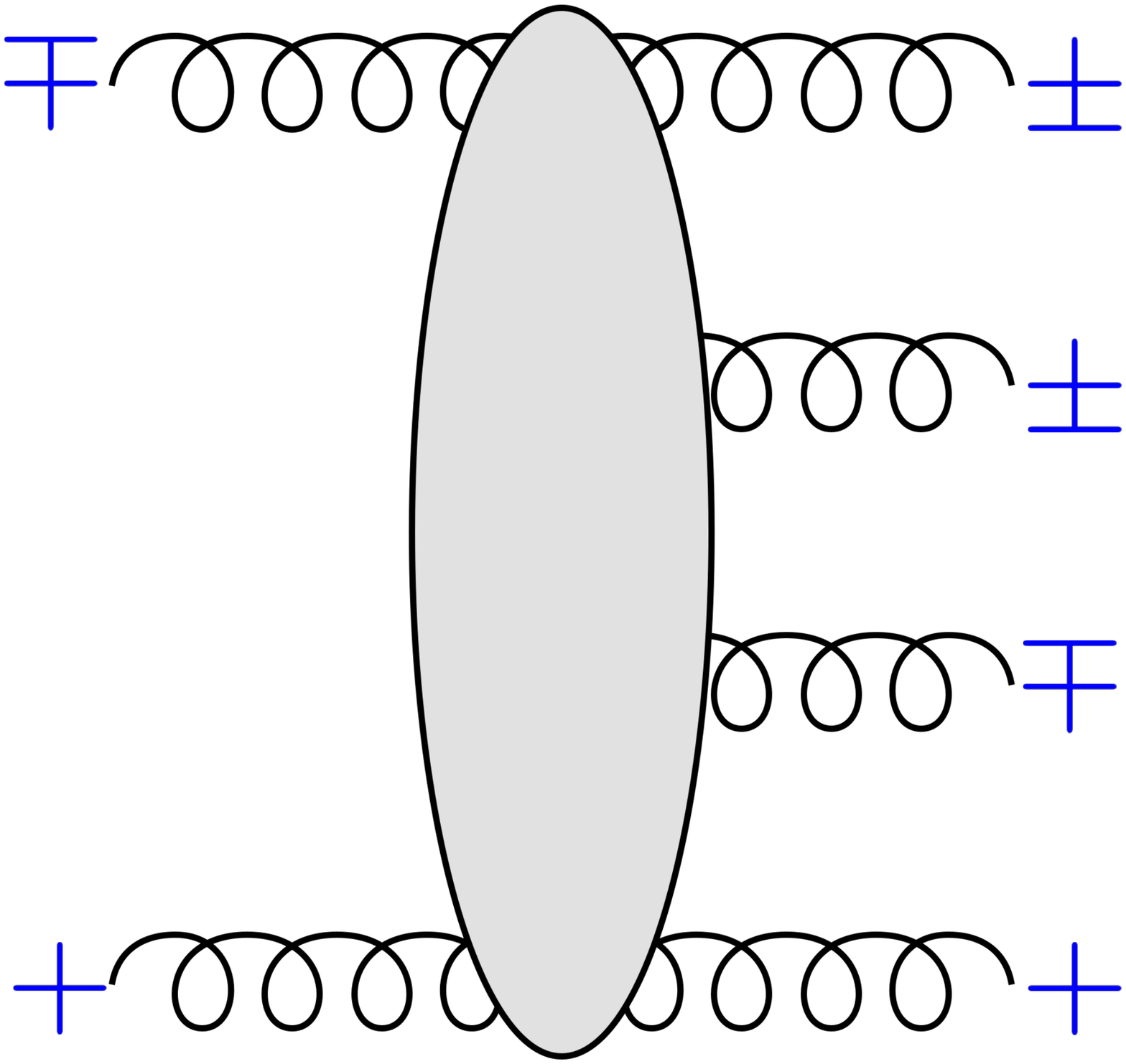}}
    \hspace*{20pt}
    \subfigure[]{\label{fig:HelicityFlipMHVe}\includegraphics[scale=0.15]{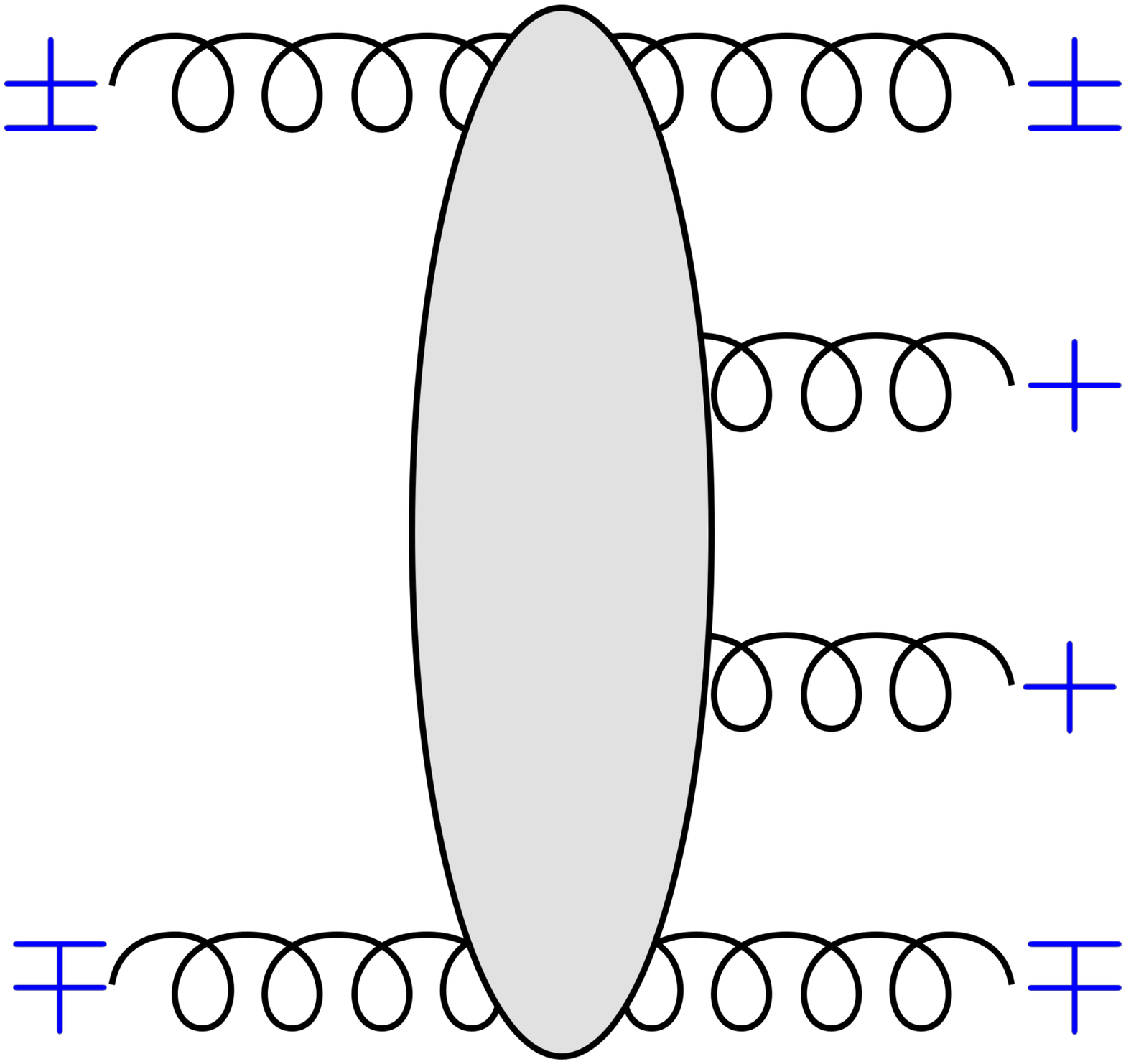}}
    \hspace*{20pt}
    \subfigure[]{\label{fig:HelicityFlipMHVf}\includegraphics[scale=0.15]{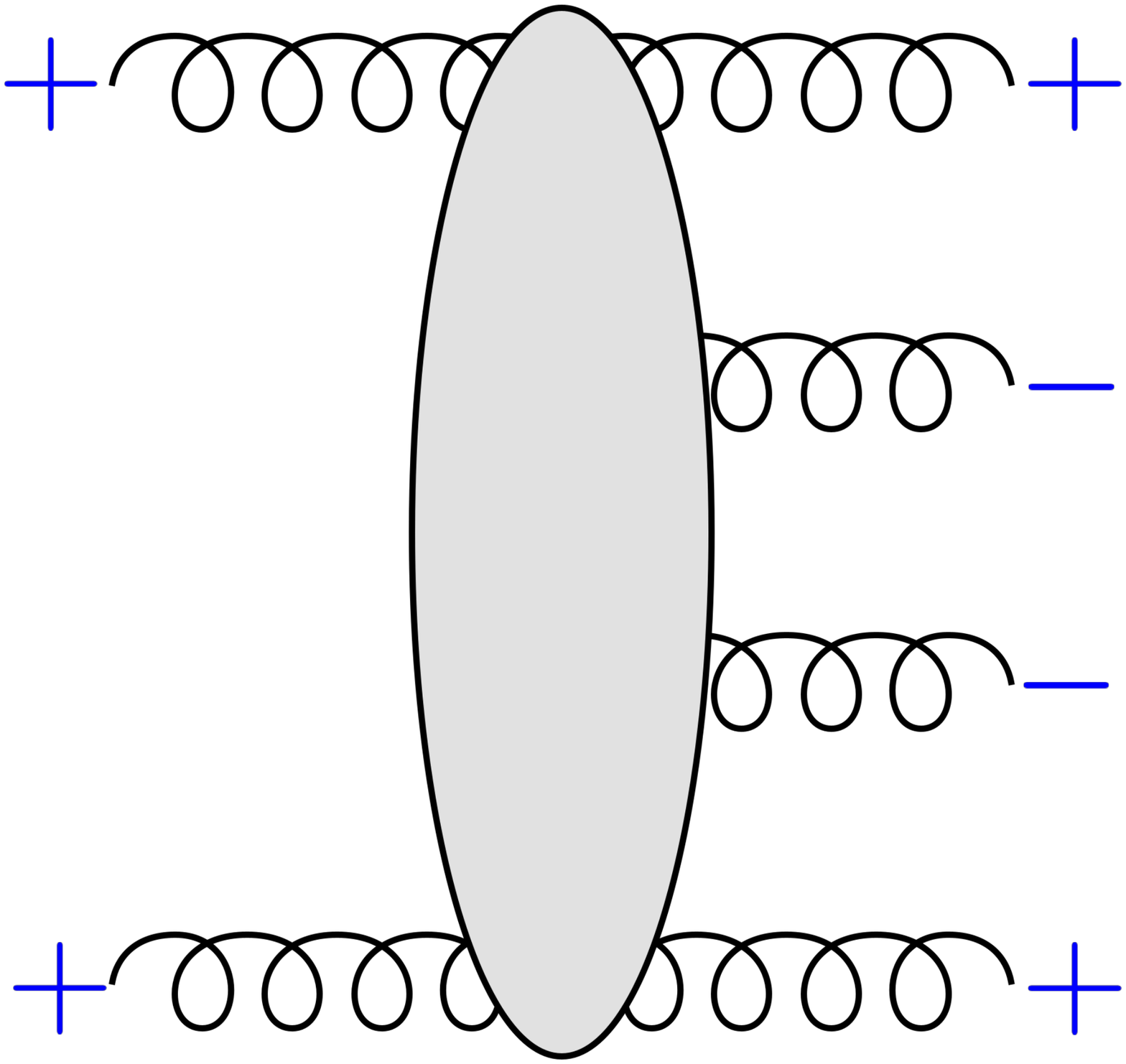}}
    \caption{Eleven of the MHV helicity configurations that are power suppressed in the central NMRK limit. The remaining eleven power suppressed MHV configurations can be obtained by flipping all helicities. The labelling of the gluons follows figure~\ref{fig:momenta}.}
    \label{fig:HelFlipMHV}
\end{figure}

The tree-level six gluon amplitude~(\ref{NLOfactorization}) describes a total of 50 helicity configurations. However, the helicity-flip impact factors $C^{g(0)}(p_2^\pm, p_3^\pm)$ and $C^{g(0)}(p_1^\pm, p_6^\pm)$ are power suppressed in $t_1/s$ and $t_3/s$,
respectively. Of the 20 NMHV configurations described by eq.~(\ref{NLOfactorization}), 12 are power suppressed in this manner, owing to one or two helicity-flip impact factors, as depicted in figure~\ref{fig:HelFlipNMHV}.
Similarly, of the 30 MHV configurations described by eq.~(\ref{NLOfactorization}), 22 are power suppressed due to one or two helicity-flip impact factors and are depicted in figure~\ref{fig:HelFlipMHV}.

The remaining 16 helicity configurations of \eqn{NLOfactorization} are all associated to amplitudes which do not have a helicity-flip impact factor, and these are the only configurations that we consider in this paper. 
\vspace{50 pt}

\FloatBarrier

\section{NMHV amplitudes in general kinematics}
\label{sec:nmhv6gamp}

In this appendix we list the tree-level and one-loop six-gluon amplitudes in $\cN=4$ SYM which are used in the main text to obtain the two-gluon  opposite-helicity CEV. For these amplitudes we use the shorthand notation for spinors listed in \eqn{eq:short_spinor}.

\subsection{Tree-level six-gluon NMHV amplitudes}
\label{sec:treeNMHV}

Here we list a representation of the tree-level six-gluon NMHV colour-ordered amplitudes obtained through the BCFW on-shell recursion relations~\cite{Britto:2004ap,Britto:2005fq},
\begin{align}
	\begin{split}
		M^{(0)}\left(p_1^\ominus, p_2^\ominus, p_3^\ominus, p_4^\oplus, p_5^\oplus, p_6^\oplus\right)=&\ \frac{\langle 1|2+3| 4]^{4}}{D_2^*}+\frac{\langle 3|4+5| 6]^{4}}{D_3},
		\label{eq:nmhvmmmppp}
	\end{split}
	\\
	\begin{split}
		M^{(0)}\left(p_1^\ominus, p_2^\ominus, p_3^\oplus, p_4^\ominus, p_5^\oplus, p_6^\oplus\right) =&\ \frac{\langle 4|1+2| 3]^{4}}{D_1^*}
		+\frac{\langle2 \, 4\rangle^{4}[5 \, 6]^{4}}{D_2}
		+\frac{\langle 1 \, 2\rangle^{4}[3 \, 5]^{4}}{D_3^*},
		\label{eq:nmhvmmpmpp}
	\end{split}
	\\
	\begin{split}
		M^{(0)}\left(p_1^\ominus, p_2^\oplus, p_3^\ominus, p_4^\oplus, p_5^\ominus, p_6^\oplus\right) =&\ \frac{\langle1 \, 3\rangle^{4}[4 \, 6]^{4}}{D_1} 
		+\frac{[2 \, 4]^{4}\langle 1 \, 5\rangle^{4}}{D_2^*}
		+\frac{[2 \, 6]^{4}\langle3 \, 5\rangle^{4}}{D_3}.
	\label{eq:nmhvmpmpmp}
	\end{split}
\end{align}
The functions in the denominators are
\begin{equation}
    D_1(1,2,3,4,5,6)=-\langle 1\, 2 \rangle \langle 2 \, 3 \rangle \left[4 \, 5\right] \left[5 \, 6\right] s_{123} \langle 1|2+3|4] \langle 3|4+5|6]\,, \label{eq:NMHV_D}
\end{equation}
and cyclic permutations thereof. We use the shorthand notation
\begin{equation}
    D_i(\sigma)=D_i(\sigma_{i},\sigma_{i+1},\sigma_{i+2},\sigma_{i+3},\sigma_{i+4},\sigma_{i+5}),
\end{equation}
where for the amplitudes in this section it is assumed  -- and left implicit -- that the ordering $\sigma$ is always taken to be that of the parent $M^{(0)}$, i.e. $\left\{1,2,3,4,5,6 \right\}$.
$D_{i+1}(\sigma)$ is obtained from $D_{i}(\sigma)$ by replacing $\sigma_i\to \sigma_{i+1}$ in the latter, with all indices taken modulo 6. However, there are only three independent cyclic orderings due to the relation $D_{i+3}(\sigma)=D_i^*(\sigma)$.

We will see that the individual rational terms that occur in eqs.~(\ref{eq:nmhvmmmppp}--\ref{eq:nmhvmpmpmp}) will appear also in the one-loop amplitudes listed in the following subsection. The coefficient of the $\epsilon^{-2}$ pole of these one-loop amplitudes must be proportional to the tree-level amplitude, thus, using the one-loop amplitudes, we obtain an alternative set of representations for the tree-level amplitudes:
\begin{align}
		M^{(0)}\left(p_1^\ominus, p_2^\ominus, p_3^\ominus, p_4^\oplus, p_5^\oplus, p_6^\oplus\right)=&\
		\frac{(s_{123})^4}{D_1^*}
		+\frac{\langle 2 \, 3 \rangle^4[5 \, 6]^4}{D_2}
		+
		\frac{\langle1 \, 2\rangle^4 [4 \, 5]^4}{D_3^*},
	\label{eq:NMHVbmmmppp}\\
		M^{(0)}\left(p_1^\ominus, p_2^\ominus, p_3^\oplus, p_4^\ominus, p_5^\oplus, p_6^\oplus\right)=&\
		\frac{\langle1 \, 2\rangle^4[5 \, 6]^4}{D_1}
		+\frac{\langle 1|2+4| 3]^{4}}{D_2^*}
		+\frac{\langle 4|1+2| 6]^{4}}{D_3},
	\label{eq:NMHVbmmpmpp}
\\
		M^{(0)}\left(p_1^\ominus, p_2^\oplus, p_3^\ominus, p_4^\oplus, p_5^\ominus, p_6^\oplus\right)=&\
		\frac{\langle 5|1+3| 2]^4}{D_1^*}
		+\frac{\langle 3|2+4|6]^4}{D_2}
		+\frac{\langle 1|2+6| 4]^4}{D_3^*}.	
	\label{eq:NMHVbmpmpmp}
\end{align}
The representations eqs.~(\ref{eq:nmhvmmmppp}--\ref{eq:nmhvmpmpmp}) are complementary to the representations eqs.~(\ref{eq:NMHVbmmmppp}--\ref{eq:NMHVbmpmpmp}) in the sense that where the NMRK limit of the rational terms of the former give rise directly to $R_{xy}^I$ (defined in eqs.~(\ref{eq:RA}) and~(\ref{eq:RB})), the rational terms of the latter give rise to ${R}_{\bar{x}\bar{y}}^I$ (defined in eqs.~(\ref{eq:RAb}) and~(\ref{eq:RBb})), and vice versa.

\subsection{One-loop six-gluon NMHV amplitudes in \texorpdfstring{$\cN=4$}{N=4} SYM}
\label{sec:1loopNMHV}
 
We will need to study the NMRK limit of the colour-ordered amplitudes derived in ref.~\cite{Bern:1994cg}, namely:
\begin{align}
	\begin{split}
		\frac{(4\pi)^2}{\cg}M^{(1)}_{\mathcal{N}=4}\left(p_1^\ominus, p_2^\ominus, p_3^\ominus, p_4^\oplus, p_5^\oplus, p_6^\oplus\right)=
		&\left(
		\frac{(s_{123})^4}{D_1^*}
		\right)W_1\\
		+&\left(
		\frac{\langle 1|2+3| 4]^{4}}{D_2^*}
		+\frac{\langle 2 \, 3 \rangle^4[5 \, 6]^4}{D_2}
		\right)W_2
		\\+&
		\left(
		\frac{\langle 3|4+5| 6]^{4}}{D_3}
		+\frac{\langle1 \, 2\rangle^4 [4 \, 5]^4}{D_3^*}
		\right)W_3\,,
	\end{split}
	\label{eq:N4mmmppp}
\end{align}
\begin{align}
	\begin{split}
		\frac{(4\pi)^2}{\cg}M^{(1)}_{\mathcal{N}=4}\left(p_1^\ominus, p_2^\ominus, p_3^\oplus, p_4^\ominus, p_5^\oplus, p_6^\oplus\right)=
		&\left(
		\frac{\langle 4 |1+2|3]^4}{D_1^*}
		+\frac{\langle1 \, 2\rangle^4[5 \,6]^4}{D_1}
		\right)W_1\\
		+&\left(
		\frac{\langle 1|2+4| 3]^{4}}{D_2^*}
		+\frac{\langle 2 \, 4 \rangle^4[5 \, 6]^4}{D_2}
		\right)W_2
		\\+&
		\left(
		\frac{\langle 4|1+2| 6]^{4}}{D_3}
		+\frac{\langle1 \, 2\rangle^4 [3 \, 5]^4}{D_3^*}
		\right)W_3\,,
	\end{split}
	\label{eq:N4mmpmpp}
\end{align}
\begin{align}
	\begin{split}
		\frac{(4\pi)^2}{\cg}M^{(1)}_{\mathcal{N}=4}\left(p_1^\ominus, p_2^\oplus, p_3^\ominus, p_4^\oplus, p_5^\ominus, p_6^\oplus\right)=
		&\left(
		\frac{\langle 1 \, 3\rangle^{4}[4 \, 6]^{4}}{D_1}
		+\frac{\langle 5|1+3| 2]^4}{D_1^*}
		\right)W_1\\
		+&\left(
		\frac{[2 \, 4]^{4}\langle 1 \, 5\rangle^{4}}{D_2^*}
		+\frac{\langle 3|2+4|6]^4}{D_2}
		\right)W_2\\
		+&\left(
		\frac{[2 \, 6]^{4}\langle 3 \, 5\rangle^{4}}{D_3}
		+\frac{\langle 1|2+6| 4]^4}{D_3^*}	
		\right)W_3\,.
	\end{split}
	\label{eq:N4mpmpmp}
\end{align}
The transcendental functions are given by
\begin{align}
	\begin{split}
		W_i(\sigma)&=-\frac{1}{2 \epsilon^{2}} \sum_{j=1}^{6}\left(\frac{\mu^{2}}{-t^{[2]}_{j}}\right)^{\epsilon}
		-\log \left(\frac{-t^{[3]}_{i}}{-t^{[2]}_{i}}\right) 
		\log \left(\frac{-t^{[3]}_{i}}{-t^{[2]}_{i+1}}\right)
		-\log \left(\frac{-t^{[3]}_{i}}{-t^{[2]}_{i+3}}\right) 
		\log \left(\frac{-t^{[3]}_{i}}{-t^{[2]}_{i+4}}\right)
		\\
		&\phantom{\frac{1}{2}}+\log \left(\frac{-t^{[3]}_{i}}{-t^{[2]}_{i+2}}\right) 
		\log \left(\frac{-t^{[3]}_{i}}{-t^{[2]}_{i+5}}\right)
		+\frac{1}{2}
		\log \left(\frac{-t^{[2]}_{i}}{-t^{[2]}_{i+3}}\right) 
		\log \left(\frac{-t^{[2]}_{i+1}}{-t^{[2]}_{i+4}}\right)
		\\&+
		\frac{1}{2}
		\log \left(\frac{-t^{[2]}_{i+5}}{-t^{[2]}_{i}}\right) 
		\log \left(\frac{-t^{[2]}_{i+1}}{-t^{[2]}_{i+2}}\right)
		+
		\frac{1}{2}\log \left(\frac{-t^{[2]}_{i+2}}{-t^{[2]}_{i+3}}\right) 
		\log \left(\frac{-t^{[2]}_{i+4}}{-t^{[2]}_{i+5}}\right)+\frac{\pi^{2}}{3}
		,
	\end{split}
	\label{eq:Wi}
\end{align}
where we use the same notation for the arguments and cyclic permutations as for the function $D_i$, discussed below \eqn{eq:NMHV_D}. While in 
eqs.~(\ref{eq:N4mmmppp}--\ref{eq:N4mpmpmp}) the ordering argument is left implicit, it is useful to display the ordering explicitly when the transcendental functions of different colour orderings appear in the same expression, e.g. in \eqn{eq:N4NMHVReW}.
We note that the function $W_i$ has cyclic symmetry in the form $W_{i+3}=W_{i}$ so there are only three independent cyclic permutations. In \eqn{eq:Wi} we use the compact notation for Mandelstam invariants introduced in \eqn{eq:ti}, and we use the prescription of~\eqn{eq:log_prescription}. We emphasise that the function $W_i$ does not depend on the individual helicities of the particles. 
\section{Additional forms of the opposite-helicity central-emission vertex}
\label{sec:altern}

In ref.~\cite{Duhr:2009uxa} an alternative form of the opposite-helicity two-gluon central-emission vertex was found by using the CSW rules~\cite{Cachazo:2004kj}, 
\begin{align}
\begin{split}
\left.A^{gg(0)}(q_1,p_4^\oplus,p_5^\ominus,q_3)\right|_4=&\sqrt{\frac{x_4}{x_5}}\frac{\left|q_{1 \perp}\right|^{2} x_{4} p_{5 \perp}^{3}}{\langle 4 \, 5\rangle\left(x_{5}\left|p_{4 \perp}\right|^{2}+x_{4}\left|p_{5 \perp}\right|^{2}\right) p_{4 \perp}\left(p_{4 \perp}+p_{5 \perp}\right)} \\
&+\sqrt{\frac{x_4}{x_5}}\frac{q_{1 \perp}^{*} x_{4}\left(q_{3 \perp}+p_{5 \perp}\right)^{3}}{\left(q_{3 \perp} \sqrt{\frac{x_4}{x_5}}-\langle 4 \, 5\rangle\right)\left(x_{5}\left|p_{4 \perp}\right|^{2}+x_{4}\left|q_{2 \perp}\right|^{2}\right) p_{4 \perp}} \\
&+\frac{q_{1 \perp} q_{3 \perp} \sqrt{x_{4}x_{5}} x_{5}}{\langle 4 \, 5 \rangle\left(x_{5} p_{4 \perp}-x_{4} q_{2 \perp}\right)} -\sqrt{\frac{x_4}{x_5}}\frac{q_{1 \perp}^{*} q_{3 \perp} x_{4}}{\left(p_{4 \perp}+p_{5 \perp}\right)[4 \, 5]} \,.
\end{split}
\label{eq:AggpmCSW}
\end{align}
In the minimal variables of section~\ref{sec:minset} this is
\begin{align}
\begin{split}
\left.A^{gg(0)}(q_1,p_4^\oplus,p_5^\ominus,q_3)\right|_4=&
-\frac{\bar{w}z^3|1-z|^2X^2}{(w-z)(w + X z)(|w|^2+X|z|^2)}
+\frac{X^2z^3(1-\bar{z})}{(1+Xz)(1+X|z|^2)}
\\&+\frac{(1-w)(1-z)z X}{(1+X)(1+Xz)(w+Xz)}
+\frac{(1-w)(1-\bar{z})\bar{w}z X^2}{(1+X)(w-z)(\bar{w}+X \bar{z})} \,,
\end{split}
\label{eq:AggpmCSWmin}
\end{align}
which has the expected physical singularities of \eqns{eq:physsing}{eq:physsing2}, and also spurious poles at $P_{u_A}=0$ and $\bar{P}_{v_A}=0$.
\par
We also list a compact representation we have found
\beqa
\left. A^{gg(0)}(q_1, p^\oplus_4, p^\ominus_5, q_3)\right|_5 &=& \frac{X z (\bar{w}-1) (Xz+w+z-1)}{(X+1) (w+Xz)} 
- \frac{X \bar{w} (w-1) (\bar{w}+X) z}{(\bar{w}+X \bar{z}) (w-z) (X+1)} \nn\\
&&+ \frac{X \bar{w} (z-1) z^2 (X z+|w|^2)}{(|w|^2+X |z|^2) (w-z) (w+Xz)}
+ \frac{X z^2}{X |z|^2+1} \,,
\label{eq:take5}
\eeqa
which has only a single spurious pole at $P_{u_A}=0$. We finally list a representation which is compact when expressed in terms of light-cone coordinates, which was obtained by partial fractioning the third term of \eqn{kosc3},
	\beqa
	\lefteqn{ \left. A^{gg(0)}(q_1, p^\oplus_4, p^\ominus_5, q_3)\right|_6 }\nn\\
	&& \qquad = x_5 - y_5\, \frac{|q_{1\perp}|^2}{p_{4\perp} p_{5\perp}^*} 
	- \frac{ q_{1\perp}^* ( q_{1\perp} - p_{4\perp})^2}{p_{4\perp} s_{234}} - \frac{x_5 p_{4\perp}^* ( q_{1\perp} - p_{4\perp})}{x_4 s_{234}}
	\nn\\
	&& \qquad \quad + \frac{q_{1\perp}^*}{p_{4\perp} \langle4 \, 5\rangle^*} x_4 \sqrt{\frac{x_4}{x_5}} ( y_4 q_{1\perp} - p_{4\perp})
	+ \frac{q_{1\perp}}{p_{5\perp}^* \langle4 \, 5\rangle} x_5 \sqrt{\frac{x_5}{x_4}} ( y_5 q_{1\perp}^* - p_{5\perp}^*) \,.
	\label{kosc4}
	\eeqa
	
\subsection{BCFW representation for the ordering \texorpdfstring{$\sigma_{A'}$}{A'}}
\label{sec:BCFWAx}

In this appendix we present the BCFW representations of the opposite-helicity vertex for the ordering $\sigma_{A'}$. These expressions 
can be obtained from the results for the ordering $\sigma_{A}$ in 
section~\ref{sec:other} by applying the rules listed in~\eqn{eq:AtoAx}. 

For the $\sigma_{A'}=\{1,2,3,5,4,6\}$ ordering we list the NMRK limits of the cross ratios,
\begin{equation}
	\begin{array}{ccccc}
		u_{A'} & = &\!\!\!\! u_A,\\
		v_{A'} & = & {\displaystyle\frac{s_{23}s_{46}}{s_{235}s_{123}}} & \toNMRK & {\displaystyle \frac{|w|^2|z-1|^2X}{(1+X)(|w+z-wz|^2+X|z|^2)} }\,,\\
		w_{A'} & = & {\displaystyle \frac{s_{35}s_{61}}{s_{345}s_{235}} }& \toNMRK &{\displaystyle \frac{|z|^4|w-1|^2X}{(|w|^2+X|z|^2)(|w+z-wz|^2+X|z|^2)}} \,.
		\label{eq:uvwAx}
	\end{array}
\end{equation}
From the BCFW representations of the NMHV amplitudes of eqs.~(\ref{eq:nmhvmmmppp}, \ref{eq:nmhvmpmpmp}, \ref{eq:NMHVbmmpmpp})
we obtain the opposite-helicity central-emission vertex,
\begin{align}
	\begin{split}
		A^{gg(0)}(q_1, p^\ominus_5, p^\oplus_4, q_3) = R^{A'}_{uv}+R^{A'}_{vw}+R^{A'}_{wu}\, ,
	\end{split}
	\label{kosc2p}
\end{align}
with
\begin{align}
	\begin{split}
		R^{A'}_{uv}&=\frac{|w-1|^2 \bar{w} X^2 |z|^2 (\bar{z}-1)}{(\bar{w}+X \bar{z})(X+1)(w-z)  (\bar{w}+\bar{z}-\bar{w}\bar{z}+X \bar{z})}\,,\\
		R^{A'}_{vw}&=\frac{wz(w-1)  (z-1)  (\bar{w}+\bar{z}-\bar{w} \bar{z})^4}{\bar{z}  (|w+z-wz|^2+X|z|^2)(\bar{w}+\bar{z}-\bar{w}\bar{z}+X \bar{z})
		(w (\bar{w}+\bar{z}-\bar{w} \bar{z})+X |z|^2)}\,, \\
		R^{A'}_{wu}&=-\frac{ (\bar{w}-1) |w|^2 X^2 |z-1|^2 z^2 |z|^2}{ (w+X z)  (|w|^2+X |z|^2)(w (\bar{w}+\bar{z}-\bar{w} \bar{z})+X |z|^2)(w-z)}\,.
	\end{split}
	\label{eq:RAxb}
\end{align}
Instead starting from the BCFW representations of eqs.~(\ref{eq:nmhvmmpmpp}, \ref{eq:NMHVbmmmppp}, \ref{eq:NMHVbmpmpmp}), we obtain the representation,
\begin{align}
	\begin{split}
		A^{gg(0)}(q_1, p^\ominus_5, p^\oplus_4, q_3) =  R^{A'}_{\bar{u}\bar{v}}+R^{A'}_{\bar{v}\bar{w}}+R^{A'}_{\bar{w}\bar{u}}\, ,
	\end{split}
	\label{kosc2p2}
\end{align}
with
\begin{align}
	\begin{split}
		R^{A'}_{\bar{u}\bar{v}}&=\frac{|w-1|^2 w  (z-1) |z|^2}{(X+1) (\bar{w}-\bar{z}) (w+X z) (w+z-w z+X z)}\,,\\
		R^{A'}_{\bar{v}\bar{w}}&=\frac{\bar{w}(\bar{w}-1)(\bar{z}-1)  z^2  |z|^2X^2 }{(|w+z-wz|^2+X|z|^2)(w+z-wz+Xz) (\bar{w}(w+z-w z)+X |z|^2) }\,, \\
		R^{A'}_{\bar{w}\bar{u}}&=-\frac{(w-1) |w|^2\bar{w}^4 |z-1|^2 z }{\bar{z} (|w|^2+X |z|^2)   (\bar{w}+X \bar{z})(\bar{w}(w+z-w z)+X |z|^2)(\bar{w}-\bar{z})}\,.
	\end{split}
	\label{eq:RAx}
\end{align}
The physical-singularity surfaces of this ordering are
\begin{equation}
	\begin{array}{ccccccc}
		\langle 4 \, 5 \rangle= 0  & \leftrightarrow & 
		p_{4\perp}\sqrt{\displaystyle \frac{p_4^+}{p_5^+}}-p_{5\perp}\sqrt{\displaystyle\frac{p_5^+}{p_4^+}}=0
		& \leftrightarrow & w+Xz  = 0\,, \\
		s_{123} = 0  & \leftrightarrow & p_4^++p_5^+=0 
		& \leftrightarrow & 1+X  = 0 \,,\\
		s_{235} = 0  & \leftrightarrow & 
		|q_{2\perp}'|^2+p_4^+p_5^-=0
		& \leftrightarrow & |w+z-wz|^2+X|z|^2 = 0 \,, \\
		s_{354} = 0  & \leftrightarrow & p_4^-+p_5^-=0  
		& \leftrightarrow & |w|^2+X|z|^2 = 0 \,,
	\end{array}
	\label{eq:physsingA'}
\end{equation}
while the unphysical-singularity surfaces are
\begin{equation}
	\begin{array}{ccccccc}
		\langle 1 |2+3|5]=0 & \leftrightarrow & P^{A'}_{v}=0 & \leftrightarrow & q^{\prime \ast }_{2\perp}-p_{5\perp}^*{\displaystyle \frac{p_4^+}{p_5^+}}=0 & \leftrightarrow  & \bar{w}+\bar{z}-\bar{w} \bar{z}+X \bar{z}=0 \,,\\
		\langle 3 |5+4|6]=0 & \leftrightarrow & P^{A'}_{u}=0 & \leftrightarrow & p_{4\perp}+p_{5\perp}=0 & \leftrightarrow  & w-z=0 \,,\\
		\langle 4 |6+1|2]=0 & \leftrightarrow & P^{A'}_{w}=0 & \leftrightarrow & q^{\prime \ast }_{2\perp}+p_{5\perp}^*
		{\displaystyle \frac{p_{5\perp}}{p_{4\perp}}\frac{p_4^+}{p_5^+}}=0 & \leftrightarrow  & w (\bar{w}+\bar{z}-\bar{w} \bar{z})+X |z|^2=0 \,.\\
	\end{array}
	\label{eq:unphysAx}
\end{equation}
We note that  
\begin{align}
	\begin{split}
		R^A_{xy}\ \ {\underset{p_4\leftrightarrow p_5}\leftrightarrow}\ \ \left(R^{A'}_{\bar{x}\bar{y}}\right)^* \,.
	\end{split}
	\label{eq:RAswitch45}
\end{align}

\boldmath
\section{Absorptive part of the one-loop six-gluon amplitude in NMRK}
\unboldmath
\label{sec:absorp}

In \eqn{eq:Msig-sig} we decomposed the \texorpdfstring{$\cN=4$}{N=4} SYM one-loop six-gluon amplitude in NMRK into a basis of symmetric $(+)$ or antisymmetric $(-)$ representations exchanged in the $t_1$ and $t_3$ channels. For one-loop amplitudes where an adjoint representation is circulating in the loop, the $(-,-)$ component is given by \eqn{eq:Modd-odd}. We list here analogous expressions for the remaining components:
\begin{align}
    \begin{split}
    &\mathcal{M}^{(1)(+,+)}=\frac{\gs^6}{4}\\
    &\times \bigg\{
    \tr\left(\left\{F^{a_2},F^{a_3}\right\}F^{a_4}F^{a_5}\left\{F^{a_6},F^{a_1}\right\}\right) \left(M^{(1)}(\sigma_A)+M^{(1)}(\overset{\leftrightarrow}{\sigma_A})+M^{(1)}(\underset{\leftrightarrow}{\sigma_A})+M^{(1)}(\overset{\leftrightarrow}{\underset{\leftrightarrow}{\sigma_A}})\right)\\
    &+\tr\left(\left\{F^{a_2},F^{a_3}\right\}F^{a_5}F^{a_4}\left\{F^{a_6},F^{a_1}\right\}\right) \left(M^{(1)}(\sigma_{A'})+M^{(1)}(\overset{\leftrightarrow}{\sigma_{A'}})+M^{(1)}(\underset{\leftrightarrow}{\sigma_{A'}})+M^{(1)}(\overset{\leftrightarrow}{\underset{\leftrightarrow}{\sigma_{A'}}})\right)\\
    &+\tr\left(\left\{F^{a_2},F^{a_3}\right\}F^{a_4}\left\{F^{a_6},F^{a_1}\right\}F^{a_5}\right) \left(M^{(1)}(\sigma_B)+M^{(1)}(\overset{\leftrightarrow}{\sigma_B})+M^{(1)}(\underset{\leftrightarrow}{\sigma_B})+M^{(1)}(\overset{\leftrightarrow}{\underset{\leftrightarrow}{\sigma_B}})\right)\bigg\}
    .
    \end{split}
    \label{eq:Meven-even}
\end{align} 

\begin{align}
    \begin{split}
    &\mathcal{M}^{(1)(+,-)}=\frac{\gs^6}{4}
    \\
    &\times \bigg\{
    \tr\left(\left\{F^{a_2},F^{a_3}\right\}F^{a_4}F^{a_5}\left[F^{a_6},F^{a_1}\right]\right) \left(M^{(1)}(\sigma_A)+M^{(1)}(\overset{\leftrightarrow}{\sigma_A})-M^{(1)}(\underset{\leftrightarrow}{\sigma_A})-M^{(1)}(\overset{\leftrightarrow}{\underset{\leftrightarrow}{\sigma_A}})\right)\\
    &+\tr\left(\left\{F^{a_2},F^{a_3}\right\}F^{a_5}F^{a_4}\left[F^{a_6},F^{a_1}\right]\right) \left(M^{(1)}(\sigma_{A'})+M^{(1)}(\overset{\leftrightarrow}{\sigma_{A'}})-M^{(1)}(\underset{\leftrightarrow}{\sigma_{A'}})-M^{(1)}(\overset{\leftrightarrow}{\underset{\leftrightarrow}{\sigma_{A'}}})\right)\\
    &+\tr\left(\left\{F^{a_2},F^{a_3}\right\}F^{a_4}\left[F^{a_6},F^{a_1}\right]F^{a_5}\right)
    \left(M^{(1)}(\sigma_B)+M^{(1)}(\overset{\leftrightarrow}{\sigma_B})-M^{(1)}(\underset{\leftrightarrow}{\sigma_B})-M^{(1)}(\overset{\leftrightarrow}{\underset{\leftrightarrow}{\sigma_B}})\right)\bigg\}
    .
    \end{split}
    \label{eq:Meven-odd}
\end{align} 
\begin{align}
    \begin{split}
    &\mathcal{M}^{(1)(-,+)}=\frac{\gs^6}{4}\\&\times \bigg\{\tr\left(\left[F^{a_2},F^{a_3}\right]F^{a_4}F^{a_5}\left\{F^{a_6},F^{a_1}\right\}\right) \left(M^{(1)}(\sigma_A)-M^{(1)}(\overset{\leftrightarrow}{\sigma_A})+M^{(1)}(\underset{\leftrightarrow}{\sigma_A})-M^{(1)}(\overset{\leftrightarrow}{\underset{\leftrightarrow}{\sigma_A}})\right)\\
    &+\tr\left(\left[F^{a_2},F^{a_3}\right]F^{a_5}F^{a_4}\left\{F^{a_6},F^{a_1}\right\}\right) \left(M^{(1)}(\sigma_{A'})-M^{(1)}(\overset{\leftrightarrow}{\sigma_{A'}})+M^{(1)}(\underset{\leftrightarrow}{\sigma_{A'}})-M^{(1)}(\overset{\leftrightarrow}{\underset{\leftrightarrow}{\sigma_{A'}}})\right)\\
    &+\tr\left(\left[F^{a_2},F^{a_3}\right]F^{a_4}\left\{F^{a_6},F^{a_1}\right\}F^{a_5}\right)
    \left(M^{(1)}(\sigma_B)-M^{(1)}(\overset{\leftrightarrow}{\sigma_B})+M^{(1)}(\underset{\leftrightarrow}{\sigma_B})-M^{(1)}(\overset{\leftrightarrow}{\underset{\leftrightarrow}{\sigma_B}})\right)\bigg\}
    .
    \end{split}
    \label{eq:Modd-even}
\end{align} 
In \sec{sec:6g1lamp} we found that in the central NMRK, the dispersive amplitude received a contribution from the $(-,-)$ component alone. The same is not true for the absorptive part of the amplitude, which is the object of study of this appendix. To this end we list the NMRK limit of the imaginary parts of the transcendental functions \eqn{eq:V6g} and \eqn{eq:Wi}, for the 12 colour orderings that appear in  eqs.~(\ref{eq:Meven-even})--(\ref{eq:Modd-even}).  We begin by giving the results for the $\sigma_A$ orderings:
\begin{align}
	\begin{split}
	\IM\Big[V_6(\sigma_A)\Big]=2\IM\Big[W_2(\sigma_A)\Big]
	\toNMRK&-\frac{4\pi}{\epsilon}+\pi\log\left(\frac{t_1p_4^-s_{45}p_5^+t_3}{(\mu^2)^4}\right) \,,\\
	\IM\Big[V_6(\overset{\leftrightarrow}{\sigma_A})\Big]=2\IM\Big[W_2(\overset{\leftrightarrow}{\sigma_A})\Big]
	\toNMRK&-\frac{2\pi}{\epsilon}+\pi\log\left(\frac{p_4^-s_{45}p_5^+t_3}{t_1(\mu^2)^2}\right) \,,\\
	\IM\Big[V_6(\underset{\leftrightarrow}{\sigma_A})\Big]=2\IM\Big[W_2(\underset{\leftrightarrow}{\sigma_A})\Big]
	\toNMRK&-\frac{2\pi}{\epsilon}+\pi\log\left(\frac{t_1p_4^-s_{45}p_5^+}{t_3(\mu^2)^2}\right) \,,\\
	\IM\Big[V_6(\overset{\leftrightarrow}{\underset{\leftrightarrow}{\sigma_A}})\Big]
	\toNMRK&-\frac{2\pi}{\epsilon}+\pi\log\left(\frac{t_1p_4^-p_5^+t_3}{s_{45}(\mu^2)^2}\right)+2\pi\log\left(\frac{u_A}{1-u_A}\right) \,,\\
	2\IM\Big[W_2(\overset{\leftrightarrow}{\underset{\leftrightarrow}{\sigma_A}})\Big]
	\toNMRK&-\frac{2\pi}{\epsilon}+\pi\log\left(\frac{t_1p_4^-p_5^+t_3}{s_{45}(\mu^2)^2}\right)-2\pi\log\left(\frac{v_A w_A}{u_A}\right) \,.
	\end{split}
\end{align}
We provide the results for $W_1$ and $W_3$ as differences from those for $W_2$. Interestingly, $\sigmaAtb$ is the only ordering which leads to differing imaginary parts for different cyclic permutations of $W_i$:
\begin{equation}
	\begin{array}{cclcccl}
		\IM\Big[W_1(\sigma_{A})-W_2(\sigma_{A})\Big] & = & 0\,, & \quad & \IM\Big[W_3(\sigma_{A})-W_2(\sigma_{A})\Big] & = & 0\,,\\
		\IM\Big[W_1(\sigmaAt)-W_2(\sigmaAt)\Big] & = & 0\,, & \quad & \IM\Big[W_3(\sigmaAt)-W_2(\sigmaAt)\Big] & = & 0\,,\\
		\IM\Big[W_1(\sigmaAb)-W_2(\sigmaAb)\Big] & = & 0\,, & \quad & \IM\Big[W_3(\sigmaAb)-W_2(\sigmaAb)\Big] & = & 0\,,\\
		\IM\Big[W_1(\sigmaAtb)-W_2(\sigmaAtb)\Big] & = & 2\pi\log(v_A)\,, & \quad & \IM\Big[W_3(\sigmaAtb)-W_2(\sigmaAtb)\Big] & = & 2\pi\log(w_A)\,.\\
	\end{array}
\end{equation}
We further note that only for the ordering $\sigmaAtb$ does the analytic continuation of the dilogarithms in \eqn{eq:V6g} to our physical region produce an imaginary part. For this colour ordering, the cross ratio 
$$\frac{s_{63}s_{45}}{s_{632}s_{245}},$$
which is related to $u_A$ by interchange of $p_1 \leftrightarrow p_6$ and $p_2 \leftrightarrow p_3$, has two positive invariants in the numerator and two negative invariants in the denominator.  The analytic continuation of the dilogarithm that depends on this cross ratio leads to factor of $i\pi \log(1-u_A)$ which is related to the factorisation breaking term discovered in ref.~\cite{Bartels:2008ce}.  The interpretation of this quantity is different in our paper; ref.~\cite{Bartels:2008ce} considers the single planar colour ordering of the amplitude, analytically continued to the physical region where $s_{12},s_{45}>0$, $s_{23},s_{34},s_{56},s_{61},s_{234},s_{345}<0$, 
while we consider a certain colour-ordered amplitude in the physical region where $s_{12},s_{34},s_{45},s_{56},s_{345}>0$, $s_{23},s_{61},s_{234}<0$.  However, the consequence is the same, namely that this term is not compatible with Regge-pole factorisation of the amplitude, which only applies to its dispersive part. Instead it contributes to a Regge cut.

Expressions for the $\sigma_{A'}$ orderings are related by an interchange of $p_4 \leftrightarrow p_5$. 

For the $\sigma_{B}$ orderings we obtain
\begin{align}
	\begin{split}
	\IM\Big[V_6(\sigma_B)\Big]=2\IM\Big[W_1(\sigma_B)\Big]
	\toNMRK&-\frac{2\pi}{\epsilon}+\pi\log\left(
	\frac{t_1p_5^+p_5^-t_3}{p_4^+p_4^-(\mu^2)^2}
	\right) \,, \\
	\IM\Big[V_6(\overset{\leftrightarrow}{\sigma_B})\Big]
	\toNMRK&-\frac{2\pi}{\epsilon}+\pi\log\left(
	\frac{t_1p_4^+p_5^-t_3}{p_4^-p_5^+(\mu^2)^2}
	\right)+2\pi\log\left(\frac{v_B}{1-v_B}\right) \,,\\
	2\IM\Big[W_1(\overset{\leftrightarrow}{\sigma_B})\Big]
	\toNMRK&-\frac{2\pi}{\epsilon}+\pi\log\left(
	\frac{t_1p_4^+p_5^-t_3}{p_4^-p_5^+(\mu^2)^2}
	\right)+2\pi\log\left(\frac{u_B v_B}{w_B}\right) \,,\\
	\IM\Big[V_6(\underset{\leftrightarrow}{\sigma_B})\Big]
	\toNMRK&-\frac{2\pi}{\epsilon}+\pi\log\left(
	\frac{t_1p_4^-p_5^+t_3}{p_4^+p_5^-(\mu^2)^2}
	\right)+2\pi\log\left(\frac{u_B}{1-u_B}\right) \,,\\
	2\IM\Big[W_1(\underset{\leftrightarrow}{\sigma_B})\Big]
	\toNMRK&-\frac{2\pi}{\epsilon}+\pi\log\left(
	\frac{t_1p_4^-p_5^+t_3}{p_4^+p_5^-(\mu^2)^2}
	\right)+2\pi \log\left(\frac{u_B v_B}{w_B}\right) \,,\\
	\IM\Big[V_6(\overset{\leftrightarrow}{\underset{\leftrightarrow}{\sigma_B}})\Big]=2\IM\Big[W_1(\overset{\leftrightarrow}{\underset{\leftrightarrow}{\sigma_B}})\Big]
	\toNMRK&-\frac{2\pi}{\epsilon}+\pi\log\left(
	\frac{t_1p_4^+p_4^-t_3}{p_5^+p_5^-(\mu^2)^2}
	\right)\,.
	\end{split}
\end{align}
Again the orderings that lead to imaginary contributions from the dilogarithms (namely $\overset{\leftrightarrow}{\sigma_B}$, $\underset{\leftrightarrow}{\sigma_B}$) are the same orderings for which the imaginary parts of the distinct cyclic permutations of $W_i$ differ:
\begin{equation}
	\begin{array}{cclcccl}
		\IM\Big[W_2(\sigma_{B})-W_1(\sigma_{B})\Big] & = & 0\,, & \quad & 
		\IM\Big[W_3(\sigma_{B})-W_1(\sigma_{B})\Big] & = & 0\,,\\
		\IM\Big[W_2(\sigmaBt)-W_1(\sigmaBt)\Big] & = & 2\pi\log\left(\frac{w_B}{u_B}\right)\,, & \quad & \IM\Big[W_3(\sigmaBt)-W_1(\sigmaBt)\Big] & = & -2\pi\log\left(u_B\right)\,,\\
		\IM\Big[W_2(\sigmaBb)-W_1(\sigmaBb)\Big] & = & -2\pi\log\left(v_B\right) \,, & \quad & \IM\Big[W_3(\sigmaBb)-W_1(\sigmaBb)\Big] & = & 2\pi\log\left(\frac{w_B}{v_B}\right)\,,\\
		\IM\Big[W_2(\sigmaBtb)-W_1(\sigmaBtb)\Big] & = & 0\,, & \quad & \IM\Big[W_3(\sigmaBtb)-W_1(\sigmaBtb)\Big] & = & 0\,.\\
	\end{array}
\end{equation}
\subsection*{Absorptive part of the same-helicity amplitude}
From the above lists of imaginary parts we can now assemble the absorptive parts of the colour-dressed amplitude for the case where $\nu_4=\nu_5=\oplus$:
\begin{align}
	\begin{split}
	\absorp{\mathcal{M}^{(1)(-,-)}_{\mathcal{N}=4}(p_4^\oplus,p_5^\oplus)}&\toNMRK \gs^6\frac{\kappa_{\Gamma}}{(4\pi)^2}s_{12}F^{a_3 a_2 c_1}C^{g(0)}\left(p_2^{\nu_2}, p_3^{-\nu_2}\right)\frac{1}{t_1}\\
	\times \frac{\pi}{2} \Bigg\{  
	\tr \left(F^{c_1}F^{a_4}F^{a_5}F^{c_3}\right)&A^{gg(0)}(q_1, p^\oplus_4, p^\oplus_5, q_3)\left(\frac{-5}{\epsilon}+\log\left(\frac{t_1s_{45}(p_4^-p_5^+)^2t_3}{(\mu^2)^5}\frac{u_A}{1-u_A}\right)\right)\\
	+\tr \left(F^{c_1}F^{a_5}F^{a_4}F^{c_3}\right)&A^{gg(0)}(q_1, p^\oplus_5, p^\oplus_4, q_3)\left(\frac{-5}{\epsilon}+\log\left(\frac{t_1s_{45}(p_4^+p_5^-)^2t_3}{(\mu^2)^5}\frac{u_{A'}}{1-u_{A'}}\right)\right)\\
	+\tr \left(F^{c_1}F^{a_5}F^{c_3}F^{a_4}\right)&B^{gg(0)}(q_1, p^\oplus_4, p^\oplus_5, q_3)
	\left(\frac{-4}{\epsilon}+ \log \left( \frac{t_1^2t_3^2}{(\mu^2)^4}\frac{v_B}{1-v_B} \frac{u_B}{1-u_B} \right)\right)
	\Bigg\} 
	\\
	&\hspace{-11 em}\times\frac{1}{t_3}F^{a_6 a_1 c_3}C^{g(0)}\left(p_1^{\nu_1}, p_6^{-\nu_1}\right)\,,
\end{split}
\label{eq:abN41loop}
\end{align}
\begin{align}
    \begin{split}
    \absorp{\mathcal{M}^{(1)(+,-)}_{\mathcal{N}=4}(p_4^\oplus,p_5^\oplus)}\toNMRK & \frac{\pi}{2}  \gs^6\frac{\kappa_{\Gamma}}{(4\pi)^2}s_{12}C^{g(0)}\left(p_2^{\nu_2}, p_3^{-\nu_2}\right)\frac{1}{t_1}
    \frac{1}{t_3}C^{g(0)}\left(p_1^{\nu_1}, p_6^{-\nu_1}\right)\\\times\Bigg\{\tr\left(\left\{F^{a_2},F^{a_3}\right\}F^{a_4}F^{a_5}\left[F^{a_6},F^{a_1}\right]\right)&A^{gg(0)}(q_1, p^\oplus_4, p^\oplus_5, q_3)\left(\frac{-1}{\epsilon}+\log\left(\frac{t_3s_{45}}{t_1\mu^2}\frac{1-u_A}{u_A}\right)\right)\\
    +\tr\left(\left\{F^{a_2},F^{a_3}\right\}F^{a_5}F^{a_4}\left[F^{a_6},F^{a_1}\right]\right)&A^{gg(0)}(q_1, p^\oplus_5, p^\oplus_4, q_3)\left(\frac{-1}{\epsilon}+\log\left(\frac{t_3s_{45}}{t_1\mu^2}\frac{1-u_{A'}}{u_{A'}}\right)\right)\\
    +\tr\left(\left\{F^{a_2},F^{a_3}\right\}F^{a_4}\left[F^{a_6},F^{a_1}\right]F^{a_5}\right) &B^{gg(0)}(q_1, p^\oplus_4, p^\oplus_5, q_3) \log\left(\left(\frac{p_5^-}{p_4^+}\right)^2\frac{1-u_B}{u_B}\frac{v_B}{1-v_B}\right)\Bigg\}
    \,,
    \label{eq:absorpM+-}
    \end{split}
\end{align} 
\begin{align}
    \begin{split}
    \absorp{\mathcal{M}^{(1)(-,+)}_{\mathcal{N}=4}(p_4^\oplus,p_5^\oplus)}\toNMRK & \frac{\pi}{2}  \gs^6\frac{\kappa_{\Gamma}}{(4\pi)^2}s_{12}C^{g(0)}\left(p_2^{\nu_2}, p_3^{-\nu_2}\right)\frac{1}{t_1}\frac{1}{t_3}C^{g(0)}\left(p_1^{\nu_1}, p_6^{-\nu_1}\right)\\\times\Bigg\{\tr\left(\left[F^{a_2},F^{a_3}\right]F^{a_4}F^{a_5}\left\{F^{a_6},F^{a_1}\right\}\right)&A^{gg(0)}(q_1, p^\oplus_4, p^\oplus_5, q_3)\left(\frac{-1}{\epsilon}+\log\left(\frac{t_1s_{45}}{t_3\mu^2}\frac{1-u_A}{u_A}\right)\right)\\
    +\tr\left(\left[F^{a_2},F^{a_3}\right]F^{a_5}F^{a_4}\left\{F^{a_6},F^{a_1}\right\}\right)&A^{gg(0)}(q_1, p^\oplus_5, p^\oplus_4, q_3)\left(\frac{-1}{\epsilon}+\log\left(\frac{t_1s_{45}}{t_3\mu^2}\frac{1-u_{A'}}{u_{A'}}\right)\right)\\
    +\tr\left(\left[F^{a_2},F^{a_3}\right]F^{a_4}\left\{F^{a_6},F^{a_1}\right\}F^{a_5}\right) &B^{gg(0)}(q_1, p^\oplus_4, p^\oplus_5, q_3) \log\left(\left(\frac{p_5^+}{p_4^-}\right)^2\frac{u_B}{1-u_B}\frac{1-v_B}{v_B}\right)\Bigg\}
    \,,
    \label{eq:absorpM-+}
    \end{split}
\end{align} 
\begin{align}
    \begin{split}
    \absorp{\mathcal{M}^{(1)(+,+)}_{\mathcal{N}=4}(p_4^\oplus,p_5^\oplus)}\toNMRK  \frac{\pi}{2} & \gs^6\frac{\kappa_{\Gamma}}{(4\pi)^2}s_{12}C^{g(0)}\left(p_2^{\nu_2}, p_3^{-\nu_2}\right)\frac{1}{t_1}\frac{1}{t_3}C^{g(0)}\left(p_1^{\nu_1}, p_6^{-\nu_1}\right)\\\times\Bigg\{\tr\left(\left\{F^{a_2},F^{a_3}\right\}F^{a_4}F^{a_5}\left\{F^{a_6},F^{a_1}\right\}\right)&A^{gg(0)}(q_1, p^\oplus_4, p^\oplus_5, q_3)\left(\frac{-1}{\epsilon}+\log\left(\frac{t_1t_3}{s_{45}\mu^2}\frac{u_A}{1-u_A}\right)\right)\\
    +\tr\left(\left\{F^{a_2},F^{a_3}\right\}F^{a_5}F^{a_4}\left\{F^{a_6},F^{a_1}\right\}\right)&A^{gg(0)}(q_1, p^\oplus_5, p^\oplus_4, q_3)\left(\frac{-1}{\epsilon}+\log\left(\frac{t_1t_3}{s_{45}\mu^2}\frac{u_{A'}}{1-u_{A'}}\right)\right)\\
    +\tr\left(\left\{F^{a_2},F^{a_3}\right\}F^{a_4}\left\{F^{a_6},F^{a_1}\right\}F^{a_5}\right) &B^{gg(0)}(q_1, p^\oplus_4, p^\oplus_5, q_3) \log\left(\frac{1-u_B}{u_B}\frac{1-v_B}{v_B}\right)\Bigg\}
    \,.
    \end{split}
    \label{eq:absorpM++}
\end{align} 
In section~\ref{sec:NLO_xs} we show that the absorptive parts listed in eqs.~(\ref{eq:abN41loop}--\ref{eq:absorpM++}) do not contribute to the NLO $gg\to gggg$ squared matrix element. These terms will only contribute to the squared matrix element at higher orders in the coupling.

\subsection*{Absorptive part of the opposite-helicity amplitude}
Here we similarly list the absorptive parts of the colour-dressed amplitude for the case where $\nu_4=-\nu_5=\oplus$:
\begin{align}
	\begin{split}
	&\hspace{-3 em}
	\absorp{\mathcal{M}^{(1)(-,-)}_{\mathcal{N}=4}(p_4^\oplus,p_5^\ominus)}\toNMRK \gs^6\frac{\kappa_{\Gamma}}{(4\pi)^2}s_{12}F^{a_3 a_2 c_1}C^{g(0)}\left(p_2^\ominus, p_3^\oplus\right)\frac{1}{t_1}\\
	\times \frac{\pi}{2}\Bigg\{  
	&\tr \left(F^{c_1}F^{a_4}F^{a_5}F^{c_3}\right)
	\bigg[
	A^{gg(0)}(q_1, p^\oplus_4, p^\ominus_5, q_3)
	\left(
	\frac{-5}{\epsilon}+\log\left(\frac{t_1s_{45}(p_4^-p_5^+)^2t_3}{(\mu^2)^5}\frac{u_A}{v_Aw_A}\right)
	\right)\\
	&\hspace{22 em} +S^A_{uv}\log(v_A)+S^A_{wu}\log(w_A)
	\bigg]\\
	+&\tr \left(F^{c_1}F^{a_5}F^{a_4}F^{c_3}\right)
	\bigg[
	A^{gg(0)}(q_1, p^\ominus_5, p^\oplus_4, q_3)\left(\frac{-5}{\epsilon}+\log\left(\frac{t_1s_{45}(p_4^+p_5^-)^2t_3}{(\mu^2)^5}\frac{u_{A'}}{v_{A'}w_{A'}}\right)\right)\\
	&\hspace{22 em} +S^{A'}_{uv}\log(v_{A'})+S^{A'}_{wu}\log(w_{A'})
	\bigg]\\
	+&\tr \left(F^{c_1}F^{a_5}F^{c_3}F^{a_4}\right)
	\bigg[
	B^{gg(0)}(q_1, p^\oplus_4, p^\ominus_5, q_3)
	\left(\frac{-4}{\epsilon}+ \log \left( \frac{t_1^2t_3^2}{(\mu^2)^4}\right)\right)\\
	&\hspace{10 em}-S^B_{uv}\log\left(\frac{w_B}{u_Bv_B}\right)\bigg]
	\Bigg\}\times\frac{1}{t_3}F^{a_6 a_1 c_3}C^{g(0)} \left(p_6^\oplus, p_1^\ominus\right) \,,
	\label{eq:absorpM--nmhv}
\end{split}
\end{align}
\begin{align}
    \begin{split}
    &\hspace{-3 em}\absorp{\mathcal{M}^{(1)(+,-)}_{\mathcal{N}=4}(p_4^\oplus,p_5^\ominus)}\toNMRK \frac{\pi}{2} \gs^6\frac{\kappa_{\Gamma}}{(4\pi)^2}s_{12}C^{g(0)}\left(p_2^{\nu_2}, p_3^{-\nu_2}\right)\frac{1}{t_1}\frac{1}{t_3}C^{g(0)}\left(p_1^{\nu_1}, p_6^{-\nu_1}\right)\\\times \Bigg\{&\tr\left(\left\{F^{a_2},F^{a_3}\right\}F^{a_4}F^{a_5}\left[F^{a_6},F^{a_1}\right]\right)
    \\&\times \bigg[A^{gg(0)}(q_1, p^\oplus_4, p^\ominus_5, q_3)\left(\frac{-1}{\epsilon}+\log\left(\frac{t_3s_{45}}{t_1\mu^2}\frac{v_A w_A}{u_A}\right)\right)
    -S^A_{uv}\log(v_A)-S^A_{wu}\log(w_A)
    \bigg]\\
    +&\tr\left(\left\{F^{a_2},F^{a_3}\right\}F^{a_5}F^{a_4}\left[F^{a_6},F^{a_1}\right]\right)
    \\&\times \bigg[A^{gg(0)}(q_1, p^\ominus_5, p^\oplus_4, q_3)\left(\frac{-1}{\epsilon}+\log\left(\frac{t_3s_{45}}{t_1\mu^2}\frac{v_{A'} w_{A'}}{u_{A'}}\right)\right)
    -S^{A'}_{uv}\log(v_{A'})-S^{A'}_{wu}\log(w_{A'})\bigg]\\
    +&\tr\left(\left\{F^{a_2},F^{a_3}\right\}F^{a_4}\left[F^{a_6},F^{a_1}\right]F^{a_5}\right)
    \\&\times \bigg[ 2B^{gg(0)}(q_1, p^\oplus_4, p^\ominus_5, q_3)\log\left(\frac{p_5^-}{p_4^+}\right)
    +S^B_{vw}\log\left(\frac{u_B}{v_Bw_B}\right)-S^B_{wu}\log\left(\frac{v_B}{w_B u_B}\right)
    \bigg]\Bigg\}
    \,,
    \label{eq:absorpM+-nmhv}
    \end{split}
\end{align} 
\begin{align}
    \begin{split}
    &\hspace{-2 em}\absorp{\mathcal{M}^{(1)(-,+)}_{\mathcal{N}=4}(p_4^\oplus,p_5^\ominus)}\toNMRK \frac{\pi}{2}
    \gs^6\frac{\kappa_{\Gamma}}{(4\pi)^2}s_{12}C^{g(0)}\left(p_2^{\nu_2}, p_3^{-\nu_2}\right)\frac{1}{t_1}\frac{1}{t_3}C^{g(0)}\left(p_1^{\nu_1}, p_6^{-\nu_1}\right)\\\times \Bigg\{&\tr\left(\left[F^{a_2},F^{a_3}\right]F^{a_4}F^{a_5}\left\{F^{a_6},F^{a_1}\right\}\right)
    \\
    &\times \bigg[A^{gg(0)}(q_1, p^\oplus_4, p^\ominus_5, q_3)\left(\frac{-1}{\epsilon}+\log\left(\frac{t_1s_{45}}{t_3\mu^2}\frac{v_A w_A}{u_A}\right)\right) -S^A_{uv}\log(v_A)-S^A_{wu}\log(w_A)
    \bigg]\\
    +&\tr\left(\left[F^{a_2},F^{a_3}\right]F^{a_5}F^{a_4}\left\{F^{a_6},F^{a_1}\right\}\right)
    \\
    &\times \bigg[A^{gg(0)}(q_1, p^\ominus_5, p^\oplus_4, q_3)\left(\frac{-1}{\epsilon}+\log\left(\frac{t_1s_{45}}{t_3\mu^2}\frac{v_{A'} w_{A'}}{u_{A'}}\right)\right) -S^{A'}_{uv}\log(v_{A'})-S^{A'}_{wu}\log(w_{A'})\bigg]\\
    +&\tr\left(\left[F^{a_2},F^{a_3}\right]F^{a_4}\left\{F^{a_6},F^{a_1}\right\}F^{a_5}\right)
    \\
    & \times \bigg[ 2B^{gg(0)}(q_1, p^\oplus_4, p^\ominus_5, q_3)\log\left(\frac{p_5^-}{p_4^+}\right)
    -S^B_{vw}\log\left(\frac{u_B}{v_Bw_B}\right)+S^B_{wu}\log\left(\frac{v_B}{w_B u_B}\right)
    \bigg]\Bigg\} \,,
    \end{split}
    \label{eq:absorpM-+nmhv}
\end{align} 
\begin{align}
    \begin{split}
    &\hspace{-2 em}\absorp{\mathcal{M}^{(1)(+,+)}_{\mathcal{N}=4}(p_4^\oplus,p_5^\ominus)}\toNMRK \frac{\pi}{2}
    \gs^6\frac{\kappa_{\Gamma}}{(4\pi)^2}s_{12}C^{g(0)}\left(p_2^{\nu_2}, p_3^{-\nu_2}\right)\frac{1}{t_1}\frac{1}{t_3}C^{g(0)}\left(p_1^{\nu_1}, p_6^{-\nu_1}\right)
    \\ \times\Bigg\{&\tr\left(\left\{F^{a_2},F^{a_3}\right\}F^{a_4}F^{a_5}\left\{F^{a_6},F^{a_1}\right\}\right)
    \\
    &\times \bigg[
    A^{gg(0)}(q_1, p^\oplus_4, p^\ominus_5, q_3)\left(\frac{-1}{\epsilon}+\log\left(\frac{t_1t_3}{s_{45}\mu^2}\frac{u_A}{v_A w_A}\right)\right)
    +S^A_{uv}\log(v_A)+S^A_{wu}\log(w_A)\bigg]\\
    +&\tr\left(\left\{F^{a_2},F^{a_3}\right\}F^{a_5}F^{a_4}\left\{F^{a_6},F^{a_1}\right\}\right)
    \\
    &
    \times 
    \bigg[
    A^{gg(0)}(q_1, p^\ominus_5, p^\oplus_4, q_3)\left(\frac{-1}{\epsilon}+\log\left(\frac{t_1t_3}{s_{45}\mu^2}\frac{u_{A'}}{v_{A'}w_{A'}}\right)\right) +S^{A'}_{uv}\log(v_{A'})+S^{A'}_{wu}\log(w_{A'})\bigg]\\
    +&\tr\left(\left\{F^{a_2},F^{a_3}\right\}F^{a_4}\left\{F^{a_6},F^{a_1}\right\}F^{a_5}\right)\times \bigg[S^B_{uv}\log\left(\frac{w_B}{u_Bv_B}\right)
    \bigg]\Bigg\} \,.
    \end{split}
    \label{eq:absorpM++nmhv}
\end{align} 
In section~\ref{sec:NLO_xs} we show that at the level of the NLO $gg\to gggg$ squared matrix element, \eqn{eq:absorpM--nmhv} is the only absorptive component of the amplitude which can contribute for a specific helicity configuration. This contribution vanishes in the sum over helicities $\nu_4 $ and $\nu_5$. All components eq.~(\ref{eq:abN41loop})--(\ref{eq:absorpM++nmhv}) will contribute at higher orders in the coupling.

\section{Kinematic limits of the two-gluon central-emission vertex}
\label{sec:lim}
Starting from the central NMRK region, it is possible to take further kinematic limits to regions which are contained within NMRK. In these more restricted kinematic regions further simplifications occur. A key example are the regions where we additionally require $p_4^+\gg p_5^+$ or $p_5^+\gg p_4^+$ which we refer to as the MRK and $\MRKx$ limits respectively. These limits of the two-gluon CEV are studied in appendix \ref{sec:mrklim}. From the central NMRK region we may alternatively take the limit where one of the central gluons becomes soft, which is the subject of appendix \ref{sec:softlim}, or the limit where the two central gluons become collinear to each other, which is the subject of appendix \ref{sec:collim}. These further limits act as checks on the results presented in this paper, and also provide intuition for the kinematic dependence of the cross ratios and the two-gluon CEV. One may also investigate the limits where both of the central gluons become soft, but this is not pursued in this appendix.
\subsection{MRK limit}
\label{sec:mrklim}
Recall that the MRK (MRK$'$) limit is obtained by taking $X\to\infty$ ($X\to0$).
In the MRK or MRK$'$ limit, the tree-level two-gluon CEV factorises into a product of Lipatov vertices times a $t$-channel pole. This factorization property applies for any helicity configuration, for example, for opposite helicities
\begin{align}
\begin{split}
	\left.A^{gg(0)}(q_1, p^\oplus_4, p^\ominus_5, q_3)\right|_{\rm MRK}&=V^{g(0)}(q_1,p^\oplus_4,q_2)\frac{1}{t_2}V^{g(0)}(q_2,p^\ominus_5,q_3)\\
	&=\frac{z}{\bar{z}}(\bar{w}-1)(\bar{z}-1),\\
\end{split}
\label{eq:A4p5mMRK}
\end{align}
\begin{align}
\begin{split}
	\left.A^{gg(0)}(q_1, p^\ominus_5,p^\oplus_4,  q_3)\right|_{\MRKx}&=V^{g(0)}(q_1,p^\ominus_5,q'_2)\frac{1}{t'_2}V^{g(0)}(q'_2,p^\oplus_4,q_3)\\
	&=\frac{z}{\bar{z}}\frac{(\bar{w} (\bar{z}-1)-\bar{z})}{(w (z-1)-z)}(w-1)(z-1) \,,
\end{split}
\label{eq:A5mp4pMRKx}
\end{align}
with Lipatov vertex $V^{g(0)}$ in \eqn{eq:centrv}.
Considering the opposite helicity configuration at one loop, we saw in section~\ref{sec:N4opp-hel} that different transcendental terms multiply rational terms which are not simply the tree amplitudes. This prompts an analysis of these individual rational terms in the MRK and $\MRKx$ limits.
As a consistency check on the MRK and $\MRKx$ limits of these individual rational terms, the relevant sums of these rational terms given in \eqn{kosc2}, \eqn{kosc3}, \eqn{kosc2p} and \eqn{kosc2p2} should reproduce the limits \eqn{eq:A4p5mMRK} and \eqn{eq:A5mp4pMRKx}.

For the $\sigma_A$ ordering in the MRK, from \eqns{eq:RA}{eq:RAb} we obtain
\begin{align}
	\begin{split}
		R^A_{uv}&\toMRK\frac{w-1}{w-z}\left.A^{gg(0)}(q_1, p^\oplus_4, p^\ominus_5, q_3)\right|_{\rm MRK}\\
		R_{\bar{v}\bar{w}}^{A}&\toMRK\left.A^{gg(0)}(q_1, p^\oplus_4, p^\ominus_5, q_3)\right|_{\rm MRK}\\
		R_{wu}^{A}&\toMRK-\frac{z-1}{w-z}\left.A^{gg(0)}(q_1, p^\oplus_4, p^\ominus_5, q_3)\right|_{\rm MRK}\\
	\end{split}
	\label{eq:RAMRK}
\end{align}
while $R_{\bar{u}\bar{v}}^{A}$, $R_{vw}^{A}$ and $R_{\bar{w}\bar{u}}^{A}$ vanish in this limit. All six of these rational terms vanish in the $\MRKx$ limit. The limits listed in \eqn{eq:RAMRK} can be used to show \eqn{eq:A4p5mMRK}, starting from either of the representations \eqn{kosc2} or \eqn{kosc3}. On the other hand, for the $\sigma_{A'}$ permutation, all rational terms vanish in MRK. In $\MRKx$, $R^{A'}_{uv}$, $R_{\bar{v}\bar{w}}^{A'}$ and $R_{wu}^{A'}$ vanish, while the remaining terms tend to
\begin{align}
	\begin{split}
		R_{\bar{u}\bar{v}}^{A'}&\toMRKx -\frac{(\bar{w}-1) \bar{z}^2}{(\bar{w}-\bar{z}) (\bar{w} (\bar{z}-1)-\bar{z})}\left.A^{gg(0)}(q_1, p^\ominus_5,p^\oplus_4,  q_3)\right|_{\MRKx}\\
		R_{vw}^{A'}&\toMRKx \left.A^{gg(0)}(q_1, p^\ominus_5,p^\oplus_4,  q_3)\right|_{\MRKx}\\
		R_{\bar{w}\bar{u}}^{A'}&\toMRKx \frac{\bar{w}^2 (\bar{z}-1)}{(\bar{w}-\bar{z}) (\bar{w} (\bar{z}-1)-\bar{z})}\left.A^{gg(0)}(q_1, p^\ominus_5,p^\oplus_4,  q_3)\right|_{\MRKx}.
	\end{split}
\end{align}
Using the representation \eqn{kosc2p} or \eqn{kosc2p2} we see that these limits are consistent with the limit of the tree-level amplitude \eqn{eq:A5mp4pMRKx}.
Finally for the $B$ configurations, we get terms that survive in the MRK limit,
\begin{align}
	\begin{split}
		R^B_{uv}&\toMRK-\frac{(w-1) (z-1)}{w (z-1)-z}\left.A^{gg(0)}(q_1, p^\oplus_4, p^\ominus_5, q_3)\right|_{\rm MRK}\\
		R_{\bar{u}\bar{v}}^{B}&\toMRK0\\
		R_{vw}^{B}&\toMRK0\\
		R_{\bar{v}\bar{w}}^{B}&\toMRK\left.-A^{gg(0)}(q_1, p^\oplus_4, p^\ominus_5, q_3)\right|_{\rm MRK}\\
		R_{wu}^{B}&\toMRK\frac{1}{w(z-1)-z}\left.A^{gg(0)}(q_1, p^\oplus_4, p^\ominus_5, q_3)\right|_{\rm MRK}\\
		R_{\bar{w}\bar{u}}^{B}&\toMRK0,
	\end{split}
\end{align}
and also terms that survive in the $\MRKx$ limit,
\begin{align}
	\begin{split}
		R^B_{uv}&\toMRKx -\frac{(\bar{w}-1) (\bar{z}-1)}{\bar{w} (\bar{z}-1)-\bar{z}}\left.A^{gg(0)}(q_1, p^\ominus_5,p^\oplus_4,  q_3)\right|_{\MRKx}\\
		R_{\bar{u}\bar{v}}^{B}&\toMRKx 0\\
		R_{vw}^{B}&\toMRKx \frac{1}{\bar{w} (\bar{z}-1)-\bar{z}}\left.A^{gg(0)}(q_1, p^\ominus_5,p^\oplus_4,  q_3)\right|_{\MRKx}\\
		R_{\bar{v}\bar{w}}^{B}&\toMRKx 0\\
		R_{wu}^{B}&\toMRKx 0\\
		R_{\bar{w}\bar{u}}^{B}&\toMRKx-\left.A^{gg(0)}(q_1, p^\ominus_5,p^\oplus_4,  q_3)\right|_{\MRKx}.
	\end{split}
\end{align}
The MRK and $\MRKx$ limits of the one-loop two gluon CEV in $\cN=4$ SYM are discussed in section~\ref{sec:6g1lamp}. The notation introduced in that section makes it trivial to obtain the MRK and $\MRKx$ limits of the transcendental terms, with the exception of terms that can be compactly described in terms of the cross ratios of eqs.~(\ref{eq:uvwA}), (\ref{eq:uvwAx}) and (\ref{eq:uvwB}). To facilitate the investigation of these latter transcendental functions, we list the cross ratios in the MRK and $\MRKx$ limits:
{ \everymath={\displaystyle}
\begin{equation}
	\begin{array}{ccc}
		u_A\toMRK 1,  & v_A\toMRK \frac{(-t_1)p_5^+}{(-t_2)p_4^+}, &  w_A\toMRK \frac{(-t_3)p_4^-}{(-t_2)p_5^-},\\
		u_{A'}\toMRKx 1,  & v_{A'}\toMRKx \frac{(-t_1)p_4^+}{(-t_2')p_5^+}, &  w_{A'}\toMRKx \frac{(-t_1)p_5^-}{(-t_2')p_4^-},\\
		u_B\toMRK 1,  & v_B\toMRK \frac{p_4^-p_5^+}{(-t_2)}, &  w_B\toMRK \frac{(-t_1)(-t_3)}{(-t_2)p_4^+ p_5^-},\\
		u_{B}\toMRKx \frac{p_4^+p_5^-}{(-t_2')},  & v_{B}\toMRKx 1, &  w_{B}\toMRKx \frac{(-t_1)(-t_3)}{(-t_2')p_4^- p_5^+}.\\
	\end{array}
\label{eq:uvwMRK}
\end{equation}
\subsection{Soft limits}
\label{sec:softlim}
In the limit that either of the gluons of the tree two-gluon CEV become soft, $p_4\to 0$ or $p_5\to 0$,
the amplitude given in  \eqn{NLOfactorization2} is singular. It factorises into an eikonal factor, which behaves like $1/p_4$ or $1/p_5$,
and a five-gluon amplitude in MRK. For example, taking the limit where all components of $p_4$ tend to zero at the same rate, \eqn{NLOfactorization2} factorises as
\beqa
\lefteqn{ \left. M^{(0)}_{6g}(p_1^{\nu_1},p_2^{\nu_2}, p_3^{\nu_3}, p^{\nu_4}_4, p^{\nu_5}_5, p_6^{\nu_6})\right|_{\mathrm{NMRK}} } \nn\\
&\tosoft&
\left[ \Soft^{(0)}(p_3, p^{\nu_4}_4, p_5)\right]_{\mathrm{NMRK}}\,\left[ M^{(0)}_{5g}(p_1^{\nu_1},p_2^{\nu_2}, p_3^{\nu_3}, p^{\nu_5}_5, p_6^{\nu_6})\right]_{\mathrm{MRK}}\,,
\label{eq:soft}
\eeqa
where the five-gluon amplitude in MRK is
\begin{equation}
     \left. M^{(0)}_{5g}(p_1^{\nu_1},p_2^{\nu_2}, p_3^{\nu_3}, p^{\nu_5}_5, p_6^{\nu_6})\right|_{\mathrm{MRK}} =s\, C^{g(0)}(p_2^{\nu_2}, p_3^{\nu_3})\, \frac1{t_1}\, V^{g(0)}(q_1, p^{\nu_5}_5, q_3)\, \frac1{t_3}\, C^{g(0)}(p_1^{\nu_1}, p_6^{\nu_6})\,,
\label{eq:mrk5g}
\end{equation}
with $q_3= q_1-p_5$ and $V^{g(0)}$ given in \eqn{eq:centrv}.
In \eqn{eq:soft}, the positive-helicity soft factor, or eikonal factor is
\beq
\Soft^{(0)}(p_3, p^\oplus_4, p_5) = \frac{\langle 3\, 5\rangle}{\langle 3\, 4\rangle \langle 4\, 5\rangle}\,.
\label{eq:eik}
\eeq
Note that this eikonal factor does not depend on the helicities of particles 3 and 5, and in fact not even on their parton flavour. The negative-helicity eikonal factor is obtained by complex conjugation. In the NMRK limit the positive-helicity eikonal factor becomes
\beq
\left. \Soft^{(0)}(p_3, p^\oplus_4, p_5)\right|_{\mathrm{NMRK}} = \Soft^{(0)}_A(p_4^{\oplus},p_5)= \frac{p_{5\perp}}{p_{4\perp}} \sqrt{\frac{p_4^+}{p_5^+}}\, \frac{1}{\langle 4\, 5\rangle} \,,
\label{eq:eikmrk}
\eeq
where in the first equality we have introduced a notation which makes it clear that this function no longer depends on $p_3$ once we have taken the NMRK limit. Using this notation, \eqn{eq:soft} implies that the two-gluon CEV factorises as
\beq
A^{gg(0)}(q_1, p^{\nu_4}_4, p^{\nu_5}_5, q_3)\tosoft \Soft^{(0)}_A(p_4^{\nu_4},p_5)\, V^{g(0)}(q_1, p^{\nu_5}_5, q_3)\,.
\label{eq:softvert}
\eeq
For the equal-helicity vertex \eqn{eq:kosc_same-hel}, which comes from a MHV amplitude, this limit is straightforward to verify. In fact, for such an MHV amplitude, the relation at the amplitude level, \eqn{eq:soft}, holds
exactly in general kinematics without even requiring that $p_4\to 0$.
For the opposite-helicity vertex, in order to check \eqn{eq:softvert} we have found it most convenient to use the representation in \eqn{kosc2}.

As a step towards obtaining the soft limit of the one-loop amplitude, we first consider the rational coefficients.  Specifically, for the limit where all components of $p_4$ tend to zero at the same rate, the minimal variables of \eqn{eq:wzparam2} and \eqn{eq:xparam} satisfy
\begin{align}
  \label{eq:softminvars}
  X\to 0, \qquad z\to -\frac{q_{1\perp}}{p_{4\perp}}, \qquad w\to
  \frac{q_{1\perp}}{p_{5\perp}}, \qquad Xz\to\ {\rm finite}.
\end{align}
In this limit $R^A_{uv}$, $R^A_{vw}$ and $R^{A}_{\bar{w}\bar{u}}$ tend to zero.  The others are given by
\begin{align}
  \begin{split}
  &R^A_{wu} \to \frac{Xz^2(\bar{w}-1)}{w+Xz}, \\
  &R^A_{\bar{u}\bar{v}}\to -\frac{Xz^2|w-1|^2}{(1+Xz)(w+Xz)},  \\
    \qquad &R^A_{\bar{v}\bar{w}}\to \frac{Xz^2(\bar{w}-1)}{1+Xz}.
  \end{split}
\end{align}
Therefore $R^A_{uv}+R^A_{vw}+R^A_{wu}$ and $R^A_{\bar{u}\bar{v}}+R^A_{\bar{v}\bar{w}}+R^A_{\bar{w}\bar{u}}$ tend to the same limit (that of \eqn{eq:softvert}). The results for the $A^\prime$ ordering are no longer related to these by $p_4\leftrightarrow p_5$ because the specific choice of taking $p_4\to0$ breaks the symmetry. We therefore list them here. The rational coefficients $R^{A'}_{uv}$, $R^{A'}_{\bar{v}\bar{w}}$ and $R^{A'}_{\bar{w}\bar{u}}$ tend to zero.  The remaining non-zero coefficients are
\begin{align}
  \label{eq:softApcoeffs}
  \begin{split}
    &R^{A'}_{\bar{u}\bar{v}}\to -\frac{wz(1-\bar{w})}{w+Xz},\\
    &R^{A'}_{vw}\to \frac{wz(1-\bar{w})^2}{w\bar{w}-w-Xz}, \\
    &R^{A'}_{wu}\to  \frac{w\bar{w}Xz^2(1-\bar{w})}{(w+Xz)(w\bar{w}-w-Xz)}, 
\end{split}
\end{align}
such that again $R^{A'}_{uv}+R^{A'}_{vw}+R^{A'}_{wu}$ and $R^{A'}_{\bar{u}\bar{v}}+R^{A'}_{\bar{v}\bar{w}}+R^{A'}_{\bar{w}\bar{u}}$ give the expected tree-level limit:
\begin{align}
  \label{eq:softApamp}
    -\frac{wz(1-\bar{w})}{w+Xz} 
    =\ \Soft^{(0)}_{A'}(p_4^{\nu_4},p_5) \,V^{g(0)}(q_1,p_5^\ominus,q_3),
\end{align}
where, analogous to \eqn{eq:eikmrk}, we have introduced the notation
\beq
\left. \Soft^{(0)}(p_5,p_4^\oplus,p_6)\right|_{\mathrm{NMRK}} = \Soft^{(0)}_{A'}(p_4^{\nu_4},p_5)= -\sqrt{\frac{p_5^+}{p_4^+}}\, \frac{1}{\langle 4 \,5\rangle} \,.
\label{eq:eikmrkAx}
\eeq
Finally, for the $B$ colour ordering $R^B_{\bar{u}\bar{v}}$, $R^B_{vw}$ and $R^B_{wu}$ tend to zero, leaving
\begin{align}
  \label{eq:Bcoeffs}
  \begin{split}
    &R^B_{uv}\to -(\bar{w}-1)z, \\
    &R^B_{\bar{v}\bar{w}}\to -\frac{Xz^2(\bar{w}-1)}{1+Xz-\bar{w}}, \\
    &R^B_{\bar{w}\bar{u}}\to -\frac{z(\bar{w}-1)^2}{\bar{w}-Xz-1}.
  \end{split}
\end{align}
These then lead to the expected tree-level result where $R^{B}_{uv}+R^{B}_{vw}+R^{B}_{wu}$ and $R^{B}_{\bar{u}\bar{v}}+R^{B}_{\bar{v}\bar{w}}+R^{B}_{\bar{w}\bar{u}}$ both sum to
\begin{align}
  \label{eq:softBamp}
    -(\bar{w}-1)z 
    =\Soft^{(0)}_{B}(p_4) \, V^{g(0)}(q_1,p_5^\ominus,q_3),
\end{align}
where as for the other colour orderings we have introduced
\beq
\left. \Soft^{(0)}(p_1,p_4^\oplus,p_2)\right|_{\mathrm{NMRK}} = \Soft^{(0)}_{B}(p_4)= \frac{1}{p_{4\perp}},
\label{eq:eikmrkB}
\eeq
which does not even depend on the helicity of the soft particle. The three relevant eikonal functions in the NMRK obey
\begin{equation}
\Soft^{(0)}_{A}(p_4^{\nu_4},p_5)+ \Soft^{(0)}_{A'}(p_4^{\nu_4},p_5)+\Soft^{(0)}_{B}(p_4)=0,
\end{equation}
consistent with the more general \eqn{eq:ABrelation}, and the eikonal identity. 

We now turn to the colour-ordered amplitudes at one-loop.  Beginning with the same-helicity vertex, taking the limit $p_4\to0$ in eqs.~(\ref{eq:V6Aexact}), (\ref{eq:ReVAx}) and (\ref{eq:V6Bexact}), we find
\begin{align}
  \label{eq:1loopresults}
  \begin{split}
N_c\kappa_\Gamma \real{\left.V_6(\sigma_{A})\right|_{\rm NMRK}} \tosoft\,\, & E(t_1,s_{35};t_3,s_{56}) + v^{g(1)}(t_1,|p_{5\perp}|^2,t_3)+N_c \, s^{(1)}_{A}(p_4^{\nu_4},p_5)\\
N_c\kappa_\Gamma \real{\left.V_6(\sigma_{A^\prime})\right|_{\rm NMRK}} \tosoft\,\, &E(t_1,s_{35};t_3,s_{56}) + v^{g(1)}(t_1,|p_{5\perp}|^2,t_3) 
+ \, N_c \,  s^{(1)}_{A'}(p_4,p_5), \\
N_c\kappa_\Gamma \real{\left.V_6(\sigma_{B})\right|_{\rm NMRK}} \tosoft\,\, &E(t_1,s_{35};t_3,s_{56}) + v^{g(1)}(t_1,|p_{5\perp}|^2,t_3)
+ N_c \, s^{(1)}_{B}(p_4),
  \end{split}
\end{align}
where $E(t_1,s_{35};t_3,s_{56})$ is the natural collection of impact factors and Regge trajectories in the absence of $p_4$ and $v^{g(1)}$ is the one-loop Lipatov vertex correction given in \eqn{eq:cev1n4}. The one-loop soft factor~\cite{Bern:1998sc} is
\begin{align}
    s^{(1)}(p_3,p_4^{\nu_4},p_5)=-\kappa_\Gamma \left(\frac{\mu^2(-s_{35})}{(-s_{34})(-s_{45})} \right)^\epsilon \left(\frac{1}{\epsilon^2} +\frac{\pi^2}{6}\right)+\mathcal{O}(\epsilon^2),
    \label{eq:soft1}
\end{align}
which must first be analytically continued to our physical region, discussed in section~\ref{sec:appa}, and then taken in the NMRK limit. In \eqn{eq:1loopresults} we have used the notation
\begin{align}
s^{(1)}_{A}(p_4,p_5)&=\real{s^{(1)}(p_3,p_4,p_5)\Big|_{\mathrm{NMRK}}}\,,\\
s^{(1)}_{A'}(p_4,p_5)&=\real{s^{(1)}(p_4,p_5,p_6)\Big|_{\mathrm{NMRK}}}\,,\\
s^{(1)}_{B}(p_4)&=\real{s^{(1)}(p_1,p_4,p_2)\Big|_{\mathrm{NMRK}}}\,,
\end{align}
which, as for the analogous tree-level functions, more clearly indicate the simpler kinematic dependence of these one-loop soft functions in the NMRK limit.

Eq.~(\ref{eq:1loopresults}) exactly matches the expected sum of the one-loop corrections to the non-central pieces, the Lipatov vertex for the single emission of $p_5$ and the one-loop soft function.  From \eqn{eq:disp_same-hel}, these are multiplied by the soft limit of the corresponding tree-level amplitude for that ordering.

Turning now to the opposite-helicity case in \eqn{eq:N4NMHVReWA+R}, and starting with the $A$ colour ordering, we find
\begin{align}
    \label{eq:SWlimits}
    \begin{split}
    S_{wu}^A &\tosoft \Soft^{(0)}_A(p_4^{\nu_4},p_5)
    V^{g(0)}(q_1,p_5^\ominus,q_3), \\
    \real{W_1(\sigma_A)-W_2(\sigma_A)}&\tosoft 0,\\
    \real{W_3(\sigma_A)-W_2(\sigma_A)}&\tosoft \log\left(\frac{s_{34}t_3}{s_{35}t_1}\right) \log\left( \frac{|p_{5\perp}|^2}{s_{45}} \right).
    \end{split}
\end{align}
This allows us to extract a common factor of the tree amplitude multiplying $$ \real{2W_2(\sigma_A) + (W_3(\sigma_A)-W_2(\sigma_A))}.$$ Using \eqns{eq:W2Aexact}{eq:W2A}, we find
\begin{align}
\begin{split}
      &\left.
      \real{
      2W_2(\sigma_A)
      + (W_3(\sigma_A)-W_2(\sigma_A))
      }
      \right|_{\rm NMRK}\\
      =& \left.\real{  V_6(\sigma_A)-\Delta V_A(u_A,v_A,w_A) + \Delta W_A(u_A,v_A,w_A) + (W_3(\sigma_A)-W_2(\sigma_A)) }\right|_{\rm NMRK}\\
      \tosoft&
      \left. \real{ V_6(\sigma_{A})}\right|_{{\rm NMRK},\, p_4\to0},
\end{split}
\end{align}
where the cancellation in the last line arises from the soft limit of eqs.~(\ref{eq:deltaVA}), (\ref{eq:deltaWA}) and (\ref{eq:SWlimits}). Eq.~(\ref{eq:1loopresults}) then immediately gives the expected one-loop structure which equals that found for the same-helicity vertex.  The helicity-dependence in the soft limit is contained only in the tree-level factors.

Similarly for the $A'$-ordering we find
\begin{align}
    \label{eq:SWlimitsAx}
    \begin{split}
    S_{uv}^{A'} &\tosoft \Soft^{(0)}_{A'}(p_4^{\nu_4},p_5)
    V^{g(0)}(q_1,p_5^\ominus,q_3), \\
    \left.\real{
    W_1(\sigma_{A'})-W_2(\sigma_{A'})
    }\right|_{\rm NMRK}&\tosoft \log\left(\frac{s_{46}t_1}{s_{56}t_3}\right) \log\left( \frac{|p_{5\perp}|^2}{s_{45}} \right),\\
    \left.\real{
    W_3(\sigma_{A'})-W_2(\sigma_{A'})}\right|_{\rm NMRK}
    &\tosoft 0.
    \end{split}
\end{align}
Extracting a common coefficient of the tree amplitude leaves
\begin{align}
    \left.\real{2W_2(\sigma_{A'}) + (W_1(\sigma_{A'})-W_2(\sigma_{A'}))}\right|_{\rm NMRK} \tosoft \left. \real{ V_6(\sigma_{A'})}\right|_{{\rm NMRK},\, p_4\to0},
\end{align}
as expected.  Finally for the $B$ ordering,
\begin{align}
    \label{eq:SWlimitsB}
    \begin{split}
    S_{vw}^B + S_{wu}^B &\tosoft \Soft^{(0)}_{B}(p_4)
V^{g(0)}(q_1,p_5^\ominus,q_3), \\
    \left.\real{
    W_2(\sigma_B)-W_1(\sigma_B)
    }\right|_{\rm NMRK}
    &\tosoft \log\left(\frac{s_{24}s_{56}}{s_{12}t_1}\right) \log\left( \frac{s_{14}s_{35}}{s_{12}t_3} \right), \\
    \left.\real{
    W_3(\sigma_B)-W_1(\sigma_B) 
    }\right|_{\rm NMRK}
    &\tosoft \log\left(\frac{s_{24}s_{56}}{s_{12}t_1}\right) \log\left( \frac{s_{14}s_{35}}{s_{12}t_3} \right).
    \end{split}
\end{align}
As the final two limits are equal, one again factors out the tree amplitude leaving
\begin{align}
    \left.
    \real{
    2W_1(\sigma_B) + (W_2(\sigma_B)-W_1(\sigma_B))
     }\right|_{\rm NMRK}
     \tosoft  
     \left.
     \real{ V_6(\sigma_B)}\right|_{\mathrm{NMRK},\, p_4\to 0}.
\end{align}
Finally, performing the change of basis of colour structures in \eqn{colour-rotation}, we collect the results of this appendix in the following expression for general helicities,
\begin{align}
\begin{split}
&\mathrm{Disp}\left[
\mathcal{M}^{(1)}_{\cN=4}(p_4^{\nu_4}, p_5^{\nu_5})\right]\xrightarrow[\mathrm{NMRK,} \ p_4\to 0]{}
\\&
\frac{\gs^6}{(4\pi)^2}\, 
s_{12}F^{a_3 a_2 c_1}
C^{g(0)}\left(p_2^{\nu_2}, p_3^{-\nu_2}\right)
\frac{1}{t_1} V^{g(0)}(q_1,p_5^{\nu_5},q_3)  \times  \\
&\Bigg\{  
		(F^{a_4}F^{a_5})_{c_1c_3}
		\bigg[
		\Soft_A(p_4^{\nu_4},p_5)\left(
		3 \,v^{g(1)}(t_1,|p_{5\perp}|^2,t_3)
		+N_c\left(2s^{(1)}_{A}(p_4,p_5)+s^{(1)}_{B}(p_4)\right) \right.
		\\&\left. +3 \, E(t_1,s_{35};t_3,s_{56})
		\right)
		+N_c\Soft_{A'}(p_4^{\nu_4},p_5)\left(s^{(1)}_{A'}(p_4,p_5)-s^{(1)}_{B}(p_4)\right)
		\bigg]
		\\
	&+(F^{a_5}F^{a_4})_{c_1c_3}
		\bigg[
		\Soft_{A'}(p_4^{\nu_4},p_5)\left(
		3 \,v^{g(1)}(t_1,|p_{5\perp}|^2,t_3)
		+N_c\left(2s^{(1)}_{A'}(p_4,p_5)+s^{(1)}_{B}(p_4)\right) \right.
		\\&\left. +3 \, E(t_1,s_{35};t_3,s_{56})
		\right)
		+N_c\Soft_{A}(p_4^{\nu_4},p_5)\left(s^{(1)}_{A}(p_4,p_5)-s^{(1)}_{B}(p_4)\right)
		\bigg]
\\
	&+d_{A}^{c_1 a_{\sigma_4} a_{\sigma_5} c_3}
		\bigg[
		\Soft_A(p_4^{\nu_4},p_5)\left(s^{(1)}_{A}(p_4,p_5)-s^{(1)}_{B}(p_4)\right)
		\\& +\Soft_{A'}(p_4^{\nu_4},p_5)\left(s^{(1)}_{A'}(p_4,p_5)-s^{(1)}_{B}(p_4)\right)
		\bigg]
		\Bigg\}\,\times\frac{1}{t_3}
F^{a_6 a_1 c_3}
C^{g(0)}\left(p_1^{\nu_1},p_6^{-\nu_1}\right)\,.
\label{eq:DispFFd_soft}
\end{split}
\end{align}

\subsection{Collinear limits}
\label{sec:collim}
In this appendix we investigate the behaviour of the one-loop amplitude \eqn{eq:DispFFdgen} in the collinear limit,
\beq
p_4 \to z_4 P\,, \qquad p_5 \to z_5 P\,, \qquad z_4+z_5 = 1\,.
\label{eq:coll45}
\eeq
We begin by recalling the collinear limit of the colour ordered six-gluon tree-level amplitude in general kinematics.
It factorises into a divergent splitting amplitude, which behaves like $1/\langle 4 5\rangle$ or $1/[ 4 5 ]$,
and a five-gluon amplitude,
\begin{equation}
M^{(0)}_{6g}(p_1^{\nu_1},p_2^{\nu_2}, p_3^{\nu_3}, p^{\nu_4}_4, p^{\nu_5}_5, p_6^{\nu_6})
\xrightarrow[]{p_4 || p_5}\sum_{\nu=\pm}\mathrm{Split}^{(0)}_{-\nu}(p^{\nu_4}_4, p^{\nu_5}_5)\, M^{(0)}_{5g}(p_1^{\nu_1}, p_2^{\nu_2}, p_3^{\nu_3}, P^\nu, p_6^{\nu_6}),
\label{eq:coll_bulk}
\end{equation}
where we sum over the helicities of the intermediate state $P^\nu$. For an equal-helicity pair, the tree-level splitting amplitudes are
\beq
{\rm{Split}}^{(0)}_{\ominus}(p^\oplus_4, p^\oplus_5) = \frac{1}{\sqrt{z_4z_5} \langle 4 \, 5\rangle}\,, \qquad {\rm{Split}}^{(0)}_{\oplus}(p^\oplus_4, p^\oplus_5) = 0 \,,
\eeq
while for an opposite-helicity pair, they are
\beq
{\rm{Split}}^{(0)}_{\oplus}(p^\oplus_4, p^\ominus_5) = \frac{z^2_5}{\sqrt{z_4z_5} \langle 4 \, 5\rangle}\,, \qquad
{\rm{Split}}^{(0)}_{\ominus}(p^\oplus_4, p^\ominus_5) = - \frac{z^2_4}{\sqrt{z_4z_5} [4 \, 5]}\,.
\eeq
In the central NMRK limit, \eqn{eq:coll_bulk} becomes
\begin{align}
&\left. M^{(0)}_{6g}(p_1^{\nu_1}, p_2^{\nu_2}, p_3^{\nu_3}, p^{\nu_4}_4, p^{\nu_5}_5, p_6^{\nu_6}) \right|_{\mathrm{NMRK}} \nn\\
&\xrightarrow[]{p_4 || p_5}\sum_{\nu=\pm}\mathrm{Split}^{(0)}_{-\nu}(p^{\nu_4}_4, p^{\nu_5}_5)\,  \left. M^{(0)}_{5g}(p_1^{\nu_1}, p_2^{\nu_2}, p_3^{\nu_3}, P^\nu, p_6^{\nu_6})\right|_{\mathrm{MRK}},
\label{eq:coll_nmrk}
\end{align}
where the five-gluon amplitude in MRK is given in \eqn{eq:mrk5g}, with $q_3= q_1-P$.
In particular, the two-gluon CEV factorises as
\beq
A^{gg(0)}(q_1, p^{\nu_4}_4, p^{\nu_5}_5, q_3) \stackrel{p_4 || p_5}{=} 
\sum_{\nu=\pm} \mathrm{Split}^{(0)}_{-\nu}(p^{\nu_4}_4, p^{\nu_5}_5)\, V^{g(0)}(q_1, P^{\nu}, q_3)\,.
\label{eq:collvert}
\eeq
For the equal-helicity vertex \eqn{eq:kosc_same-hel}, it is straightforward to verify \eqn{eq:collvert}.
For the opposite-helicity vertex, we have found it more convenient to check \eqn{eq:collvert} by starting from \eqn{kosc2} or \eqn{kosc3} rather than \eqn{kosc}, where as we have mentioned in section~\ref{sec:bcfw}, the BCFW representations make clear the collinear singularity structure of the amplitudes. Working in the minimal variables, the collinear limit can be expressed as
\begin{align}
    \begin{split}
        w&\tocol-\frac{z_4}{z_5}z \,,\\
        X&\tocol \frac{z_4}{z_5} \,.
    \end{split}
\end{align}
For the $\sigma_A$ ordering we find that four of the six rational terms factorise into a splitting function and Lipatov vertex, while two of the rational terms do not diverge in this limit (unless we move to the further soft-collinear limits of $z_4\to0$ or $z_4\to1$),
\begin{align}
	\begin{split}
		R^A_{uv}&\tocol V^{g(0)}(q_1,P^\oplus,q_3)\mathrm{Split}^{(0)}_\ominus(p_4^\oplus,p_5^\ominus),\\
		R_{\bar{u}\bar{v}}^{A}&\tocol V^{g(0)}(q_1,P^\ominus,q_3)\mathrm{Split}^{(0)}_\oplus(p_4^\oplus,p_5^\ominus),\\
		R_{vw}^{A}&\tocol-\frac{(z-1)  (z z_4+z_5)z_5}{(\bar{z}-1) (\bar{z}z_4 +z_5) (|z|^2z_4+z_5)},\\
		R_{\bar{v}\bar{w}}^{A}&\tocol -\frac{z^4 z_4^2 (\bar{z}-1) (\bar{z}z_4 +z_5)}{z_5(z-1)(z z_4+z_5) (|z|^2z_4+z_5)},\\
		R_{wu}^{A}&\tocol V^{g(0)}(q_1,P^\ominus,q_3)\mathrm{Split}^{(0)}_\oplus(p_4^\oplus,p_5^\ominus),\\
		R_{\bar{w}\bar{u}}^{A}&\tocol V^{g(0)}(q_1,P^\oplus,q_3)\mathrm{Split}^{(0)}_\ominus(p_4^\oplus,p_5^\ominus).
	\end{split}
	\label{eq:RAcoll}
\end{align}
Analogous results for the $\sigma_{A'}$ ordering can be found by exchanging $p_4\leftrightarrow p_5$ and taking the complex conjugate, as in \eqn{eq:RAswitch45}.
For the $\sigma_{B}$ ordering, as expected, none of the rational terms diverge in this limit,
\begin{align}
	\begin{split}
		R^B_{uv}&\tocol\frac{|z-1|^2|z z_4+z_5|^2}{z_5((z-1)z_4+z_5) \bar{z}},\\
		R_{\bar{u}\bar{v}}^{B}&\tocol0,\\
		R_{vw}^{B}&\tocol\frac{z_5(z-1) (z z_4+z_5)}{\bar{z} (|z|^2 z_4+z_5) ((z+\bar{z}-1)z_4 +z_5)},\\
		R_{\bar{v}\bar{w}}^{B}&\tocol\frac{z^3 z_4^2 (\bar{z}-1) (\bar{z}z_4+z_5)}
		{z_5(|z|^2z_4+z_5) ((z+\bar{z}-1)z_4+z_5)},\\
		R_{wu}^{B}&\tocol\frac{z_5 z_4^2 (\bar{z}-1) (\bar{z} z_4+z_5)}
		{((z-1)z_4+z_5) 
		((z+\bar{z}-1)z_4+z_5)
		\left(|z_5+(z-1)z_4|^2+z_4z_5\right) },\\
		R_{\bar{w}\bar{u}}^{B}&\tocol\frac{(z-1) (z z_4+z_5) ((\bar{z}-1)z_4+z_5)^3}
		{z_5
		((z+\bar{z}-1)z_4+z_5)
		\left(|z_5+(z-1)z_4|^2+z_4z_5\right) 
		}.
	\end{split}
\end{align}
From these lists of rational terms we can immediately verify that the divergent pieces satisfy the condition in \eqn{eq:collvert} from the representations eq.~(\ref{kosc2}) or eq.~(\ref{kosc3}). We note that in this limit the divergent pieces also satisfy
\begin{equation}
    \left. A^{gg(0)}(q_1,p^{\nu_4}_4,p^{\nu_5}_5,q_3)\right|_{p_4 || p_5}=-\left.A^{gg(0)}(q_1,p^{\nu_5}_5,p^{\nu_4}_4,q_3)\right|_{p_4 || p_5}.
\end{equation}

In addition to \eqn{eq:coll45}, the amplitude in \eqn{NLOfactorization2} is divergent in the collinear regions $p_{3\perp} = - q_{1\perp} \rightarrow 0$ or
$p_{6\perp} = q_{3\perp} \rightarrow 0$. In those regions, \eqn{NLOfactorization2} must not diverge more rapidly than $1/|q_{i\perp}|$,
with $i=1,3$, in order for the related cross section not to diverge more than logarithmically.
Since \eqn{NLOfactorization2} displays poles as $|q_{1\perp}|^2$ and $|q_{3\perp}|^2$
go to zero, the central-emission vertex must be at least linear in $|q_{i\perp}|$ as $|q_{i\perp}|\rightarrow 0$,
\begin{equation}
A^{gg(0)}(q_1, p^{\nu_4}_4, p^{\nu_5}_5, q_3) \stackrel{|q_{i\perp}|\rightarrow 0}{=} \ord(|q_{i\perp}|)\,,
\label{alim}
\end{equation}
with $i=1,3$, which is indeed fulfilled by \eqn{kosc}.
\par
Let us now consider the collinear limit of the one-loop amplitudes in $\cN=4$ SYM in the NMRK limit. It is most convenient if we work at the level of colour-ordered amplitudes, where in general kinematics the colour-ordered one-loop six gluon amplitude has the collinear limit~\cite{Bern:1994zx},
\begin{align}
\begin{split}
M^{(1)}_{6g}(p_2^{\nu_2}, p_3^{\nu_3}, p^{\nu_4}_4, p^{\nu_5}_5, p_6^{\nu_6}, p_1^{\nu_1})
\xrightarrow[]{p_4 || p_5}\sum_{\nu=\pm}\Big(&\mathrm{Split}_{-\nu}^{(0)}(p^{\nu_4}_4, p^{\nu_5}_5)\, M^{(1)}_{5g}(p_2^{\nu_2}, p_3^{\nu_3}, P^\nu, p_6^{\nu_6}, p_1^{\nu_1})\\
+&\mathrm{Split}_{-\nu}^{(1)}(p^{\nu_4}_4, p^{\nu_5}_5)\, M^{(0)}_{5g}(p_2^{\nu_2}, p_3^{\nu_3}, P^\nu, p_6^{\nu_6}, p_1^{\nu_1}) \Big) \,,
\label{eq:coll_bulk_loop}
\end{split}
\end{align}
where the one-loop splitting functions were obtained in~\cite{Bern:1994zx}. In $\mathcal{N}=4$ SYM, these splitting functions are proportional to the tree-level splitting function,
\begin{equation}
    \mathrm{Split}_{-\nu}^{(1)}(p^{\nu_4}_4, p^{\nu_5}_5)=\kappa_\Gamma\, \mathrm{Split}_{-\nu}^{(0)}(p^{\nu_4}_4, p^{\nu_5}_5)\, r_S^{\mathrm{SUSY}}(z_4,s_{45}),
\end{equation}
where the loop factor,
\begin{equation}
r_S^{\mathrm{SUSY}}(z_4,s_{45})=-\frac{1}{\epsilon^2}\left(\frac{\mu^2}{z_4(1-z_4)(-s_{45})}\right)^\epsilon+2\log(z_4)\log(1-z_4)-\frac{\pi^2}{6},
\label{eq:rSUSY}
\end{equation}
is independent of the helicities. In the central NMRK limit we expect to obtain
\begin{align}
\begin{split}
\left.M^{(1)}_{6g}(p_2^{\nu_2}, p_3^{\nu_3}, p^{\nu_4}_4, p^{\nu_5}_5, p_6^{\nu_6}, p_1^{\nu_1})\right|&_{\mathrm{NMRK}}
\xrightarrow[]{p_4 || p_5}\\
\sum_{\nu=\pm}\bigg(&\mathrm{Split}_{-\nu}^{(0)}(p^{\nu_4}_4, p^{\nu_5}_5)\, \left.M^{(1)}_{5g}(p_2^{\nu_2}, p_3^{\nu_3}, P^\nu, p_6^{\nu_6}, p_1^{\nu_1})\right|_{\mathrm{MRK}}\\
+&\mathrm{Split}_{-\nu}^{(1)}(p^{\nu_4}_4, p^{\nu_5}_5)\, \left.M^{(0)}_{5g}(p_2^{\nu_2}, p_3^{\nu_3}, P^\nu, p_6^{\nu_6}, p_1^{\nu_1})\right|_{\mathrm{MRK}} \bigg).
\label{eq:coll_nmrk_loop}
\end{split}
\end{align}
Let us check this first for the same-helicity case. As we are working at the level of colour-ordered amplitudes, we consider the three representative orderings of \eqn{eq:disp_same-hel}. For the $\sigma_A$ colour ordering, we find 
\begin{align}
\begin{split}
    s_{234}&\tocol z_5 t_1 + z_4 t_3\, ,\\
\end{split}
\end{align}
so that in the collinear limit,
\begin{equation}
    v_A+w_A\tocol 1\, .
\end{equation}
In particular, this allows us to use the identity,
\begin{equation}
    \Li\left(1-x\right)+\Li\left(x\right)=\frac{\pi^2}{6}-\log\left( 1-x\right)\log\left(x\right).
\end{equation}
to write
\begin{align}
\begin{split}
    \Delta V_A(u_A,v_A,w_A)\tocol N_c \cg \bigg(&
    -\frac{1}{4}\log^2\left(\frac{s_{45}}{|P_{\perp}|^2}\right)
    -\frac{1}{2}\log\left(\frac{s_{45}}{|P_{\perp}|^2}\right)\log\left(\frac{t_1 t_3 z_4 z_5}{s_{234}^2} \right)
    \\
    &+\log\left(\frac{z_4 t_3}{s_{234}}\right)
    \log\left(\frac{z_5 t_1}{s_{234}}\right)
     \bigg) \,.
    \end{split}
\end{align}
For the full transcendental function of this colour ordering we find
\begin{align}
	\begin{split}
		N_c \cg \real{\left.V_6(\sigma_A)\right|_{\mathrm{NMRK}}}
		\tocol
		&E(t_1,p_3^+P^-;t_3,P^+p_6^-)   +v^{g(1)}_{\mathcal{N}=4}\left(t_1,|P_{\perp}|^2,t_3\right)\\
    +&N_c\real{r^{\mathrm{SUSY}}(z_4,s_{45})} \,,
		\label{eq:V6gtaColl}
	\end{split}
\end{align}
which is invariant under $p_4 \leftrightarrow p_5$.
After manipulating the colour factors to obtain the natural association of one adjoint generator with each three-point tree-level factor, we obtain
\begin{align}
	\begin{split}
		\real{\mathcal{M}^{(1)}_{\mathcal{N}=4}(p_4^\oplus,p_5^\oplus)}
		&\xrightarrow[\mathrm{NMRK},\,p_4||p_5]{}\\
		\frac{\gs^6}{(4\pi)^2}\kappa_{\Gamma}s_{12}&\left[F^{a_3 a_2 c_1}C^{g(0)}\left(p_2^{\nu_2}, p_3^{-\nu_2}\right)\right]\frac{1}{t_1}\\
		\times&\left[F^{c_1c_3c_2} V^{g(0)}(q_1,P^\oplus,q_3)\right]  \left[F^{c_2a_5a_4}\mathrm{Split}^{(0)}_{\ominus}(p_4^\oplus,p_5^\oplus)\right]
		\\\times
		&\frac{1}{t_3}\left[F^{a_6 a_1 c_3}C^{g(0)}\left(p_1^{\nu_1},p_6^{-\nu_1} \right)\right]
		\times \bigg\{  
		N_c \left. \real{V_6(\sigma_A)} \right|_{\mathrm{NMRK},\, p_4||p_5}
		\bigg\} \,.
	\end{split}
	\label{eq:M1Coll_eq}
\end{align}
The opposite-helicity case proceeds similarly.
From \eqn{eq:RAcoll} we see that in the collinear limit, both $S^A_{uv}$ and $S^A_{wu}$ are equal to the tree-level amplitude, and these terms combine with the $\Delta W_A$ term to play the same role as $\Delta V_A$ in this limit,
\begin{align}
A^{gg(0)}(q_1,p_4^\oplus,p_4^\ominus,q_3)\Delta W_A(u_A,v_A,w_A)+ S^A_{uv}\log(v_A)\left(\frac{w_A}{u_A}\right)+S^A_{wu}\log(w_A)\left(\frac{v_A}{u_A}\right) \nn\\
\tocol 
\left(\sum_{\nu=\pm}V^{g(0)}(q_1,P^\nu,q_3) \mathrm{Split}^{(0)}_{-\nu}(p_4^\oplus,p_5^\ominus)\right)
\left. \Delta V_A(u_A,v_A,w_A)\right|_{p_4||p_5}
\end{align}
This means we can straightforwardly write
\begin{align}
	\begin{split}
		\real{\mathcal{M}^{(1)}_{\mathcal{N}=4}(p_4^\oplus,p_5^\ominus)}&\xrightarrow[\mathrm{NMRK},p_4||p_5]{}\\
		\frac{\gs^6}{(4\pi)^2}\kappa_{\Gamma}s_{12}&\left[F^{a_3 a_2 c_1}C^{g(0)}\left(p_2^{\nu_2}, p_3^{-\nu_2}\right)\right]\frac{1}{t_1}\\
		\times&\sum_{\nu=\pm}\left(\left[F^{c_1c_3c_2} V^{g(0)}(q_1,P^\nu,q_3)\right]  \left[F^{c_2a_5a_4}\mathrm{Split}^{(0)}_{-\nu}(p_4^\oplus,p_5^\ominus)\right]\right)
		\\\times
		&\frac{1}{t_3}\left[F^{a_6 a_1 c_3}C^{g(0)}\left(p_1^{\nu_1},p_6^{-\nu_1} \right)\right]
		\times \bigg\{  
		N_c \left. \real{V_6(\sigma_A)}\right|_{\mathrm{NMRK},\, p_4||p_5}
		\bigg\}.
	\end{split}
	\label{eq:M1Coll_opp}
\end{align}

\bibliography{refs.bib}
\end{document}